\documentclass{jaa}
\usepackage{natbib}
\usepackage{xcolor}
\usepackage{hyperref}
\usepackage{amsmath}
\usepackage{amssymb}
\usepackage{aas_macros}

\usepackage{graphicx}
\usepackage{xspace}
\usepackage{multirow}

\hypersetup{
    colorlinks=true,
    linkcolor=red,
    citecolor=blue,
    filecolor=magenta, 
    urlcolor=blue,
}

\newcommand{\rficlean}{\texttt{RFIClean}\xspace}

\newcommand{\dmcalc}{\texttt{DMcalc}}

\begin{document}\sloppy

\title{Nanohertz Gravitational Wave Astronomy during the SKA Era: An InPTA perspective}

\author{Bhal Chandra Joshi\textsuperscript{1,*}, 
Achamveedu Gopakumar\textsuperscript{2}, 
Arul Pandian\textsuperscript{3}, 
Thiagaraj Prabu\textsuperscript{3}, 
Lankeswar Dey\textsuperscript{2}, 
Manjari Bagchi\textsuperscript{4,5}, 
Shantanu Desai\textsuperscript{6}, 
Pratik Tarafdar\textsuperscript{4}, 
Prerna Rana\textsuperscript{2}, 
Yogesh Maan\textsuperscript{1}, 
Neelam Dhanda Batra\textsuperscript{7}, 
Raghav Girgaonkar\textsuperscript{8},
Nikita Agarwal\textsuperscript{9}, 
Paramasivan Arumugam\textsuperscript{10},
Sarmistha Banik\textsuperscript{21}
Avishek Basu\textsuperscript{11}, 
Adarsh Bathula\textsuperscript{12}, 
Subhajit Dandapat\textsuperscript{2}, 
Yashwant Gupta\textsuperscript{1}, 
Shinnosuke Hisano\textsuperscript{13}, 
Ryo Kato\textsuperscript{14,15}, 
Divyansh Kharbanda\textsuperscript{6},
Tomonosuke Kikunaga\textsuperscript{13}, 
Neel Kolhe\textsuperscript{16}, 
M. A. Krishnakumar\textsuperscript{17,18}, 
P. K. Manoharan\textsuperscript{19}, 
Piyush Marmat\textsuperscript{10}, 
Arun Naidu\textsuperscript{20}, 
K. Nobleson\textsuperscript{21}, 
Avinash Kumar Paladi\textsuperscript{22}, 
Dhruv Pathak\textsuperscript{23}, 
Jaikhomba Singha\textsuperscript{10}, 
Aman Srivastava\textsuperscript{6}, 
Mayuresh Surnis\textsuperscript{11}, 
Sai Chaitanya Susarla\textsuperscript{24},
Abhimanyu Susobhanan\textsuperscript{25,1}, \and 
Keitaro Takahashi\textsuperscript{26,27}}

\affilOne{\textsuperscript{1}National Centre for Radio Astrophysics (TIFR), Post Bag 3, Ganeshkhind, Pune - 411007, India.\\}
\affilTwo{\textsuperscript{2}Tata Institute of Fundamental Research, Mumbai 400005, INDIA.\\}
\affilThree{\textsuperscript{3}Raman Research Institute, Bengaluru, Karnataka, India.\\}
\affilFour{\textsuperscript{4}The Institute of Mathetical Sciences, CIT Campus, Taramani, Chennai 600113, Tamil Nadu, India \\}
\affilFive{\textsuperscript{5} Homi Bhabha National Institute, Training School Complex, Anushakti Nagar, Mumbai 400094, India.\\}
\affilSix{\textsuperscript{6}Dept of Physics, IIT Hyderabad, Kandi. Telangana 5022085 India.\\}
\affilSeven{\textsuperscript{7}Department of Physics and Astrophysics, University of Delhi, Delhi.\\}
\affilEight{\textsuperscript{8}Amity Centre of Excellence in Astrobiology, Amity University Mumbai 410206, Maharashtra, India.\\}
\affilNine{\textsuperscript{9} Manipal Institute of Technology, Manipal 576104, Karnataka, India.\\}
\affilTen{\textsuperscript{10}Department of Physics, Indian Insitute of Technology Roorkee, Roorkee 247667, Uttarakhand, India.\\}
\affilEleven{\textsuperscript{11}Jodrell Bank Centre for Astrophysics, University of Manchester, Manchester M13 9PL, UK.\\}
\affilTwelve{\textsuperscript{12}The Indian Institute of Science Education and Research, Mohali, India.\\}
\affilThirteen{\textsuperscript{13}Kumamoto University, Graduate School of Science and Technology, Kumamoto, 860-8555, Japan.\\}
\affilFourteen{\textsuperscript{14} Faculty of Advanced Science and Technology, Kumamoto University, Kumamoto, 8608555, Japan.\\}
\affilFifteen{\textsuperscript{15}Osaka City University Advanced Mathematical Institute, Osaka, 5588585, Japan.\\}
\affilSixteen{\textsuperscript{16}Department of Physics, St. Xavier’s College (Autonomous), Mumbai 400001, Maharashtra, India.\\}
\affilSeventeen{\textsuperscript{17}Max-Planck-Institut für Radioastronomie, Auf dem Hügel 69, 53121 Bonn (Germany).\\}
\affilEighteen{\textsuperscript{18}Universität Bielefeld, Fakultät für Physik, Universitätsstr. 25, D-33615 Bielefeld (Germany).\\}
\affilNineteen{\textsuperscript{19}Arecibo Observatory, University of Central Florida, Arecibo 00612, USA.\\}
\affilTwenty{\textsuperscript{20}University of Oxford, Department of Physics, Deny's Wilkinson Building, Keble Road, Oxford, UK OX1 3RH.\\}
\affilTwentyone{\textsuperscript{21}Department of Physics, BITS Pilani Hyderabad Campus, Hyderabad 500078, Telangana, India.\\}
\affilTwentytwo{\textsuperscript{22}Indian Institute of Space Science and Technology, Thiruvananthapuram, Kerala 695547, India.\\}
\affilTwentythree{\textsuperscript{23}Inter-University Centre for Astronomy and Astrophysics (IUCAA), Pune, India .\\}
\affilTwentyfour{\textsuperscript{24}School of Mathematics, National University of Ireland, Galway, University Road, Galway H91TK33, Ireland.\\}
\affilTwentyfive{\textsuperscript{25} National Astronomical Observatories, Chinese Academy of Sciences, Beijing 100101, China.\\}
\affilTwentysix{\textsuperscript{26}Faculty of Advanced Science and Technology, Kumamoto University, Japan.\\}
\affilTwentyseven{\textsuperscript{27}International Research Organization for Advanced Science and Technology, Kumamoto University, Japan.\\}

\twocolumn[{

\maketitle

\corres{bcj@ncra.tifr.res.in}

\msinfo{01 April 2022}{XXXXX}

}]

\newpage

\twocolumn[{

\begin{abstract}

Decades long monitoring of millisecond pulsars, which exhibit highly stable rotational periods, in pulsar timing array experiments is on the threshold of discovering nanohertz  stochastic gravitational wave background. This paper describes the Indian Pulsar timing array (InPTA) experiment, which employs the upgraded Giant Metrewave Radio Telescope (uGMRT) for timing an ensemble of millisecond pulsars for this purpose.  We highlight InPTA's observation strategies and analysis methods, which are relevant for a future PTA experiment with the more sensitive  Square Kilometer Array (SKA) telescope. We show that the unique multi-sub-array multi-band wide-bandwidth frequency coverage of the  InPTA provides Dispersion Measure estimates with  unprecedented precision for PTA pulsars, e.g., $\sim 2 \times$ 10$^{-5}$  pc\,cm$^{-3}$ for PSR J1909$-$3744. Configuring the SKA-low and SKA-mid as two and four sub-arrays respectively, it is shown that comparable  precision is achievable, using observation strategies similar to those pursued by the InPTA, for a larger sample of 62 pulsars requiring about 26 and 7 hours per epoch for the SKA-mid and the SKA-low telescopes  respectively. We also review the ongoing efforts to develop PTA-relevant general relativistic constructs that will be required to search for nanohertz gravitational waves from isolated super-massive black hole binary systems like blazar OJ~287. These efforts should be relevant to pursue persistent multi-messenger gravitational wave astronomy during the forthcoming era of the SKA telescope, the Thirty Meter Telescope, and the next-generation Event Horizon Telescope.

\end{abstract}


}]


\doinum{12.3456/s78910-011-012-3}
\artcitid{\#\#\#\#}
\volnum{000}
\year{0000}
\pgrange{1--}
\setcounter{page}{1}
\lp{1}

\newpage
\section{Introduction}

\par
Gravitational waves (GWs) are propagating ripples in the space-time curvature  as predicted by Einstein's General Theory of Relativity \citep{ein18}.  The routine detection of high-frequency GWs (30$-$800 Hz) by the LIGO-Virgo-KAGRA collaboration\footnote{\url{https://www.ligo.org/science/Publication-O3aCatalog/index.php} } from around 100 merger events that involve black holes (BHs) and neutron stars (NSs) in the last 6 years has opened a new window to the universe which is complementary to the traditional electromagnetic window \citep{Abbot2019_GWTC1,Abbot2020_GWTC-2,Abbot2021_GWTC3}. These observations are providing key insights into a number of astrophysical puzzles that are difficult to tackle by traditional electromagnetic astronomy \citep{Arimoto21}. 
This includes the formation channels for stellar-mass BHs, Equation of State for NSs, measuring the Hubble constant by employing standard and dark sirens, and testing general relativity in ultra-strong regimes,  as well as constraining alternative theories which  dispense with dark energy and dark matter    \citep{Arimoto21,Mapelli20,BB17,StdSiren17,Boran18,DrkSiren19,TGR21}. 
Therefore, it is reasonable to expect that the other GW frequency windows should provide opportunities to explore many aspects of physics, astrophysics and cosmology \citep{Bailes21}. 
Specifically, nanohertz stochastic gravitational wave background (SGWB) formed by the incoherent superposition of GWs from supermassive black hole binaries (SMBHB) is expected to be detected in the coming years \citep{IPTA_GWB_2022}. The discovery of such nanohertz (nHz) GWs will provide unique insights into the formation and evolution of galaxies and their constituent SMBHs, a precise description of solar system, cosmological standard sirens and  fundamental tests of  gravitation \citep{Burke-Spolaor2019}.

\par
These nHz GWs can be detected by employing the precision timing of an ensemble of radio millisecond pulsars \citep[MSPs : ][]{Detweiler1979,S78}. 
A pulsar timing array (PTA) is a dedicated  experiment  to time pulsars with the aim of detecting low-frequency GWs \citep[PTAs:][]{FosterBacker1990}. At present, there exist four such established efforts that  employ the world's best radio telescopes along with three more emerging experiments\footnote{\url{http://ipta4gw.org/}}. PTAs require long-term high-cadence monitoring of a large ensemble of MSPs using large highly sensitive radio telescopes. The high sensitivity of the Square Kilometer Array (SKA) telescope  with its multiple-beam design will not only help to discover nHz GW sources, but also will strengthen the  post-discovery nHz GW science with sensitive high-cadence  observations. This is one of the main goals of the pulsar key science project of the SKA \citep{ks15,jhm+15}.
\par
While pulsar astronomy has traditionally been carried out with large single-dish telescopes, SKA is designed to be an interferometer where the signal collected over its large collecting area is recorded using a beam-former. Out of the currently operating four PTAs, only the Indian Pulsar Timing Array (InPTA) primarily uses  an interferometer, the upgraded Giant Metre-wave Radio Telescope (uGMRT), for its observations\footnote{It may be noted that Large European Array for pulsars \citep[LEAP:][]{Bassa2016} also uses a phased array of multi-element telescopes, but these form a subset of European Pulsar Timing Array experiment, which is largely based on single dish observations. Another interferometer, which contributes data to European Pulsar Timing Array experiment is the Westerbok Synthesis Radio Telescope. Recently, the MeerTime experiment \citep{Bailes2018_MeerTime} has started collecting data with MeerKat, which is also an interferometer. It may be noted that these telescopes have not been used as sub-arrays unlike the uGMRT}. The uGMRT is also a pathfinder telescope for the SKA with features similar to the SKA. Hence, the observation strategies uniquely adopted in the InPTA program are very much relevant for initiating the SKA-era PTA program.

\par
One of the unique features of InPTA is characterisation of propagation effects with unprecedented precision, improving the overall precision of the pulsar timing required for eventual nHz GW detection. The techniques developed for the analysis of the low radio frequency data by the InPTA astronomers would allow the Indian community to meaningfully contribute to  PTA work with the SKA. Additionally, InPTA researchers are developing approaches to model nHz GW sources  and associated detection techniques with a particular emphasis on orbital eccentricities. These efforts should provide another pathway for us to contribute to the SKA science efforts, particularly from the multi-messenger perspectives.

\par
We now present the relevance of the InPTA program for the discovery and characterisation of nHz GW sources with the SKA telescope, motivated by the above-mentioned considerations. The structure of the paper is as  follows. A brief review of the current state-of-art in the detection of nHz GWs and the existing pulsar timing array experiments is provided in Section \ref{pta} followed by a description of the  InPTA experiment in Section \ref{inpta}. A comparison of the uGMRT and the SKA from this perspective is discussed in Section \ref{inptanska} outlining features of the InPTA experiment which are relevant to a future SKA program. A discussion of analysis techniques and methods developed for the InPTA program,  which are likely to be carried forward to the SKA program, is presented in Section \ref{method}. Possibilities for  multi-messenger astronomy for individual sources in the context of the SKA are outlined in Section \ref{mmess}.  After a discussion on  contributions of the InPTA on development of the Indian and Japanese neutron star community as future SKA ready users in Section \ref{ucd}, we conclude by listing future directions in Section \ref{conc}.

\section{Pulsar Timing Arrays for Nanohertz GW Astronomy}
\label{pta}

\par
SMBHBs with total masses in the $10^{8}-10^{10}\, M_{\odot}$ range  are the primary sources of nHz GWs \citep{Detweiler1979}. Such binaries emit nHz GWs during their GW emission-dominated orbital evolution phases and are expected to have orbital periods of years. This is because the natural (orbital) frequency of such self-gravitating BH binary systems  roughly follow $\omega^2\,r^3 = G\,M$, where $\omega, r$ and $M$ are the orbital angular frequency, the associated separation and the total mass of the system respectively  \citep{SS2009}. For a typical circular SMBHB system,  the frequency of the emitted GWs is given by  \citep{Detweiler1979}

\begin{equation}
    f_{\rm GW} \sim 20\,\text{nHz}\,\left(\frac{200\,M}{r}\right)^{3/2}\,\left(\frac{10^{10}\,M_{\odot}}{M}\right) \,,
\end{equation}
where we have used the fact that the frequency of GWs emitted from a circular binary is twice its orbital frequency.

\par
As noted earlier, the existing PTAs monitor an ensemble of MSPs to detect  nHz GWs emitted by SMBHBs \citep{FosterBacker1990, HobbsDai2017_PTA}. This is because the times of arrival (TOAs) of MSP pulses can be modelled and predicted with very high accuracy due to the extremely stable rotational periods of MSPs \citep{Taylor1992}. It turns out that such predictions require the use of an accurate  \textit{timing model} that incorporates various astrophysical and instrumental delays \citep{Edwards2006_tempo2}. Typically, PTAs compare the observed TOAs from MSPs  with predictions for their arrival times based on a pulsar timing model. The differences between the predictions and the actual measurements are usually referred to as the pulsar `timing residuals'. When a GW passes between the Earth and an MSP, it affects the geodesics along which the pulses arrive at the Earth and therefore modulates the pulsar TOAs \citep{EstabrookWahlquist1975, Anholm2009}. The resulting GW-induced timing residuals at a given epoch depends on the two GW polarization states $h_{+,\times}$ such that \citep{Susobhanan2020}
{\small
\begin{align}
R(t)=&\begin{bmatrix}F_{+} & F_{\times}\end{bmatrix}\begin{bmatrix}\cos2\psi & -\sin2\psi\\
\sin2\psi & \cos2\psi
\end{bmatrix}\nonumber\\
&\qquad\int_{0}^{t}dt'\,\begin{bmatrix}h_{+}(t')-h_{+}(t'+\Delta_{p})\\
h_{\times}(t')-h_{\times}(t'+\Delta_{p})
\end{bmatrix}\,,
\end{align}
}
where $\Delta_p$ is a geometric delay, $F_{+,\times}$ are known as the antenna pattern functions, $\psi$ is the GW polarization angle, and the integration is over the coordinate time measured at the solar system barycentre. The expressions for $\Delta_p$ and $F_{+,\times}$ may be found in, e.g., \citet{Anholm2009}.

\par
The existing PTAs are expected to detect first a stochastic GW background (SGWB) created by the incoherent superposition of  GWs from the whole cosmic population of in-spiralling MBHBs \citep{ptk+21}. Moreover, it is customary to treat the resulting SGWB in the $10^{-9}-10^{-7}$ Hz (GW) frequency window  to be an isotropic Gaussian  stationary signal which is described by a characteristic amplitude $h_c$ that follows a power-law spectrum, $h_c \propto f^{-2/3}$ \citep{Phinney2001}. Further, due to the quadrupolar nature of GWs, the above SGWB is expected to induce the following correlation between the timing residuals of each pulsar pair of PTAs \citep{HellingsDowns1983}:
\begin{align}
c(\theta) &= \frac{3}{2} x \, \ln{x} - \frac{x}{4} + \frac{1}{2} + \frac{1}{2} \delta(x)\,,
\end{align}
where $x = (1 - \cos{\theta} )/2$ for an angle $\theta$ on the sky between two pulsars, and $\delta(x)$ is the  regular Dirac delta function (the resulting curve is usually referred to as the Hellings-and-Downs curve). Therefore, the actual detection of the nHz SGWB, characterised by certain $h_c$, involves detailed investigations that determine how close the observed correlations between the timing residuals of each pulsar pair of a given PTA follow the Hellings-and-Downs curve. An incontrovertible detection of the SGWB due to inspiralling SMBHBs will be made if the timing residuals of all pulsar-pairs in a PTA are shown to follow the Hellings-and-Downs curve with a common red spectrum. In order to achieve these goals,  PTA experiments usually monitor MSPs with a cadence of few weeks and the resulting timing data set now spans a few decades for the three established PTAs \citep{IPTA_DR2}. This ensures that the existing PTA data sets should be sensitive to GWs with periods from weeks to years  corresponding to a nHz GW frequency window typically in the $10^{-8} - 10^{-9}$ Hz range.

\par
Currently, four such experiments are operational with a time-baseline of more than a few years. While Parkes Pulsar Timing Array \citep[PPTA :][]{mhb+13}, North American Nano-hertz Observatory for Gravitational Waves \citep[NANOGrav :][]{abb+18} and European Pulsar Timing Array \citep[EPTA :][]{dcl+16} have been gathering data for more than a decade, an Indian effort that employs the uGMRT, called the Indian pulsar timing array (InPTA),  has been in operation since 2015 \citep{jab+18}. Additionally, PTA efforts are being initiated by employing  two more facilities, namely the Five hundred meter radio telescope (FAST) and the MeerKat \citep{CPTA2016,SAPTA2016}. The PTAs periodically pool their data together to create the International Pulsar Timing Array (IPTA) data releases. Two such data releases have happened during the last decade \citep{IPTA_DR1, IPTA_DR2}. IPTA is  a consortium of consortia that is governed by a steering committee with members from all the above mentioned PTAs.

\par
All the well-established PTAs regularly make public their data sets along with their GW astronomy-related analysis results \citep{Alam2020_NANOGrav12.5_narrowband,Kerr2020_PPTA_DR2,Desvignes2016_EPTA, Chen2021_EPTA_6psr}.
Within the last two years, detailed and independent investigations of the most up-to-date NANOGrav, PPTA and EPTA data sets reveal the presence of  a common red noise process in their pulsars with spectral signature consistent with an SGWB due to in-spiralling SMBHBs \citep{Arzoumanian2020_SGWB,Goncharov2021,Chen2021_EPTA_6psr}.
However, these data sets do not strongly support the presence of the  Hellings-and-Downs spatial correlation, crucial to establish the presence of nHz SGWB that arises from merging SMBHBs. This is being attributed to the presence of various statistical uncertainties in the TOA measurements.  Similar conclusions were drawn from a detailed analysis of more sensitive IPTA DR2 data set formed from a combination of older public data releases from the individual PTAs  \citep{IPTA_DR2,IPTA_GWB_2022}.
At present, the constituent members of IPTA are working in tandem by updating their data sets and employing  multiple analysis algorithms to increase the confidence in a possible detection of SGWB in the coming years. 

\par
Unfortunately, the stochastic nature of our astrophysical GWB signal can be mimicked by other random processes that occur in the PTA pulsars. Specifically, the epoch to epoch variations in dispersion measure (DM)\footnote{Dispersion measure is defined as the integral of the column density of electrons over the line-of-sight to the pulsar} and scattering variations need to be accurately incorporated while obtaining TOA residuals from various PTA data sets. These propagation effects are due to the relative motion of the pulsar, the earth and the inhomogenous ionised inter-stellar and inter-planetary medium (IISM and IPM). These  are strong functions of frequency thereby dominating low radio frequency observations of the PTA pulsars \citep{dvt+20,tsb+21}. The wide-band low-frequency coverage in the InPTA experiment provides very high precision measurements of these effects \citep{kmj+21,Nobleson+2021}. A comparable precision is expected from  a similar future  experiment with the SKA. It was argued that one needs to model these variations appropriately to ensure the correctness of the astrophysical interpretation of the expected SGWB signals, extracted from   the timing residuals \citep{Lentati2016}. Work is in progress to build such models for incorporating in a future IPTA data release.

\section{Indian Pulsar Timing Array experiment}
\label{inpta}

\par
Indian Pulsar Timing Array experiment (InPTA) was started in 2015  using the Ooty Radio Telescope \citep[ORT:][]{Swarup1971} and the legacy Giant Metre-wave Radio Telescope \citep[GMRT:][]{Swarup1991} as a pilot experiment. The InPTA now utilises the unique  capabilities of the upgraded Giant Metre-wave Radio Telescope \citep[uGMRT: ][]{gak+17}, which is also a pathfinder telescope for the SKA. The salient features of this experiment are described in this section.

\begin{figure}[H]
\centering
\includegraphics[width=0.5\textwidth]{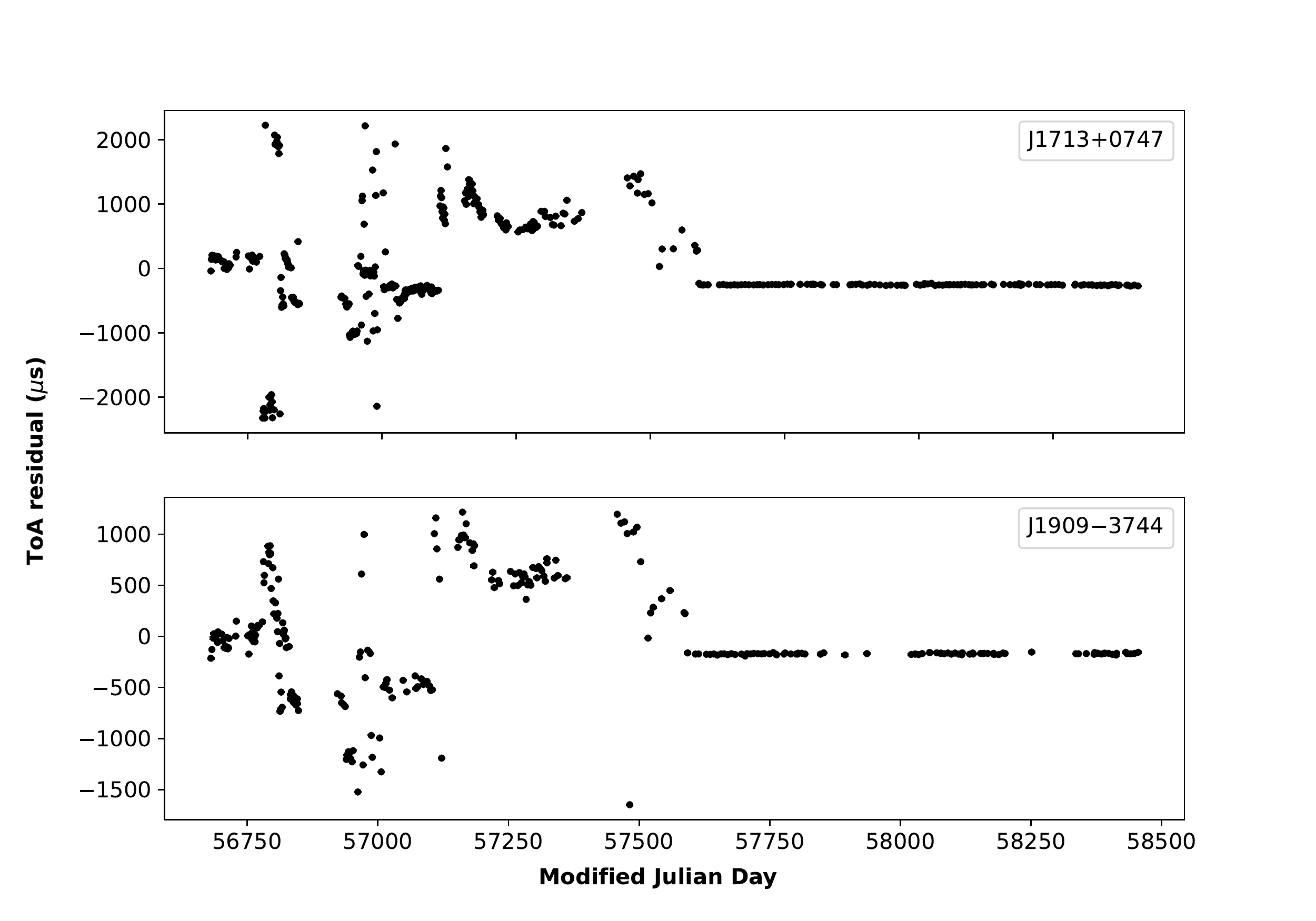}
\caption{Timing residuals for two PTA pulsars with the ORT during and after the legacy phase of the proposal. The wander in the residuals before mid-2016 was traced to instability in the observatory time and frequency standards and associated electronics and was subsequently rectified.}
\label{fig:ORTres}
\end{figure} 

\par
The experiment started as a pilot experiment in 2014, first using the ORT for observing two pulsars, PSRs J1713+0747 and J1909-3744, to test the precision timing capabilities of the Indian instruments. A newly commissioned pulsar receiver called PONDER \citep{njmk15} was employed to coherently dedisperse the data over 16 MHz band-pass at 326.5 MHz and generate time-stamped integrated pulse profiles (IPPs) using topocentric pulse frequency calculated from polynomial coefficients with the best known available timing solutions. The folded time-series were analysed using \texttt{TEMPO2}  \citep{Hobbs2006_tempo2,Edwards2006_tempo2} pulsar timing software package to obtain timing residuals. Early observations showed residuals varying over a few milliseconds and with clock jitter (Figure \ref{fig:ORTres}). These were traced to instability in the observatory time and frequency standard and associated electronics. Based on these initial observations, the observatory instrumentation was improved. Further observations showed stable timing residuals demonstrating  the  feasibility of a PTA experiment and also helped to obtain refined timing solutions for the next phase.  

\par
Encouraged by these results, a pilot program with a sample of 9 pulsars was initiated using both the legacy GMRT and the ORT in 2015. The multi-frequency data, acquired in this pilot phase, was used to test the feasibility of a PTA with both  instruments for the first time. The data were used (a) to plan improvements in the observatory instrumentation, (b) to assess the timing precision achievable with the legacy GMRT and the ORT, (c) to determine the fixed pipeline delay between data acquisition systems \citep{smj+21}, (d) to build timing solutions for a larger sample of PTA pulsars   and (d) to train new members of the collaboration in observing and data reduction.  Different data reduction techniques using a variety of pulsar off-line processing packages, such as \texttt{PRESTO}\footnote{\url{https://www.cv.nrao.edu/\~sransom/presto/}} \citep{Ransom2011_PRESTO},  \texttt{SIGPROC}\footnote{\url{http://sigproc.sourceforge.net/}} \citep{Lorimer2011_sigproc}, \texttt{PSRCHIVE}\footnote{\url{http://psrchive.sourceforge.net/}} \citep{Hotan2004_psrchive}, and \texttt{DSPSR}\footnote{\url{http://dspsr.sourceforge.net/}} \citep{vanStraten2011_dspsr}  were tried on these data to decide the correct strategy for a bigger PTA experiment in the future.

\begin{table*}
\centering
\begin{tabular}{c|c|c|c|c|c|c}
\hline
 & Name & Period & DM & Median S/N & Median S/N & Obs time \\ 
& &  (s)   & (pc\,cm$^{-3}$)  & 1460    & 500 & (min) \\ \hline
&J0645+5158   & 0.008853& 18.25& 7&52 & 30 \\
&J1024-0719   & 0.005162&  6.49& 13& 17& 30 \\
&J1455$-$3330 & 0.007987& 13.57& 9&19 & 30 \\
&J1614-2230   & 0.003151& 34.49&12 &18 & 30 \\
Exploratory InPTA &J1640+2224   & 0.003163& 18.43& 20& 34& 30 \\
& J1730$-$2304&0.008123& 9.62 & 61     & 56  & 30 \\
& J1738+0333  & 0.005850&33.77&   7   & 10  & 30 \\
& J2317+1439  & 0.003445&21.91& 8 & 21 & 30\\
& J2302+4442  &0.005192& 13.79& 9  & 24     & 30 \\
\hline
& J1643$-$1224& 0.004622& 62.41& 129  & 423 & 50\\
& J1713+0747  & 0.004570& 15.91& 254 & 174 & 50\\
& J1857+0943  & 0.005362& 13.31& 74  & 134 & 30\\
Classic InPTA& J1909$-$3744& 0.002947& 10.39& 77 & 289 & 50\\
& J1939+2134  & 0.001558& 71.02& 159 & 323 & 20 \\
& J2145$-$0750& 0.016052&  9.00&103 & 946 & 50 \\
& J2124$-$3358& 0.004931& 4.60& 39  & 208 & 50\\
\hline
& J0437$-$4715&0.005757 & 2.64& 958    & 2343   & 15 \\
& J0613$-$0200& 0.003061&38.77&  61      & 119      & 50 \\
& J0751+1807  & 0.003479&30.25&  32      & 59      & 30 \\
Expanded InPTA & J1012$+$5307& 0.005255& 9.02&  259      & 359     & 30 \\
& J1022$+$1001& 0.016452&10.25&  38     & 417     & 20 \\
& J1600$-$3053& 0.003597&52.32&  71      & 67     & 50 \\
&J1744$-$1134 &0.004075 & 3.14&    72   & 343     & 50 \\
\hline \hline
\end{tabular}
\caption{The pulsar sample of the InPTA. The sample labelled "Classic InPTA" was observed from 2018$-$2022, whereas the sample labelled "Exploratory InPTA" was observed only in 2018$-$2019. Likewise, the sample labelled "Expanded InPTA" was mostly observed only after 2021 with high sensitivity. The "Exploratory InPTA" sample also included all pulsars listed in the "Classic InPTA" and "Expanded InPTA" sample. The "Classic InPTA" and the "Expanded InPTA" pulsars are being observed using Band 3 and 5 with 200 MHz bandwidth, whereas "Exploratory InPTA" sample was observed with Band 3, 4 and 5 with 100 MHz bandwidth.}
\label{tab:sample}
\end{table*}

\par
The GMRT was upgraded between 2010 and 2017 \citep{gak+17} providing a seamless frequency coverage from 50 to 1500 MHz with five new wide-band feeds (Band 1 : 50$-$80 MHz; Band 2 : 120$-$250 MHz; Band 3: 250$-$500 MHz; Band 4 : 550$-$850 MHz and Band 5 : 1050$-$1450 MHz). The wide-band data are acquired with new wide-band receivers and back-end, capable of recording data with 400 MHz band-pass, as beam-former time-series from 4 simultaneous beams. After an initial exploratory experiment to assess the timing precision of the uGMRT with 19 pulsars, a strategy to observe each pulsar in more than one frequency band simultaneously with the uGMRT was adopted for the ongoing InPTA experiment since 2018. 

\par
The uGMRT is a multi-element interferometer much like the SKA. The large sensitivity of the uGMRT is achieved by compensating the geometric and instrumental phase delays (including ionospheric delays) across the interferometric array, consisting of 45-m diameter antennas, to form a phased array. While a phased array of 20 antennas provides a collecting area of about 180-m diameter single dish, an eight antenna phased array synthesises a 125-m single dish. In addition, the relatively smaller beam of the phased array helps in both reducing the sky background as well as in filtering uncorrelated noise as compared to a single dish telescope.  The second advantage of an interferometer like the uGMRT is a possibility of forming multiple phased arrays using different groups of antennas, called sub-arrays. If different sub-arrays are set up to observe at different frequencies, this allows simultaneous multi-frequency observations covering a large bandwidth. On the other hand, multiple sources can be covered at the same frequencies by multiple sub-arrays. Thus, the uGMRT not only provided a sensitivity similar to the large telescopes used in other PTA experiments, but an opportunity to try out new observations strategies with an interferometer. The ongoing InPTA experiment has explored this flexibility of an interferometer to optimise the precision of the experiment.

\par
In the beginning, the InPTA used three uGMRT bands (400$-$500, 650$-$750 and 1360$-$1460 MHz in Band 3, 4 and 5 of the uGMRT respectively) by grouping five to 15 antennas in three sub-arrays using three out of four beams of the uGMRT. As the uGMRT back-end \citep[GMRT Wideband Backend - GWB : ][]{rkg+17} provides a coherent dedispersion capability for a maximum of 200 MHz bandwidth \citep{dg16}, two of the Bands (Band 3 and 5) were coherently dedispersed to remove dispersive effects of the IISM with 100 MHz band-pass respectively, while Band 4 was observed with a 1024 channel digital filter-bank. While this strategy sampled the 300$-$1500 MHz with three non-contiguous bands, the smaller bandwidth as well as lesser number of antennas in each band limited the achievable sensitivity of InPTA observations. 

\par
We addressed this shortcoming since 2019 by using only two sub-arrays with 10 and 15 antennas in Band 3 and 5. As most of PTA pulsars are relatively low DM pulsars (2.6 to 30.0 pc\,cm$^{-3}$), the dispersion smear for these pulsars can be kept smaller than the sampling time employed (40.96 $\mu$s) by using a filter-bank with 1024 sub-bands or more at Band 5. This permitted using the entire 200 MHz bandwidth for coherent dedispersion at Band 3 to eliminate  the significantly larger dispersive smear at frequencies near 300 MHz. Thus, the data were observed in both bands with 200 MHz bandwidth with coherent dedispersion used only in Band 3. The total integration time for all the pulsars was also increased to one hour compared to 30 minute observations in 2018. This strategy provided 4 times larger sensitivity than earlier observations significantly reducing ToA uncertainties as well as improving DM precision as is shown later.

\par
\begin{figure}
\centering
\includegraphics[scale=0.45]{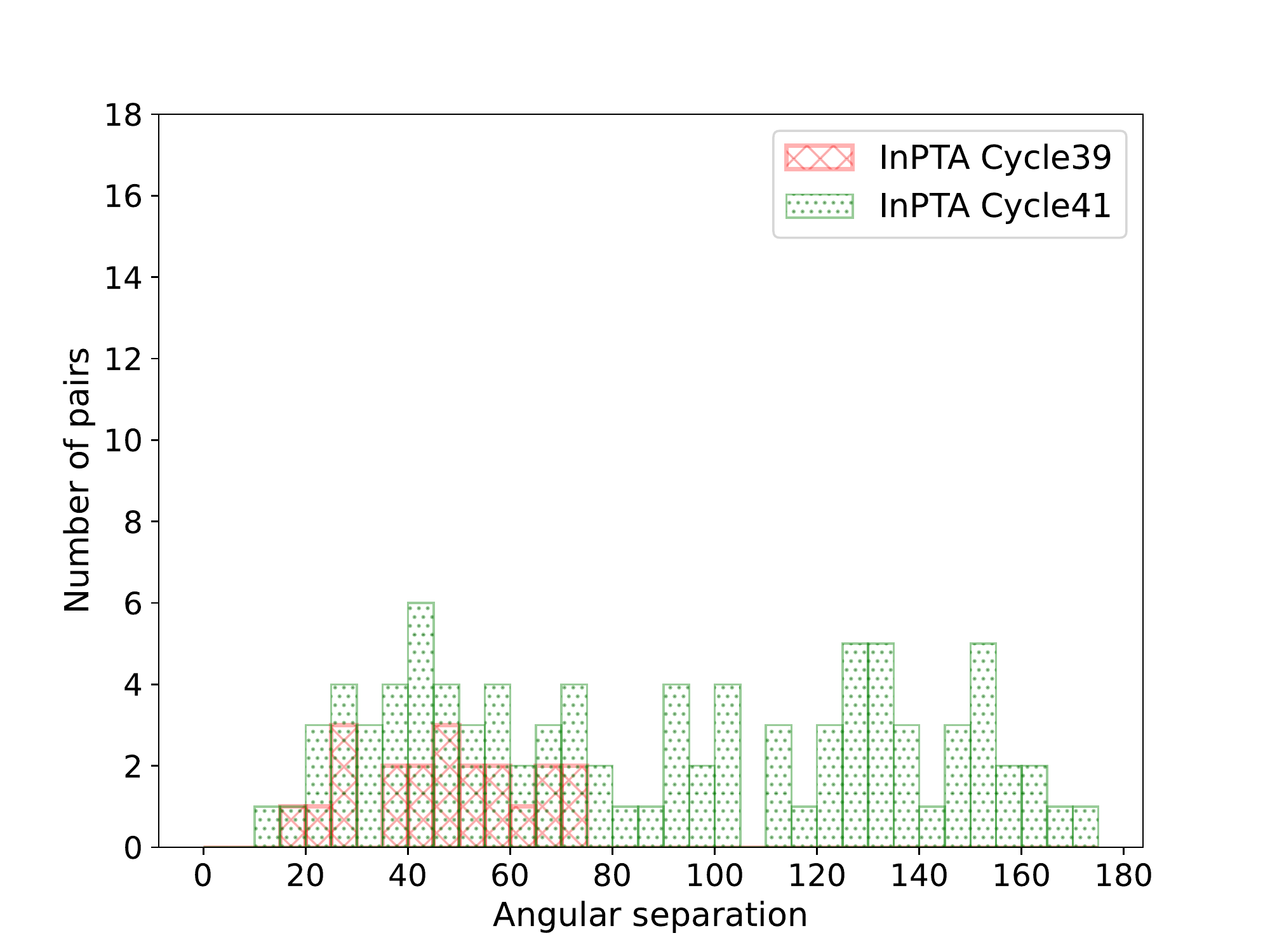}
\caption{Histograms for the distribution of pairwise angular separations between the InPTA pulsars. The sample was gradually increased to achieve a more uniform sampling of spatial separations (Hellings-Downs overlap function \citep{HellingsDowns1983}) over the course of the experiment. The red and green histograms represent the sample over two 6 month observing semesters for the uGMRT.   }
\label{fig:distinpta}
\end{figure}

In this new strategy, the monitoring observations were started with 6  bright pulsars and the sample is being gradually increased to 14  pulsars(Table \ref{tab:sample}).  New pulsars were selected from the pulsars which formed a part of the IPTA Data Release I \citep{IPTA_DR1}. Apart from choosing only the pulsars visible with the GMRT (Dec $>$ -50$^\circ$), the  signal-to-noise ratio (S/N) achieved in the exploratory phase of the experiment (labelled "Exploratory InPTA" in Table \ref{tab:sample}) with the selected configuration was used to make a shortlist. As a good sampling of Hellings-Down overlap function \citep{HellingsDowns1983} is required for detection of SGWB, new pulsars were added to optimise a good coverage of this overlap function  with pulsar pairs (Figure \ref{fig:distinpta}).

\par
The current data set for InPTA, including the pilot, legacy and exploratory uGMRT phases spans about 8 years. However, the span and cadence of individual pulsars in our sample have varied over the years due to the considerations mentioned above. While most of the recent dataset has been taken using  the upgraded GWB \citep{rkg+17}, the frequency bands and receivers employed in observations prior to 2018 have varied. All the legacy GMRT and the uGMRT data were acquired with two hands of polarisation, but the ORT data were single polarisation data due to the nature of the instrument. Lastly, the pipeline delays as well as other data acquisition features are well characterised for the uGMRT receivers. As a consequence, the collaboration has decided to limit its upcoming first public data release to observations with the uGMRT alone to maintain the homogeneity in data properties. 

\begin{figure}
\centering
\includegraphics[scale=0.5]{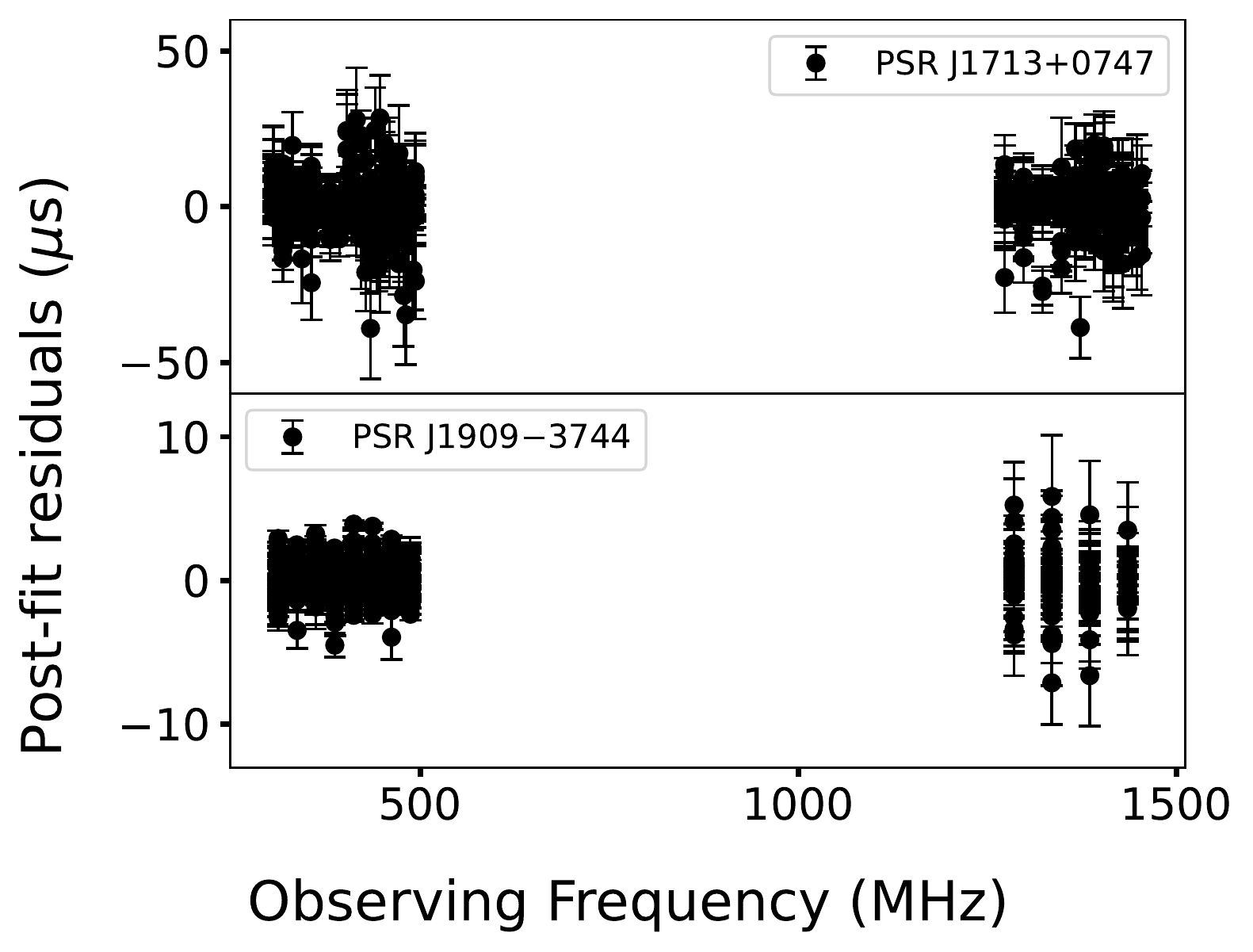}
\caption{Timing residuals obtained with data observed simultaneously over 300 to 1460 MHz for PSRs J1713+0747 and J1909$-$3744 are shown in this figure after a DM fit over this range without using any frequency dependent FD parameter or DMJUMP. Epoch by epoch fits provide very precise DMs in the InPTA experiment.  }
\label{fig:freqres}
\end{figure}

\par
The pulsar data generated by uGMRT beamformer observations are recorded in a channelised time series binary format by the GWB, together with a timestamp representing the start of the observation. A pipeline named \texttt{pinta}\footnote{\url{https://github.com/abhisrkckl/pinta}} \citep{smj+21}  was developed to process such datasets obtained using the uGMRT for the InPTA experiment. This pipeline performs RFI mitigation using two different softwares, \texttt{RFIClean}\footnote{\url{https://github.com/ymaan4/RFIClean}} \citep{Maan2020_rfiClean} and \texttt{gptool}\footnote{\url{https://github.com/chowdhuryaditya/gptool}}, and produces partially folded PSRFITS archives which can be used for further analysis using standard pulsar software such as \texttt{PSRCHIVE}. After the data reduction offline with this pipeline, the multi-band data were further analysed with post-processing pipelines, \texttt{DMCalc} \footnote{\url{https://github.com/kkma89/dmcalc.git}} and \texttt{PulsePotratiture} \citep{Pennucci+2014} to obtain precision DM measurements. These pipelines were developed and/or tuned for the uGMRT datasets and yield ToAs apart from DM measurements. These were analysed with pulsar timing software, \texttt{TEMPO2}, to obtain the final timing solutions as well as timing residuals, which form the input for GW analysis. 

\begin{figure}
\centering
\includegraphics[scale=0.5]{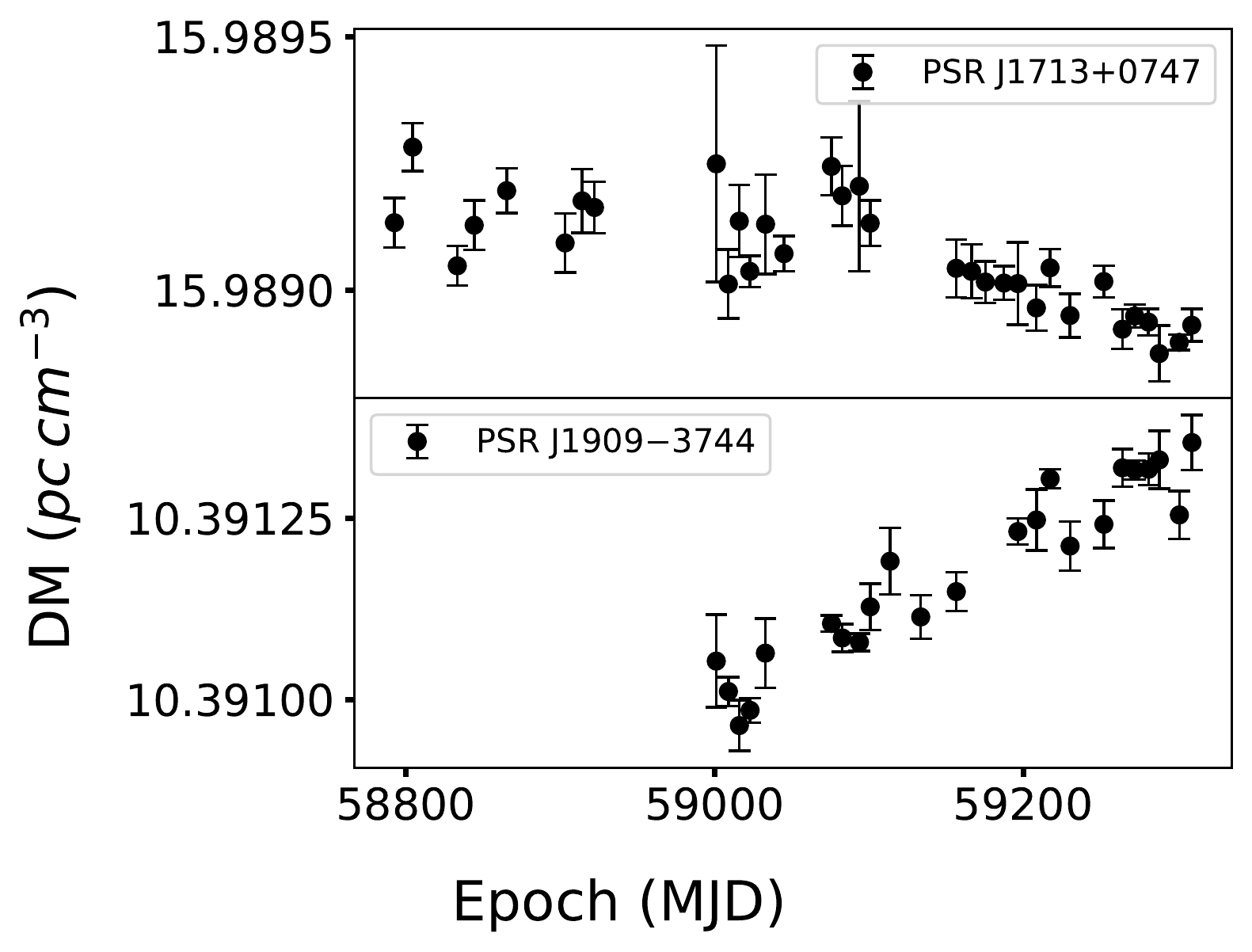}
\caption{DM measurements using the data from 300 to 1460 MHz obtained for PSRs J1713+0747 and J1909$-$3744 in the InPTA experiment.  }
\label{fig:dmtimeseries}
\end{figure}

\par
\begin{figure}
\centering
\includegraphics[scale=0.5]{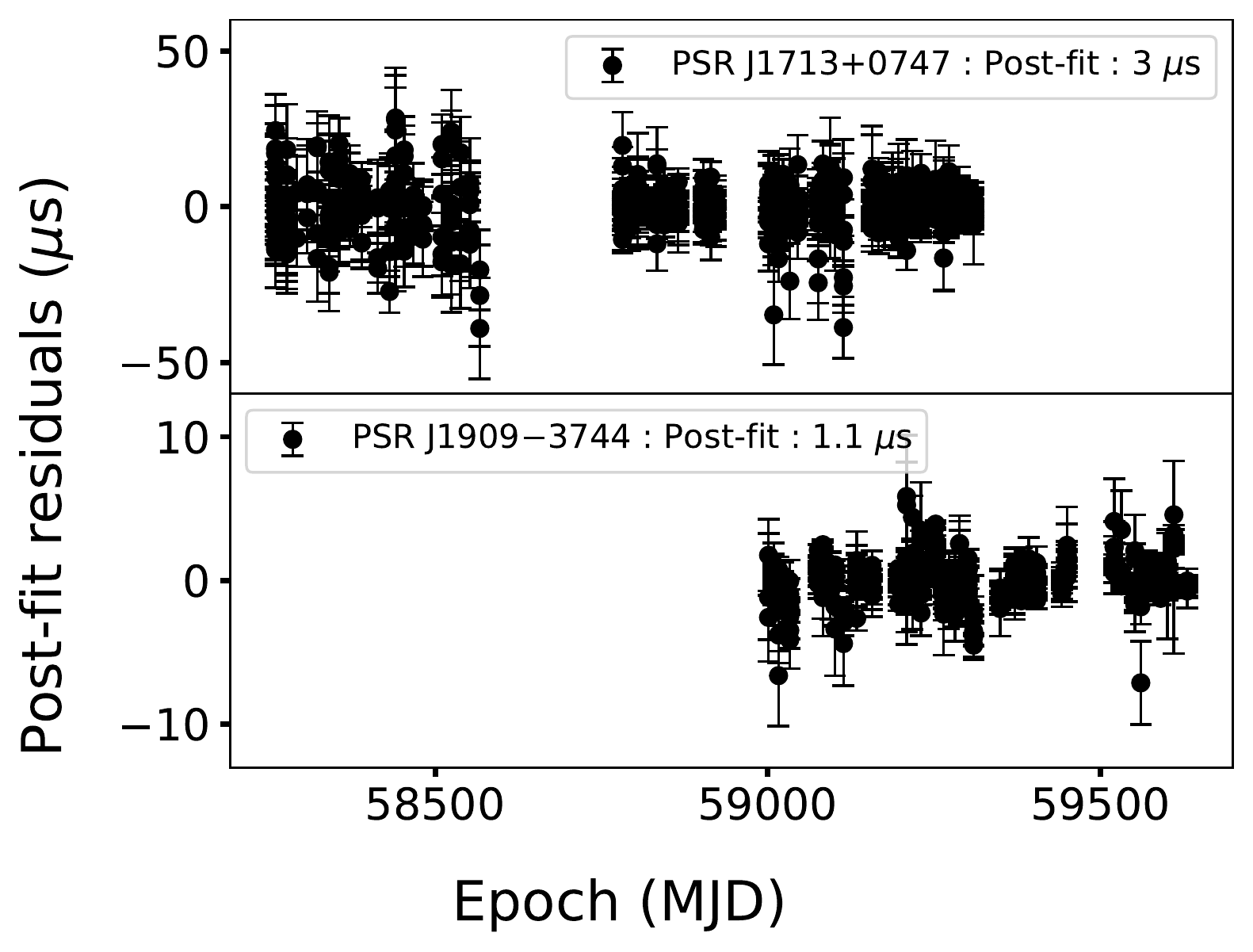}
\caption{Timing residuals for PSRs J1713+0747 and J1909$-$3744 with the data obtained in the InPTA experiment are shown as a function of observation epochs. The RMS post-fit residuals achieved are indicated in the legend  }
\label{fig:timres}
\end{figure}

The experiment has resulted in the highest precision measurements of DM reported so far by any PTA, primarily due to a combination of simultaneous Band 3 and Band 5 wide-band data covering 300$-$1460 MHz,  which is unique to our experiment. Such a combination is affected by misalignment due to profile evolution and scatter broadening over this frequency range.  As explained later (Section \ref{method}), analysis of ToAs generated using frequency resolved templates showed normally distributed residuals  without using FD parameters or DMJUMP to align Band 3 and Band 5 data   (Figure \ref{fig:freqres}). Consequently, we were able to achieve high precision with the median uncertainty as small as  2 $\times$ 10$^{-5}$ pc\,cm$^{-3}$ for PSR J1909$-$3744. The variation of DM for PSRs J1713+0747 and J1909$-$3744 are shown in Figure \ref{fig:dmtimeseries} as an illustration. The SKA has a similar frequency coverage and is likely to provide these measurements with similar or better precision using a strategy similar to the InPTA. With these measurements, we have also achieved $\mu$s post-fit residuals, comparable to those obtained with the other PTA experiments and the preliminary results from the upcoming public data release of the InPTA are shown in Figure \ref{fig:timres}.

\par
While we did not attempt observations of multiple sources using multiple beams-sub-arrays, the experience from our strategy can be extrapolated to evaluate such an observing strategy. The InPTA observations provide useful experience to understand the phasing constraints from an ionospheric turbulence point of view for different observations. This is important to decide the configurations of interferometer antennas for the sub-arrays to be used, particularly for the most distant antennas in the sub-array.

\section{Lessons from InPTA for GW science with SKA}
\label{inptanska}

\par
The multi-element interferometer nature of SKA makes it an instrument similar to the uGMRT, albeit with a much larger sensitivity. Hence, our experience with the uGMRT, which is also a recognised path-finder for the SKA, can inform the observations and analysis strategy for a future PTA with the SKA. In this section, some possibilities based on the InPTA experience are discussed.

\par
The SKA consists of two telescopes, both of which will be constructed in two phases. Here, we provide a top level description of the capabilities of the phase-1 of the telescope, which will represent about 10 percent of the final SKA collecting area. The SKA-mid will be located in Karoo, South Africa. It will consist of 197 dishes spread over three spiral arms on an area of a diameter of about 150-km. These include 133, 15-m dishes and 64, 13.5-m MeerKat dishes. It is planned to have three frequency bands operational in the first phase. While Band 2 covers 0.95 to 1.76 GHz and is likely to become available first, Band 5 (4.6 $-$ 15.3 GHz) and Band 1 (0.35 $-$ 1.05 GHz) will be commissioned subsequently. The planned back-end consists of a beam-former providing 16 simultaneous beams with coherent dedispersion carried over 300 to 800 MHz band-pass. About 120 antennas, consisting of 46 MeerKat antennas, are located in a cluster of dimensions 2 km by 2 km, whereas the rest of the  77 antennas are spread beyond these baselines. Further, about 20 antennas each from the spiral arms (15 SKA dishes and 5 MeerKat dishes) can  be used as two 3 km by 3 km clusters. While further arm antennas can also be used as phased arrays, the baselines between these antennas are too long for a stable phase solution at a frequency below 2 GHz. These baselines for the outer spiral arms are probably useful as phased arrays at Band 5, where ionospheric propagation effects are much smaller. However, the sensitivity of a Band 5 tied array is smaller than Band 2 for the same number of antennas as the SKA-mid antennas are less sensitive at Band 5. Moreover, pulsars are steep spectrum  sources and fainter at Band 5. Hence, only 120 to 140 antennas are considered here  for high sensitivity PTA observations.

\begin{table}[ht]
\begin{tabular}{|c|c|c|}
\hline
{\bf GMRT } & {\bf Single antenna} & {\bf Subarray }   \\
\hline
Band 2  & & 8 antennas \\
 Aeff/Tsys ($m^2/K$) &	 3.8	& 30 \\
SEFD (Jy) &	727	& 91	    \\
\hline
Band 3 	& &	8 antennas	\\		
Aeff / Tsys ($m^2/K$)	& 10.5  & 	84 \\		
SEFD (Jy)&	263	&33 	\\	
\hline
Band 4 &  & 5 antennas\\		
Aeff / Tsys ($m^2/K$)&	9.7	& 48 \\
SEFD (Jy) &	286 & 57		 \\	
\hline
Band 5 &  &	15 antennas\\		
Aeff / Tsys ($m^2/K$)&	9.2 & 	138   \\		
SEFD (Jy) &	300	& 20   \\	
\hline
\end{tabular}
\caption{ The system equivalent flux density estimated for the uGMRT sub-array configurations employed in the InPTA experiment}
\label{tab:sefd_gmrt}
\end{table}


\par
The SKA-low telescope consists of  131,072 dual polarised log periodic antenna-based aperture arrays, located  in Western Australia in Murchinson Radio-astronomy Observatory. These dipoles are arranged in 512 stations in the central core and three spiral arms located approximately 65 km apart with each station consisting of 512 dipoles. The core consists of 224 stations with the largest baseline of 1 km with 13 more clusters, each with 6 stations, located within 2 km of the core. The central core can be used as two phased arrays, each proving a physical aperture of 127020 m$^2$ or as a single phased array of two times that area.

\par
SKA-mid uses a real-time digital back-end for pulsar timing, called Pulsar Timing Engine \citep[PST : ][]{k2018}. The PST helps to record high time resolution full-Stokes pulsar data. It receives 16 tied-array beams, with each beam pointing on a pulsar to be timed concurrently. Observation durations can be from 180 to 1800 seconds. Each beam can be coherently dedispersed over 300 to 800 MHz in real-time to provide high time resolution data, which can be further processed to provide frequency-resolved sub-integrated time-stamped IPP in the PSRFITS format \citep{Hotan2004_psrchive}. The multiple beams in both SKA-mid and SKA-low allow multiple targets/bands  to be observed concurrently making both these instruments very flexible for different observing strategies.

\par
The uGMRT, which is a pathfinder telescope for the SKA, is similar to the SKA due to its multi-band, multi-beam and multi-sub-array nature. The 30 antennas of the GMRT can be grouped into 4 sub-arrays to operate over 5 frequency bands using 4 beams provided by the GMRT back-end (See Section \ref{inpta}). Using our experience for such a strategy, a PTA program with the SKA can utilise at least 4 sub-arrays in the SKA-mid and  two sub-arrays in the SKA-low for simultaneous multi-frequency observations as well as concurrent observations of multiple targets. Apart from optimal utilisation of telescope time for the pulsar Key Science Project (KSP), such a strategy will also provide a wide-frequency coverage for separating ISM noise from the precision timing in a way similar to that in the InPTA. Some possibilities are explored below for a suitable strategy for a PTA program with the SKA.

\begin{table*}[ht]
\centering
\begin{tabular}{|c|c|c|c|}
\hline
{\bf } & {\bf 	Single} & {\bf Single}  & {\bf All }   \\
{\bf SKA Low} & {\bf 	station} & {\bf subarray }  & {\bf subarrays  }   \\
{\bf } & {\bf 	 } & {\bf   }  & {\bf  within 1 km }   \\
\hline
&   & & \\	
Low Band  & - &	100	& 200\\
(50-350 MHz) & - &	stations	& stations\\
Aeff / Tsys ($m^2/K$)&		1.10  &	110.00 &	220.00	  \\	
SEFD (Jy) &	2,510.27 &	25.10 &	12.55	   \\
\hline
\hline
{\bf } & {\bf 	Single} & {\bf Single}  & {\bf All }   \\
{\bf SKA MID} & {\bf 	antenna} & {\bf subarray }  & {\bf subarrays  }   \\
{\bf } & {\bf 	 } & {\bf   }  & {\bf  within 2 km }   \\
\hline
Mid Band 1a  &  & & \\
(0.35-0.65 GHz) & - & 30 antennas & 120\\
 Aeff/Tsys ($m^2/K$) &		2.10 &	63.00 &	252.00 \\
SEFD (Jy) &		1,314.90 &	43.83 &	10.96  \\
\hline
Mid Band 1b &  & & \\
(0.65-1.05 GHz) & - & 30 antennas & 120\\		
Aeff / Tsys ($m^2/K$)	& 4.20	& 126.00 &	504.00\\		
SEFD (Jy)&	657.45 &	21.92 &	5.48 	\\	
\hline
Mid Band 2 &   & & \\	
(0.95-1.76 GHz)& - &  30 antennas & 120\\
Aeff / Tsys ($m^2/K$)&		11.00 &	330.0 &	1,320.00  \\	
SEFD (Jy) &	251.03 &	8.37 &	2.09   \\
\hline
\end{tabular}
\caption{The system equivalent flux density estimates for the proposed sub-array configurations for a PTA experiment using the SKA Low and SKA Mid telescopes}
\label{tab:sefd_ska}
\end{table*}

\par
\begin{figure*}
 \centering
 \includegraphics[width=0.6\textwidth]{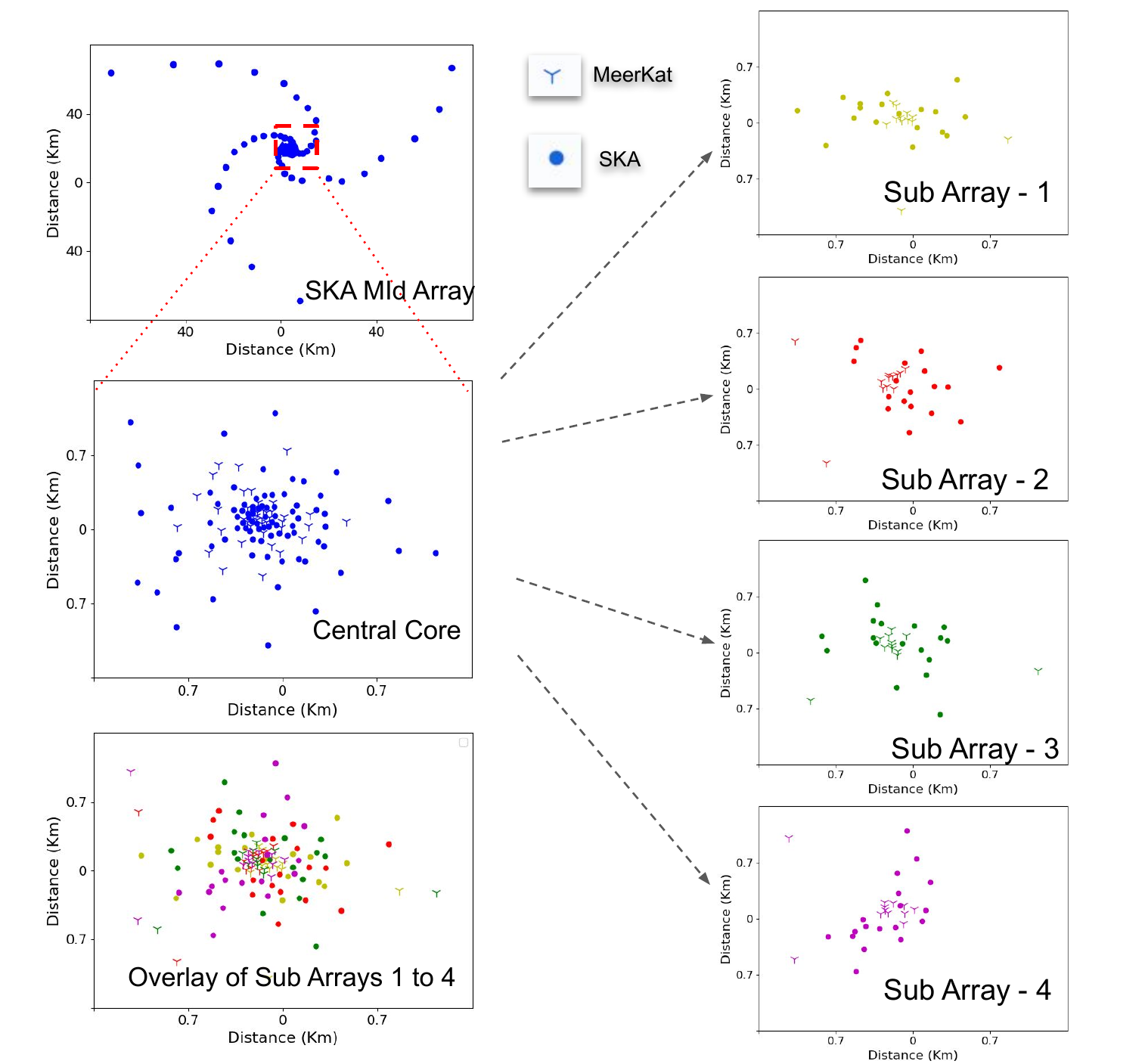}
 \caption{ The proposed sub-array configuration for a pulsar timing experiment with the SKA-mid Phase-1 Telescope }
 \label{fig:skamid_sub}
 \end{figure*}
\begin{figure*}
 \centering
 \includegraphics[width=0.6\textwidth]{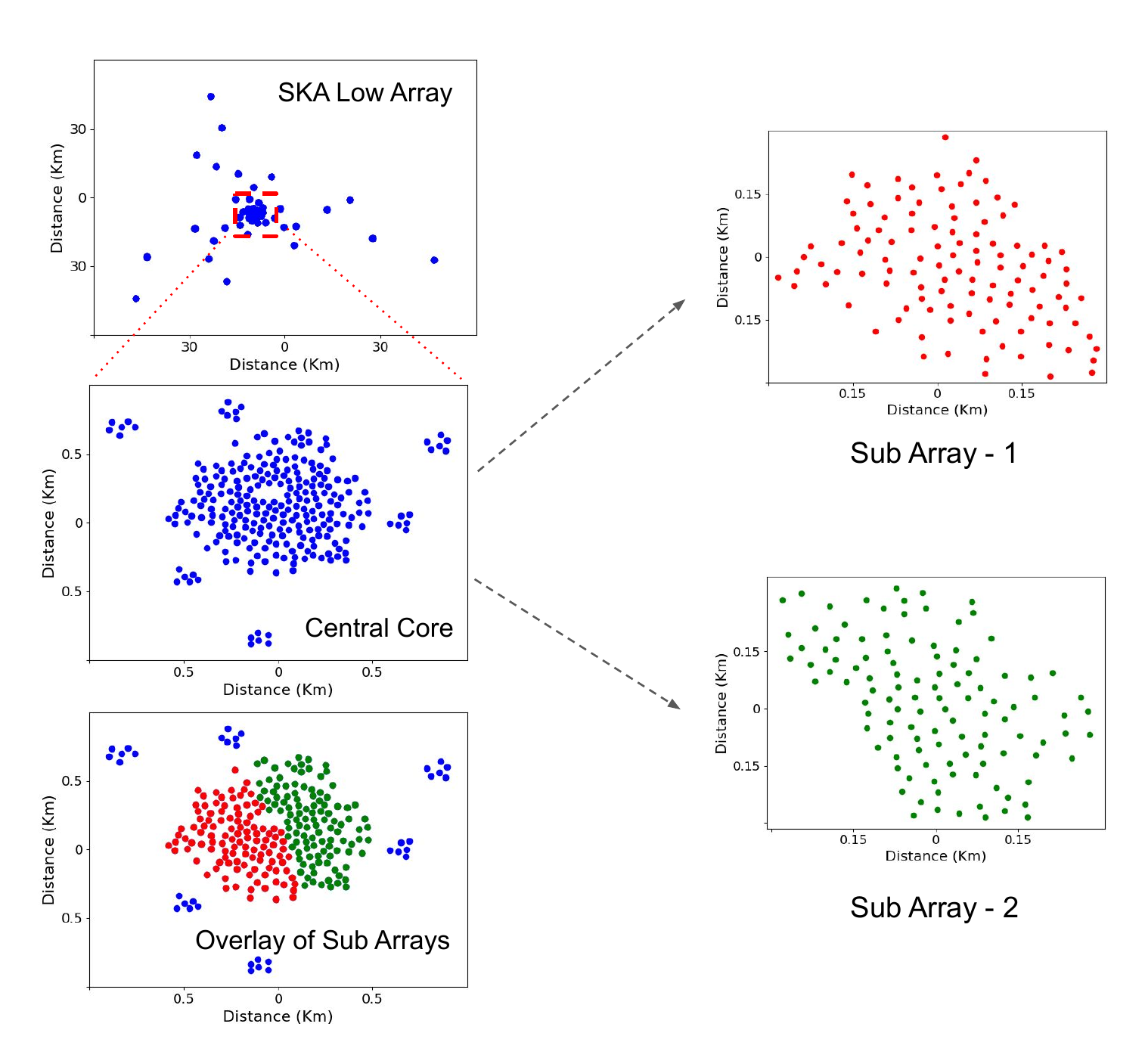}
 \caption{ The proposed sub-array configuration for a pulsar timing experiment with the SKA-low Phase-1 Telescope }
 \label{fig:skalow_sub}
 \end{figure*}

A comparison of the system equivalent flux density (SEFD) for the uGMRT arrays employed in the InPTA experiment and those for the SKA-mid and SKA-low sub-arrays, 
shown in Tables \ref{tab:sefd_gmrt} and \ref{tab:sefd_ska}, suggests that a 30 antenna SKA sub-array is equivalent to 6, 12 and 35  antenna uGMRT sub-array at band 3, 4 and 5 of the uGMRT respectively. While such  SKA sub-arrays provide marginally smaller sensitivity than the uGMRT Band 3 sub-array used in the InPTA experiment, a higher sensitivity is indicated  for sub-arrays at other bands. With about 120 antennas in the central part of the SKA-mid, this suggests the  use of a maximum 4 sub-arrays using 4 beams with sensitivity similar to or greater than the uGMRT. For weaker pulsars, one could instead use only two or even a single SKA-mid sub-array to achieve almost four times  the uGMRT sensitivity. We recommend a possible configuration of 4 sub-arrays with the SKA-mid telescope as shown in Figure \ref{fig:skamid_sub}. Given that the available bandwidth is 300/800 MHz compared to 200 MHz used in the InPTA program, one would achieve much larger sensitivity than the uGMRT sub-arrays in practice. While 16 sub-arrays are in principle possible in the SKA, this comes at the cost of the sensitivity of each sub-array and is not recommended based on the InPTA experience, where  only 2 sub-arrays are used instead of possible 4 in the current ongoing observations. A similar comparison with the SKA-low configuration suggests that two sub-arrays using each half of SKA-low core stations provide significantly larger sensitivity than Band 2 of the uGMRT (See Tables  \ref{tab:sefd_ska} and \ref{tab:sefd_gmrt}).  We recommend a possible configuration of 2 sub-arrays with the SKA-low telescope as shown in Figure \ref{fig:skalow_sub}. Since SKA-mid and SKA-low are two different telescopes with independent correlator and beam-former, coordinated observations between the telescopes can provide the low-frequency support for higher frequency observations with a SKA-PTA covering an unprecedented frequency range of 50 $-$ 1800 MHz to provide the most sensitive PTA dataset ever.

\par
The two to four sub-array configurations, discussed above, can be employed in a variety of ways.  One option would be to use the same observing strategy as InPTA, where the different sub-arrays are used for simultaneous multi-frequency observations of the same target pulsar (hereby referred to as OPTION I). This provides a wide frequency coverage, useful for characterising the ISM noise, the jitter noise and bias due to profile evolution across the frequency. The other option would be to use each sub-array to observe a different target pulsar concurrently (OPTION II). This option optimises the use of telescope time to cover a much larger sample of PTA pulsars. A third option is to use a hybrid strategy combining both multi-frequency and multi-target strategies (OPTION III). Finally, fainter pulsars can be covered with the full 120/200 antenna array, one source at a time, with three 
observations employing a different frequency band (OPTION IV). It may be noted that the brightness of each pulsar (or equivalently the achievable S/N) determines the precision of ToAs. Consequently, the number of sub-arrays possible depends on the distribution of flux density of the PTA pulsars, which also is a critical factor in deciding the most useful observing strategy. Thus, given the pulsar sample, all four of the observing strategies may be required.

\begin{table*}[ht]
\begin{tabular}{|c|c|c|}
\hline
List of pulsars & Number of Pulsars & Observing time required \\
& & Option I/II/III/IV  \\
& & (hrs) ) \\
\hline 
\hline
J0437-4715, J0621+1002, J0636+5128, J0900-3144, J1022+1001 & & \\
J1045-4509, J1600-3053, J1603-7202, J1640+2224, J1643-1224 & 24 & 4/2/2 \\ 
J1713+0747,J1730-2304, J1732-5049, J1747-4036, J1744-1134& &  \\
J1832-0836, B1855+09, J1903+0327, J1909-3744, B1937+21 & & \\
J1944+0907, J2124-3358, J2145-0750, J2234+0944& &  \\
\hline
\hline
J0023+0923, J0030+0451, J0034-0534, J0218+4232, J0340+4130& & \\
J0610-2100, J0613-0200, J0711-6830, J0751+1807, J0931-1902& & \\
J1024-0719, J1453+1902, J1455-3330, J1614-2230, J1721-2457& & \\
J1738+0333, J1741+1351, J1802-2124, J1804-2717, B1821-24A & 38 & 22 \\
J1853+1303, J1910+1256, J1911-1114, J1911+1347, J1918-0642 &&\\ 
J1923+2515, J1946+3417, J1949+3106, B1953+29, J2010-1323 &&\\ 
J2017+0603, J2033+1734, J2043+1711, J2129-5721, J2234+0611 && \\ 
J2214+3000, J2302+4442 J2317+1439& & \\
\hline
\hline
\end{tabular}
\caption{Recommended observation strategy for a PTA program with the SKA telescope. The pulsar sample was divided into two categories. The bright pulsars could be observed for 10 minutes each using Option I (Single Target simultaneous observations with three multi-band  sub-arrays), Option II (Multiple targets with 4 sub-arrays in three non-simultaneous multi-band observations) or Option III(two targets with two band sub-arrays each). For the fainter pulsars in the second category, option IV is recommended with the full array utilised in three multi-band observations)} 
\label{tab:obsstrat}
\end{table*}


\par
With the above considerations in mind, we classified the PTA sample into two different categories. The first category consists of the brightest pulsars, where we adopt OPTION I, II or III using the maximum number of sub-arrays available for each telescope optimising the telescope time as well as frequency coverage. The second  category of  pulsars consists of fainter pulsars, where option IV is appropriate. Given the advertised parameters of SKA telescopes, the required observation time  for each pulsar in the respective categories using these options was estimated from the radiometer equation using the flux density estimates from the ATNF pulsar catalog \citep{mhth05} \footnote{\url{https://www.atnf.csiro.au/research/pulsar/psrcat/}}. These calculations provide an estimate for the required telescope time. While details of some parameters of SKA-telescopes are still being worked out in the construction phase and a detailed optimisation is planned in future, some recommendations are made below based on InPTA experience and available information. 

\par
A list of target pulsars to be observed using the suggested four options for the SKA-mid telescope is provided in Table \ref{tab:obsstrat}.  The total observing time required for each epoch of observations is also indicated in this table. We find that a sample of 62 pulsars, visible with the SKA telescope, can be observed with about 26 hours with the SKA-mid telescope. Likewise, seven hours are needed for each epoch with the  SKA-low telescope.  Assuming a 10-day cadence, the SKA key science program will require 936 hours for a PTA program with the SKA. This estimate can vary depending on the number of available antennas in the early construction phase of the SKA telescope (which increases progressively from the current 64 MeerKat antennas during AA1 to AA* phase) as well as on achievable SEFD and backend configuration and evolving radio Frequency Interference environment. Finally, the observing proposed here can be further optimised by reserving the SKA observations to weaker pulsars or pulsars which are affected by significant inter-stellar propagation effects. While a more detailed study is planned in future incorporating this information, the recommendation in this section already provides a blue-print for an SKA PTA, probably for the first time.

\par
In conclusion, the experience gained with different configurations of the uGMRT interferometer for frequencies between 300 to 1500 MHz in the InPTA experiment is useful to propose observation configurations that can be used once the SKA phase 1 telescope becomes available. Suggestions for the required observing strategies are described in some detail in this section and can form the basis of future SKA-PTA program of the SKA pulsar key science project.

\section{Analysis methodology from InPTA relevant to SKA}
\label{method} 

\par
The reduction and the analysis of the InPTA data, particularly that obtained at the lower frequency band, required the development of new techniques or tuning  the techniques available in the literature. As mentioned in Section \ref{inpta}, the uGMRT raw data were reduced by using an off-line pipeline, called \texttt{pinta}, to obtain  partially folded PSRFITS archives.  While the SKA will employ a real-time pipeline to process pulsar observations, the RFI mitigation and excision techniques (\texttt{RFIClean} and \texttt{gptool}) will be relevant for the SKA pipeline as well, particularly for Band 1 data from SKA-mid and data from SKA-low. The final product of SKA real-time pipeline  is also expected to be in the PSRFITS format, which will allow the InPTA techniques for determination and application of precision DMs to be incorporated in the SKA Science Data processor easily.  The experience relevant to the analysis of the data from the SKA beam-former is described in this section.

\subsection{Online RFI-excision before partially folding the data}
\label{rfi_excision}

\par
Radio frequency interference (RFI) remains a growing concern to obtain quality data and achieve the designed sensitivity. For most of the PTA experiments, the data are processed in real-time to output data that are partially folded over the pulsar spin period. While this real-time processing results in highly reduced data rates as well as requirements of disk-space and computing resources for off-line processing, it is much harder to identify and excise RFI from the resultant partially integrated data. Moreover, a variety of RFI, e.g., the periodic RFI, faint and short time-scale RFI, and temporally varying RFI that mostly causes baseline variations, is not possible to mitigate from the partially integrated data as it requires analysing contiguous temporal sections of data to identify such RFI. The above understanding is reinforced by our InPTA data processing experience, wherein mitigation of RFI in the filterbank data using the excision tools \rficlean \citep{Maan2020_rfiClean} and \texttt{gptool} greatly aid in the detection significance. In fact, in some particular cases, the pulsars are detected only after RFI is excised using these tools. The SKA will also employ a real-time pipeline to process pulsar observations. To fully utilise the telescope's sensitivity, it is desirable that the above or similar RFI excision tools are employed in the SKA's real-time data reduction pipeline to mitigate RFI before the data are partially folded. The computation time estimates based on the InPTA experiment \citep{smj+21} suggest that these can run in real-time, particularly on GPGPU based hardware as the source code is easily portable to  GPGPU kernels. 

\subsection{Narrow-band technique for times-of-arrival and dispersion measure estimation}
\label{narrowband}

\par
Precise estimation of times-of-arrival (ToA) and DM at lower frequencies with the help of traditional narrow-band technique is impeded by the following factors: (1) the measured DM and frequency-dependent profile evolution are intertwined, (2) uncorrected DM variations can introduce smear while frequency-collapsing the data to degrade the S/N, and (3) systematic biases can also get introduced in the estimated DM due to interstellar scatter broadening. A frequency resolved narrow-band analysis method was developed for the post-processing of the InPTA data. This method is called DMcalc and utilises python interface of PSRCHIVE package \citep{kmj+21}.

\par
The ToAs are obtained in a timing experiment by cross-correlating the data profiles with a noise-free template. Traditionally, a noise-free template is obtained by averaging  frequency-collapsed IPP from several epochs. Epoch to epoch DM variations can introduce significant smear when multiple widely separated epochs are averaged to obtain a noise-free template. Additionally, an incorrect alignment of the pulse across the frequencies for pulsars showing significant evolution of profiles with frequency leads to DM offsets. It may be noted that both these effects are significantly larger at lower frequencies.  Indeed, our work shows that DM estimates using a single frequency-collapsed template or an offset fiducial DM used to dedisperse the template itself, can have  constant offsets from the actual DMs \citep{kmj+21}. Traditionally, this is handled by using FD parameters or DMJUMPS in the timing analysis. 

\par
In the InPTA narrow-band analysis, we use a frequency-resolved template and an iterative procedure to estimate DM of each epoch. The number of sub-bands used in the frequency resolved template is a trade-off between the profile evolution and the achievable S/N in each sub-band and all DM measurements use frequency resolved ToAs obtained by cross-correlating the data with frequency resolved noise-free templates. The sub-band profiles in the templates for each epoch are aligned by determining the epoch dependent DM in an iterative procedure. Here, two or more sub-integrations Band 3 data of each epoch are analysed using ToAs obtained by cross-correlating time-collapsed  profiles of the epoch aligned first with an assumed DM obtained from \texttt{pdmp} analysis \citep{kmj+21}. This analysis is repeated in subsequent iterations progressively improving the  DM estimate. A final frequency resolved noise-free template is then formed by averaging the epochs aligned using the DM obtained in the final iteration.

\par
The main conclusion from the InPTA analysis with the procedure described here is that FD parameters or DMJUMPS are not needed for any of the pulsars in the InPTA sample. Considering that the sample includes pulsars which show significant profile evolution from 300 to 1460 MHz as well as significant scatter broadening at frequencies around 400 MHz (e.g. PSRs J1643$-$1224 and J2145$-$0750), this suggests that using such frequency resolved analysis appears to be more robust. Please note that our simultaneous multi-frequency observations avoid any possibility of small variations in DM between near-simultaneous high and low frequency observations adopted in other PTA experiments. Additionally, we use  higher precision DM measurements from 300 to 500 MHz. Thus, we recommend following an observing strategy and analysis similar to the InPTA for a PTA experiment with the SKA. Thus, OPTION I and III (See Section \ref{inptanska}) are preferred options provided the required S/N is achievable for the target pulsars.

\par
The post processing pipeline \dmcalc{} was developed for application on the combined multi-band data with the noise-free template updated with the fiducial DM obtained in the previous steps. It is straight-forward to incorporate this in the SKA science data processor with a possibility of providing real-time measurements. In turn, these will be useful for a quick follow-up of DM events, such as those reported in PSR J1713+0747 \citep{Lam2018}, as well as updating DM for coherent dedispersion for the subsequent observations.

\subsection{Wideband technique for simultaneous measurement of times-of-arrival and dispersion measure}
\label{wideband}

\par
Pulsar observations are increasingly using wideband receivers (defined as receivers with a fractional band-pass of 100 percent or greater) to acquire more precise TOAs \citep{gak+17,Johnston+2021}. Pulse profiles can evolve significantly over such wide bandwidths. This introduces a bias in arrival times as well as DM measurements when such data are analysed with a single band-collapsed 1-dimensional template.  \citet{Pennucci+2014} devised an algorithm for generating a 2-dimensional template and for using these for simultaneous measurement of DMs and ToAs from wideband pulsar data. They proposed pulsar timing with such wide-band ToAs\citep{Pennucci2019ApJ}. This "wideband timing technique" has been applied on various datasets, such as the NANOGrav 12.5-year data \citep{Alam+2021ApJS} and the CHIME and GBT-L data in the  400$-$800 MHz frequency range \citep{Fonseca+2021ApJ}.  Recently, we applied this technique to the uGMRT data in the 300$-$500 MHz frequency range \citep{Nobleson+2021}. As the profile evolution is much more dominant for many pulsars below 400 MHz apart from a significant smearing due to pulse broadening, this study demonstrated the effectiveness of this technique below 400 MHz for the first time.

\par
The wideband timing technique uses frequency-resolved high S/N IPP to decompose this 2-dimensional data into principal component eigen-profiles which capture the variation of the profile with the observing frequency. A noise-free 2-dimensional \texttt{PulsePortraiture} is generated by fitting spline functions to the coefficients of eigen-profiles  required for modeling the 2-dimensional data \footnote{\url{https://github.com/pennucci/PulsePortraiture}} \citep{2016ascl.soft06013P}.  The phase offset of the wide-band data with the pulse portrait is obtained by least-square minimization of the phase offset of each sub-band pulse profile incorporating the cold-plasma dispersion law. This yields both a high precision ToA and DM estimate. The timing solution and residuals for a PTA  pulsar are then obtained by maximising a likelihood function, implemented in the software package TEMPO \citep{Nice+2015ascl}, with these ToAs and DM estimates as inputs.

\par
Pulsars, such as PSRs J1643$-$1224 and J1939+2134, show significant profile evolution due to scatter-broadening in the InPTA sample, whereas PSR J2145$-$0750 shows an evolving profile. Analysis of Band 3 InPTA data needed three eigen-profiles to model the pulse-portrait of the former two pulsars, while two eigen-profiles were required for the latter \citep{Nobleson+2021}. We also noticed that RFI contributes dominant eigenvalues and corresponding noisy eigen-profiles, so it is important to remove RFI  before the principal component analysis for this technique to work well. The wavelet and  spline fits needed to be appropriately tuned for the technique to work  at our frequency. This experience can easily be adopted for SKA data.

\par
We are currently developing the technique to include the likelihood function used in the TEMPO2 \citep{Hobbs2006_tempo2} software package which will use the wideband DM measurements from the ToAs as priors to estimate the DM model parameters. We also plan to develop a technique to combine multi-band data in the regime of wideband analysis. The standardisation of techniques as part of the InPTA analysis is  directly relevant  to the SKA due to the wide multi-band frequency coverage of the SKA in view of the observations strategy proposed in this paper.

\section{Persistent Multi-Messenger GW Astronomy with the SKA }
\label{mmess}

\par
The LIGO-Virgo-KAGRA collaboration has inaugurated the era of GW astronomy due to their routine detection of hecto-hertz GWs from around 100 stellar-mass BH binaries since 2015 \citep{GWTC2_2020, GWTC3_2021}.
Further, this collaboration heralded the era of multi-messenger GW astronomy by the observations of transient hectohertz GWs from a binary neutron star merger (GW170817) and its counterparts (EM170817) across the electromagnetic spectrum \citep{GW170817_2017, GW170817_EM}.
Strikingly, these multi-messenger observations of 
such a $100$ second-long transient GW event 
are already providing fundamental contributions to the areas of  physics, astrophysics, and cosmology
\citep{GW170817_2017}.
Recall that multi-messenger astronomy usually involves combining data from  various messengers that arise from spectacular astrophysical events.
\par
During the SKA era, we should be able to practice what we term  the persistent multi-messenger GW astronomy \citep{VDG2021}.
This is influenced by the fact that the rapidly maturing PTAs should be detecting a stochastic nHz GW background created by a population of merging SMBHBs in the coming years \citep{IPTA_GWB_2022}.
Thereafter, continuous nHz GWs from bright individual SMBHBs that stand above this background should be routinely detected with the help of the SKA-era PTA \citep{Burke-Spolaor2019}. 
The eventual detection of continuous nHz GWs from individual SMBHBs should allow us to pursue multi-wavelength electromagnetic observations of such GW sources as they are expected to reside in the active galactic nuclei \citep{Burke-Spolaor2019,Xin2021}.
In other words, persistent multi-wavelength electromagnetic observations of such continuous GW sources should allow us to pursue multi-messenger GW astronomy during the SKA era \citep{VDG2021}.
It turns out that the resulting persistent multi-messenger nHz GW astronomy can make profound contributions to the emerging field of multi-messenger GW astronomy \citep{Burke-Spolaor2019}.
\par
The InPTA researchers are well-placed to contribute to the various aspects of this emerging area of astronomy during the SKA era.
This is partly due to our involvement in the general relativity-based predictions of certain  Bremsstrahlung flares from the unique bright blazar OJ~287 and the successful observational campaigns of these predicted flares \citep{Dey2018, Dey2019, Laine2020}.
These investigations and the associated campaigns provided firm observational evidence 
for the presence of a nHz GW-emitting SMBH binary in the blazar OJ~287.
The underlying model, developed by Mauri Valtonen and his collaborators, prescribes the observed double-peaked Bremsstrahlung flares from OJ~287 to the repeated impacts of a secondary BH with the accretion disk of the primary BH twice during its general relativistic eccentric orbit. 
Notably,  the 2019 {Eddington flare} from OJ~287, predicted in \cite{Dey2018}, was successfully observed with the {Spitzer space telescope} and this campaign even allowed us to provide a parametric constraint on the celebrated BH no-hair theorem \citep{Laine2020}.
\par
These successful observational campaigns of a number of predicted impact flares in optical wavelengths prompted us to explore the possibility of using other electromagnetic wavelengths to probe the SMBH binary central engine description for OJ~287 \citep{VDG2021}. 
Fortunately, several high-resolution Very Long Baseline Interferometry (VLBI) observational campaigns have tried to image the radio jet of OJ~287 during the past three decades \citep{Hodgson2017, Cohen2017}.
These observational campaigns reveal that the position angle (PA) of the projected jet on the sky plane shows systematic variations at both millimeter and centimeter wavelengths \citep{Dey2021}.
Therefore, some of us pursued detailed investigations that allowed us to explain the observed PA variations of OJ~287’s radio jet while employing the now established SMBH binary central engine description \citep{Dey2021}.
Further, we have provided specific predictions that should be testable by the GMVA and EHT campaigns on OJ~287, planned for the coming years.
These efforts are ensuring that multi-band EM observations of OJ~287 will be pursued in the coming years to probe various aspects of its SMBH binary central engine description \citep{VDG2021}.
It is expected that 
such multi-wavelength and multidisciplinary efforts will be required to pursue multi-messenger nHz GW astronomy with OJ 287 during the SKA era.

\par
We are also developing general relativistic approaches that will be required for extracting nHz GWs from isolated SMBH binaries from the SKA era PTA data sets.
Specifically, efforts are ongoing to develop ready-to-use computationally efficient routines that model the expected pulsar timing residuals induced by GWs from isolated SMBH binaries inspiraling along general relativistic eccentric orbits \citep{Susobhanan2020}.
These approaches employ post-Newtonian(PN) approximation for describing the general relativistic SMBH binary dynamics and this approximation provides general relativity based  corrections to Newtonian  dynamics in terms of a dimensionless parameter that involves SMBH binary orbital velocity \citep{Blanchet2014}.
In \citet{Susobhanan2020}, we employed 3PN-accurate Keplerian parametric solution to model general relativistic eccentric orbits of SMBH binaries while an improved version of GW phasing approach was employed to incorporate the effects of GW emission \citep{MGS2004,DGI2004}.
It may be recalled that these two elements, namely PN-accurate Keplerian type parametric solution and the GW phasing approach,  are employed in the widely used Damour-Deruelle timing formula to time MSPs in general relativistic eccentric orbits \citep{DD1986}.
Further, we are incorporating the spin effects to model pulsar timing residuals induced by inspiral GWs from spinning SMBH binaries in PN-accurate eccentric orbits with the help of \citet{KG2005} 

\par
\begin{figure*}
    \centering
    \includegraphics[width=0.75\textwidth]{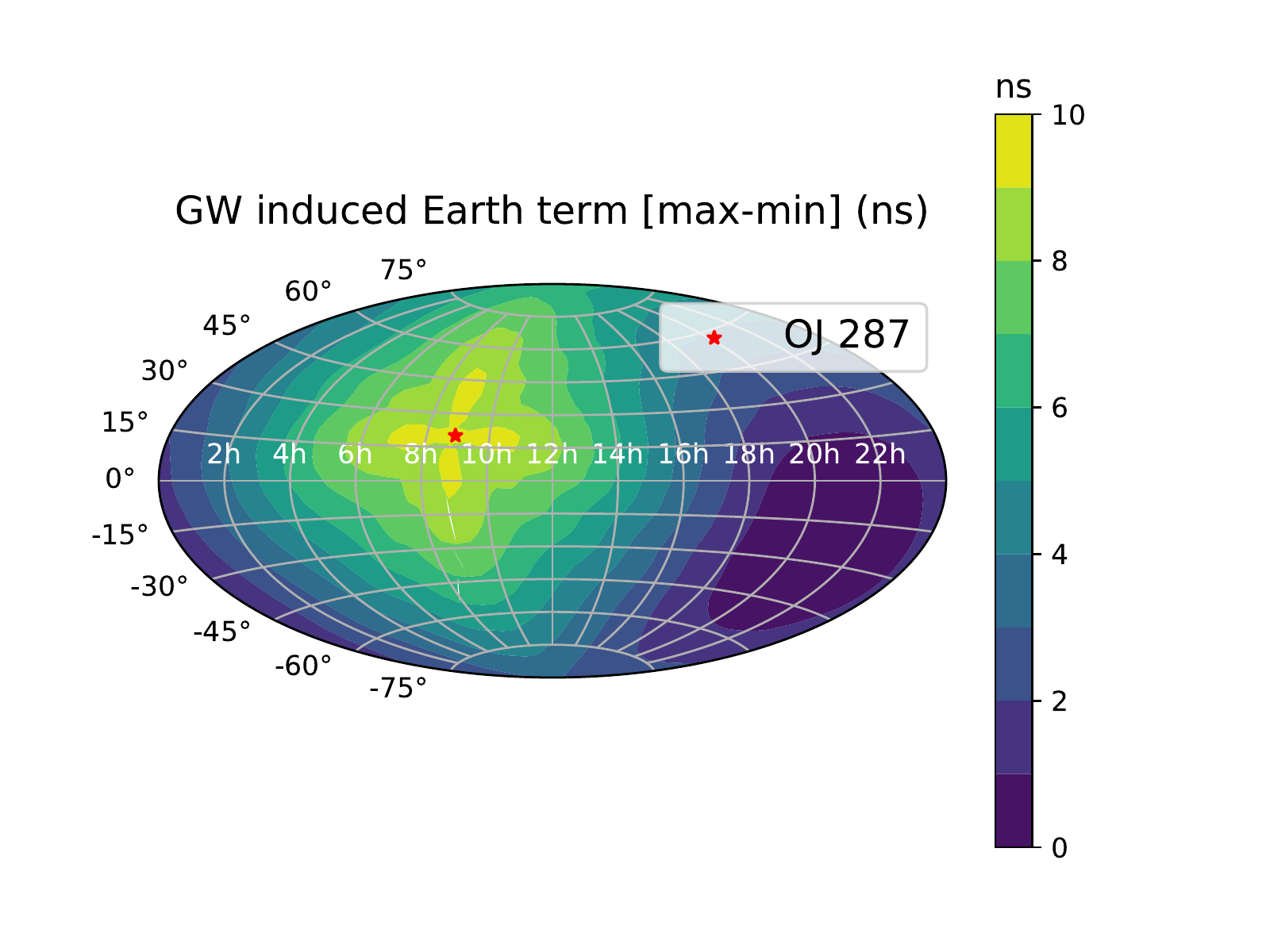}\\
    \includegraphics[width=0.75\textwidth]{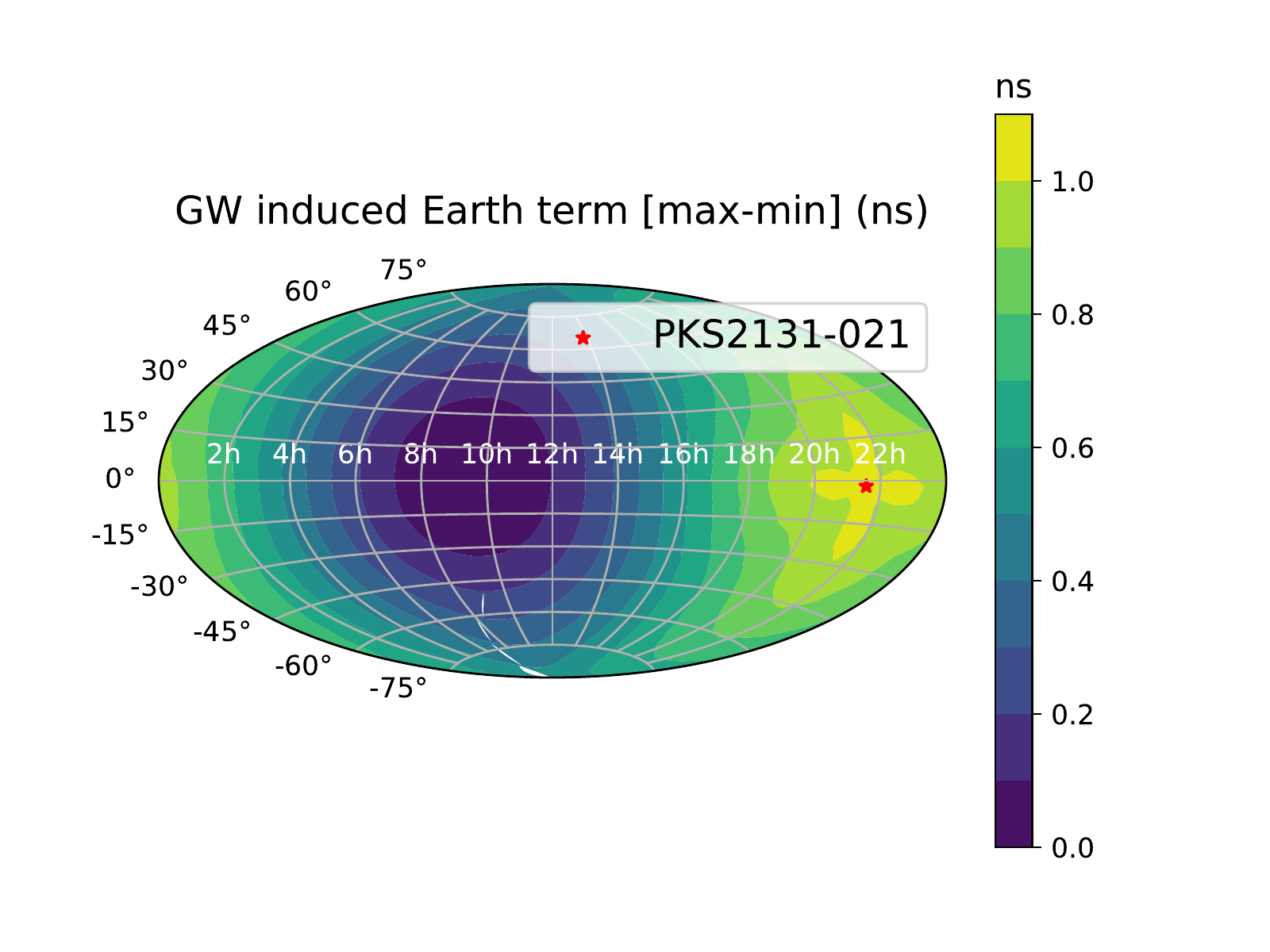}
    \caption{Distribution of the GW-induced Earth term amplitudes, namely $max - min$ of the Earth term contributions to the timing residuals, for pulsars around two different sky locations associated with two SMBHB candidates: OJ~287 and PKS2131-021. The binary parameters of OJ~287 are taken from \citet{Dey2018}
    while we employ \citet{ONeill2022} for
    PKS2131-021.
    Influenced by \citet{ONeill2022}, we let the total mass $M = 2 \times 10^9$ and mass ratio $q=1$ for the second candidate. The GW-induced residuals are calculated by extending the method of \citet{Susobhanan2020} while incorporating PN-accurate GW emission effects for  tracking the evolution of the SMBHBs. }
    \label{fig:sky_maps}
\end{figure*}

In Figure~\ref{fig:sky_maps}, we show the amplitude of the expected GW-induced pulsar residuals for pulsars across the sky for two SKA era nHz GW emitting 
SMBH binaries, namely blazars OJ~287 and PKS~2131-021 
while focusing only on the earth term contributions
\citep{ONeill2022}.
The inference of a nHz GW emitting SMBH binary in PKS~2131-021 is attributed to a very recent effort that demonstrated two epochs of strong sinusoidal variation with essentially the same period and phase in the 45.1-year radio light curve \citep{ONeill2022}.
A natural explanation is that this blazar is hosting an SMBH binary with an orbital period of $\sim 4.75$ years (redshifted) in contrast to $\sim 12$ years for OJ~287.
The heat maps in Figure~\ref{fig:sky_maps} reveal that SKA-era PTA efforts should be able to pursue persistent multi-messenger nHz GW astronomy with these two sources that may involve other facilities like the  Thirty Meter Telescope and The Next Generation Event Horizon Telescope \citep{VDG2021}.
\par
 With the help of such general relativistic constructs, 
we are working to constrain the presence of isolated eccentric SMBH binaries in PTA data sets, influenced by \citet{N11_isolated}.
 Additionally, we are modelling general relativistic burst with memory PTA signals due to SMBH binaries in PN-accurate hyperbolic orbits with the help of \citet{Cho2018}.
These efforts should allow us to contribute to the ongoing efforts to constrain the presence of such isolated SMBH binaries in the existing IPTA data sets and their updates.
Clearly, such efforts will be crucial to detect routinely  isolated GW sources in the SKA era PTA data sets and therefore pursue persistent multi-messenger GW astronomy during that era.

\section{InPTA and the development of SKA user community in Japan and India}
\label{ucd}

The InPTA consortium is playing an active role in building a scientific community that will have expertise to use the SKA products to pursue its key science projects. 
Taking a cue from the LIGO collaboration, which recruited many high energy physicists into  gravitational wave astronomy (after the cancellation of the SSC), we have tried to build a community of researchers from diverse areas such as  Cosmology, General Relativity, Nuclear and High Energy Physics, Astronomical Instrumentation, Neutron Star Physics, Solar Physics to list a few examples. These researchers have nicely complemented the original core group, which consisted of 
radio astronomers 
working on various aspects of pulsar astronomy.

At present, InPTA hosts seven undergrad students, nine graduate students, and ten postdoctoral researchers and we expect this collaboration to grow further as we head into the SKA era. All these young researchers are now well  experienced in performing pulsar observations simultaneously at different uGMRT bands in the sub-array mode, which will be a necessity in the SKA era. There were some other undergraduate students, who were members of InPTA earlier and then moved to different research fields for their Ph.D,   as well as some former postdocs who have moved on to non-academic jobs. 
We gather that the training they received as members of InPTA remains  invaluable to them.  Furthermore,  several InPTA members who started with InPTA projects while being based at Indian institutes have continued to remain active after received their under-graduate and graduate degrees, in spite of taking on new positions abroad.

To train new members, InPTA organises training schools and busy weeks (where the whole collaboration members gather for in person interactions). 
Further, young InPTA members are encouraged to present research results in national and international levels and they participate in outreach activities.
InPTA members have organised a conference series titled ``Asia SKA Initiative On NS'' twice, first time in 2016 in India and the second time in 2019 in Thailand. Both of the meetings were attended by scientists from India, China, Japan, Thailand, etc. The participants were from various stages of their careers.
InPTA members also played a key role to form a consortium called `Gravitational Radiation and Science with Pulsars' (GRASP) that involve researchers from South Africa and China in addition to InPTA members. Monthly meetings of GRASP is playing a significant role in exchange of knowledge and will help in future international collaborations in the SKA era.
Finally, 
InPTA hosted the annual meeting of the IPTA during June 2019 in Pune, India. This meeting included a student training week followed by a conference week. In addition to the 2019 meeting, InPTA members have acted as as members of the organising committees of other IPTA meetings too, as well as delivered lectures in student weeks and presented research works. 

In what follows, we provide a brief timeline of our consortium.
InPTA started in 2015 as a pilot program consisting of only five members using the legacy GMRT. It was accompanied by another pilot program using uGMRT during 2016$-$2017. Thereafter, the two programs merged and InPTA program using uGMRT emerged with 19 members during 2018. We were an associate member of the IPTA consortium  until 2021, when we were given the full membership. In 2021, four researchers from Japan (based at Kumamoto University) joined the InPTA consortium. Hence the word ``InPTA'' is now a moniker for an Indo-Japanese  pulsar timing array program using an Indian telescope, but is not confined to only Indian researchers.

\vspace{-2em}
\section{Conclusion}
\label{conc}

The observing strategy and analysis methods used in the InPTA are useful in informing the design of a PTA experiment with the SKA telescope. A precision better than 2 $\times$ 10$^{-5}$ pc\,cm$^{-3}$ for pulsars such as PSR J1909$-$3744, already demonstrated in our analysis with the InPTA in this paper, is possible with much larger sensitivity and frequency coverage provided by the SKA. We recommend using the SKA-low and SKA-mid as one to four sub-arrays respectively to cover frequencies from 50 to 1800 MHz concurrently based on the InPTA experience. We show that such observing can be used for a 62 pulsar sample three times a month requiring 936 hours similar to observing time required by currently on-going PTA experiments. The development of general relativistic constructs by the InPTA astronomers promise better tuned methods for search for isolated SMBHB GW sources in the PTA data-sets in the SKA  era as well as constrain the astrophysics of these sources with multi-messenger observations. Last, but not the least the InPTA experiment is developing a vibrant Indo-Japanese user community for GW physics with the SKA telescope when this telescope comes online later in the decade.

As the SKA-construction plan gets underway, it is possible to devise an optimised observing plan complimentary to the four major PTA experiments (the EPTA, the InPTA, the NANOgrav and the PPTA) for a higher cadence coverage of an extended sample of PTA pulsars. This involves an optimisation in sensitivity and frequency coverage using different combination of available antennas as they become available in a phased manner from 2024 on-wards. A design of such an optimal strategy is in progress and will be reported in future.

A potential discovery of the nano-Hertz GW  is likely in near future. This can further inform the strategy best suited for post-discovery GW science in the SKA-era. The discussion presented here will hopefully motivate a deeper discussion amongst the major PTA experiments and the IPTA for future directions of exciting GW astrophysics with the SKA, which is likely to revolutionise the radio astronomy with its Square kilo-meter collecting area in the next two decades.




\section*{Acknowledgements}

This work is carried out by InPTA, which is part of the International Pulsar Timing Array consortium. We thank the staff of the GMRT who made our observations possible. GMRT is run by the National Centre for Radio Astrophysics of the Tata Institute of Fundamental Research.  BCJ, PR, AS, SD, LD, and YG acknowledge the support of the Department of Atomic Energy, Government  of India, under project identification \# RTI4002. BCJ and YG acknowledge support from the Department of Atomic Energy, Government of India, under project \# 12-R\&D-TFR-5.02-0700.  
AS is supported in part by the National Natural Science Foundation of China grant No. 11988101.

\vspace{-1em}



\bibliography{journals_apj,references}

\begin{thebibliography}{}
\expandafter\ifx\csname natexlab\endcsname\relax\def\natexlab#1{#1}\fi

\bibitem[{{Abbott} {$et~al$.}(2017{\natexlab{a}}){Abbott}, {Abbott}, {Abbott},
  {Acernese}, {Ackley}, {Adams}, {Adams}, {Addesso}, {Adhikari}, {Adya},
  {Affeldt}, {Afrough}, {Agarwal}, {Agathos}, {Agatsuma}, {Aggarwal}, {Aguiar},
  {Aiello}, {Ain}, {Ajith}, {Allen}, {Allen}, {Allocca}, {Altin}, {Amato},
  {Ananyeva}, {Anderson}, {Anderson}, {Angelova}, {Antier}, {Appert}, {Arai},
  {Araya}, {Areeda}, {Arnaud}, {Arun}, {Ascenzi}, {Ashton}, {Ast}, {Aston},
  {Astone}, {Atallah}, {Aufmuth}, {Aulbert}, {Aultoneal}, {Austin},
  {Avila-Alvarez}, {Babak}, {Bacon}, {Bader}, {Bae}, {Baker}, {Baldaccini},
  {Ballardin}, {Ballmer}, {Banagiri}, {Barayoga}, {Barclay}, {Barish},
  {Barker}, {Barkett}, {Barone}, {Barr}, {Barsotti}, {Barsuglia}, {Barta},
  {Bartlett}, {Bartos}, {Bassiri}, {Basti}, {Batch}, {Bawaj}, {Bayley},
  {Bazzan}, {B{\'e}csy}, {Beer}, {Bejger}, {Belahcene}, {Bell}, {Berger},
  {Bergmann}, {Bero}, {Berry}, {Bersanetti}, {Bertolini}, {Betzwieser},
  {Bhagwat}, {Bhandare}, {Bilenko}, {Billingsley}, {Billman}, {Birch},
  {Birney}, {Birnholtz}, {Biscans}, {Biscoveanu}, {Bisht}, {Bitossi}, {Biwer},
  {Bizouard}, {Blackburn}, {Blackman}, {Blair}, {Blair}, {Blair}, {Bloemen},
  {Bock}, {Bode}, {Boer}, {Bogaert}, {Bohe}, {Bondu}, {Bonilla}, {Bonnand},
  {Boom}, {Bork}, {Boschi}, {Bose}, {Bossie}, {Bouffanais}, {Bozzi},
  {Bradaschia}, {Brady}, {Branchesi}, {Brau}, {Briant}, {Brillet}, {Brinkmann},
  {Brisson}, {Brockill}, {Broida}, {Brooks}, {Brown}, {Brown}, {Brunett},
  {Buchanan}, {Buikema}, {Bulik}, {Bulten}, {Buonanno}, {Buskulic}, {Buy},
  {Byer}, {Cabero}, {Cadonati}, {Cagnoli}, {Cahillane}, {Bustillo},
  {Callister}, {Calloni}, {Camp}, {Canepa}, {Canizares}, {Cannon}, {Cao},
  {Cao}, {Capano}, {Capocasa}, {Carbognani}, {Caride}, {Carney}, {Diaz},
  {Casentini}, {Caudill}, {Cavagli{\`a}}, {Cavalier}, {Cavalieri}, {Cella},
  {Cepeda}, {Cerd{\'a}-Dur{\'a}n}, {Cerretani}, {Cesarini}, {Chamberlin},
  {Chan}, {Chao}, {Charlton}, {Chase}, {Chassande-Mottin}, {Chatterjee},
  {Chatziioannou}, {Cheeseboro}, {Chen}, {Chen}, {Chen}, {Cheng}, {Chia},
  {Chincarini}, {Chiummo}, {Chmiel}, {Cho}, {Cho}, {Chow}, {Christensen},
  {Chu}, {Chua}, {Chua}, {Chung}, {Chung}, {Ciani}, {Ciolfi}, {Cirelli},
  {Cirone}, {Clara}, {Clark}, {Clearwater}, {Cleva}, {Cocchieri}, {Coccia},
  {Cohadon}, {Cohen}, {Colla}, {Collette}, {Cominsky}, {Constancio}, {Conti},
  {Cooper}, {Corban}, {Corbitt}, {Cordero-Carri{\'o}n}, {Corley}, {Cornish},
  {Corsi}, {Cortese}, {Costa}, {Coughlin}, {Coughlin}, {Coulon}, {Countryman},
  {Couvares}, {Covas}, {Cowan}, {Coward}, {Cowart}, {Coyne}, {Coyne},
  {Creighton}, {Creighton}, {Cripe}, {Crowder}, {Cullen}, {Cumming},
  {Cunningham}, {Cuoco}, {Dal Canton}, {D{\'a}lya}, {Danilishin}, {D'Antonio},
  {Danzmann}, {Dasgupta}, {da Silva Costa}, {Datrier}, {Dattilo}, {Dave},
  {Davier}, {Davis}, {Daw}, {Day}, {de}, {Debra}, {Degallaix}, {de Laurentis},
  {Del{\'e}glise}, {Del Pozzo}, {Demos}, {Denker}, {Dent}, {de Pietri},
  {Dergachev}, {De Rosa}, {Derosa}, {de Rossi}, {Desalvo}, {de Varona},
  {Devenson}, {Dhurandhar}, {D{\'\i}az}, {di Fiore}, {di Giovanni}, {di
  Girolamo}, {di Lieto}, {di Pace}, {di Palma}, {di Renzo}, {Doctor},
  {Dolique}, {Donovan}, {Dooley}, {Doravari}, {Dorrington}, {Douglas}, {Dovale
  {\'A}lvarez}, {Downes}, {Drago}, {Dreissigacker}, {Driggers}, {Du}, {Ducrot},
  {Dupej}, {Dwyer}, {Edo}, {Edwards}, {Effler}, {Eggenstein}, {Ehrens},
  {Eichholz}, {Eikenberry}, {Eisenstein}, {Essick}, {Estevez}, {Etienne},
  {Etzel}, {Evans}, {Evans}, {Factourovich}, {Fafone}, {Fair}, {Fairhurst},
  {Fan}, {Farinon}, {Farr}, {Farr}, {Fauchon-Jones}, {Favata}, {Fays}, {Fee},
  {Fehrmann}, {Feicht}, {Fejer}, {Fernandez-Galiana}, {Ferrante}, {Ferreira},
  {Ferrini}, {Fidecaro}, {Finstad}, {Fiori}, {Fiorucci}, {Fishbach}, {Fisher},
  {Fitz-Axen}, {Flaminio}, {Fletcher}, {Fong}, {Font}, {Forsyth}, {Forsyth},
  {Fournier}, {Frasca}, {Frasconi}, {Frei}, {Freise}, {Frey}, {Frey}, {Fries},
  {Fritschel}, {Frolov}, {Fulda}, {Fyffe}, {Gabbard}, {Gadre}, {Gaebel},
  {Gair}, {Gammaitoni}, {Ganija}, {Gaonkar}, {Garcia-Quiros}, {Garufi},
  {Gateley}, {Gaudio}, {Gaur}, {Gayathri}, {Gehrels}, {Gemme}, {Genin},
  {Gennai}, {George}, {George}, {Gergely}, {Germain}, {Ghonge}, {Ghosh},
  {Ghosh}, {Ghosh}, {Giaime}, {Giardina}, {Giazotto}, {Gill}, {Glover},
  {Goetz}, {Goetz}, {Gomes}, {Goncharov}, {Gonz{\'a}lez}, {Castro},
  {Gopakumar}, {Gorodetsky}, {Gossan}, {Gosselin}, {Gouaty}, {Grado}, {Graef},
  {Granata}, {Grant}, {Gras}, {Gray}, {Greco}, {Green}, {Gretarsson}, {Groot},
  {Grote}, {Grunewald}, {Gruning}, {Guidi}, {Guo}, {Gupta}, {Gupta}, {Gushwa},
  {Gustafson}, {Gustafson}, {Halim}, {Hall}, {Hall}, {Hamilton}, {Hammond},
  {Haney}, {Hanke}, {Hanks}, {Hanna}, {Hannam}, {Hannuksela}, {Hanson},
  {Hardwick}, {Harms}, {Harry}, {Harry}, {Hart}, {Haster}, {Haughian}, {Healy},
  {Heidmann}, {Heintze}, {Heitmann}, {Hello}, {Hemming}, {Hendry}, {Heng},
  {Hennig}, {Heptonstall}, {Heurs}, {Hild}, {Hinderer}, {Hoak}, {Hofman},
  {Holt}, {Holz}, {Hopkins}, {Horst}, {Hough}, {Houston}, {Howell}, {Hreibi},
  {Hu}, {Huerta}, {Huet}, {Hughey}, {Husa}, {Huttner}, {Huynh-Dinh}, {Indik},
  {Inta}, {Intini}, {Isa}, {Isac}, {Isi}, {Iyer}, {Izumi}, {Jacqmin}, {Jani},
  {Jaranowski}, {Jawahar}, {Jim{\'e}nez-Forteza}, {Johnson}, {Jones}, {Jones},
  {Jonker}, {Ju}, {Junker}, {Kalaghatgi}, {Kalogera}, {Kamai}, {Kandhasamy},
  {Kang}, {Kanner}, {Kapadia}, {Karki}, {Karvinen}, {Kasprzack}, {Katolik},
  {Katsavounidis}, {Katzman}, {Kaufer}, {Kawabe}, {K{\'e}f{\'e}lian}, {Keitel},
  {Kemball}, {Kennedy}, {Kent}, {Key}, {Khalili}, {Khan}, {Khan}, {Khan},
  {Khazanov}, {Kijbunchoo}, {Kim}, {Kim}, {Kim}, {Kim}, {Kim}, {Kim},
  {Kimbrell}, {King}, {King}, {Kinley-Hanlon}, {Kirchhoff}, {Kissel},
  {Kleybolte}, {Klimenko}, {Knowles}, {Koch}, {Koehlenbeck}, {Koley},
  {Kondrashov}, {Kontos}, {Korobko}, {Korth}, {Kowalska}, {Kozak},
  {Kr{\"a}mer}, {Kringel}, {Krishnan}, {Kr{\'o}lak}, {Kuehn}, {Kumar}, {Kumar},
  {Kumar}, {Kuo}, {Kutynia}, {Kwang}, {Lackey}, {Lai}, {Landry}, {Lang},
  {Lange}, {Lantz}, {Lanza}, {Lartaux-Vollard}, {Lasky}, {Laxen}, {Lazzarini},
  {Lazzaro}, {Leaci}, {Leavey}, {Lee}, {Lee}, {Lee}, {Lee}, {Lee}, {Lehmann},
  {Lenon}, {Leonardi}, {Leroy}, {Letendre}, {Levin}, {Li}, {Linker},
  {Littenberg}, {Liu}, {Liu}, {Lo}, {Lockerbie}, {London}, {Lord}, {Lorenzini},
  {Loriette}, {Lormand}, {Losurdo}, {Lough}, {Lousto}, {Lovelace}, {L{\"u}ck},
  {Lumaca}, {Lundgren}, {Lynch}, {Ma}, {Macas}, {Macfoy}, {Machenschalk},
  {Macinnis}, {MacLeod}, {Hernandez}, {Maga{\~n}a-Sandoval}, {Zertuche},
  {Magee}, {Majorana}, {Maksimovic}, {Man}, {Mandic}, {Mangano}, {Mansell},
  {Manske}, {Mantovani}, {Marchesoni}, {Marion}, {M{\'a}rka}, {M{\'a}rka},
  {Markakis}, {Markosyan}, {Markowitz}, {Maros}, {Marquina}, {Martelli},
  {Martellini}, {Martin}, {Martin}, {Martynov}, {Mason}, {Massera}, {Masserot},
  {Massinger}, {Masso-Reid}, {Mastrogiovanni}, {Matas}, {Matichard}, {Matone},
  {Mavalvala}, {Mazumder}, {McCarthy}, {McClelland}, {McCormick}, {McCuller},
  {McGuire}, {McIntyre}, {McIver}, {McManus}, {McNeill}, {McRae}, {McWilliams},
  {Meacher}, {Meadors}, {Mehmet}, {Meidam}, {Mejuto-Villa}, {Melatos},
  {Mendell}, {Mercer}, {Merilh}, {Merzougui}, {Meshkov}, {Messenger},
  {Messick}, {Metzdorff}, {Meyers}, {Miao}, {Michel}, {Middleton}, {Mikhailov},
  {Milano}, {Miller}, {Miller}, {Miller}, {Millhouse}, {Milovich-Goff},
  {Minazzoli}, {Minenkov}, {Ming}, {Mishra}, {Mitra}, {Mitrofanov},
  {Mitselmakher}, {Mittleman}, {Moffa}, {Moggi}, {Mogushi}, {Mohan},
  {Mohapatra}, {Montani}, {Moore}, {Moraru}, {Moreno}, {Morriss}, {Mours},
  {Mow-Lowry}, {Mueller}, {Muir}, {Mukherjee}, {Mukherjee}, {Mukherjee},
  {Mukund}, {Mullavey}, {Munch}, {Mu{\~n}iz}, {Muratore}, {Murray}, {Napier},
  {Nardecchia}, {Naticchioni}, {Nayak}, {Neilson}, {Nelemans}, {Nelson},
  {Nery}, {Neunzert}, {Nevin}, {Newport}, {Newton}, {Ng}, {Nguyen}, {Nichols},
  {Nielsen}, {Nissanke}, {Nitz}, {Noack}, {Nocera}, {Nolting}, {North},
  {Nuttall}, {Oberling}, {O'Dea}, {Ogin}, {Oh}, {Oh}, {Ohme}, {Okada},
  {Oliver}, {Oppermann}, {Oram}, {O'Reilly}, {Ormiston}, {Ortega},
  {O'Shaughnessy}, {Ossokine}, {Ottaway}, {Overmier}, {Owen}, {Pace}, {Page},
  {Page}, {Pai}, {Pai}, {Palamos}, {Palashov}, {Palomba}, {Pal-Singh}, {Pan},
  {Pan}, {Pang}, {Pang}, {Pankow}, {Pannarale}, {Pant}, {Paoletti}, {Paoli},
  {Papa}, {Parida}, {Parker}, {Pascucci}, {Pasqualetti}, {Passaquieti},
  {Passuello}, {Patil}, {Patricelli}, {Pearlstone}, {Pedraza}, {Pedurand},
  {Pekowsky}, {Pele}, {Penn}, {Perez}, {Perreca}, {Perri}, {Pfeiffer},
  {Phelps}, {Piccinni}, {Pichot}, {Piergiovanni}, {Pierro}, {Pillant},
  {Pinard}, {Pinto}, {Pirello}, {Pitkin}, {Poe}, {Poggiani}, {Popolizio},
  {Porter}, {Post}, {Powell}, {Prasad}, {Pratt}, {Pratten}, {Predoi},
  {Prestegard}, {Prijatelj}, {Principe}, {Privitera}, {Prodi}, {Prokhorov},
  {Puncken}, {Punturo}, {Puppo}, {P{\"u}rrer}, {Qi}, {Quetschke}, {Quintero},
  {Quitzow-James}, {Raab}, {Rabeling}, {Radkins}, {Raffai}, {Raja}, {Rajan},
  {Rajbhandari}, {Rakhmanov}, {Ramirez}, {Ramos-Buades}, {Rapagnani},
  {Raymond}, {Razzano}, {Read}, {Regimbau}, {Rei}, {Reid}, {Reitze}, {Ren},
  {Reyes}, {Ricci}, {Ricker}, {Rieger}, {Riles}, {Rizzo}, {Robertson}, {Robie},
  {Robinet}, {Rocchi}, {Rolland}, {Rollins}, {Roma}, {Romano}, {Romano},
  {Romel}, {Romie}, {Rosi{\'n}ska}, {Ross}, {Rowan}, {R{\"u}diger}, {Ruggi},
  {Rutins}, {Ryan}, {Sachdev}, {Sadecki}, {Sadeghian}, {Sakellariadou},
  {Salconi}, {Saleem}, {Salemi}, {Samajdar}, {Sammut}, {Sampson}, {Sanchez},
  {Sanchez}, {Sanchis-Gual}, {Sandberg}, {Sanders}, {Sassolas},
  {Sathyaprakash}, {Saulson}, {Sauter}, {Savage}, {Sawadsky}, {Schale},
  {Scheel}, {Scheuer}, {Schmidt}, {Schmidt}, {Schnabel}, {Schofield},
  {Sch{\"o}nbeck}, {Schreiber}, {Schuette}, {Schulte}, {Schutz}, {Schwalbe},
  {Scott}, {Scott}, {Seidel}, {Sellers}, {Sengupta}, {Sentenac}, {Sequino},
  {Sergeev}, {Shaddock}, {Shaffer}, {Shah}, {Shahriar}, {Shaner}, {Shao},
  {Shapiro}, {Shawhan}, {Sheperd}, {Shoemaker}, {Shoemaker}, {Siellez},
  {Siemens}, {Sieniawska}, {Sigg}, {Silva}, {Singer}, {Singh}, {Singhal},
  {Sintes}, {Slagmolen}, {Smith}, {Smith}, {Smith}, {Somala}, {Son},
  {Sonnenberg}, {Sorazu}, {Sorrentino}, {Souradeep}, {Spencer}, {Srivastava},
  {Staats}, {Staley}, {Steer}, {Steinke}, {Steinlechner}, {Steinlechner},
  {Steinmeyer}, {Stevenson}, {Stone}, {Stops}, {Strain}, {Stratta}, {Strigin},
  {Strunk}, {Sturani}, {Stuver}, {Summerscales}, {Sun}, {Sunil}, {Suresh},
  {Sutton}, {Swinkels}, {Szczepa{\'n}czyk}, {Tacca}, {Tait}, {Talbot},
  {Talukder}, {Tanner}, {T{\'a}pai}, {Taracchini}, {Tasson}, {Taylor},
  {Taylor}, {Tewari}, {Theeg}, {Thies}, {Thomas}, {Thomas}, {Thomas}, {Thorne},
  {Thrane}, {Tiwari}, {Tiwari}, {Tokmakov}, {Toland}, {Tonelli}, {Tornasi},
  {Torres-Forn{\'e}}, {Torrie}, {T{\"o}yr{\"a}}, {Travasso}, {Traylor},
  {Trinastic}, {Tringali}, {Trozzo}, {Tsang}, {Tse}, {Tso}, {Tsukada}, {Tsuna},
  {Tuyenbayev}, {Ueno}, {Ugolini}, {Unnikrishnan}, {Urban}, {Usman},
  {Vahlbruch}, {Vajente}, {Valdes}, {van Bakel}, {van Beuzekom}, {van den
  Brand}, {van den Broeck}, {Vander-Hyde}, {van der Schaaf}, {van Heijningen},
  {van Veggel}, {Vardaro}, {Varma}, {Vass}, {Vas{\'u}th}, {Vecchio},
  {Vedovato}, {Veitch}, {Veitch}, {Venkateswara}, {Venugopalan}, {Verkindt},
  {Vetrano}, {Vicer{\'e}}, {Viets}, {Vinciguerra}, {Vine}, {Vinet}, {Vitale},
  {Vo}, {Vocca}, {Vorvick}, {Vyatchanin}, {Wade}, {Wade}, {Wade}, {Walet},
  {Walker}, {Wallace}, {Walsh}, {Wang}, {Wang}, {Wang}, {Wang}, {Wang}, {Ward},
  {Warner}, {Was}, {Watchi}, {Weaver}, {Wei}, {Weinert}, {Weinstein}, {Weiss},
  {Wen}, {Wessel}, {We{\ss}els}, {Westerweck}, {Westphal}, {Wette}, {Whelan},
  {Whitcomb}, {Whiting}, {Whittle}, {Wilken}, {Williams}, {Williams},
  {Williamson}, {Willis}, {Willke}, {Wimmer}, {Winkler}, {Wipf}, {Wittel},
  {Woan}, {Woehler}, {Wofford}, {Wong}, {Worden}, {Wright}, {Wu}, {Wysocki},
  {Xiao}, {Yamamoto}, {Yancey}, {Yang}, {Yap}, {Yazback}, {Yu}, {Yu}, {Yvert},
  {Zadro{\.z}ny}, {Zanolin}, {Zelenova}, {Zendri}, {Zevin}, {Zhang}, {Zhang},
  {Zhang}, {Zhang}, {Zhao}, {Zhou}, {Zhou}, {Zhu}, {Zhu}, {Zimmerman},
  {Zucker}, {Zweizig}, {Foley}, {Coulter}, {Drout}, {Kasen}, {Kilpatrick},
  {Madore}, {Murguia-Berthier}, {Pan}, {Piro}, {Prochaska}, {Ramirez-Ruiz},
  {Rest}, {Rojas-Bravo}, {Shappee}, {Siebert}, {Simon}, {Ulloa}, {Annis},
  {Soares-Santos}, {Brout}, {Scolnic}, {Diehl}, {Frieman}, {Berger},
  {Alexander}, {Allam}, {Balbinot}, {Blanchard}, {Butler}, {Chornock}, {Cook},
  {Cowperthwaite}, {Drlica-Wagner}, {Drout}, {Durret}, {Eftekhari}, {Finley},
  {Fong}, {Fryer}, {Garc{\'\i}a-Bellido}, {Gill}, {Gruendl}, {Hanna},
  {Hartley}, {Herner}, {Huterer}, {Kasen}, {Kessler}, {Li}, {Lin}, {Lopes},
  {Louren{\c{c}}o}, {Margutti}, {Marriner}, {Marshall}, {Matheson}, {Medina},
  {Metzger}, {Mu{\~n}oz}, {Muir}, {Nicholl}, {Nugent}, {Palmese},
  {Paz-Chinch{\'o}n}, {Quataert}, {Sako}, {Sauseda}, {Schlegel}, {Secco},
  {Smith}, {Sobreira}, {Stebbins}, {Villar}, {Vivas}, {Wester}, {Williams},
  {Yanny}, {Zenteno}, {Abbott}, {Abdalla}, {Bechtol}, {Benoit-L{\'e}vy},
  {Bertin}, {Bridle}, {Brooks}, {Buckley-Geer}, {Burke}, {Rosell}, {Kind},
  {Carretero}, {Castander}, {Cunha}, {D'Andrea}, {da Costa}, {Davis}, {Depoy},
  {Desai}, {Dietrich}, {Estrada}, {Fernandez}, {Flaugher}, {Fosalba},
  {Gaztanaga}, {Gerdes}, {Giannantonio}, {Goldstein}, {Gruen}, {Gutierrez},
  {Hartley}, {Honscheid}, {Jain}, {James}, {Jeltema}, {Johnson}, {Kent},
  {Krause}, {Kron}, {Kuehn}, {Kuhlmann}, {Kuropatkin}, {Lahav}, {Lima}, {Maia},
  {March}, {Miller}, {Miquel}, {Neilsen}, {Nord}, {Ogando}, {Plazas}, {Romer},
  {Roodman}, {Rykoff}, {Sanchez}, {Scarpine}, {Schubnell}, {Sevilla-Noarbe},
  {Smith}, {Smith}, {Suchyta}, {Tarle}, {Thomas}, {Thomas}, {Troxel}, {Tucker},
  {Vikram}, {Walker}, {Weller}, {Zhang}, {Haislip}, {Kouprianov}, {Reichart},
  {Tartaglia}, {Sand}, {Valenti}, {Yang}, {Arcavi}, {Hosseinzadeh}, {Howell},
  {McCully}, {Poznanski}, {Vasylyev}, {Tanvir}, {Levan}, {Hjorth}, {Cano},
  {Copperwheat}, {de Ugarte-Postigo}, {Evans}, {Fynbo},
  {Gonz{\'a}lez-Fern{\'a}ndez}, {Greiner}, {Irwin}, {Lyman}, {Mandel},
  {McMahon}, {Milvang-Jensen}, {O'Brien}, {Osborne}, {Perley}, {Pian},
  {Palazzi}, {Rol}, {Rosetti}, {Rosswog}, {Rowlinson}, {Schulze}, {Steeghs},
  {Th{\"o}ne}, {Ulaczyk}, {Watson}, {Wiersema}, {Lipunov}, {Gorbovskoy},
  {Kornilov}, {Tyurina}, {Balanutsa}, {Vlasenko}, {Gorbunov}, {Podesta},
  {Levato}, {Saffe}, {Buckley}, {Budnev}, {Gress}, {Yurkov}, {Rebolo}, \&
  {Serra-Ricart}}]{StdSiren17}
{Abbott}, B.~P., {Abbott}, R., {Abbott}, T.~D., {$et~al$.} 2017{\natexlab{a}},
  \nat, 551, 85

\bibitem[{{Abbott} {$et~al$.}(2017{\natexlab{b}}){Abbott}, {Abbott}, {Abbott},
  {Acernese}, {Ackley}, {Adams}, {Adams}, {Addesso}, {Adhikari}, {Adya},
  {Affeldt}, {Afrough}, {Agarwal}, {Agathos}, {Agatsuma}, {Aggarwal}, {Aguiar},
  {Aiello}, {Ain}, {Ajith}, {Allen}, {Allen}, {Allocca}, {Altin}, {Amato},
  {Ananyeva}, {Anderson}, {Anderson}, {Angelova}, {Antier}, {Appert}, {Arai},
  {Araya}, {Areeda}, {Arnaud}, {Arun}, {Ascenzi}, {Ashton}, {Ast}, {Aston},
  {Astone}, {Atallah}, {Aufmuth}, {Aulbert}, {AultONeal}, {Austin},
  {Avila-Alvarez}, {Babak}, {Bacon}, {Bader}, {Bae}, {Bailes}, {Baker},
  {Baldaccini}, {Ballardin}, {Ballmer}, {Banagiri}, {Barayoga}, {Barclay},
  {Barish}, {Barker}, {Barkett}, {Barone}, {Barr}, {Barsotti}, {Barsuglia},
  {Barta}, {Barthelmy}, {Bartlett}, {Bartos}, {Bassiri}, {Basti}, {Batch},
  {Bawaj}, {Bayley}, {Bazzan}, {B{\'e}csy}, {Beer}, {Bejger}, {Belahcene},
  {Bell}, {Berger}, {Bergmann}, {Bernuzzi}, {Bero}, {Berry}, {Bersanetti},
  {Bertolini}, {Betzwieser}, {Bhagwat}, {Bhandare}, {Bilenko}, {Billingsley},
  {Billman}, {Birch}, {Birney}, {Birnholtz}, {Biscans}, {Biscoveanu}, {Bisht},
  {Bitossi}, {Biwer}, {Bizouard}, {Blackburn}, {Blackman}, {Blair}, {Blair},
  {Blair}, {Bloemen}, {Bock}, {Bode}, {Boer}, {Bogaert}, {Bohe}, {Bondu},
  {Bonilla}, {Bonnand}, {Boom}, {Bork}, {Boschi}, {Bose}, {Bossie},
  {Bouffanais}, {Bozzi}, {Bradaschia}, {Brady}, {Branchesi}, {Brau}, {Briant},
  {Brillet}, {Brinkmann}, {Brisson}, {Brockill}, {Broida}, {Brooks}, {Brown},
  {Brown}, {Brunett}, {Buchanan}, {Buikema}, {Bulik}, {Bulten}, {Buonanno},
  {Buskulic}, {Buy}, {Byer}, {Cabero}, {Cadonati}, {Cagnoli}, {Cahillane},
  {Calder{\'o}n Bustillo}, {Callister}, {Calloni}, {Camp}, {Canepa},
  {Canizares}, {Cannon}, {Cao}, {Cao}, {Capano}, {Capocasa}, {Carbognani},
  {Caride}, {Carney}, {Carullo}, {Casanueva Diaz}, {Casentini}, {Caudill},
  {Cavagli{\`a}}, {Cavalier}, {Cavalieri}, {Cella}, {Cepeda},
  {Cerd{\'a}-Dur{\'a}n}, {Cerretani}, {Cesarini}, {Chamberlin}, {Chan}, {Chao},
  {Charlton}, {Chase}, {Chassande-Mottin}, {Chatterjee}, {Chatziioannou},
  {Cheeseboro}, {Chen}, {Chen}, {Chen}, {Cheng}, {Chia}, {Chincarini},
  {Chiummo}, {Chmiel}, {Cho}, {Cho}, {Chow}, {Christensen}, {Chu}, {Chua},
  {Chua}, {Chung}, {Chung}, {Ciani}, {Ciolfi}, {Cirelli}, {Cirone}, {Clara},
  {Clark}, {Clearwater}, {Cleva}, {Cocchieri}, {Coccia}, {Cohadon}, {Cohen},
  {Colla}, {Collette}, {Cominsky}, {Constancio}, {Conti}, {Cooper}, {Corban},
  {Corbitt}, {Cordero-Carri{\'o}n}, {Corley}, {Cornish}, {Corsi}, {Cortese},
  {Costa}, {Coughlin}, {Coughlin}, {Coulon}, {Countryman}, {Couvares}, {Covas},
  {Cowan}, {Coward}, {Cowart}, {Coyne}, {Coyne}, {Creighton}, {Creighton},
  {Cripe}, {Crowder}, {Cullen}, {Cumming}, {Cunningham}, {Cuoco}, {Dal Canton},
  {D{\'a}lya}, {Danilishin}, {D'Antonio}, {Danzmann}, {Dasgupta}, {Da Silva
  Costa}, {Dattilo}, {Dave}, {Davier}, {Davis}, {Daw}, {Day}, {De}, {DeBra},
  {Degallaix}, {De Laurentis}, {Del{\'e}glise}, {Del Pozzo}, {Demos}, {Denker},
  {Dent}, {De Pietri}, {Dergachev}, {De Rosa}, {DeRosa}, {De Rossi}, {DeSalvo},
  {de Varona}, {Devenson}, {Dhurandhar}, {D{\'\i}az}, {Dietrich}, {Di Fiore},
  {Di Giovanni}, {Di Girolamo}, {Di Lieto}, {Di Pace}, {Di Palma}, {Di Renzo},
  {Doctor}, {Dolique}, {Donovan}, {Dooley}, {Doravari}, {Dorrington},
  {Douglas}, {Dovale {\'A}lvarez}, {Downes}, {Drago}, {Dreissigacker},
  {Driggers}, {Du}, {Ducrot}, {Dudi}, {Dupej}, {Dwyer}, {Edo}, {Edwards},
  {Effler}, {Eggenstein}, {Ehrens}, {Eichholz}, {Eikenberry}, {Eisenstein},
  {Essick}, {Estevez}, {Etienne}, {Etzel}, {Evans}, {Evans}, {Factourovich},
  {Fafone}, {Fair}, {Fairhurst}, {Fan}, {Farinon}, {Farr}, {Farr},
  {Fauchon-Jones}, {Favata}, {Fays}, {Fee}, {Fehrmann}, {Feicht}, {Fejer},
  {Fernandez-Galiana}, {Ferrante}, {Ferreira}, {Ferrini}, {Fidecaro},
  {Finstad}, {Fiori}, {Fiorucci}, {Fishbach}, {Fisher}, {Fitz-Axen},
  {Flaminio}, {Fletcher}, {Fong}, {Font}, {Forsyth}, {Forsyth}, {Fournier},
  {Frasca}, {Frasconi}, {Frei}, {Freise}, {Frey}, {Frey}, {Fries}, {Fritschel},
  {Frolov}, {Fulda}, {Fyffe}, {Gabbard}, {Gadre}, {Gaebel}, {Gair},
  {Gammaitoni}, {Ganija}, {Gaonkar}, {Garcia-Quiros}, {Garufi}, {Gateley},
  {Gaudio}, {Gaur}, {Gayathri}, {Gehrels}, {Gemme}, {Genin}, {Gennai},
  {George}, {George}, {Gergely}, {Germain}, {Ghonge}, {Ghosh}, {Ghosh},
  {Ghosh}, {Giaime}, {Giardina}, {Giazotto}, {Gill}, {Glover}, {Goetz},
  {Goetz}, {Gomes}, {Goncharov}, {Gonz{\'a}lez}, {Gonzalez Castro},
  {Gopakumar}, {Gorodetsky}, {Gossan}, {Gosselin}, {Gouaty}, {Grado}, {Graef},
  {Granata}, {Grant}, {Gras}, {Gray}, {Greco}, {Green}, {Gretarsson}, {Groot},
  {Grote}, {Grunewald}, {Gruning}, {Guidi}, {Guo}, {Gupta}, {Gupta}, {Gushwa},
  {Gustafson}, {Gustafson}, {Halim}, {Hall}, {Hall}, {Hamilton}, {Hammond},
  {Haney}, {Hanke}, {Hanks}, {Hanna}, {Hannam}, {Hannuksela}, {Hanson},
  {Hardwick}, {Harms}, {Harry}, {Harry}, {Hart}, {Haster}, {Haughian}, {Healy},
  {Heidmann}, {Heintze}, {Heitmann}, {Hello}, {Hemming}, {Hendry}, {Heng},
  {Hennig}, {Heptonstall}, {Heurs}, {Hild}, {Hinderer}, {Ho}, {Hoak}, {Hofman},
  {Holt}, {Holz}, {Hopkins}, {Horst}, {Hough}, {Houston}, {Howell}, {Hreibi},
  {Hu}, {Huerta}, {Huet}, {Hughey}, {Husa}, {Huttner}, {Huynh-Dinh}, {Indik},
  {Inta}, {Intini}, {Isa}, {Isac}, {Isi}, {Iyer}, {Izumi}, {Jacqmin}, {Jani},
  {Jaranowski}, {Jawahar}, {Jim{\'e}nez-Forteza}, {Johnson},
  {Johnson-McDaniel}, {Jones}, {Jones}, {Jonker}, {Ju}, {Junker}, {Kalaghatgi},
  {Kalogera}, {Kamai}, {Kandhasamy}, {Kang}, {Kanner}, {Kapadia}, {Karki},
  {Karvinen}, {Kasprzack}, {Kastaun}, {Katolik}, {Katsavounidis}, {Katzman},
  {Kaufer}, {Kawabe}, {K{\'e}f{\'e}lian}, {Keitel}, {Kemball}, {Kennedy},
  {Kent}, {Key}, {Khalili}, {Khan}, {Khan}, {Khan}, {Khazanov}, {Kijbunchoo},
  {Kim}, {Kim}, {Kim}, {Kim}, {Kim}, {Kim}, {Kimbrell}, {King}, {King},
  {Kinley-Hanlon}, {Kirchhoff}, {Kissel}, {Kleybolte}, {Klimenko}, {Knowles},
  {Koch}, {Koehlenbeck}, {Koley}, {Kondrashov}, {Kontos}, {Korobko}, {Korth},
  {Kowalska}, {Kozak}, {Kr{\"a}mer}, {Kringel}, {Krishnan}, {Kr{\'o}lak},
  {Kuehn}, {Kumar}, {Kumar}, {Kumar}, {Kuo}, {Kutynia}, {Kwang}, {Lackey},
  {Lai}, {Landry}, {Lang}, {Lange}, {Lantz}, {Lanza}, {Larson},
  {Lartaux-Vollard}, {Lasky}, {Laxen}, {Lazzarini}, {Lazzaro}, {Leaci},
  {Leavey}, {Lee}, {Lee}, {Lee}, {Lee}, {Lee}, {Lehmann}, {Lenon}, {Leon},
  {Leonardi}, {Leroy}, {Letendre}, {Levin}, {Li}, {Linker}, {Littenberg},
  {Liu}, {Liu}, {Lo}, {Lockerbie}, {London}, {Lord}, {Lorenzini}, {Loriette},
  {Lormand}, {Losurdo}, {Lough}, {Lousto}, {Lovelace}, {L{\"u}ck}, {Lumaca},
  {Lundgren}, {Lynch}, {Ma}, {Macas}, {Macfoy}, {Machenschalk}, {MacInnis},
  {Macleod}, {Maga{\~n}a Hernandez}, {Maga{\~n}a-Sandoval}, {Maga{\~n}a
  Zertuche}, {Magee}, {Majorana}, {Maksimovic}, {Man}, {Mandic}, {Mangano},
  {Mansell}, {Manske}, {Mantovani}, {Marchesoni}, {Marion}, {M{\'a}rka},
  {M{\'a}rka}, {Markakis}, {Markosyan}, {Markowitz}, {Maros}, {Marquina},
  {Marsh}, {Martelli}, {Martellini}, {Martin}, {Martin}, {Martynov}, {Marx},
  {Mason}, {Massera}, {Masserot}, {Massinger}, {Masso-Reid}, {Mastrogiovanni},
  {Matas}, {Matichard}, {Matone}, {Mavalvala}, {Mazumder}, {McCarthy},
  {McClelland}, {McCormick}, {McCuller}, {McGuire}, {McIntyre}, {McIver},
  {McManus}, {McNeill}, {McRae}, {McWilliams}, {Meacher}, {Meadors}, {Mehmet},
  {Meidam}, {Mejuto-Villa}, {Melatos}, {Mendell}, {Mercer}, {Merilh},
  {Merzougui}, {Meshkov}, {Messenger}, {Messick}, {Metzdorff}, {Meyers},
  {Miao}, {Michel}, {Middleton}, {Mikhailov}, {Milano}, {Miller}, {Miller},
  {Miller}, {Millhouse}, {Milovich-Goff}, {Minazzoli}, {Minenkov}, {Ming},
  {Mishra}, {Mitra}, {Mitrofanov}, {Mitselmakher}, {Mittleman}, {Moffa},
  {Moggi}, {Mogushi}, {Mohan}, {Mohapatra}, {Molina}, {Montani}, {Moore},
  {Moraru}, {Moreno}, {Morisaki}, {Morriss}, {Mours}, {Mow-Lowry}, {Mueller},
  {Muir}, {Mukherjee}, {Mukherjee}, {Mukherjee}, {Mukund}, {Mullavey}, {Munch},
  {Mu{\~n}iz}, {Muratore}, {Murray}, {Nagar}, {Napier}, {Nardecchia},
  {Naticchioni}, {Nayak}, {Neilson}, {Nelemans}, {Nelson}, {Nery}, {Neunzert},
  {Nevin}, {Newport}, {Newton}, {Ng}, {Nguyen}, {Nguyen}, {Nichols}, {Nielsen},
  {Nissanke}, {Nitz}, {Noack}, {Nocera}, {Nolting}, {North}, {Nuttall},
  {Oberling}, {O'Dea}, {Ogin}, {Oh}, {Oh}, {Ohme}, {Okada}, {Oliver},
  {Oppermann}, {Oram}, {O'Reilly}, {Ormiston}, {Ortega}, {O'Shaughnessy},
  {Ossokine}, {Ottaway}, {Overmier}, {Owen}, {Pace}, {Page}, {Page}, {Pai},
  {Pai}, {Palamos}, {Palashov}, {Palomba}, {Pal-Singh}, {Pan}, {Pan}, {Pang},
  {Pang}, {Pankow}, {Pannarale}, {Pant}, {Paoletti}, {Paoli}, {Papa}, {Parida},
  {Parker}, {Pascucci}, {Pasqualetti}, {Passaquieti}, {Passuello}, {Patil},
  {Patricelli}, {Pearlstone}, {Pedraza}, {Pedurand}, {Pekowsky}, {Pele},
  {Penn}, {Perez}, {Perreca}, {Perri}, {Pfeiffer}, {Phelps}, {Piccinni},
  {Pichot}, {Piergiovanni}, {Pierro}, {Pillant}, {Pinard}, {Pinto}, {Pirello},
  {Pitkin}, {Poe}, {Poggiani}, {Popolizio}, {Porter}, {Post}, {Powell},
  {Prasad}, {Pratt}, {Pratten}, {Predoi}, {Prestegard}, {Prijatelj},
  {Principe}, {Privitera}, {Prix}, {Prodi}, {Prokhorov}, {Puncken}, {Punturo},
  {Puppo}, {P{\"u}rrer}, {Qi}, {Quetschke}, {Quintero}, {Quitzow-James},
  {Raab}, {Rabeling}, {Radkins}, {Raffai}, {Raja}, {Rajan}, {Rajbhandari},
  {Rakhmanov}, {Ramirez}, {Ramos-Buades}, {Rapagnani}, {Raymond}, {Razzano},
  {Read}, {Regimbau}, {Rei}, {Reid}, {Reitze}, {Ren}, {Reyes}, {Ricci},
  {Ricker}, {Rieger}, {Riles}, {Rizzo}, {Robertson}, {Robie}, {Robinet},
  {Rocchi}, {Rolland}, {Rollins}, {Roma}, {Romano}, {Romano}, {Romel}, {Romie},
  {Rosi{\'n}ska}, {Ross}, {Rowan}, {R{\"u}diger}, {Ruggi}, {Rutins}, {Ryan},
  {Sachdev}, {Sadecki}, {Sadeghian}, {Sakellariadou}, {Salconi}, {Saleem},
  {Salemi}, {Samajdar}, {Sammut}, {Sampson}, {Sanchez}, {Sanchez},
  {Sanchis-Gual}, {Sandberg}, {Sanders}, {Sassolas}, {Sathyaprakash},
  {Saulson}, {Sauter}, {Savage}, {Sawadsky}, {Schale}, {Scheel}, {Scheuer},
  {Schmidt}, {Schmidt}, {Schnabel}, {Schofield}, {Sch{\"o}nbeck}, {Schreiber},
  {Schuette}, {Schulte}, {Schutz}, {Schwalbe}, {Scott}, {Scott}, {Seidel},
  {Sellers}, {Sengupta}, {Sentenac}, {Sequino}, {Sergeev}, {Shaddock},
  {Shaffer}, {Shah}, {Shahriar}, {Shaner}, {Shao}, {Shapiro}, {Shawhan},
  {Sheperd}, {Shoemaker}, {Shoemaker}, {Siellez}, {Siemens}, {Sieniawska},
  {Sigg}, {Silva}, {Singer}, {Singh}, {Singhal}, {Sintes}, {Slagmolen},
  {Smith}, {Smith}, {Smith}, {Somala}, {Son}, {Sonnenberg}, {Sorazu},
  {Sorrentino}, {Souradeep}, {Spencer}, {Srivastava}, {Staats}, {Staley},
  {Steinke}, {Steinlechner}, {Steinlechner}, {Steinmeyer}, {Stevenson},
  {Stone}, {Stops}, {Strain}, {Stratta}, {Strigin}, {Strunk}, {Sturani},
  {Stuver}, {Summerscales}, {Sun}, {Sunil}, {Suresh}, {Sutton}, {Swinkels},
  {Szczepa{\'n}czyk}, {Tacca}, {Tait}, {Talbot}, {Talukder}, {Tanner},
  {T{\'a}pai}, {Taracchini}, {Tasson}, {Taylor}, {Taylor}, {Tewari}, {Theeg},
  {Thies}, {Thomas}, {Thomas}, {Thomas}, {Thorne}, {Thorne}, {Thrane},
  {Tiwari}, {Tiwari}, {Tokmakov}, {Toland}, {Tonelli}, {Tornasi},
  {Torres-Forn{\'e}}, {Torrie}, {T{\"o}yr{\"a}}, {Travasso}, {Traylor},
  {Trinastic}, {Tringali}, {Trozzo}, {Tsang}, {Tse}, {Tso}, {Tsukada}, {Tsuna},
  {Tuyenbayev}, {Ueno}, {Ugolini}, {Unnikrishnan}, {Urban}, {Usman},
  {Vahlbruch}, {Vajente}, {Valdes}, {Vallisneri}, {van Bakel}, {van Beuzekom},
  {van den Brand}, {Van Den Broeck}, {Vander-Hyde}, {van der Schaaf}, {van
  Heijningen}, {van Veggel}, {Vardaro}, {Varma}, {Vass}, {Vas{\'u}th},
  {Vecchio}, {Vedovato}, {Veitch}, {Veitch}, {Venkateswara}, {Venugopalan},
  {Verkindt}, {Vetrano}, {Vicer{\'e}}, {Viets}, {Vinciguerra}, {Vine}, {Vinet},
  {Vitale}, {Vo}, {Vocca}, {Vorvick}, {Vyatchanin}, {Wade}, {Wade}, {Wade},
  {Walet}, {Walker}, {Wallace}, {Walsh}, {Wang}, {Wang}, {Wang}, {Wang},
  {Wang}, {Ward}, {Warner}, {Was}, {Watchi}, {Weaver}, {Wei}, {Weinert},
  {Weinstein}, {Weiss}, {Wen}, {Wessel}, {We{\ss}els}, {Westerweck},
  {Westphal}, {Wette}, {Whelan}, {Whitcomb}, {Whiting}, {Whittle}, {Wilken},
  {Williams}, {Williams}, {Williamson}, {Willis}, {Willke}, {Wimmer},
  {Winkler}, {Wipf}, {Wittel}, {Woan}, {Woehler}, {Wofford}, {Wong}, {Worden},
  {Wright}, {Wu}, {Wysocki}, {Xiao}, {Yamamoto}, {Yancey}, {Yang}, {Yap},
  {Yazback}, {Yu}, {Yu}, {Yvert}, {Zadro{\.Z}ny}, {Zanolin}, {Zelenova},
  {Zendri}, {Zevin}, {Zhang}, {Zhang}, {Zhang}, {Zhang}, {Zhao}, {Zhou},
  {Zhou}, {Zhu}, {Zhu}, {Zimmerman}, {Zucker}, {Zweizig}, {LIGO Scientific
  Collaboration}, \& {Virgo Collaboration}}]{GW170817_2017}
---. 2017{\natexlab{b}}, \prl, 119, 161101

\bibitem[{{Abbott} {$et~al$.}(2017{\natexlab{c}}){Abbott}, {Abbott}, {Abbott},
  {Acernese}, {Ackley}, {Adams}, {Adams}, {Addesso}, {Adhikari}, {Adya},
  {Affeldt}, {Afrough}, {Agarwal}, {Agathos}, {Agatsuma}, {Aggarwal}, {Aguiar},
  {Aiello}, {Ain}, {Ajith}, {Allen}, {Allen}, {Allocca}, {Altin}, {Amato},
  {Ananyeva}, {Anderson}, {Anderson}, {Angelova}, {Antier}, {Appert}, {Arai},
  {Araya}, {Areeda}, {Arnaud}, {Arun}, {Ascenzi}, {Ashton}, {Ast}, {Aston},
  {Astone}, {Atallah}, {Aufmuth}, {Aulbert}, {AultONeal}, {Austin},
  {Avila-Alvarez}, {Babak}, {Bacon}, {Bader}, {Bae}, {Baker}, {Baldaccini},
  {Ballardin}, {Ballmer}, {Banagiri}, {Barayoga}, {Barclay}, {Barish},
  {Barker}, {Barkett}, {Barone}, {Barr}, {Barsotti}, {Barsuglia}, {Barta},
  {Barthelmy}, {Bartlett}, {Bartos}, {Bassiri}, {Basti}, {Batch}, {Bawaj},
  {Bayley}, {Bazzan}, {B{\'e}csy}, {Beer}, {Bejger}, {Belahcene}, {Bell},
  {Berger}, {Bergmann}, {Bero}, {Berry}, {Bersanetti}, {Bertolini},
  {Betzwieser}, {Bhagwat}, {Bhandare}, {Bilenko}, {Billingsley}, {Billman},
  {Birch}, {Birney}, {Birnholtz}, {Biscans}, {Biscoveanu}, {Bisht}, {Bitossi},
  {Biwer}, {Bizouard}, {Blackburn}, {Blackman}, {Blair}, {Blair}, {Blair},
  {Bloemen}, {Bock}, {Bode}, {Boer}, {Bogaert}, {Bohe}, {Bondu}, {Bonilla},
  {Bonnand}, {Boom}, {Bork}, {Boschi}, {Bose}, {Bossie}, {Bouffanais}, {Bozzi},
  {Bradaschia}, {Brady}, {Branchesi}, {Brau}, {Briant}, {Brillet}, {Brinkmann},
  {Brisson}, {Brockill}, {Broida}, {Brooks}, {Brown}, {Brown}, {Brunett},
  {Buchanan}, {Buikema}, {Bulik}, {Bulten}, {Buonanno}, {Buskulic}, {Buy},
  {Byer}, {Cabero}, {Cadonati}, {Cagnoli}, {Cahillane}, {Calder{\'o}n
  Bustillo}, {Callister}, {Calloni}, {Camp}, {Canepa}, {Canizares}, {Cannon},
  {Cao}, {Cao}, {Capano}, {Capocasa}, {Carbognani}, {Caride}, {Carney},
  {Casanueva Diaz}, {Casentini}, {Caudill}, {Cavagli{\`a}}, {Cavalier},
  {Cavalieri}, {Cella}, {Cepeda}, {Cerd{\'a}-Dur{\'a}n}, {Cerretani},
  {Cesarini}, {Chamberlin}, {Chan}, {Chao}, {Charlton}, {Chase},
  {Chassande-Mottin}, {Chatterjee}, {Chatziioannou}, {Cheeseboro}, {Chen},
  {Chen}, {Chen}, {Cheng}, {Chia}, {Chincarini}, {Chiummo}, {Chmiel}, {Cho},
  {Cho}, {Chow}, {Christensen}, {Chu}, {Chua}, {Chua}, {Chung}, {Chung},
  {Ciani}, {Ciolfi}, {Cirelli}, {Cirone}, {Clara}, {Clark}, {Clearwater},
  {Cleva}, {Cocchieri}, {Coccia}, {Cohadon}, {Cohen}, {Colla}, {Collette},
  {Cominsky}, {Constancio}, {Conti}, {Cooper}, {Corban}, {Corbitt},
  {Cordero-Carri{\'o}n}, {Corley}, {Cornish}, {Corsi}, {Cortese}, {Costa},
  {Coughlin}, {Coughlin}, {Coulon}, {Countryman}, {Couvares}, {Covas}, {Cowan},
  {Coward}, {Cowart}, {Coyne}, {Coyne}, {Creighton}, {Creighton}, {Cripe},
  {Crowder}, {Cullen}, {Cumming}, {Cunningham}, {Cuoco}, {Dal Canton},
  {D{\'a}lya}, {Danilishin}, {D'Antonio}, {Danzmann}, {Dasgupta}, {Da Silva
  Costa}, {Dattilo}, {Dave}, {Davier}, {Davis}, {Daw}, {Day}, {De}, {DeBra},
  {Degallaix}, {De Laurentis}, {Del{\'e}glise}, {Del Pozzo}, {Demos}, {Denker},
  {Dent}, {De Pietri}, {Dergachev}, {De Rosa}, {DeRosa}, {De Rossi}, {DeSalvo},
  {de Varona}, {Devenson}, {Dhurandhar}, {D{\'\i}az}, {Di Fiore}, {Di
  Giovanni}, {Di Girolamo}, {Di Lieto}, {Di Pace}, {Di Palma}, {Di Renzo},
  {Doctor}, {Dolique}, {Donovan}, {Dooley}, {Doravari}, {Dorrington},
  {Douglas}, {Dovale {\'A}lvarez}, {Downes}, {Drago}, {Dreissigacker},
  {Driggers}, {Du}, {Ducrot}, {Dupej}, {Dwyer}, {Edo}, {Edwards}, {Effler},
  {Ehrens}, {Eichholz}, {Eikenberry}, {Eisenstein}, {Essick}, {Estevez},
  {Etienne}, {Etzel}, {Evans}, {Evans}, {Factourovich}, {Fafone}, {Fair},
  {Fairhurst}, {Fan}, {Farinon}, {Farr}, {Farr}, {Fauchon-Jones}, {Favata},
  {Fays}, {Fee}, {Fehrmann}, {Feicht}, {Fejer}, {Fernandez-Galiana},
  {Ferrante}, {Ferreira}, {Ferrini}, {Fidecaro}, {Finstad}, {Fiori},
  {Fiorucci}, {Fishbach}, {Fisher}, {Fitz-Axen}, {Flaminio}, {Fletcher},
  {Fong}, {Font}, {Forsyth}, {Forsyth}, {Fournier}, {Frasca}, {Frasconi},
  {Frei}, {Freise}, {Frey}, {Frey}, {Fries}, {Fritschel}, {Frolov}, {Fulda},
  {Fyffe}, {Gabbard}, {Gadre}, {Gaebel}, {Gair}, {Gammaitoni}, {Ganija},
  {Gaonkar}, {Garcia-Quiros}, {Garufi}, {Gateley}, {Gaudio}, {Gaur},
  {Gayathri}, {Gehrels}, {Gemme}, {Genin}, {Gennai}, {George}, {George},
  {Gergely}, {Germain}, {Ghonge}, {Ghosh}, {Ghosh}, {Ghosh}, {Giaime},
  {Giardina}, {Giazotto}, {Gill}, {Glover}, {Goetz}, {Goetz}, {Gomes},
  {Goncharov}, {Gonz{\'a}lez}, {Gonzalez Castro}, {Gopakumar}, {Gorodetsky},
  {Gossan}, {Gosselin}, {Gouaty}, {Grado}, {Graef}, {Granata}, {Grant}, {Gras},
  {Gray}, {Greco}, {Green}, {Gretarsson}, {Griswold}, {Groot}, {Grote},
  {Grunewald}, {Gruning}, {Guidi}, {Guo}, {Gupta}, {Gupta}, {Gushwa},
  {Gustafson}, {Gustafson}, {Halim}, {Hall}, {Hall}, {Hamilton}, {Hammond},
  {Haney}, {Hanke}, {Hanks}, {Hanna}, {Hannam}, {Hannuksela}, {Hanson},
  {Hardwick}, {Harms}, {Harry}, {Harry}, {Hart}, {Haster}, {Haughian}, {Healy},
  {Heidmann}, {Heintze}, {Heitmann}, {Hello}, {Hemming}, {Hendry}, {Heng},
  {Hennig}, {Heptonstall}, {Heurs}, {Hild}, {Hinderer}, {Hoak}, {Hofman},
  {Holt}, {Holz}, {Hopkins}, {Horst}, {Hough}, {Houston}, {Howell}, {Hreibi},
  {Hu}, {Huerta}, {Huet}, {Hughey}, {Husa}, {Huttner}, {Huynh-Dinh}, {Indik},
  {Inta}, {Intini}, {Isa}, {Isac}, {Isi}, {Iyer}, {Izumi}, {Jacqmin}, {Jani},
  {Jaranowski}, {Jawahar}, {Jim{\'e}nez-Forteza}, {Johnson}, {Jones}, {Jones},
  {Jonker}, {Ju}, {Junker}, {Kalaghatgi}, {Kalogera}, {Kamai}, {Kandhasamy},
  {Kang}, {Kanner}, {Kapadia}, {Karki}, {Karvinen}, {Kasprzack}, {Katolik},
  {Katsavounidis}, {Katzman}, {Kaufer}, {Kawabe}, {K{\'e}f{\'e}lian}, {Keitel},
  {Kemball}, {Kennedy}, {Kent}, {Key}, {Khalili}, {Khan}, {Khan}, {Khan},
  {Khazanov}, {Kijbunchoo}, {Kim}, {Kim}, {Kim}, {Kim}, {Kim}, {Kim},
  {Kimbrell}, {King}, {King}, {Kinley-Hanlon}, {Kirchhoff}, {Kissel},
  {Kleybolte}, {Klimenko}, {Knowles}, {Koch}, {Koehlenbeck}, {Koley},
  {Kondrashov}, {Kontos}, {Korobko}, {Korth}, {Kowalska}, {Kozak},
  {Kr{\"a}mer}, {Kringel}, {Krishnan}, {Kr{\'o}lak}, {Kuehn}, {Kumar}, {Kumar},
  {Kumar}, {Kuo}, {Kutynia}, {Kwang}, {Lackey}, {Lai}, {Landry}, {Lang},
  {Lange}, {Lantz}, {Lanza}, {Larson}, {Lartaux-Vollard}, {Lasky}, {Laxen},
  {Lazzarini}, {Lazzaro}, {Leaci}, {Leavey}, {Lee}, {Lee}, {Lee}, {Lee}, {Lee},
  {Lehmann}, {Lenon}, {Leonardi}, {Leroy}, {Letendre}, {Levin}, {Li}, {Linker},
  {Littenberg}, {Liu}, {Lo}, {Lockerbie}, {London}, {Lord}, {Lorenzini},
  {Loriette}, {Lormand}, {Losurdo}, {Lough}, {Lousto}, {Lovelace}, {L{\"u}ck},
  {Lumaca}, {Lundgren}, {Lynch}, {Ma}, {Macas}, {Macfoy}, {Machenschalk},
  {MacInnis}, {Macleod}, {Maga{\~n}a Hernandez}, {Maga{\~n}a-Sandoval},
  {Maga{\~n}a Zertuche}, {Magee}, {Majorana}, {Maksimovic}, {Man}, {Mandic},
  {Mangano}, {Mansell}, {Manske}, {Mantovani}, {Marchesoni}, {Marion},
  {M{\'a}rka}, {M{\'a}rka}, {Markakis}, {Markosyan}, {Markowitz}, {Maros},
  {Marquina}, {Marsh}, {Martelli}, {Martellini}, {Martin}, {Martin},
  {Martynov}, {Mason}, {Massera}, {Masserot}, {Massinger}, {Masso-Reid},
  {Mastrogiovanni}, {Matas}, {Matichard}, {Matone}, {Mavalvala}, {Mazumder},
  {McCarthy}, {McClelland}, {McCormick}, {McCuller}, {McGuire}, {McIntyre},
  {McIver}, {McManus}, {McNeill}, {McRae}, {McWilliams}, {Meacher}, {Meadors},
  {Mehmet}, {Meidam}, {Mejuto-Villa}, {Melatos}, {Mendell}, {Mercer}, {Merilh},
  {Merzougui}, {Meshkov}, {Messenger}, {Messick}, {Metzdorff}, {Meyers},
  {Miao}, {Michel}, {Middleton}, {Mikhailov}, {Milano}, {Miller}, {Miller},
  {Miller}, {Millhouse}, {Milovich-Goff}, {Minazzoli}, {Minenkov}, {Ming},
  {Mishra}, {Mitra}, {Mitrofanov}, {Mitselmakher}, {Mittleman}, {Moffa},
  {Moggi}, {Mogushi}, {Mohan}, {Mohapatra}, {Montani}, {Moore}, {Moraru},
  {Moreno}, {Morriss}, {Mours}, {Mow-Lowry}, {Mueller}, {Muir}, {Mukherjee},
  {Mukherjee}, {Mukherjee}, {Mukund}, {Mullavey}, {Munch}, {Mu{\~n}iz},
  {Muratore}, {Murray}, {Napier}, {Nardecchia}, {Naticchioni}, {Nayak},
  {Neilson}, {Nelemans}, {Nelson}, {Nery}, {Neunzert}, {Nevin}, {Newport},
  {Newton}, {Ng}, {Nguyen}, {Nguyen}, {Nichols}, {Nielsen}, {Nissanke}, {Nitz},
  {Noack}, {Nocera}, {Nolting}, {North}, {Nuttall}, {Oberling}, {O'Dea},
  {Ogin}, {Oh}, {Oh}, {Ohme}, {Okada}, {Oliver}, {Oppermann}, {Oram},
  {O'Reilly}, {Ormiston}, {Ortega}, {O'Shaughnessy}, {Ossokine}, {Ottaway},
  {Overmier}, {Owen}, {Pace}, {Page}, {Page}, {Pai}, {Pai}, {Palamos},
  {Palashov}, {Palomba}, {Pal-Singh}, {Pan}, {Pan}, {Pang}, {Pang}, {Pankow},
  {Pannarale}, {Pant}, {Paoletti}, {Paoli}, {Papa}, {Parida}, {Parker},
  {Pascucci}, {Pasqualetti}, {Passaquieti}, {Passuello}, {Patil}, {Patricelli},
  {Pearlstone}, {Pedraza}, {Pedurand}, {Pekowsky}, {Pele}, {Penn}, {Perez},
  {Perreca}, {Perri}, {Pfeiffer}, {Phelps}, {Piccinni}, {Pichot},
  {Piergiovanni}, {Pierro}, {Pillant}, {Pinard}, {Pinto}, {Pirello}, {Pitkin},
  {Poe}, {Poggiani}, {Popolizio}, {Porter}, {Post}, {Powell}, {Prasad},
  {Pratt}, {Pratten}, {Predoi}, {Prestegard}, {Price}, {Prijatelj}, {Principe},
  {Privitera}, {Prodi}, {Prokhorov}, {Puncken}, {Punturo}, {Puppo},
  {P{\"u}rrer}, {Qi}, {Quetschke}, {Quintero}, {Quitzow-James}, {Raab},
  {Rabeling}, {Radkins}, {Raffai}, {Raja}, {Rajan}, {Rajbhandari}, {Rakhmanov},
  {Ramirez}, {Ramos-Buades}, {Rapagnani}, {Raymond}, {Razzano}, {Read},
  {Regimbau}, {Rei}, {Reid}, {Reitze}, {Ren}, {Reyes}, {Ricci}, {Ricker},
  {Rieger}, {Riles}, {Rizzo}, {Robertson}, {Robie}, {Robinet}, {Rocchi},
  {Rolland}, {Rollins}, {Roma}, {Romano}, {Romel}, {Romie}, {Rosi{\'n}ska},
  {Ross}, {Rowan}, {R{\"u}diger}, {Ruggi}, {Rutins}, {Ryan}, {Sachdev},
  {Sadecki}, {Sadeghian}, {Sakellariadou}, {Salconi}, {Saleem}, {Salemi},
  {Samajdar}, {Sammut}, {Sampson}, {Sanchez}, {Sanchez}, {Sanchis-Gual},
  {Sandberg}, {Sanders}, {Sassolas}, {Sathyaprakash}, {Saulson}, {Sauter},
  {Savage}, {Sawadsky}, {Schale}, {Scheel}, {Scheuer}, {Schmidt}, {Schmidt},
  {Schnabel}, {Schofield}, {Sch{\"o}nbeck}, {Schreiber}, {Schuette}, {Schulte},
  {Schutz}, {Schwalbe}, {Scott}, {Scott}, {Seidel}, {Sellers}, {Sengupta},
  {Sentenac}, {Sequino}, {Sergeev}, {Shaddock}, {Shaffer}, {Shah}, {Shahriar},
  {Shaner}, {Shao}, {Shapiro}, {Shawhan}, {Sheperd}, {Shoemaker}, {Shoemaker},
  {Siellez}, {Siemens}, {Sieniawska}, {Sigg}, {Silva}, {Singer}, {Singh},
  {Singhal}, {Sintes}, {Slagmolen}, {Smith}, {Smith}, {Smith}, {Somala}, {Son},
  {Sonnenberg}, {Sorazu}, {Sorrentino}, {Souradeep}, {Spencer}, {Srivastava},
  {Staats}, {Staley}, {Steinke}, {Steinlechner}, {Steinlechner}, {Steinmeyer},
  {Stevenson}, {Stone}, {Stops}, {Strain}, {Stratta}, {Strigin}, {Strunk},
  {Sturani}, {Stuver}, {Summerscales}, {Sun}, {Sunil}, {Suresh}, {Sutton},
  {Swinkels}, {Szczepa{\'n}czyk}, {Tacca}, {Tait}, {Talbot}, {Talukder},
  {Tanner}, {T{\'a}pai}, {Taracchini}, {Tasson}, {Taylor}, {Taylor}, {Tewari},
  {Theeg}, {Thies}, {Thomas}, {Thomas}, {Thomas}, {Thorne}, {Thorne}, {Thrane},
  {Tiwari}, {Tiwari}, {Tokmakov}, {Toland}, {Tonelli}, {Tornasi},
  {Torres-Forn{\'e}}, {Torrie}, {T{\"o}yr{\"a}}, {Travasso}, {Traylor},
  {Trinastic}, {Tringali}, {Trozzo}, {Tsang}, {Tse}, {Tso}, {Tsukada}, {Tsuna},
  {Tuyenbayev}, {Ueno}, {Ugolini}, {Unnikrishnan}, {Urban}, {Usman},
  {Vahlbruch}, {Vajente}, {Valdes}, {van Bakel}, {van Beuzekom}, {van den
  Brand}, {Van Den Broeck}, {Vander-Hyde}, {van der Schaaf}, {van Heijningen},
  {van Veggel}, {Vardaro}, {Varma}, {Vass}, {Vas{\'u}th}, {Vecchio},
  {Vedovato}, {Veitch}, {Veitch}, {Venkateswara}, {Venugopalan}, {Verkindt},
  {Vetrano}, {Vicer{\'e}}, {Viets}, {Vinciguerra}, {Vine}, {Vinet}, {Vitale},
  {Vo}, {Vocca}, {Vorvick}, {Vyatchanin}, {Wade}, {Wade}, {Wade}, {Walet},
  {Walker}, {Wallace}, {Walsh}, {Wang}, {Wang}, {Wang}, {Wang}, {Wang}, {Ward},
  {Warner}, {Was}, {Watchi}, {Weaver}, {Wei}, {Weinert}, {Weinstein}, {Weiss},
  {Wen}, {Wessel}, {Wessels}, {Westerweck}, {Westphal}, {Wette}, {Whelan},
  {Whitcomb}, {Whiting}, {Whittle}, {Wilken}, {Williams}, {Williams},
  {Williamson}, {Willis}, {Willke}, {Wimmer}, {Winkler}, {Wipf}, {Wittel},
  {Woan}, {Woehler}, {Wofford}, {Wong}, {Worden}, {Wright}, {Wu}, {Wysocki},
  {Xiao}, {Yamamoto}, {Yancey}, {Yang}, {Yap}, {Yazback}, {Yu}, {Yu}, {Yvert},
  {Zadro{\.z}ny}, {Zanolin}, {Zelenova}, {Zendri}, {Zevin}, {Zhang}, {Zhang},
  {Zhang}, {Zhang}, {Zhao}, {Zhou}, {Zhou}, {Zhu}, {Zhu}, {Zimmerman},
  {Zucker}, {Zweizig}, {LIGO Scientific Collaboration}, {Virgo Collaboration},
  {Wilson-Hodge}, {Bissaldi}, {Blackburn}, {Briggs}, {Burns}, {Cleveland},
  {Connaughton}, {Gibby}, {Giles}, {Goldstein}, {Hamburg}, {Jenke}, {Hui},
  {Kippen}, {Kocevski}, {McBreen}, {Meegan}, {Paciesas}, {Poolakkil}, {Preece},
  {Racusin}, {Roberts}, {Stanbro}, {Veres}, {von Kienlin}, {GBM}, {Savchenko},
  {Ferrigno}, {Kuulkers}, {Bazzano}, {Bozzo}, {Brandt}, {Chenevez},
  {Courvoisier}, {Diehl}, {Domingo}, {Hanlon}, {Jourdain}, {Laurent}, {Lebrun},
  {Lutovinov}, {Martin-Carrillo}, {Mereghetti}, {Natalucci}, {Rodi}, {Roques},
  {Sunyaev}, {Ubertini}, {INTEGRAL}, {Aartsen}, {Ackermann}, {Adams},
  {Aguilar}, {Ahlers}, {Ahrens}, {Samarai}, {Altmann}, {Andeen}, {Anderson},
  {Ansseau}, {Anton}, {Arg{\"u}elles}, {Auffenberg}, {Axani}, {Bagherpour},
  {Bai}, {Barron}, {Barwick}, {Baum}, {Bay}, {Beatty}, {Becker Tjus},
  {Bernardini}, {Besson}, {Binder}, {Bindig}, {Blaufuss}, {Blot}, {Bohm},
  {B{\"o}rner}, {Bos}, {Bose}, {B{\"o}ser}, {Botner}, {Bourbeau}, {Bourbeau},
  {Bradascio}, {Braun}, {Brayeur}, {Brenzke}, {Bretz}, {Bron},
  {Brostean-Kaiser}, {Burgman}, {Carver}, {Casey}, {Casier}, {Cheung},
  {Chirkin}, {Christov}, {Clark}, {Classen}, {Coenders}, {Collin}, {Conrad},
  {Cowen}, {Cross}, {Day}, {de Andr{\'e}}, {De Clercq}, {DeLaunay},
  {Dembinski}, {De Ridder}, {Desiati}, {de Vries}, {de Wasseige}, {de With},
  {DeYoung}, {D{\'\i}az-V{\'e}lez}, {di Lorenzo}, {Dujmovic}, {Dumm},
  {Dunkman}, {Dvorak}, {Eberhardt}, {Ehrhardt}, {Eichmann}, {Eller}, {Evenson},
  {Fahey}, {Fazely}, {Felde}, {Filimonov}, {Finley}, {Flis}, {Franckowiak},
  {Friedman}, {Fuchs}, {Gaisser}, {Gallagher}, {Gerhardt}, {Ghorbani}, {Giang},
  {Glauch}, {Gl{\"u}senkamp}, {Goldschmidt}, {Gonzalez}, {Grant}, {Griffith},
  {Haack}, {Hallgren}, {Halzen}, {Hanson}, {Hebecker}, {Heereman}, {Helbing},
  {Hellauer}, {Hickford}, {Hignight}, {Hill}, {Hoffman}, {Hoffmann},
  {Hokanson-Fasig}, {Hoshina}, {Huang}, {Huber}, {Hultqvist}, {H{\"u}nnefeld},
  {In}, {Ishihara}, {Jacobi}, {Japaridze}, {Jeong}, {Jero}, {Jones},
  {Kalaczynski}, {Kang}, {Kappes}, {Karg}, {Karle}, {Kauer}, {Keivani},
  {Kelley}, {Kheirandish}, {Kim}, {Kim}, {Kintscher}, {Kiryluk}, {Kittler},
  {Klein}, {Kohnen}, {Koirala}, {Kolanoski}, {K{\"o}pke}, {Kopper}, {Kopper},
  {Koschinsky}, {Koskinen}, {Kowalski}, {Krings}, {Kroll}, {Kr{\"u}ckl},
  {Kunnen}, {Kunwar}, {Kurahashi}, {Kuwabara}, {Kyriacou}, {Labare},
  {Lanfranchi}, {Larson}, {Lauber}, {Lesiak-Bzdak}, {Leuermann}, {Liu}, {Lu},
  {L{\"u}nemann}, {Luszczak}, {Madsen}, {Maggi}, {Mahn}, {Mancina}, {Maruyama},
  {Mase}, {Maunu}, {McNally}, {Meagher}, {Medici}, {Meier}, {Menne}, {Merino},
  {Meures}, {Miarecki}, {Micallef}, {Moment{\'e}}, {Montaruli}, {Moore},
  {Moulai}, {Nahnhauer}, {Nakarmi}, {Naumann}, {Neer}, {Niederhausen},
  {Nowicki}, {Nygren}, {Obertacke Pollmann}, {Olivas}, {O'Murchadha},
  {Palczewski}, {Pandya}, {Pankova}, {Peiffer}, {Pepper}, {P{\'e}rez de los
  Heros}, {Pieloth}, {Pinat}, {Price}, {Przybylski}, {Raab}, {R{\"a}del},
  {Rameez}, {Rawlins}, {Rea}, {Reimann}, {Relethford}, {Relich}, {Resconi},
  {Rhode}, {Richman}, {Robertson}, {Rongen}, {Rott}, {Ruhe}, {Ryckbosch},
  {Rysewyk}, {S{\"a}lzer}, {Sanchez Herrera}, {Sandrock}, {Sandroos},
  {Santander}, {Sarkar}, {Sarkar}, {Satalecka}, {Schlunder}, {Schmidt},
  {Schneider}, {Schoenen}, {Sch{\"o}neberg}, {Schumacher}, {Seckel},
  {Seunarine}, {Soedingrekso}, {Soldin}, {Song}, {Spiczak}, {Spiering},
  {Stachurska}, {Stamatikos}, {Stanev}, {Stasik}, {Stettner}, {Steuer},
  {Stezelberger}, {Stokstad}, {St{\"o}ssl}, {Strotjohann}, {Stuttard},
  {Sullivan}, {Sutherland}, {Taboada}, {Tatar}, {Tenholt}, {Ter-Antonyan},
  {Terliuk}, {Te{\v{s}}i{\'c}}, {Tilav}, {Toale}, {Tobin}, {Toscano}, {Tosi},
  {Tselengidou}, {Tung}, {Turcati}, {Turley}, {Ty}, {Unger}, {Usner},
  {Vandenbroucke}, {Van Driessche}, {van Eijndhoven}, {Vanheule}, {van Santen},
  {Vehring}, {Vogel}, {Vraeghe}, {Walck}, {Wallace}, {Wallraff}, {Wandler},
  {Wandkowsky}, {Waza}, {Weaver}, {Weiss}, {Wendt}, {Werthebach}, {Whelan},
  {Wiebe}, {Wiebusch}, {Wille}, {Williams}, {Wills}, {Wolf}, {Wood}, {Woolsey},
  {Woschnagg}, {Xu}, {Xu}, {Xu}, {Yanez}, {Yodh}, {Yoshida}, {Yuan}, {Zoll},
  {IceCube Collaboration}, {Balasubramanian}, {Mate}, {Bhalerao},
  {Bhattacharya}, {Vibhute}, {Dewangan}, {Rao}, {Vadawale}, {AstroSat Cadmium
  Zinc Telluride Imager Team}, {Svinkin}, {Hurley}, {Aptekar}, {Frederiks},
  {Golenetskii}, {Kozlova}, {Lysenko}, {Oleynik}, {Tsvetkova}, {Ulanov},
  {Cline}, {IPN Collaboration}, {Li}, {Xiong}, {Zhang}, {Lu}, {Song}, {Cao},
  {Chang}, {Chen}, {Chen}, {Chen}, {Chen}, {Chen}, {Chen}, {Cui}, {Cui},
  {Deng}, {Dong}, {Du}, {Fu}, {Gao}, {Gao}, {Gao}, {Ge}, {Gu}, {Guan}, {Guo},
  {Han}, {Hu}, {Huang}, {Huo}, {Jia}, {Jiang}, {Jiang}, {Jin}, {Jin}, {Li},
  {Li}, {Li}, {Li}, {Li}, {Li}, {Li}, {Li}, {Li}, {Li}, {Li}, {Liang}, {Liao},
  {Liu}, {Liu}, {Liu}, {Liu}, {Liu}, {Liu}, {Liu}, {Lu}, {Lu}, {Luo}, {Ma},
  {Meng}, {Nang}, {Nie}, {Ou}, {Qu}, {Sai}, {Sun}, {Tan}, {Tao}, {Tao}, {Tuo},
  {Wang}, {Wang}, {Wang}, {Wang}, {Wang}, {Wen}, {Wu}, {Wu}, {Xiao}, {Xu},
  {Xu}, {Yan}, {Yang}, {Yang}, {Yang}, {Zhang}, {Zhang}, {Zhang}, {Zhang},
  {Zhang}, {Zhang}, {Zhang}, {Zhang}, {Zhang}, {Zhang}, {Zhang}, {Zhang},
  {Zhang}, {Zhang}, {Zhang}, {Zhang}, {Zhang}, {Zhang}, {Zhao}, {Zhao}, {Zhao},
  {Zheng}, {Zhu}, {Zhu}, {Zou}, {Insight-HXMT Collaboration}, {Albert},
  {Andr{\'e}}, {Anghinolfi}, {Ardid}, {Aubert}, {Aublin}, {Avgitas}, {Baret},
  {Barrios-Mart{\'\i}}, {Basa}, {Belhorma}, {Bertin}, {Biagi}, {Bormuth},
  {Bourret}, {Bouwhuis}, {Br{\^a}nza{\c{s}}}, {Bruijn}, {Brunner}, {Busto},
  {Capone}, {Caramete}, {Carr}, {Celli}, {Cherkaoui El Moursli}, {Chiarusi},
  {Circella}, {Coelho}, {Coleiro}, {Coniglione}, {Costantini}, {Coyle},
  {Creusot}, {D{\'\i}az}, {Deschamps}, {De Bonis}, {Distefano}, {Di Palma},
  {Domi}, {Donzaud}, {Dornic}, {Drouhin}, {Eberl}, {El Bojaddaini}, {El
  Khayati}, {Els{\"a}sser}, {Enzenh{\"o}fer}, {Ettahiri}, {Fassi}, {Felis},
  {Fusco}, {Gay}, {Giordano}, {Glotin}, {Gr{\'e}goire}, {Ruiz}, {Graf},
  {Hallmann}, {van Haren}, {Heijboer}, {Hello}, {Hern{\'a}ndez-Rey},
  {H{\"o}ssl}, {Hofest{\"a}dt}, {Hugon}, {Illuminati}, {James}, {de Jong},
  {Jongen}, {Kadler}, {Kalekin}, {Katz}, {Kiessling}, {Kouchner}, {Kreter},
  {Kreykenbohm}, {Kulikovskiy}, {Lachaud}, {Lahmann}, {Lef{\`e}vre}, {Leonora},
  {Lotze}, {Loucatos}, {Marcelin}, {Margiotta}, {Marinelli},
  {Mart{\'\i}nez-Mora}, {Mele}, {Melis}, {Michael}, {Migliozzi}, {Moussa},
  {Navas}, {Nezri}, {Organokov}, {P{\u{a}}v{\u{a}}la{\c{s}}}, {Pellegrino},
  {Perrina}, {Piattelli}, {Popa}, {Pradier}, {Quinn}, {Racca}, {Riccobene},
  {S{\'a}nchez-Losa}, {Salda{\~n}a}, {Salvadori}, {Samtleben}, {Sanguineti},
  {Sapienza}, {Sieger}, {Spurio}, {Stolarczyk}, {Taiuti}, {Tayalati},
  {Trovato}, {Turpin}, {T{\"o}nnis}, {Vallage}, {Van Elewyck}, {Versari},
  {Vivolo}, {Vizzoca}, {Wilms}, {Zornoza}, {Z{\'u}{\~n}iga}, {ANTARES
  Collaboration}, {Beardmore}, {Breeveld}, {Burrows}, {Cenko}, {Cusumano},
  {D'A{\`\i}}, {de Pasquale}, {Emery}, {Evans}, {Giommi}, {Gronwall}, {Kennea},
  {Krimm}, {Kuin}, {Lien}, {Marshall}, {Melandri}, {Nousek}, {Oates},
  {Osborne}, {Pagani}, {Page}, {Palmer}, {Perri}, {Siegel}, {Sbarufatti},
  {Tagliaferri}, {Tohuvavohu}, {Swift Collaboration}, {Tavani}, {Verrecchia},
  {Bulgarelli}, {Evangelista}, {Pacciani}, {Feroci}, {Pittori}, {Giuliani},
  {Del Monte}, {Donnarumma}, {Argan}, {Trois}, {Ursi}, {Cardillo}, {Piano},
  {Longo}, {Lucarelli}, {Munar-Adrover}, {Fuschino}, {Labanti}, {Marisaldi},
  {Minervini}, {Fioretti}, {Parmiggiani}, {Gianotti}, {Trifoglio}, {Di Persio},
  {Antonelli}, {Barbiellini}, {Caraveo}, {Cattaneo}, {Costa}, {Colafrancesco},
  {D'Amico}, {Ferrari}, {Morselli}, {Paoletti}, {Picozza}, {Pilia}, {Rappoldi},
  {Soffitta}, {Vercellone}, {AGILE Team}, {Foley}, {Coulter}, {Kilpatrick},
  {Drout}, {Piro}, {Shappee}, {Siebert}, {Simon}, {Ulloa}, {Kasen}, {Madore},
  {Murguia-Berthier}, {Pan}, {Prochaska}, {Ramirez-Ruiz}, {Rest},
  {Rojas-Bravo}, {1M2H Team}, {Berger}, {Soares-Santos}, {Annis}, {Alexander},
  {Allam}, {Balbinot}, {Blanchard}, {Brout}, {Butler}, {Chornock}, {Cook},
  {Cowperthwaite}, {Diehl}, {Drlica-Wagner}, {Drout}, {Durret}, {Eftekhari},
  {Finley}, {Fong}, {Frieman}, {Fryer}, {Garc{\'\i}a-Bellido}, {Gruendl},
  {Hartley}, {Herner}, {Kessler}, {Lin}, {Lopes}, {Louren{\c{c}}o}, {Margutti},
  {Marshall}, {Matheson}, {Medina}, {Metzger}, {Mu{\~n}oz}, {Muir}, {Nicholl},
  {Nugent}, {Palmese}, {Paz-Chinch{\'o}n}, {Quataert}, {Sako}, {Sauseda},
  {Schlegel}, {Scolnic}, {Secco}, {Smith}, {Sobreira}, {Villar}, {Vivas},
  {Wester}, {Williams}, {Yanny}, {Zenteno}, {Zhang}, {Abbott}, {Banerji},
  {Bechtol}, {Benoit-L{\'e}vy}, {Bertin}, {Brooks}, {Buckley-Geer}, {Burke},
  {Capozzi}, {Carnero Rosell}, {Carrasco Kind}, {Castander}, {Crocce}, {Cunha},
  {D'Andrea}, {da Costa}, {Davis}, {DePoy}, {Desai}, {Dietrich}, {Eifler},
  {Fernandez}, {Flaugher}, {Fosalba}, {Gaztanaga}, {Gerdes}, {Giannantonio},
  {Goldstein}, {Gruen}, {Gschwend}, {Gutierrez}, {Honscheid}, {James},
  {Jeltema}, {Johnson}, {Johnson}, {Kent}, {Krause}, {Kron}, {Kuehn}, {Lahav},
  {Lima}, {Maia}, {March}, {Martini}, {McMahon}, {Menanteau}, {Miller},
  {Miquel}, {Mohr}, {Nichol}, {Ogando}, {Plazas}, {Romer}, {Roodman}, {Rykoff},
  {Sanchez}, {Scarpine}, {Schindler}, {Schubnell}, {Sevilla-Noarbe}, {Sheldon},
  {Smith}, {Smith}, {Stebbins}, {Suchyta}, {Swanson}, {Tarle}, {Thomas},
  {Troxel}, {Tucker}, {Vikram}, {Walker}, {Wechsler}, {Weller}, {Carlin},
  {Gill}, {Li}, {Marriner}, {Neilsen}, {Dark Energy Camera GW-EM
  Collaboration}, {DES Collaboration}, {Haislip}, {Kouprianov}, {Reichart},
  {Sand}, {Tartaglia}, {Valenti}, {Yang}, {DLT40 Collaboration}, {Benetti},
  {Brocato}, {Campana}, {Cappellaro}, {Covino}, {D'Avanzo}, {D'Elia}, {Getman},
  {Ghirlanda}, {Ghisellini}, {Limatola}, {Nicastro}, {Palazzi}, {Pian},
  {Piranomonte}, {Possenti}, {Rossi}, {Salafia}, {Tomasella}, {Amati},
  {Antonelli}, {Bernardini}, {Bufano}, {Capaccioli}, {Casella}, {Dadina}, {De
  Cesare}, {Di Paola}, {Giuffrida}, {Giunta}, {Israel}, {Lisi}, {Maiorano},
  {Mapelli}, {Masetti}, {Pescalli}, {Pulone}, {Salvaterra}, {Schipani},
  {Spera}, {Stamerra}, {Stella}, {Testa}, {Turatto}, {Vergani}, {Aresu},
  {Bachetti}, {Buffa}, {Burgay}, {Buttu}, {Caria}, {Carretti}, {Casasola},
  {Castangia}, {Carboni}, {Casu}, {Concu}, {Corongiu}, {Deiana}, {Egron},
  {Fara}, {Gaudiomonte}, {Gusai}, {Ladu}, {Loru}, {Leurini}, {Marongiu},
  {Melis}, {Melis}, {Migoni}, {Milia}, {Navarrini}, {Orlati}, {Ortu}, {Palmas},
  {Pellizzoni}, {Perrodin}, {Pisanu}, {Poppi}, {Righini}, {Saba}, {Serra},
  {Serrau}, {Stagni}, {Surcis}, {Vacca}, {Vargiu}, {Hunt}, {Jin}, {Klose},
  {Kouveliotou}, {Mazzali}, {M{\o}ller}, {Nava}, {Piran}, {Selsing}, {Vergani},
  {Wiersema}, {Toma}, {Higgins}, {Mundell}, {di Serego Alighieri}, {G{\'o}tz},
  {Gao}, {Gomboc}, {Kaper}, {Kobayashi}, {Kopac}, {Mao}, {Starling}, {Steele},
  {van der Horst}, {GRAWITA: GRAvitational Wave Inaf TeAm}, {Acero}, {Atwood},
  {Baldini}, {Barbiellini}, {Bastieri}, {Berenji}, {Bellazzini}, {Bissaldi},
  {Blandford}, {Bloom}, {Bonino}, {Bottacini}, {Bregeon}, {Buehler}, {Buson},
  {Cameron}, {Caputo}, {Caraveo}, {Cavazzuti}, {Chekhtman}, {Cheung}, {Chiang},
  {Ciprini}, {Cohen-Tanugi}, {Cominsky}, {Costantin}, {Cuoco}, {D'Ammando}, {de
  Palma}, {Digel}, {Di Lalla}, {Di Mauro}, {Di Venere}, {Dubois}, {Fegan},
  {Focke}, {Franckowiak}, {Fukazawa}, {Funk}, {Fusco}, {Gargano}, {Gasparrini},
  {Giglietto}, {Giordano}, {Giroletti}, {Glanzman}, {Green}, {Grondin},
  {Guillemot}, {Guiriec}, {Harding}, {Horan}, {J{\'o}hannesson}, {Kamae},
  {Kensei}, {Kuss}, {La Mura}, {Latronico}, {Lemoine-Goumard}, {Longo},
  {Loparco}, {Lovellette}, {Lubrano}, {Magill}, {Maldera}, {Manfreda},
  {Mazziotta}, {McEnery}, {Meyer}, {Michelson}, {Mirabal}, {Monzani},
  {Moretti}, {Morselli}, {Moskalenko}, {Negro}, {Nuss}, {Ojha}, {Omodei},
  {Orienti}, {Orlando}, {Palatiello}, {Paliya}, {Paneque}, {Pesce-Rollins},
  {Piron}, {Porter}, {Principe}, {Rain{\`o}}, {Rando}, {Razzano}, {Razzaque},
  {Reimer}, {Reimer}, {Reposeur}, {Rochester}, {Saz Parkinson}, {Sgr{\`o}},
  {Siskind}, {Spada}, {Spandre}, {Suson}, {Takahashi}, {Tanaka}, {Thayer},
  {Thayer}, {Thompson}, {Tibaldo}, {Torres}, {Torresi}, {Troja}, {Venters},
  {Vianello}, {Zaharijas}, {Fermi Large Area Telescope Collaboration},
  {Allison}, {Bannister}, {Dobie}, {Kaplan}, {Lenc}, {Lynch}, {Murphy},
  {Sadler}, {Australia Telescope Compact Array}, {Hotan}, {James}, {Oslowski},
  {Raja}, {Shannon}, {Whiting}, {Australian SKA Pathfinder}, {Arcavi},
  {Howell}, {McCully}, {Hosseinzadeh}, {Hiramatsu}, {Poznanski}, {Barnes},
  {Zaltzman}, {Vasylyev}, {Maoz}, {Las Cumbres Observatory Group}, {Cooke},
  {Bailes}, {Wolf}, {Deller}, {Lidman}, {Wang}, {Gendre}, {Andreoni}, {Ackley},
  {Pritchard}, {Bessell}, {Chang}, {M{\"o}ller}, {Onken}, {Scalzo},
  {Ridden-Harper}, {Sharp}, {Tucker}, {Farrell}, {Elmer}, {Johnston},
  {Venkatraman Krishnan}, {Keane}, {Green}, {Jameson}, {Hu}, {Ma}, {Sun}, {Wu},
  {Wang}, {Shang}, {Hu}, {Ashley}, {Yuan}, {Li}, {Tao}, {Zhu}, {Zhang},
  {Suntzeff}, {Zhou}, {Yang}, {Orange}, {Morris}, {Cucchiara}, {Giblin},
  {Klotz}, {Staff}, {Thierry}, {Schmidt}, {OzGrav}, {(Deeper}, {Wider},
  {program}, {AST3}, {CAASTRO Collaborations}, {Tanvir}, {Levan}, {Cano}, {de
  Ugarte-Postigo}, {Gonz{\'a}lez-Fern{\'a}ndez}, {Greiner}, {Hjorth}, {Irwin},
  {Kr{\"u}hler}, {Mandel}, {Milvang-Jensen}, {O'Brien}, {Rol}, {Rosetti},
  {Rosswog}, {Rowlinson}, {Steeghs}, {Th{\"o}ne}, {Ulaczyk}, {Watson}, {Bruun},
  {Cutter}, {Figuera Jaimes}, {Fujii}, {Fruchter}, {Gompertz}, {Jakobsson},
  {Hodosan}, {J{\`e}rgensen}, {Kangas}, {Kann}, {Rabus}, {Schr{\o}der},
  {Stanway}, {Wijers}, {VINROUGE Collaboration}, {Lipunov}, {Gorbovskoy},
  {Kornilov}, {Tyurina}, {Balanutsa}, {Kuznetsov}, {Vlasenko}, {Podesta},
  {Lopez}, {Podesta}, {Levato}, {Saffe}, {Mallamaci}, {Budnev}, {Gress},
  {Kuvshinov}, {Gorbunov}, {Vladimirov}, {Zimnukhov}, {Gabovich}, {Yurkov},
  {Sergienko}, {Rebolo}, {Serra-Ricart}, {Tlatov}, {Ishmuhametova}, {MASTER
  Collaboration}, {Abe}, {Aoki}, {Aoki}, {Asakura}, {Baar}, {Barway}, {Bond},
  {Doi}, {Finet}, {Fujiyoshi}, {Furusawa}, {Honda}, {Itoh}, {Kanda},
  {Kawabata}, {Kawabata}, {Kim}, {Koshida}, {Kuroda}, {Lee}, {Liu},
  {Matsubayashi}, {Miyazaki}, {Morihana}, {Morokuma}, {Motohara}, {Murata},
  {Nagai}, {Nagashima}, {Nagayama}, {Nakaoka}, {Nakata}, {Ohsawa}, {Ohshima},
  {Ohta}, {Okita}, {Saito}, {Saito}, {Sako}, {Sekiguchi}, {Sumi}, {Tajitsu},
  {Takahashi}, {Takayama}, {Tamura}, {Tanaka}, {Tanaka}, {Terai}, {Tominaga},
  {Tristram}, {Uemura}, {Utsumi}, {Yamaguchi}, {Yasuda}, {Yoshida}, {Zenko},
  {J-GEM}, {Adams}, {Anupama}, {Bally}, {Barway}, {Bellm}, {Blagorodnova},
  {Cannella}, {Chandra}, {Chatterjee}, {Clarke}, {Cobb}, {Cook}, {Copperwheat},
  {De}, {Emery}, {Feindt}, {Foster}, {Fox}, {Frail}, {Fremling}, {Frohmaier},
  {Garcia}, {Ghosh}, {Giacintucci}, {Goobar}, {Gottlieb}, {Grefenstette},
  {Hallinan}, {Harrison}, {Heida}, {Helou}, {Ho}, {Horesh}, {Hotokezaka}, {Ip},
  {Itoh}, {Jacobs}, {Jencson}, {Kasen}, {Kasliwal}, {Kassim}, {Kim}, {Kiran},
  {Kuin}, {Kulkarni}, {Kupfer}, {Lau}, {Madsen}, {Mazzali}, {Miller},
  {Miyasaka}, {Mooley}, {Myers}, {Nakar}, {Ngeow}, {Nugent}, {Ofek},
  {Palliyaguru}, {Pavana}, {Perley}, {Peters}, {Pike}, {Piran}, {Qi}, {Quimby},
  {Rana}, {Rosswog}, {Rusu}, {Sadler}, {Van Sistine}, {Sollerman}, {Xu}, {Yan},
  {Yatsu}, {Yu}, {Zhang}, {Zhao}, {GROWTH}, {JAGWAR}, {Caltech-NRAO},
  {TTU-NRAO}, {NuSTAR Collaborations}, {Chambers}, {Huber}, {Schultz},
  {Bulger}, {Flewelling}, {Magnier}, {Lowe}, {Wainscoat}, {Waters}, {Willman},
  {Pan-STARRS}, {Ebisawa}, {Hanyu}, {Harita}, {Hashimoto}, {Hidaka}, {Hori},
  {Ishikawa}, {Isobe}, {Iwakiri}, {Kawai}, {Kawai}, {Kawamuro}, {Kawase},
  {Kitaoka}, {Makishima}, {Matsuoka}, {Mihara}, {Morita}, {Morita}, {Nakahira},
  {Nakajima}, {Nakamura}, {Negoro}, {Oda}, {Sakamaki}, {Sasaki}, {Serino},
  {Shidatsu}, {Shimomukai}, {Sugawara}, {Sugita}, {Sugizaki}, {Tachibana},
  {Takao}, {Tanimoto}, {Tomida}, {Tsuboi}, {Tsunemi}, {Ueda}, {Ueno}, {Yamada},
  {Yamaoka}, {Yamauchi}, {Yatabe}, {Yoneyama}, {Yoshii}, {MAXI Team}, {Coward},
  {Crisp}, {Macpherson}, {Andreoni}, {Laugier}, {Noysena}, {Klotz}, {Gendre},
  {Thierry}, {Turpin}, {Consortium}, {Im}, {Choi}, {Kim}, {Yoon}, {Lim}, {Lee},
  {Lee}, {Kim}, {Ko}, {Joe}, {Kwon}, {Kim}, {Lim}, {Choi}, {KU Collaboration},
  {Fynbo}, {Malesani}, {Xu}, {Optical Telescope}, {Smartt}, {Jerkstrand},
  {Kankare}, {Sim}, {Fraser}, {Inserra}, {Maguire}, {Leloudas}, {Magee},
  {Shingles}, {Smith}, {Young}, {Kotak}, {Gal-Yam}, {Lyman}, {Homan},
  {Agliozzo}, {Anderson}, {Angus}, {Ashall}, {Barbarino}, {Bauer}, {Berton},
  {Botticella}, {Bulla}, {Cannizzaro}, {Cartier}, {Cikota}, {Clark}, {De Cia},
  {Della Valle}, {Dennefeld}, {Dessart}, {Dimitriadis}, {Elias-Rosa}, {Firth},
  {Fl{\"o}rs}, {Frohmaier}, {Galbany}, {Gonz{\'a}lez-Gait{\'a}n}, {Gromadzki},
  {Guti{\'e}rrez}, {Hamanowicz}, {Harmanen}, {Heintz}, {Hernandez}, {Hodgkin},
  {Hook}, {Izzo}, {James}, {Jonker}, {Kerzendorf}, {Kostrzewa-Rutkowska},
  {Kromer}, {Kuncarayakti}, {Lawrence}, {Manulis}, {Mattila}, {McBrien},
  {M{\"u}ller}, {Nordin}, {O'Neill}, {Onori}, {Palmerio}, {Pastorello},
  {Patat}, {Pignata}, {Podsiadlowski}, {Razza}, {Reynolds}, {Roy}, {Ruiter},
  {Rybicki}, {Salmon}, {Pumo}, {Prentice}, {Seitenzahl}, {Smith}, {Sollerman},
  {Sullivan}, {Szegedi}, {Taddia}, {Taubenberger}, {Terreran}, {Van Soelen},
  {Vos}, {Walton}, {Wright}, {Wyrzykowski}, {Yaron}, {pre=''(''>ePESSTO},
  {Chen}, {Kr{\"u}hler}, {Schady}, {Wiseman}, {Greiner}, {Rau}, {Schweyer},
  {Klose}, {Nicuesa Guelbenzu}, {GROND}, {Palliyaguru}, {Tech University},
  {Shara}, {Williams}, {Vaisanen}, {Potter}, {Romero Colmenero}, {Crawford},
  {Buckley}, {Mao}, {SALT Group}, {D{\'\i}az}, {Macri}, {Garc{\'\i}a Lambas},
  {Mendes de Oliveira}, {Nilo Castell{\'o}n}, {Ribeiro}, {S{\'a}nchez},
  {Schoenell}, {Abramo}, {Akras}, {Alcaniz}, {Artola}, {Beroiz}, {Bonoli},
  {Cabral}, {Camuccio}, {Chavushyan}, {Coelho}, {Colazo}, {Costa-Duarte},
  {Cuevas Larenas}, {Dom{\'\i}nguez Romero}, {Dultzin}, {Fern{\'a}ndez},
  {Garc{\'\i}a}, {Girardini}, {Gon{\c{c}}alves}, {Gon{\c{c}}alves}, {Gurovich},
  {Jim{\'e}nez-Teja}, {Kanaan}, {Lares}, {Lopes de Oliveira}, {L{\'o}pez-Cruz},
  {Melia}, {Molino}, {Padilla}, {Pe{\~n}uela}, {Placco}, {Qui{\~n}ones},
  {Ram{\'\i}rez Rivera}, {Renzi}, {Riguccini}, {R{\'\i}os-L{\'o}pez},
  {Rodriguez}, {Sampedro}, {Schneiter}, {Sodr{\'e}}, {Starck}, {Torres-Flores},
  {Tornatore}, {Zadro{\.z}ny}, {Castillo}, {TOROS: Transient Robotic
  Observatory of South Collaboration}, {Castro-Tirado}, {Tello}, {Hu}, {Zhang},
  {Cunniffe}, {Castell{\'o}n}, {Hiriart}, {Caballero-Garc{\'\i}a},
  {Jel{\'\i}nek}, {Kub{\'a}nek}, {P{\'e}rez del Pulgar}, {Park}, {Jeong},
  {Castro Cer{\'o}n}, {Pandey}, {Yock}, {Querel}, {Fan}, {Wang}, {BOOTES
  Collaboration}, {Beardsley}, {Brown}, {Crosse}, {Emrich}, {Franzen},
  {Gaensler}, {Horsley}, {Johnston-Hollitt}, {Kenney}, {Morales}, {Pallot},
  {Sokolowski}, {Steele}, {Tingay}, {Trott}, {Walker}, {Wayth}, {Williams},
  {Wu}, {Murchison Widefield Array}, {Yoshida}, {Sakamoto}, {Kawakubo},
  {Yamaoka}, {Takahashi}, {Asaoka}, {Ozawa}, {Torii}, {Shimizu}, {Tamura},
  {Ishizaki}, {Cherry}, {Ricciarini}, {Penacchioni}, {Marrocchesi}, {CALET
  Collaboration}, {Pozanenko}, {Volnova}, {Mazaeva}, {Minaev}, {Krugov},
  {Kusakin}, {Reva}, {Moskvitin}, {Rumyantsev}, {Inasaridze}, {Klunko},
  {Tungalag}, {Schmalz}, {Burhonov}, {IKI-GW Follow-up Collaboration},
  {Abdalla}, {Abramowski}, {Aharonian}, {Ait Benkhali}, {Ang{\"u}ner},
  {Arakawa}, {Arrieta}, {Aubert}, {Backes}, {Balzer}, {Barnard}, {Becherini},
  {Becker Tjus}, {Berge}, {Bernhard}, {Bernl{\"o}hr}, {Blackwell},
  {B{\"o}ttcher}, {Boisson}, {Bolmont}, {Bonnefoy}, {Bordas}, {Bregeon},
  {Brun}, {Brun}, {Bryan}, {B{\"u}chele}, {Bulik}, {Capasso}, {Caroff},
  {Carosi}, {Casanova}, {Cerruti}, {Chakraborty}, {Chaves}, {Chen},
  {Chevalier}, {Colafrancesco}, {Condon}, {Conrad}, {Davids}, {Decock}, {Deil},
  {Devin}, {deWilt}, {Dirson}, {Djannati-Ata{\"\i}}, {Donath}, {O'C. Drury},
  {Dutson}, {Dyks}, {Edwards}, {Egberts}, {Emery}, {Ernenwein}, {Eschbach},
  {Farnier}, {Fegan}, {Fernandes}, {Fiasson}, {Fontaine}, {Funk},
  {F{\"u}ssling}, {Gabici}, {Gallant}, {Garrigoux}, {Gat{\'e}}, {Giavitto},
  {Giebels}, {Glawion}, {Glicenstein}, {Gottschall}, {Grondin}, {Hahn},
  {Haupt}, {Hawkes}, {Heinzelmann}, {Henri}, {Hermann}, {Hinton}, {Hofmann},
  {Hoischen}, {Holch}, {Holler}, {Horns}, {Ivascenko}, {Iwasaki},
  {Jacholkowska}, {Jamrozy}, {Jankowsky}, {Jankowsky}, {Jingo}, {Jouvin},
  {Jung-Richardt}, {Kastendieck}, {Katarzy{\'n}ski}, {Katsuragawa},
  {Kerszberg}, {Khangulyan}, {Kh{\'e}lifi}, {King}, {Klepser}, {Klochkov},
  {Klu{\'z}niak}, {Komin}, {Kosack}, {Krakau}, {Kraus}, {Kr{\"u}ger}, {Laffon},
  {Lamanna}, {Lau}, {Lees}, {Lefaucheur}, {Lemi{\`e}re}, {Lemoine-Goumard},
  {Lenain}, {Leser}, {Lohse}, {Lorentz}, {Liu}, {Lypova}, {Malyshev},
  {Marandon}, {Marcowith}, {Mariaud}, {Marx}, {Maurin}, {Maxted}, {Mayer},
  {Meintjes}, {Meyer}, {Mitchell}, {Moderski}, {Mohamed}, {Mohrmann},
  {Mor{\r{a}}}, {Moulin}, {Murach}, {Nakashima}, {de Naurois}, {Ndiyavala},
  {Niederwanger}, {Niemiec}, {Oakes}, {O'Brien}, {Odaka}, {Ohm}, {Ostrowski},
  {Oya}, {Padovani}, {Panter}, {Parsons}, {Pekeur}, {Pelletier}, {Perennes},
  {Petrucci}, {Peyaud}, {Piel}, {Pita}, {Poireau}, {Poon}, {Prokhorov},
  {Prokoph}, {P{\"u}hlhofer}, {Punch}, {Quirrenbach}, {Raab}, {Rauth},
  {Reimer}, {Reimer}, {Renaud}, {de los Reyes}, {Rieger}, {Rinchiuso},
  {Romoli}, {Rowell}, {Rudak}, {Rulten}, {Sahakian}, {Saito}, {Sanchez},
  {Santangelo}, {Sasaki}, {Schlickeiser}, {Sch{\"u}ssler}, {Schulz},
  {Schwanke}, {Schwemmer}, {Seglar-Arroyo}, {Settimo}, {Seyffert}, {Shafi},
  {Shilon}, {Shiningayamwe}, {Simoni}, {Sol}, {Spanier}, {Spir-Jacob},
  {Stawarz}, {Steenkamp}, {Stegmann}, {Steppa}, {Sushch}, {Takahashi},
  {Tavernet}, {Tavernier}, {Taylor}, {Terrier}, {Tibaldo}, {Tiziani},
  {Tluczykont}, {Trichard}, {Tsirou}, {Tsuji}, {Tuffs}, {Uchiyama}, {van der
  Walt}, {van Eldik}, {van Rensburg}, {van Soelen}, {Vasileiadis}, {Veh},
  {Venter}, {Viana}, {Vincent}, {Vink}, {Voisin}, {V{\"o}lk}, {Vuillaume},
  {Wadiasingh}, {Wagner}, {Wagner}, {Wagner}, {White}, {Wierzcholska},
  {Willmann}, {W{\"o}rnlein}, {Wouters}, {Yang}, {Zaborov}, {Zacharias},
  {Zanin}, {Zdziarski}, {Zech}, {Zefi}, {Ziegler}, {Zorn}, {{\.Z}ywucka},
  {H.~E.~S.~S. Collaboration}, {Fender}, {Broderick}, {Rowlinson}, {Wijers},
  {Stewart}, {ter Veen}, {Shulevski}, {LOFAR Collaboration}, {Kavic},
  {Simonetti}, {League}, {Tsai}, {Obenberger}, {Nathaniel}, {Taylor}, {Dowell},
  {Liebling}, {Estes}, {Lippert}, {Sharma}, {Vincent}, {Farella}, {Wavelength
  Array}, {Abeysekara}, {Albert}, {Alfaro}, {Alvarez}, {Arceo},
  {Arteaga-Vel{\'a}zquez}, {Avila Rojas}, {Ayala Solares}, {Barber}, {Becerra
  Gonzalez}, {Becerril}, {Belmont-Moreno}, {BenZvi}, {Berley}, {Bernal},
  {Braun}, {Brisbois}, {Caballero-Mora}, {Capistr{\'a}n}, {Carrami{\~n}ana},
  {Casanova}, {Castillo}, {Cotti}, {Cotzomi}, {Couti{\~n}o de Le{\'o}n}, {De
  Le{\'o}n}, {De la Fuente}, {Diaz Hernandez}, {Dichiara}, {Dingus},
  {DuVernois}, {D{\'\i}az-V{\'e}lez}, {Ellsworth}, {Engel},
  {Enr{\'\i}quez-Rivera}, {Fiorino}, {Fleischhack}, {Fraija},
  {Garc{\'\i}a-Gonz{\'a}lez}, {Garfias}, {Gerhardt}, {Gonz{\~o}lez Mu{\~n}oz},
  {Gonz{\'a}lez}, {Goodman}, {Hampel-Arias}, {Harding}, {Hernandez},
  {Hernandez-Almada}, {Hona}, {H{\"u}ntemeyer}, {Iriarte}, {Jardin-Blicq},
  {Joshi}, {Kaufmann}, {Kieda}, {Lara}, {Lauer}, {Lennarz}, {Le{\'o}n Vargas},
  {Linnemann}, {Longinotti}, {Raya}, {Luna-Garc{\'\i}a}, {L{\'o}pez-Coto},
  {Malone}, {Marinelli}, {Martinez}, {Martinez-Castellanos},
  {Mart{\'\i}nez-Castro}, {Mart{\'\i}nez-Huerta}, {Matthews},
  {Miranda-Romagnoli}, {Moreno}, {Mostaf{\'a}}, {Nellen}, {Newbold}, {Nisa},
  {Noriega-Papaqui}, {Pelayo}, {Pretz}, {P{\'e}rez-P{\'e}rez}, {Ren}, {Rho},
  {Rivi{\`e}re}, {Rosa-Gonz{\'a}lez}, {Rosenberg}, {Ruiz-Velasco}, {Salazar},
  {Salesa Greus}, {Sandoval}, {Schneider}, {Schoorlemmer}, {Sinnis}, {Smith},
  {Springer}, {Surajbali}, {Tibolla}, {Tollefson}, {Torres}, {Ukwatta},
  {Weisgarber}, {Westerhoff}, {Wisher}, {Wood}, {Yapici}, {Yodh}, {Younk},
  {Zhou}, {{\'A}lvarez}, {HAWC Collaboration}, {Aab}, {Abreu}, {Aglietta},
  {Albuquerque}, {Albury}, {Allekotte}, {Almela}, {Alvarez Castillo},
  {Alvarez-Mu{\~n}iz}, {Anastasi}, {Anchordoqui}, {Andrada}, {Andringa},
  {Aramo}, {Arsene}, {Asorey}, {Assis}, {Avila}, {Badescu}, {Balaceanu},
  {Barbato}, {Barreira Luz}, {Becker}, {Bellido}, {Berat}, {Bertaina},
  {Bertou}, {Biermann}, {Biteau}, {Blaess}, {Blanco}, {Blazek}, {Bleve},
  {Boh{\'a}{\v{c}}ov{\'a}}, {Bonifazi}, {Borodai}, {Botti}, {Brack}, {Brancus},
  {Bretz}, {Bridgeman}, {Briechle}, {Buchholz}, {Bueno}, {Buitink}, {Buscemi},
  {Caballero-Mora}, {Caccianiga}, {Cancio}, {Canfora}, {Caruso}, {Castellina},
  {Catalani}, {Cataldi}, {Cazon}, {Chavez}, {Chinellato}, {Chudoba}, {Clay},
  {Cobos Cerutti}, {Colalillo}, {Coleman}, {Collica}, {Coluccia},
  {Concei{\c{c}}{\~a}o}, {Consolati}, {Contreras}, {Cooper}, {Coutu},
  {Covault}, {Cronin}, {D'Amico}, {Daniel}, {Dasso}, {Daumiller}, {Dawson},
  {Day}, {de Almeida}, {de Jong}, {De Mauro}, {de Mello Neto}, {De Mitri}, {de
  Oliveira}, {de Souza}, {Debatin}, {Deligny}, {D{\'\i}az Castro}, {Diogo},
  {Dobrigkeit}, {D'Olivo}, {Dorosti}, {Dos Anjos}, {Dova}, {Dundovic}, {Ebr},
  {Engel}, {Erdmann}, {Erfani}, {Escobar}, {Espadanal}, {Etchegoyen}, {Falcke},
  {Farmer}, {Farrar}, {Fauth}, {Fazzini}, {Feldbusch}, {Fenu}, {Fick},
  {Figueira}, {Filip{\v{c}}i{\v{c}}}, {Freire}, {Fujii}, {Fuster},
  {Ga{\"\i}or}, {Garc{\'\i}a}, {Gat{\'e}}, {Gemmeke}, {Gherghel-Lascu}, {Ghia},
  {Giaccari}, {Giammarchi}, {Giller}, {G{\l}as}, {Glaser}, {Golup}, {G{\'o}mez
  Berisso}, {G{\'o}mez Vitale}, {Gonz{\'a}lez}, {Gorgi}, {Gottowik}, {Grillo},
  {Grubb}, {Guarino}, {Guedes}, {Halliday}, {Hampel}, {Hansen}, {Harari},
  {Harrison}, {Harvey}, {Haungs}, {Hebbeker}, {Heck}, {Heimann}, {Herve},
  {Hill}, {Hojvat}, {Holt}, {Homola}, {H{\"o}randel}, {Horvath},
  {Hrabovsk{\'y}}, {Huege}, {Hulsman}, {Insolia}, {Isar}, {Jandt}, {Johnsen},
  {Josebachuili}, {Jurysek}, {K{\"a}{\"a}p{\"a}}, {Kampert}, {Keilhauer},
  {Kemmerich}, {Kemp}, {Kieckhafer}, {Klages}, {Kleifges}, {Kleinfeller},
  {Krause}, {Krohm}, {Kuempel}, {Kukec Mezek}, {Kunka}, {Kuotb Awad}, {Lago},
  {LaHurd}, {Lang}, {Lauscher}, {Legumina}, {Leigui de Oliveira},
  {Letessier-Selvon}, {Lhenry-Yvon}, {Link}, {Lo Presti}, {Lopes}, {L{\'o}pez},
  {L{\'o}pez Casado}, {Lorek}, {Luce}, {Lucero}, {Malacari}, {Mallamaci},
  {Mandat}, {Mantsch}, {Mariazzi}, {Maris}, {Marsella}, {Martello}, {Martinez},
  {Mart{\'\i}nez Bravo}, {Mas{\'\i}as Meza}, {Mathes}, {Mathys}, {Matthews},
  {Matthiae}, {Mayotte}, {Mazur}, {Medina}, {Medina-Tanco}, {Melo},
  {Menshikov}, {Merenda}, {Michal}, {Micheletti}, {Middendorf}, {Miramonti},
  {Mitrica}, {Mockler}, {Mollerach}, {Montanet}, {Morello}, {Morlino},
  {M{\"u}ller}, {M{\"u}ller}, {Muller}, {M{\"u}ller}, {Mussa}, {Naranjo},
  {Nguyen}, {Niculescu-Oglinzanu}, {Niechciol}, {Niemietz}, {Niggemann},
  {Nitz}, {Nosek}, {Novotny}, {No{\v{z}}ka}, {N{\'u}{\~n}ez}, {Oikonomou},
  {Olinto}, {Palatka}, {Pallotta}, {Papenbreer}, {Parente}, {Parra}, {Paul},
  {Pech}, {Pedreira}, {P{\c{e}}kala}, {Pe{\~n}a-Rodriguez}, {Pereira},
  {Perlin}, {Perrone}, {Peters}, {Petrera}, {Phuntsok}, {Pierog}, {Pimenta},
  {Pirronello}, {Platino}, {Plum}, {Poh}, {Porowski}, {Prado}, {Privitera},
  {Prouza}, {Quel}, {Querchfeld}, {Quinn}, {Ramos-Pollan}, {Rautenberg},
  {Ravignani}, {Ridky}, {Riehn}, {Risse}, {Ristori}, {Rizi}, {Rodrigues de
  Carvalho}, {Rodriguez Fernandez}, {Rodriguez Rojo}, {Roncoroni}, {Roth},
  {Roulet}, {Rovero}, {Ruehl}, {Saffi}, {Saftoiu}, {Salamida}, {Salazar},
  {Saleh}, {Salina}, {S{\'a}nchez}, {Sanchez-Lucas}, {Santos}, {Santos},
  {Sarazin}, {Sarmento}, {Sarmiento-Cano}, {Sato}, {Schauer}, {Scherini},
  {Schieler}, {Schimp}, {Schmidt}, {Scholten}, {Schov{\'a}nek}, {Schr{\"o}der},
  {Schr{\"o}der}, {Schulz}, {Schumacher}, {Sciutto}, {Segreto}, {Shadkam},
  {Shellard}, {Sigl}, {Silli}, {{\v{S}}m{\'\i}da}, {Snow}, {Sommers},
  {Sonntag}, {Soriano}, {Squartini}, {Stanca}, {Stani{\v{c}}}, {Stasielak},
  {Stassi}, {Stolpovskiy}, {Strafella}, {Streich}, {Suarez},
  {Suarez-Dur{\'a}n}, {Sudholz}, {Suomij{\"a}rvi}, {Supanitsky},
  {{\v{S}}up{\'\i}k}, {Swain}, {Szadkowski}, {Taboada}, {Taborda},
  {Timmermans}, {Todero Peixoto}, {Tomankova}, {Tom{\'e}}, {Torralba Elipe},
  {Travnicek}, {Trini}, {Tueros}, {Ulrich}, {Unger}, {Urban}, {Vald{\'e}s
  Galicia}, {Vali{\~n}o}, {Valore}, {van Aar}, {van Bodegom}, {van den Berg},
  {van Vliet}, {Varela}, {Vargas C{\'a}rdenas}, {V{\'a}zquez}, {Veberi{\v{c}}},
  {Ventura}, {Vergara Quispe}, {Verzi}, {Vicha}, {Villase{\~n}or}, {Vorobiov},
  {Wahlberg}, {Wainberg}, {Walz}, {Watson}, {Weber}, {Weindl}, {Wiede{\'n}ski},
  {Wiencke}, {Wilczy{\'n}ski}, {Wirtz}, {Wittkowski}, {Wundheiler}, {Yang},
  {Yushkov}, {Zas}, {Zavrtanik}, {Zavrtanik}, {Zepeda}, {Zimmermann},
  {Ziolkowski}, {Zong}, {Zuccarello}, {Pierre Auger Collaboration}, {Kim},
  {Schulze}, {Bauer}, {Corral-Santana}, {de Gregorio-Monsalvo},
  {Gonz{\'a}lez-L{\'o}pez}, {Hartmann}, {Ishwara-Chandra}, {Mart{\'\i}n},
  {Mehner}, {Misra}, {Micha{\l}owski}, {Resmi}, {ALMA Collaboration}, {Paragi},
  {Agudo}, {An}, {Beswick}, {Casadio}, {Frey}, {Jonker}, {Kettenis}, {Marcote},
  {Moldon}, {Szomoru}, {van Langevelde}, {Yang}, {Euro VLBI Team}, {Cwiek},
  {Cwiok}, {Czyrkowski}, {Dabrowski}, {Kasprowicz}, {Mankiewicz}, {Nawrocki},
  {Opiela}, {Piotrowski}, {Wrochna}, {Zaremba}, {{\.Z}arnecki}, {Pi of the Sky
  Collaboration}, {Haggard}, {Nynka}, {Ruan}, {Chandra Team at McGill
  University}, {Bland}, {Booler}, {Devillepoix}, {de Gois}, {Hancock}, {Howie},
  {Paxman}, {Sansom}, {Towner}, {Desert Fireball Network}, {Tonry}, {Coughlin},
  {Stubbs}, {Denneau}, {Heinze}, {Stalder}, {Weiland}, {ATLAS}, {Eatough},
  {Kramer}, {Kraus}, {Time Resolution Universe Survey}, {Troja}, {Piro},
  {Becerra Gonz{\'a}lez}, {Butler}, {Fox}, {Khandrika}, {Kutyrev}, {Lee},
  {Ricci}, {Ryan}, {S{\'a}nchez-Ram{\'\i}rez}, {Veilleux}, {Watson},
  {Wieringa}, {Burgess}, {van Eerten}, {Fontes}, {Fryer}, {Korobkin},
  {Wollaeger}, {RIMAS}, {RATIR}, {Camilo}, {Foley}, {Goedhart}, {Makhathini},
  {Oozeer}, {Smirnov}, {Fender}, {Woudt}, \& {South
  Africa/MeerKAT}}]{GW170817_EM}
---. 2017{\natexlab{c}}, \apjl, 848, L12

\bibitem[{Abbott {$et~al$.}(2019)Abbott, Abbott, Abbott, Abraham, Acernese,
  Ackley, Adams, Adhikari, Adya, Affeldt, Agathos, Agatsuma, Aggarwal, Aguiar,
  Aiello, Ain, Ajith, Allen, Allocca, Aloy, Altin, Amato, Ananyeva, Anderson,
  Anderson, Angelova, Antier, Appert, Arai, Araya, Areeda, Ar{\`{e}}ne, Arnaud,
  Arun, Ascenzi, Ashton, Aston, Astone, Aubin, Aufmuth, Aultoneal, Austin,
  Avendano, Avila-Alvarez, Babak, Bacon, Badaracco, Bader, Bae, Baker,
  Baldaccini, Ballardin, Ballmer, Banagiri, Barayoga, Barclay, Barish, Barker,
  Barkett, Barnum, Barone, Barr, Barsotti, Barsuglia, Barta, Bartlett, Bartos,
  Bassiri, Basti, Bawaj, Bayley, Bazzan, B{\'{e}}csy, Bejger, Belahcene, Bell,
  Beniwal, Berger, Bergmann, Bernuzzi, Bero, Berry, Bersanetti, Bertolini,
  Betzwieser, Bhandare, Bidler, Bilenko, Bilgili, Billingsley, Birch, Birney,
  Birnholtz, Biscans, Biscoveanu, Bisht, Bitossi, Bizouard, Blackburn,
  Blackman, Blair, Blair, Blair, Bloemen, Bode, Boer, Boetzel, Bogaert, Bondu,
  Bonilla, Bonnand, Booker, Boom, Booth, Bork, Boschi, Bose, Bossie, Bossilkov,
  Bosveld, Bouffanais, Bozzi, Bradaschia, Brady, Bramley, Branchesi, Brau,
  Briant, Briggs, Brighenti, Brillet, Brinkmann, Brisson, Brockill, Brooks,
  Brown, Brunett, Buikema, Bulik, Bulten, Buonanno, Buskulic, {Bustamante
  Rosell}, Buy, Byer, Cabero, Cadonati, Cagnoli, Cahillane, {Calder{\'{o}}n
  Bustillo}, Callister, Calloni, Camp, Campbell, Canepa, Cannon, Cao, Cao,
  Capocasa, Carbognani, Caride, Carney, Carullo, {Casanueva Diaz}, Casentini,
  Caudill, Cavagli{\`{a}}, Cavalier, Cavalieri, Cella, Cerd{\'{a}}-Dur{\'{a}}n,
  Cerretani, Cesarini, Chaibi, Chakravarti, Chamberlin, Chan, Chao, Charlton,
  Chase, Chassande-Mottin, Chatterjee, Chaturvedi, Chatziioannou, Cheeseboro,
  Chen, Chen, Chen, Cheng, Cheong, Chia, Chincarini, Chiummo, Cho, Cho, Cho,
  Christensen, Chu, Chua, Chung, Chung, Ciani, Ciobanu, Ciolfi, Cipriano,
  Cirone, Clara, Clark, Clearwater, Cleva, Cocchieri, Coccia, Cohadon, Cohen,
  Colgan, Colleoni, Collette, Collins, Cominsky, Constancio, Conti, Cooper,
  Corban, Corbitt, Cordero-Carri{\'{o}}n, Corley, Cornish, Corsi, Cortese,
  Costa, Cotesta, Coughlin, Coughlin, Coulon, Countryman, Couvares, Covas,
  Cowan, Coward, Cowart, Coyne, Coyne, Creighton, Creighton, Cripe, Croquette,
  Crowder, Cullen, Cumming, Cunningham, Cuoco, Canton, D{\'{a}}lya, Danilishin,
  D'Antonio, Danzmann, Dasgupta, {Da Silva Costa}, Datrier, Dattilo, Dave,
  Davier, Davis, Daw, Debra, Deenadayalan, Degallaix, {De Laurentis},
  Del{\'{e}}glise, {Del Pozzo}, Demarchi, Demos, Dent, {De Pietri}, Derby, {De
  Rosa}, {De Rossi}, Desalvo, {De Varona}, Dhurandhar, D{\'{i}}az, Dietrich,
  {Di Fiore}, {Di Giovanni}, {Di Girolamo}, {Di Lieto}, Ding, {Di Pace}, {Di
  Palma}, {Di Renzo}, Dmitriev, Doctor, Donovan, Dooley, Doravari, Dorrington,
  Downes, Drago, Driggers, Du, Ducoin, Dupej, Dwyer, Easter, Edo, Edwards,
  Effler, Ehrens, Eichholz, Eikenberry, Eisenmann, Eisenstein, Essick,
  Estelles, Estevez, Etienne, Etzel, Evans, Evans, Fafone, Fair, Fairhurst,
  Fan, Farinon, Farr, Farr, Fauchon-Jones, Favata, Fays, Fazio, Fee, Feicht,
  Fejer, Feng, Fernandez-Galiana, Ferrante, Ferreira, Ferreira, Ferrini,
  Fidecaro, Fiori, Fiorucci, Fishbach, Fisher, Fishner, Fitz-Axen, Flaminio,
  Fletcher, Flynn, Fong, Font, Forsyth, Fournier, Frasca, Frasconi, Frei,
  Freise, Frey, Frey, Fritschel, Frolov, Fulda, Fyffe, Gabbard, Gadre, Gaebel,
  Gair, Gammaitoni, Ganija, Gaonkar, Garcia, Garc{\'{i}}a-Quir{\'{o}}s, Garufi,
  Gateley, Gaudio, Gaur, Gayathri, Gemme, Genin, Gennai, George, George,
  Gergely, Germain, Ghonge, Ghosh, Ghosh, Ghosh, Giacomazzo, Giaime, Giardina,
  Giazotto, Gill, Giordano, Glover, Godwin, Goetz, Goetz, Goncharov,
  Gonz{\'{a}}lez, {Gonzalez Castro}, Gopakumar, Gorodetsky, Gossan, Gosselin,
  Gouaty, Grado, Graef, Granata, Grant, Gras, Grassia, Gray, Gray, Greco,
  Green, Green, Gretarsson, Groot, Grote, Grunewald, Gruning, Guidi, Gulati,
  Guo, Gupta, Gupta, Gustafson, Gustafson, Haegel, Halim, Hall, Hall, Hamilton,
  Hammond, Haney, Hanke, Hanks, Hanna, Hannam, Hannuksela, Hanson, Hardwick,
  Haris, Harms, Harry, Harry, Haster, Haughian, Hayes, Healy, Heidmann,
  Heintze, Heitmann, Hello, Hemming, Hendry, Heng, Hennig, Heptonstall,
  {Hernandez Vivanco}, Heurs, Hild, Hinderer, Hoak, Hochheim, Hofman, Holgado,
  Holland, Holt, Holz, Hopkins, Horst, Hough, Howell, Hoy, Hreibi, Huang,
  Huerta, Huet, Hughey, Hulko, Husa, Huttner, Huynh-Dinh, Idzkowski, Iess,
  Ingram, Inta, Intini, Irwin, Isa, Isac, Isi, Iyer, Izumi, Jacqmin, Jadhav,
  Jani, Janthalur, Jaranowski, Jenkins, Jiang, Johnson, Johnson-Mcdaniel,
  Jones, Jones, Jones, Jonker, Ju, Junker, Kalaghatgi, Kalogera, Kamai,
  Kandhasamy, Kang, Kanner, Kapadia, Karki, Karvinen, Kashyap, Kasprzack,
  Katsanevas, Katsavounidis, Katzman, Kaufer, Kawabe, Keerthana,
  K{\'{e}}f{\'{e}}lian, Keitel, Kennedy, Key, Khalili, Khan, Khan, Khan, Khan,
  Khazanov, Khursheed, Kijbunchoo, Kim, Kim, Kim, Kim, Kim, Kim, Kimball, King,
  King, Kinley-Hanlon, Kirchhoff, Kissel, Kleybolte, Klika, Klimenko, Knowles,
  Koch, Koehlenbeck, Koekoek, Koley, Kondrashov, Kontos, Koper, Korobko, Korth,
  Kowalska, Kozak, Kringel, Krishnendu, Kr{\'{o}}lak, Kuehn, Kumar, Kumar,
  Kumar, Kumar, Kuo, Kutynia, Kwang, Lackey, Lai, Lam, Landry, Lane, Lang,
  Lange, Lantz, Lanza, Lartaux-Vollard, Lasky, Laxen, Lazzarini, Lazzaro,
  Leaci, Leavey, Lecoeuche, Lee, Lee, Lee, Lee, Lee, Lee, Lehmann, Lenon,
  Leroy, Letendre, Levin, Li, Li, Li, Li, Lin, Linde, Linker, Littenberg, Liu,
  Liu, Lo, Lockerbie, London, Longo, Lorenzini, Loriette, Lormand, Losurdo,
  Lough, Lousto, Lovelace, Lower, L{\"{u}}ck, Lumaca, Lundgren, Lynch, Ma,
  MacAs, MacFoy, MacInnis, MacLeod, MacQuet, Maga{\~{n}}a-Sandoval,
  {Maga{\~{n}}a Zertuche}, Magee, Majorana, Maksimovic, Malik, Man, Mandic,
  Mangano, Mansell, Manske, Mantovani, Marchesoni, Marion, M{\'{a}}rka,
  M{\'{a}}rka, Markakis, Markosyan, Markowitz, Maros, Marquina, Marsat,
  Martelli, Martin, Martin, Martynov, Mason, Massera, Masserot, Massinger,
  Masso-Reid, Mastrogiovanni, Matas, Matichard, Matone, Mavalvala, Mazumder,
  McCann, McCarthy, McClelland, McCormick, McCuller, McGuire, McIver, McManus,
  McRae, McWilliams, Meacher, Meadors, Mehmet, Mehta, Meidam, Melatos, Mendell,
  Mercer, Mereni, Merilh, Merzougui, Meshkov, Messenger, Messick, Metzdorff,
  Meyers, Miao, Michel, Middleton, Mikhailov, Milano, Miller, Miller,
  Millhouse, Mills, Milovich-Goff, Minazzoli, Minenkov, Mishkin, Mishra,
  Mistry, Mitra, Mitrofanov, Mitselmakher, Mittleman, Mo, Moffa, Mogushi,
  Mohapatra, Montani, Moore, Moraru, Moreno, Morisaki, Mours, Mow-Lowry,
  Mukherjee, Mukherjee, Mukherjee, Mukund, Mullavey, Munch, Mu{\~{n}}iz,
  Muratore, Murray, Nagar, Nardecchia, Naticchioni, Nayak, Neilson, Nelemans,
  Nelson, Nery, Neunzert, Ng, Ng, Nguyen, Nichols, Nielsen, Nissanke, Nitz,
  Nocera, North, Nuttall, Obergaulinger, Oberling, O'Brien, O'Dea, Ogin, Oh,
  Oh, Ohme, Ohta, Okada, Oliver, Oppermann, Oram, O'Reilly, Ormiston, Ortega,
  O'Shaughnessy, Ossokine, Ottaway, Overmier, Owen, Pace, Pagano, Page, Pai,
  Pai, Palamos, Palashov, Palomba, Pal-Singh, Pan, Pang, Pang, Pankow,
  Pannarale, Pant, Paoletti, Paoli, Papa, Parida, Parker, Pascucci,
  Pasqualetti, Passaquieti, Passuello, Patil, Patricelli, Pearlstone, Pedersen,
  Pedraza, Pedurand, Pele, Penn, Perego, Perez, Perreca, Pfeiffer, Phelps,
  Phukon, Piccinni, Pichot, Piergiovanni, Pillant, Pinard, Pirello, Pitkin,
  Poggiani, Pong, Ponrathnam, Popolizio, Porter, Powell, Prajapati, Prasad,
  Prasai, Prasanna, Pratten, Prestegard, Privitera, Prodi, Prokhorov, Puncken,
  Punturo, Puppo, P{\"{u}}rrer, Qi, Quetschke, Quinonez, Quintero,
  Quitzow-James, Raab, Radkins, Radulescu, Raffai, Raja, Rajan, Rajbhandari,
  Rakhmanov, Ramirez, Ramos-Buades, Rana, Rao, Rapagnani, Raymond, Razzano,
  Read, Regimbau, Rei, Reid, Reitze, Ren, Ricci, Richardson, Richardson,
  Ricker, Riemenschneider, Riles, Rizzo, Robertson, Robie, Robinet, Rocchi,
  Rolland, Rollins, Roma, Romanelli, Romano, Romel, Romie, Rose,
  Rosi{\'{n}}ska, Rosofsky, Ross, Rowan, R{\"{u}}diger, Ruggi, Rutins, Ryan,
  Sachdev, Sadecki, Sakellariadou, Salafia, Salconi, Saleem, Salemi, Samajdar,
  Sammut, Sanchez, Sanchez, Sanchis-Gual, Sandberg, Sanders, Santiago, Sarin,
  Sassolas, Sathyaprakash, Saulson, Sauter, Savage, Schale, Scheel, Scheuer,
  Schmidt, Schnabel, Schofield, Sch{\"{o}}nbeck, Schreiber, Schulte, Schutz,
  Schwalbe, Scott, Scott, Seidel, Sellers, Sengupta, Sennett, Sentenac,
  Sequino, Sergeev, Setyawati, Shaddock, Shaffer, Shahriar, Shaner, Shao,
  Sharma, Shawhan, Shen, Shink, Shoemaker, Shoemaker, Shyamsundar, Siellez,
  Sieniawska, Sigg, Silva, Singer, Singh, Singhal, Sintes, Sitmukhambetov,
  Skliris, Slagmolen, Slaven-Blair, Smith, Smith, Somala, Son, Sorazu,
  Sorrentino, Souradeep, Sowell, Spencer, Srivastava, Srivastava, Staats,
  Stachie, Standke, Steer, Steinke, Steinlechner, Steinlechner, Steinmeyer,
  Stevenson, Stocks, Stone, Stops, Strain, Stratta, Strigin, Strunk, Sturani,
  Stuver, Sudhir, Summerscales, Sun, Sunil, Suresh, Sutton, Swinkels,
  Szczepa{\'{n}}czyk, Tacca, Tait, Talbot, Talukder, Tanner, T{\'{a}}pai,
  Taracchini, Tasson, Taylor, Thies, Thomas, Thomas, Thondapu, Thorne, Thrane,
  Tiwari, Tiwari, Tiwari, Toland, Tonelli, Tornasi, Torres-Forn{\'{e}}, Torrie,
  T{\"{o}}yr{\"{a}}, Travasso, Traylor, Tringali, Trovato, Trozzo, Trudeau,
  Tsang, Tse, Tso, Tsukada, Tsuna, Tuyenbayev, Ueno, Ugolini, Unnikrishnan,
  Urban, Usman, Vahlbruch, Vajente, Valdes, {Van Bakel}, {Van Beuzekom}, {Van
  Den Brand}, {Van Den Broeck}, Vander-Hyde, {Van Heijningen}, {Van Der
  Schaaf}, {Van Veggel}, Vardaro, Varma, Vass, Vas{\'{u}}th, Vecchio, Vedovato,
  Veitch, Veitch, Venkateswara, Venugopalan, Verkindt, Vetrano, Vicer{\'{e}},
  Viets, Vine, Vinet, Vitale, Vo, Vocca, Vorvick, Vyatchanin, Wade, Wade, Wade,
  Walet, Walker, Wallace, Walsh, Wang, Wang, Wang, Wang, Wang, Ward, Warden,
  Warner, Was, Watchi, Weaver, Wei, Weinert, Weinstein, Weiss, Wellmann, Wen,
  Wessel, We{\ss}els, Westhouse, Wette, Whelan, White, Whiting, Whittle,
  Wilken, Williams, Williamson, Willis, Willke, Wimmer, Winkler, Wipf, Wittel,
  Woan, Woehler, Wofford, Worden, Wright, Wu, Wysocki, Xiao, Yamamoto, Yancey,
  Yang, Yap, Yazback, Yeeles, Yu, Yu, Yuen, Yvert, Zadro{\.{z}}ny, Zanolin,
  Zappa, Zelenova, Zendri, Zevin, Zhang, Zhang, Zhang, Zhao, Zhou, Zhou, Zhu,
  Zimmerman, Zlochower, Zucker, \& Zweizig}]{Abbot2019_GWTC1}
Abbott, B.~P., Abbott, R., Abbott, T.~D., {$et~al$.} 2019, Physical Review X,
  9, 031040

\bibitem[{{Abbott} {$et~al$.}(2020){Abbott}, {Abbott}, {Abraham}, {Acernese},
  {Ackley}, {Adams}, {Adams}, {Adhikari}, {Adya}, {Affeldt}, {Agathos},
  {Agatsuma}, {Aggarwal}, {Aguiar}, {Aiello}, {Ain}, {Ajith}, {Akcay}, {Allen},
  {Allocca}, {Altin}, {Amato}, {Anand}, {Ananyeva}, {Anderson}, {Anderson},
  {Angelova}, {Ansoldi}, {Antelis}, {Antier}, {Appert}, {Arai}, {Araya},
  {Areeda}, {Ar{\`e}ne}, {Arnaud}, {Aronson}, {Arun}, {Asali}, {Ascenzi},
  {Ashton}, {Aston}, {Astone}, {Aubin}, {Aufmuth}, {AultONeal}, {Austin},
  {Avendano}, {Babak}, {Badaracco}, {Bader}, {Bae}, {Baer}, {Bagnasco},
  {Baird}, {Ball}, {Ballardin}, {Ballmer}, {Bals}, {Balsamo}, {Baltus},
  {Banagiri}, {Bankar}, {Bankar}, {Barayoga}, {Barbieri}, {Barish}, {Barker},
  {Barneo}, {Barnum}, {Barone}, {Barr}, {Barsotti}, {Barsuglia}, {Barta},
  {Bartlett}, {Bartos}, {Bassiri}, {Basti}, {Bawaj}, {Bayley}, {Bazzan},
  {Becher}, {B{\'e}csy}, {Bedakihale}, {Bejger}, {Belahcene}, {Beniwal},
  {Benjamin}, {Bennett}, {Bentley}, {Bergamin}, {Berger}, {Bergmann},
  {Bernuzzi}, {Berry}, {Bersanetti}, {Bertolini}, {Betzwieser}, {Bhandare},
  {Bhandari}, {Bhattacharjee}, {Bidler}, {Bilenko}, {Billingsley}, {Birney},
  {Birnholtz}, {Biscans}, {Bischi}, {Biscoveanu}, {Bisht}, {Bitossi},
  {Bizouard}, {Blackburn}, {Blackman}, {Blair}, {Blair}, {Blair}, {Blanch},
  {Bobba}, {Bode}, {Boer}, {Boetzel}, {Bogaert}, {Boldrini}, {Bondu},
  {Bonnand}, {Bonilla}, {Booker}, {Boom}, {Bork}, {Boschi}, {Bose},
  {Bossilkov}, {Boudart}, {Bouffanais}, {Bozzi}, {Bradaschia}, {Brady},
  {Bramley}, {Branchesi}, {Brau}, {Breschi}, {Briant}, {Briggs}, {Brighenti},
  {Brillet}, {Brinkmann}, {Brockill}, {Brooks}, {Brooks}, {Brown}, {Brunett},
  {Bruno}, {Bruntz}, {Buikema}, {Bulik}, {Bulten}, {Buonanno}, {Buscicchio},
  {Buskulic}, {Byer}, {Cabero}, {Cadonati}, {Caesar}, {Cagnoli}, {Cahillane},
  {Calder{\'o}n Bustillo}, {Callaghan}, {Callister}, {Calloni}, {Camp},
  {Canepa}, {Cannon}, {Cao}, {Cao}, {Carapella}, {Carbognani}, {Carney},
  {Carpinelli}, {Carullo}, {Carver}, {Casanueva Diaz}, {Casentini}, {Caudill},
  {Cavagli{\`a}}, {Cavalier}, {Cavalieri}, {Cella}, {Cerd{\'a}-Dur{\'a}n},
  {Cesarini}, {Chaibi}, {Chakravarti}, {Chan}, {Chan}, {Chandra}, {Chanial},
  {Chao}, {Charlton}, {Chase}, {Chassande-Mottin}, {Chatterjee},
  {Chattopadhyay}, {Chaturvedi}, {Chatziioannou}, {Chen}, {Chen}, {Chen},
  {Chen}, {Cheng}, {Cheong}, {Chia}, {Chiadini}, {Chierici}, {Chincarini},
  {Chiummo}, {Cho}, {Cho}, {Cho}, {Choate}, {Christensen}, {Chu}, {Chua},
  {Chung}, {Chung}, {Ciani}, {Ciecielag}, {Cie{\'s}lar}, {Cifaldi}, {Ciobanu},
  {Ciolfi}, {Cipriano}, {Cirone}, {Clara}, {Clark}, {Clark}, {Clarke},
  {Clearwater}, {Clesse}, {Cleva}, {Coccia}, {Cohadon}, {Cohen}, {Colleoni},
  {Collette}, {Collins}, {Colpi}, {Constancio}, {Conti}, {Cooper}, {Corban},
  {Corbitt}, {Cordero-Carri{\'o}n}, {Corezzi}, {Corley}, {Cornish}, {Corre},
  {Corsi}, {Cortese}, {Costa}, {Cotesta}, {Coughlin}, {Coughlin}, {Coulon},
  {Countryman}, {Cousins}, {Couvares}, {Covas}, {Coward}, {Cowart}, {Coyne},
  {Coyne}, {Creighton}, {Creighton}, {Croquette}, {Crowder}, {Cudell},
  {Cullen}, {Cumming}, {Cummings}, {Cunningham}, {Cuoco}, {Curylo}, {Dal
  Canton}, {D{\'a}lya}, {Dana}, {DaneshgaranBajastani}, {D'Angelo}, {Danila},
  {Danilishin}, {D'Antonio}, {Danzmann}, {Darsow-Fromm}, {Dasgupta}, {Datrier},
  {Dattilo}, {Dave}, {Davier}, {Davies}, {Davis}, {Daw}, {Dean}, {DeBra},
  {Deenadayalan}, {Degallaix}, {De Laurentis}, {Del{\'e}glise}, {Del Favero},
  {De Lillo}, {De Lillo}, {Del Pozzo}, {DeMarchi}, {De Matteis}, {D'Emilio},
  {Demos}, {Denker}, {Dent}, {Depasse}, {De Pietri}, {De Rosa}, {De Rossi},
  {DeSalvo}, {de Varona}, {Dhurandhar}, {D{\'\i}az}, {Diaz-Ortiz}, {Didio},
  {Dietrich}, {Di Fiore}, {DiFronzo}, {Di Giorgio}, {Di Giovanni}, {Di
  Giovanni}, {Di Girolamo}, {Di Lieto}, {Ding}, {Di Pace}, {Di Palma}, {Di
  Renzo}, {Divakarla}, {Dmitriev}, {Doctor}, {D'Onofrio}, {Donovan}, {Dooley},
  {Doravari}, {Dorrington}, {Downes}, {Drago}, {Driggers}, {Du}, {Ducoin},
  {Dupej}, {Durante}, {D'Urso}, {Duverne}, {Dwyer}, {Easter}, {Eddolls},
  {Edelman}, {Edo}, {Edy}, {Effler}, {Eichholz}, {Eikenberry}, {Eisenmann},
  {Eisenstein}, {Ejlli}, {Errico}, {Essick}, {Estell{\'e}s}, {Estevez},
  {Etienne}, {Etzel}, {Evans}, {Evans}, {Ewing}, {Fafone}, {Fair}, {Fairhurst},
  {Fan}, {Farah}, {Farinon}, {Farr}, {Farr}, {Fauchon-Jones}, {Favata}, {Fays},
  {Fazio}, {Feicht}, {Fejer}, {Feng}, {Fenyvesi}, {Ferguson},
  {Fernandez-Galiana}, {Ferrante}, {Ferreira}, {Fidecaro}, {Figura}, {Fiori},
  {Fiorucci}, {Fishbach}, {Fisher}, {Fishner}, {Fittipaldi}, {Fitz-Axen},
  {Fiumara}, {Flaminio}, {Floden}, {Flynn}, {Fong}, {Font}, {Forsyth},
  {Fournier}, {Frasca}, {Frasconi}, {Frei}, {Freise}, {Frey}, {Frey},
  {Fritschel}, {Frolov}, {Fronz{\'e}}, {Fulda}, {Fyffe}, {Gabbard}, {Gadre},
  {Gaebel}, {Gair}, {Gais}, {Galaudage}, {Gamba}, {Ganapathy}, {Ganguly},
  {Gaonkar}, {Garaventa}, {Garc{\'\i}a-Quir{\'o}s}, {Garufi}, {Gateley},
  {Gaudio}, {Gayathri}, {Gemme}, {Gennai}, {George}, {George}, {George},
  {Gergely}, {Ghonge}, {Ghosh}, {Ghosh}, {Ghosh}, {Giacomazzo}, {Giacoppo},
  {Giaime}, {Giardina}, {Gibson}, {Gier}, {Gill}, {Giri}, {Glanzer}, {Gleckl},
  {Godwin}, {Goetz}, {Goetz}, {Gohlke}, {Goncharov}, {Gonz{\'a}lez},
  {Gopakumar}, {Gossan}, {Gosselin}, {Gouaty}, {Grace}, {Grado}, {Granata},
  {Granata}, {Grant}, {Gras}, {Grassia}, {Gray}, {Gray}, {Greco}, {Green},
  {Green}, {Gretarsson}, {Griggs}, {Grignani}, {Grimaldi}, {Grimes}, {Grimm},
  {Grote}, {Grunewald}, {Gruning}, {Guerrero}, {Guidi}, {Guimaraes},
  {Guix{\'e}}, {Gulati}, {Guo}, {Gupta}, {Gupta}, {Gupta}, {Gustafson},
  {Gustafson}, {Guzman}, {Haegel}, {Halim}, {Hall}, {Hamilton}, {Hammond},
  {Haney}, {Hanke}, {Hanks}, {Hanna}, {Hannam}, {Hannuksela}, {Hannuksela},
  {Hansen}, {Hansen}, {Hanson}, {Harder}, {Hardwick}, {Haris}, {Harms},
  {Harry}, {Harry}, {Hartwig}, {Hasskew}, {Haster}, {Haughian}, {Hayes},
  {Healy}, {Heidmann}, {Heintze}, {Heinze}, {Heinzel}, {Heitmann}, {Hellman},
  {Hello}, {Helmling-Cornell}, {Hemming}, {Hendry}, {Heng}, {Hennes}, {Hennig},
  {Hennig}, {Hernandez Vivanco}, {Heurs}, {Hild}, {Hill}, {Hines}, {Hochheim},
  {Hofgard}, {Hofman}, {Hohmann}, {Holgado}, {Holland}, {Hollows}, {Holmes},
  {Holt}, {Holz}, {Hopkins}, {Horst}, {Hough}, {Howell}, {Hoy}, {Hoyland},
  {Huang}, {H{\"u}bner}, {Huddart}, {Huerta}, {Hughey}, {Hui}, {Husa},
  {Huttner}, {Hutzler}, {Huxford}, {Huynh-Dinh}, {Idzkowski}, {Iess},
  {Imperato}, {Inchauspe}, {Ingram}, {Intini}, {Isi}, {Iyer},
  {JaberianHamedan}, {Jacqmin}, {Jadhav}, {Jadhav}, {James}, {Jani},
  {Janssens}, {Janthalur}, {Jaranowski}, {Jariwala}, {Jaume}, {Jenkins},
  {Jeunon}, {Jiang}, {Johns}, {Johnson-McDaniel}, {Jones}, {Jones}, {Jones},
  {Jones}, {Jones}, {Jonker}, {Ju}, {Junker}, {Kalaghatgi}, {Kalogera},
  {Kamai}, {Kandhasamy}, {Kang}, {Kanner}, {Kapadia}, {Kapasi}, {Karathanasis},
  {Karki}, {Kashyap}, {Kasprzack}, {Kastaun}, {Katsanevas}, {Katsavounidis},
  {Katzman}, {Kawabe}, {K{\'e}f{\'e}lian}, {Keitel}, {Key}, {Khadka},
  {Khalili}, {Khan}, {Khan}, {Khazanov}, {Khetan}, {Khursheed}, {Kijbunchoo},
  {Kim}, {Kim}, {Kim}, {Kim}, {Kim}, {Kim}, {Kimball}, {King}, {Kinley-Hanlon},
  {Kirchhoff}, {Kissel}, {Kleybolte}, {Klimenko}, {Knowles}, {Knyazev}, {Koch},
  {Koehlenbeck}, {Koekoek}, {Koley}, {Kolstein}, {Komori}, {Kondrashov},
  {Kontos}, {Koper}, {Korobko}, {Korth}, {Kovalam}, {Kozak}, {Kr{\"a}mer},
  {Kringel}, {Krishnendu}, {Kr{\'o}lak}, {Kuehn}, {Kumar}, {Kumar}, {Kumar},
  {Kumar}, {Kuns}, {Kwang}, {Lackey}, {Laghi}, {Lalande}, {Lam}, {Lamberts},
  {Landry}, {Lane}, {Lang}, {Lange}, {Lantz}, {Lanza}, {La Rosa},
  {Lartaux-Vollard}, {Lasky}, {Laxen}, {Lazzarini}, {Lazzaro}, {Leaci},
  {Leavey}, {Lecoeuche}, {Lee}, {Lee}, {Lee}, {Lee}, {Lehmann}, {Leon},
  {Leroy}, {Letendre}, {Levin}, {Li}, {Li}, {Li}, {Li}, {Li}, {Linde},
  {Linker}, {Linley}, {Littenberg}, {Liu}, {Liu}, {Llorens-Monteagudo}, {Lo},
  {Lockwood}, {London}, {Longo}, {Lorenzini}, {Loriette}, {Lormand}, {Losurdo},
  {Lough}, {Lousto}, {Lovelace}, {L{\"u}ck}, {Lumaca}, {Lundgren}, {Ma},
  {Macas}, {MacInnis}, {Macleod}, {MacMillan}, {Macquet}, {Maga{\~n}a
  Hernandez}, {Maga{\~n}a-Sandoval}, {Magazz{\`u}}, {Magee}, {Majorana},
  {Maksimovic}, {Maliakal}, {Malik}, {Man}, {Mandic}, {Mangano}, {Mansell},
  {Manske}, {Mantovani}, {Mapelli}, {Marchesoni}, {Marion}, {M{\'a}rka},
  {M{\'a}rka}, {Markakis}, {Markosyan}, {Markowitz}, {Maros}, {Marquina},
  {Marsat}, {Martelli}, {Martin}, {Martin}, {Martinez}, {Martinez}, {Martynov},
  {Masalehdan}, {Mason}, {Massera}, {Masserot}, {Massinger}, {Masso-Reid},
  {Mastrogiovanni}, {Matas}, {Mateu-Lucena}, {Matichard}, {Matiushechkina},
  {Mavalvala}, {Maynard}, {McCann}, {McCarthy}, {McClelland}, {McCormick},
  {McCuller}, {McGuire}, {McIsaac}, {McIver}, {McManus}, {McRae}, {McWilliams},
  {Meacher}, {Meadors}, {Mehmet}, {Mehta}, {Melatos}, {Melchor}, {Mendell},
  {Menendez-Vazquez}, {Mercer}, {Mereni}, {Merfeld}, {Merilh}, {Merritt},
  {Merzougui}, {Meshkov}, {Messenger}, {Messick}, {Metzdorff}, {Meyers},
  {Meylahn}, {Mhaske}, {Miani}, {Miao}, {Michaloliakos}, {Michel}, {Middleton},
  {Milano}, {Miller}, {Millhouse}, {Mills}, {Milotti}, {Milovich-Goff},
  {Minazzoli}, {Minenkov}, {Mir}, {Mishkin}, {Mishra}, {Mistry}, {Mitra},
  {Mitrofanov}, {Mitselmakher}, {Mittleman}, {Mo}, {Mogushi}, {Mohapatra},
  {Mohite}, {Molina}, {Molina-Ruiz}, {Mondin}, {Montani}, {Moore}, {Moraru},
  {Morawski}, {Moreno}, {Morisaki}, {Mours}, {Mow-Lowry}, {Mozzon},
  {Muciaccia}, {Mukherjee}, {Mukherjee}, {Mukherjee}, {Mukherjee}, {Mukund},
  {Mullavey}, {Munch}, {Mu{\~n}iz}, {Murray}, {Nadji}, {Nagar}, {Nardecchia},
  {Naticchioni}, {Nayak}, {Neil}, {Neilson}, {Nelemans}, {Nelson}, {Nery},
  {Neunzert}, {Nitz}, {Ng}, {Ng}, {Nguyen}, {Nguyen}, {Nguyen}, {Nichols},
  {Nissanke}, {Nocera}, {Noh}, {North}, {Nothard}, {Nuttall}, {Oberling},
  {O'Brien}, {O'Dell}, {Oganesyan}, {Ogin}, {Oh}, {Oh}, {Ohme}, {Ohta},
  {Okada}, {Olivetto}, {Oppermann}, {Oram}, {O'Reilly}, {Ormiston}, {Ortega},
  {O'Shaughnessy}, {Ossokine}, {Osthelder}, {Ottaway}, {Overmier}, {Owen},
  {Pace}, {Pagano}, {Page}, {Pagliaroli}, {Pai}, {Pai}, {Palamos}, {Palashov},
  {Palomba}, {Pan}, {Panda}, {Pang}, {Pankow}, {Pannarale}, {Pant}, {Paoletti},
  {Paoli}, {Paolone}, {Parker}, {Pascucci}, {Pasqualetti}, {Passaquieti},
  {Passuello}, {Patel}, {Patricelli}, {Payne}, {Pechsiri}, {Pedraza},
  {Pegoraro}, {Pele}, {Penn}, {Perego}, {Perez}, {P{\'e}rigois}, {Perreca},
  {Perri{\`e}s}, {Petermann}, {Petterson}, {Pfeiffer}, {Pham}, {Phukon},
  {Piccinni}, {Pichot}, {Piendibene}, {Piergiovanni}, {Pierini}, {Pierro},
  {Pillant}, {Pilo}, {Pinard}, {Pinto}, {Piotrzkowski}, {Pirello}, {Pitkin},
  {Placidi}, {Plastino}, {Pluchar}, {Poggiani}, {Polini}, {Pong}, {Ponrathnam},
  {Popolizio}, {Porter}, {Poverman}, {Powell}, {Pracchia}, {Prajapati},
  {Prasai}, {Prasanna}, {Pratten}, {Prestegard}, {Principe}, {Prodi},
  {Prokhorov}, {Prosposito}, {Prudenzi}, {Puecher}, {Punturo}, {Puosi},
  {Puppo}, {P{\"u}rrer}, {Qi}, {Quetschke}, {Quinonez}, {Quitzow-James},
  {Raab}, {Raaijmakers}, {Radkins}, {Radulesco}, {Raffai}, {Rafferty}, {Rail},
  {Raja}, {Rajan}, {Rajbhandari}, {Rakhmanov}, {Ramirez}, {Ramirez},
  {Ramos-Buades}, {Rana}, {Rao}, {Rapagnani}, {Rapol}, {Ratto}, {Raymond},
  {Razzano}, {Read}, {Regimbau}, {Rei}, {Reid}, {Reitze}, {Rettegno}, {Ricci},
  {Richardson}, {Richardson}, {Richardson}, {Ricker}, {Riemenschneider},
  {Riles}, {Rizzo}, {Robertson}, {Robinet}, {Rocchi}, {Rocha}, {Rodriguez},
  {Rodriguez-Soto}, {Rolland}, {Rollins}, {Roma}, {Romanelli}, {Romano},
  {Romel}, {Romero}, {Romero-Shaw}, {Romie}, {Ronchini}, {Rose}, {Rose},
  {Rose}, {Rosell}, {Rosi{\'n}ska}, {Rosofsky}, {Ross}, {Rowan}, {Rowlinson},
  {Roy}, {Roy}, {Ruggi}, {Ryan}, {Sachdev}, {Sadecki}, {Sadiq},
  {Sakellariadou}, {Salafia}, {Salconi}, {Saleem}, {Samajdar}, {Sanchez},
  {Sanchez}, {Sanchez}, {Sanchis-Gual}, {Sanders}, {Sandles}, {Santiago},
  {Santos}, {Saravanan}, {Sarin}, {Sassolas}, {Sathyaprakash}, {Sauter},
  {Savage}, {Savant}, {Sawant}, {Sayah}, {Schaetzl}, {Schale}, {Scheel},
  {Scheuer}, {Schindler-Tyka}, {Schmidt}, {Schnabel}, {Schofield},
  {Sch{\"o}nbeck}, {Schreiber}, {Schulte}, {Schutz}, {Schwarm}, {Schwartz},
  {Scott}, {Scott}, {Seglar-Arroyo}, {Seidel}, {Sellers}, {Sengupta},
  {Sennett}, {Sentenac}, {Sequino}, {Sergeev}, {Setyawati}, {Shaffer},
  {Shahriar}, {Sharifi}, {Sharma}, {Sharma}, {Shawhan}, {Shen}, {Shikauchi},
  {Shink}, {Shoemaker}, {Shoemaker}, {Shukla}, {ShyamSundar}, {Sieniawska},
  {Sigg}, {Singer}, {Singh}, {Singh}, {Singha}, {Singhal}, {Sintes}, {Sipala},
  {Skliris}, {Slagmolen}, {Slaven-Blair}, {Smetana}, {Smith}, {Smith},
  {Somala}, {Son}, {Soni}, {Soni}, {Sorazu}, {Sordini}, {Sorrentino},
  {Sorrentino}, {Soulard}, {Souradeep}, {Sowell}, {Spencer}, {Spera},
  {Srivastava}, {Srivastava}, {Staats}, {Stachie}, {Steer}, {Steinhoff},
  {Steinke}, {Steinlechner}, {Steinlechner}, {Steinmeyer}, {Stevenson},
  {Stolle-McAllister}, {Stops}, {Stover}, {Strain}, {Stratta}, {Strunk},
  {Sturani}, {Stuver}, {S{\"u}dbeck}, {Sudhagar}, {Sudhir}, {Suh},
  {Summerscales}, {Sun}, {Sun}, {Sunil}, {Sur}, {Suresh}, {Sutton}, {Swinkels},
  {Szczepa{\'n}czyk}, {Tacca}, {Tait}, {Talbot}, {Tanasijczuk}, {Tanner},
  {Tao}, {Tapia}, {Tapia San Martin}, {Tasson}, {Taylor}, {Tenorio},
  {Terkowski}, {Thirugnanasambandam}, {Thomas}, {Thomas}, {Thomas}, {Thompson},
  {Thondapu}, {Thorne}, {Thrane}, {Tiwari}, {Tiwari}, {Tiwari}, {Toland},
  {Tolley}, {Tonelli}, {Tornasi}, {Torres-Forn{\'e}}, {Torrie}, {Melo},
  {T{\"o}yr{\"a}}, {Tran}, {Trapananti}, {Travasso}, {Traylor}, {Tringali},
  {Tripathee}, {Trovato}, {Trudeau}, {Tsai}, {Tsang}, {Tse}, {Tso}, {Tsukada},
  {Tsuna}, {Tsutsui}, {Turconi}, {Ubhi}, {Udall}, {Ueno}, {Ugolini},
  {Unnikrishnan}, {Urban}, {Usman}, {Utina}, {Vahlbruch}, {Vajente}, {Vajpeyi},
  {Valdes}, {Valentini}, {Valsan}, {van Bakel}, {van Beuzekom}, {van den
  Brand}, {Van Den Broeck}, {Vander-Hyde}, {van der Schaaf}, {van Heijningen},
  {Vardaro}, {Vargas}, {Varma}, {Vass}, {Vas{\'u}th}, {Vecchio}, {Vedovato},
  {Veitch}, {Veitch}, {Venkateswara}, {Venneberg}, {Venugopalan}, {Verkindt},
  {Verma}, {Veske}, {Vetrano}, {Vicer{\'e}}, {Viets}, {Vijaykumar},
  {Villa-Ortega}, {Vinet}, {Vitale}, {Vo}, {Vocca}, {Vorvick}, {Vyatchanin},
  {Wade}, {Wade}, {Wade}, {Walet}, {Walker}, {Wallace}, {Wallace}, {Walsh},
  {Wang}, {Wang}, {Wang}, {Wang}, {Ward}, {Warner}, {Was}, {Washington},
  {Watchi}, {Weaver}, {Wei}, {Weinert}, {Weinstein}, {Weiss}, {Wellmann},
  {Wen}, {We{\ss}els}, {Westhouse}, {Wette}, {Whelan}, {White}, {White},
  {Whiting}, {Whittle}, {Wilken}, {Williams}, {Williams}, {Williamson},
  {Willis}, {Willke}, {Wilson}, {Wimmer}, {Winkler}, {Wipf}, {Woan}, {Woehler},
  {Wofford}, {Wong}, {Wrangel}, {Wright}, {Wu}, {Wysocki}, {Xiao}, {Yamamoto},
  {Yang}, {Yang}, {Yang}, {Yap}, {Yeeles}, {Yoon}, {Yu}, {Yu}, {Yuen},
  {Zadro{\.z}ny}, {Zanolin}, {Zelenova}, {Zendri}, {Zevin}, {Zhang}, {Zhang},
  {Zhang}, {Zhang}, {Zhao}, {Zhao}, {Zheng}, {Zhou}, {Zhou}, {Zhu},
  {Zimmerman}, {Zlochower}, {Zucker}, \& {Zweizig}}]{GWTC2_2020}
{Abbott}, R., {Abbott}, T.~D., {Abraham}, S., {$et~al$.} 2020, arXiv e-prints,
  arXiv:2010.14527

\bibitem[{Abbott {$et~al$.}(2021{\natexlab{a}})Abbott, Abbott, Abraham,
  Acernese, Ackley, Adams, Adams, Adhikari, Adya, Affeldt, Agathos, Agatsuma,
  Aggarwal, Aguiar, Aiello, Ain, Ajith, Akcay, Allen, Allocca, Altin, Amato,
  Anand, Ananyeva, Anderson, Anderson, Angelova, Ansoldi, Antelis, Antier,
  Appert, Arai, Araya, Areeda, Ar{\`{e}}ne, Arnaud, Aronson, Arun, Asali,
  Ascenzi, Ashton, Aston, Astone, Aubin, Aufmuth, AultONeal, Austin, Avendano,
  Babak, Badaracco, Bader, Bae, Baer, Bagnasco, Baird, Ball, Ballardin,
  Ballmer, Bals, Balsamo, Baltus, Banagiri, Bankar, Bankar, Barayoga, Barbieri,
  Barish, Barker, Barneo, Barnum, Barone, Barr, Barsotti, Barsuglia, Barta,
  Bartlett, Bartos, Bassiri, Basti, Bawaj, Bayley, Bazzan, Becher, B{\'{e}}csy,
  Bedakihale, Bejger, Belahcene, Beniwal, Benjamin, Bennett, Bentley, Bergamin,
  Berger, Bergmann, Bernuzzi, Berry, Bersanetti, Bertolini, Betzwieser,
  Bhandare, Bhandari, Bhattacharjee, Bidler, Bilenko, Billingsley, Birney,
  Birnholtz, Biscans, Bischi, Biscoveanu, Bisht, Bitossi, Bizouard, Blackburn,
  Blackman, Blair, Blair, Blair, Blanch, Bobba, Bode, Boer, Boetzel, Bogaert,
  Boldrini, Bondu, Bonilla, Bonnand, Booker, Boom, Bork, Boschi, Bose,
  Bossilkov, Boudart, Bouffanais, Bozzi, Bradaschia, Brady, Bramley, Branchesi,
  Brau, Breschi, Briant, Briggs, Brighenti, Brillet, Brinkmann, Brockill,
  Brooks, Brooks, Brown, Brunett, Bruno, Bruntz, Buikema, Bulik, Bulten,
  Buonanno, Buscicchio, Buskulic, Byer, Cabero, Cadonati, Caesar, Cagnoli,
  Cahillane, {Calder{\'{o}}n Bustillo}, Callaghan, Callister, Calloni, Camp,
  Canepa, Cannon, Cao, Cao, Carapella, Carbognani, Carney, Carpinelli, Carullo,
  Carver, {Casanueva Diaz}, Casentini, Caudill, Cavagli{\`{a}}, Cavalier,
  Cavalieri, Cella, Cerd{\'{a}}-Dur{\'{a}}n, Cesarini, Chaibi, Chakravarti,
  Chan, Chan, Chandra, Chanial, Chao, Charlton, Chase, Chassande-Mottin,
  Chatterjee, Chattopadhyay, Chaturvedi, Chatziioannou, Chen, Chen, Chen, Chen,
  Cheng, Cheong, Chia, Chiadini, Chierici, Chincarini, Chiummo, Cho, Cho, Cho,
  Choate, Christensen, Chu, Chua, Chung, Chung, Ciani, Ciecielag,
  Cie{\'{s}}lar, Cifaldi, Ciobanu, Ciolfi, Cipriano, Cirone, Clara, Clark,
  Clark, Clarke, Clearwater, Clesse, Cleva, Coccia, Cohadon, Cohen, Colleoni,
  Collette, Collins, Colpi, Constancio, Conti, Cooper, Corban, Corbitt,
  Cordero-Carri{\'{o}}n, Corezzi, Corley, Cornish, Corre, Corsi, Cortese,
  Costa, Cotesta, Coughlin, Coughlin, Coulon, Countryman, Cousins, Couvares,
  Covas, Coward, Cowart, Coyne, Coyne, Creighton, Creighton, Croquette,
  Crowder, Cudell, Cullen, Cumming, Cummings, Cunningham, Cuoco, Cury{\l}o,
  Canton, D{\'{a}}lya, Dana, DaneshgaranBajastani, D'Angelo, Danila,
  Danilishin, D'Antonio, Danzmann, Darsow-Fromm, Dasgupta, Datrier, Dattilo,
  Dave, Davier, Davies, Davis, Daw, Dean, DeBra, Deenadayalan, Degallaix, {De
  Laurentis}, Del{\'{e}}glise, {Del Favero}, {De Lillo}, {De Lillo}, {Del
  Pozzo}, DeMarchi, {De Matteis}, D'Emilio, Demos, Denker, Dent, Depasse, {De
  Pietri}, {De Rosa}, {De Rossi}, DeSalvo, de~Varona, Dhurandhar, D{\'{i}}az,
  Diaz-Ortiz, Didio, Dietrich, {Di Fiore}, DiFronzo, {Di Giorgio}, {Di
  Giovanni}, {Di Giovanni}, {Di Girolamo}, {Di Lieto}, Ding, {Di Pace}, {Di
  Palma}, {Di Renzo}, Divakarla, Dmitriev, Doctor, D'Onofrio, Donovan, Dooley,
  Doravari, Dorrington, Downes, Drago, Driggers, Du, Ducoin, Dupej, Durante,
  D'Urso, Duverne, Dwyer, Easter, Eddolls, Edelman, Edo, Edy, Effler, Eichholz,
  Eikenberry, Eisenmann, Eisenstein, Ejlli, Errico, Essick, Estell{\'{e}}s,
  Estevez, Etienne, Etzel, Evans, Evans, Ewing, Fafone, Fair, Fairhurst, Fan,
  Farah, Farinon, Farr, Farr, Fauchon-Jones, Favata, Fays, Fazio, Feicht,
  Fejer, Feng, Fenyvesi, Ferguson, Fernandez-Galiana, Ferrante, Ferreira,
  Fidecaro, Figura, Fiori, Fiorucci, Fishbach, Fisher, Fishner, Fittipaldi,
  Fitz-Axen, Fiumara, Flaminio, Floden, Flynn, Fong, Font, Forsyth, Fournier,
  Frasca, Frasconi, Frei, Freise, Frey, Frey, Fritschel, Frolov, Fronz{\'{e}},
  Fulda, Fyffe, Gabbard, Gadre, Gaebel, Gair, Gais, Galaudage, Gamba,
  Ganapathy, Ganguly, Gaonkar, Garaventa, Garc{\'{i}}a-Quir{\'{o}}s, Garufi,
  Gateley, Gaudio, Gayathri, Gemme, Gennai, George, George, George, Gergely,
  Ghonge, Ghosh, Ghosh, Ghosh, Giacomazzo, Giacoppo, Giaime, Giardina, Gibson,
  Gier, Gill, Giri, Glanzer, Gleckl, Godwin, Goetz, Goetz, Gohlke, Goncharov,
  Gonz{\'{a}}lez, Gopakumar, Gossan, Gosselin, Gouaty, Grace, Grado, Granata,
  Granata, Grant, Gras, Grassia, Gray, Gray, Greco, Green, Green, Gretarsson,
  Griggs, Grignani, Grimaldi, Grimes, Grimm, Grote, Grunewald, Gruning,
  Guerrero, Guidi, Guimaraes, Guix{\'{e}}, Gulati, Guo, Gupta, Gupta, Gupta,
  Gustafson, Gustafson, Guzman, Haegel, Halim, Hall, Hamilton, Hammond, Haney,
  Hanke, Hanks, Hanna, Hannam, Hannuksela, Hannuksela, Hansen, Hansen, Hanson,
  Harder, Hardwick, Haris, Harms, Harry, Harry, Hartwig, Hasskew, Haster,
  Haughian, Hayes, Healy, Heidmann, Heintze, Heinze, Heinzel, Heitmann,
  Hellman, Hello, Helmling-Cornell, Hemming, Hendry, Heng, Hennes, Hennig,
  Hennig, {Hernandez Vivanco}, Heurs, Hild, Hill, Hines, Hochheim, Hofgard,
  Hofman, Hohmann, Holgado, Holland, Hollows, Holmes, Holt, Holz, Hopkins,
  Horst, Hough, Howell, Hoy, Hoyland, Huang, H{\"{u}}bner, Huddart, Huerta,
  Hughey, Hui, Husa, Huttner, Hutzler, Huxford, Huynh-Dinh, Idzkowski, Iess,
  Imperato, Inchauspe, Ingram, Intini, Isi, Iyer, JaberianHamedan, Jacqmin,
  Jadhav, Jadhav, James, Jani, Janssens, Janthalur, Jaranowski, Jariwala,
  Jaume, Jenkins, Jeunon, Jiang, Johns, Johnson-McDaniel, Jones, Jones, Jones,
  Jones, Jones, Jonker, Ju, Junker, Kalaghatgi, Kalogera, Kamai, Kandhasamy,
  Kang, Kanner, Kapadia, Kapasi, Karathanasis, Karki, Kashyap, Kasprzack,
  Kastaun, Katsanevas, Katsavounidis, Katzman, Kawabe, K{\'{e}}f{\'{e}}lian,
  Keitel, Key, Khadka, Khalili, Khan, Khan, Khazanov, Khetan, Khursheed,
  Kijbunchoo, Kim, Kim, Kim, Kim, Kim, Kim, Kimball, King, Kinley-Hanlon,
  Kirchhoff, Kissel, Kleybolte, Klimenko, Knowles, Knyazev, Koch, Koehlenbeck,
  Koekoek, Koley, Kolstein, Komori, Kondrashov, Kontos, Koper, Korobko, Korth,
  Kovalam, Kozak, Kr{\"{a}}mer, Kringel, Krishnendu, Kr{\'{o}}lak, Kuehn,
  Kumar, Kumar, Kumar, Kumar, Kuns, Kwang, Lackey, Laghi, Lalande, Lam,
  Lamberts, Landry, Lane, Lang, Lange, Lantz, Lanza, {La Rosa},
  Lartaux-Vollard, Lasky, Laxen, Lazzarini, Lazzaro, Leaci, Leavey, Lecoeuche,
  Lee, Lee, Lee, Lee, Lehmann, Leon, Leroy, Letendre, Levin, Li, Li, Li, Li,
  Li, Linde, Linker, Linley, Littenberg, Liu, Liu, Llorens-Monteagudo, Lo,
  Lockwood, London, Longo, Lorenzini, Loriette, Lormand, Losurdo, Lough,
  Lousto, Lovelace, L{\"{u}}ck, Lumaca, Lundgren, Ma, Macas, MacInnis, Macleod,
  MacMillan, Macquet, {Maga{\~{n}}a Hernandez}, Maga{\~{n}}a-Sandoval,
  Magazz{\`{u}}, Magee, Majorana, Maksimovic, Maliakal, Malik, Man, Mandic,
  Mangano, Mansell, Manske, Mantovani, Mapelli, Marchesoni, Marion,
  M{\'{a}}rka, M{\'{a}}rka, Markakis, Markosyan, Markowitz, Maros, Marquina,
  Marsat, Martelli, Martin, Martin, Martinez, Martinez, Martynov, Masalehdan,
  Mason, Massera, Masserot, Massinger, Masso-Reid, Mastrogiovanni, Matas,
  Mateu-Lucena, Matichard, Matiushechkina, Mavalvala, Maynard, McCann,
  McCarthy, McClelland, McCormick, McCuller, McGuire, McIsaac, McIver, McManus,
  McRae, McWilliams, Meacher, Meadors, Mehmet, Mehta, Melatos, Melchor,
  Mendell, Menendez-Vazquez, Mercer, Mereni, Merfeld, Merilh, Merritt,
  Merzougui, Meshkov, Messenger, Messick, Metzdorff, Meyers, Meylahn, Mhaske,
  Miani, Miao, Michaloliakos, Michel, Middleton, Milano, Miller, Millhouse,
  Mills, Milotti, Milovich-Goff, Minazzoli, Minenkov, Mir, Mishkin, Mishra,
  Mistry, Mitra, Mitrofanov, Mitselmakher, Mittleman, Mo, Mogushi, Mohapatra,
  Mohite, Molina, Molina-Ruiz, Mondin, Montani, Moore, Moraru, Morawski,
  Moreno, Morisaki, Mours, Mow-Lowry, Mozzon, Muciaccia, Mukherjee, Mukherjee,
  Mukherjee, Mukherjee, Mukund, Mullavey, Munch, Mu{\~{n}}iz, Murray, Nadji,
  Nagar, Nardecchia, Naticchioni, Nayak, Neil, Neilson, Nelemans, Nelson, Nery,
  Neunzert, Nitz, Ng, Ng, Nguyen, Nguyen, Nguyen, Nichols, Nissanke, Nocera,
  Noh, North, Nothard, Nuttall, Oberling, O'Brien, O'Dell, Oganesyan, Ogin, Oh,
  Oh, Ohme, Ohta, Okada, Olivetto, Oppermann, Oram, O'Reilly, Ormiston, Ortega,
  O'Shaughnessy, Ossokine, Osthelder, Ottaway, Overmier, Owen, Pace, Pagano,
  Page, Pagliaroli, Pai, Pai, Palamos, Palashov, Palomba, Pan, Panda, Pang,
  Pankow, Pannarale, Pant, Paoletti, Paoli, Paolone, Parker, Pascucci,
  Pasqualetti, Passaquieti, Passuello, Patel, Patricelli, Payne, Pechsiri,
  Pedraza, Pegoraro, Pele, Penn, Perego, Perez, P{\'{e}}rigois, Perreca,
  Perri{\`{e}}s, Petermann, Petterson, Pfeiffer, Pham, Phukon, Piccinni,
  Pichot, Piendibene, Piergiovanni, Pierini, Pierro, Pillant, Pilo, Pinard,
  Pinto, Piotrzkowski, Pirello, Pitkin, Placidi, Plastino, Pluchar, Poggiani,
  Polini, Pong, Ponrathnam, Popolizio, Porter, Poverman, Powell, Pracchia,
  Prajapati, Prasai, Prasanna, Pratten, Prestegard, Principe, Prodi, Prokhorov,
  Prosposito, Prudenzi, Puecher, Punturo, Puosi, Puppo, P{\"{u}}rrer, Qi,
  Quetschke, Quinonez, Quitzow-James, Raab, Raaijmakers, Radkins, Radulesco,
  Raffai, Rafferty, Rail, Raja, Rajan, Rajbhandari, Rakhmanov, Ramirez,
  Ramirez, Ramos-Buades, Rana, Rao, Rapagnani, Rapol, Ratto, Raymond, Razzano,
  Read, Regimbau, Rei, Reid, Reitze, Rettegno, Ricci, Richardson, Richardson,
  Richardson, Ricker, Riemenschneider, Riles, Rizzo, Robertson, Robinet,
  Rocchi, Rocha, Rodriguez, Rodriguez-Soto, Rolland, Rollins, Roma, Romanelli,
  Romano, Romel, Romero, Romero-Shaw, Romie, Ronchini, Rose, Rose, Rose,
  Rosell, Rosi{\'{n}}ska, Rosofsky, Ross, Rowan, Rowlinson, Roy, Roy, Ruggi,
  Ryan, Sachdev, Sadecki, Sadiq, Sakellariadou, Salafia, Salconi, Saleem,
  Samajdar, Sanchez, Sanchez, Sanchez, Sanchis-Gual, Sanders, Sandles,
  Santiago, Santos, Saravanan, Sarin, Sassolas, Sathyaprakash, Sauter, Savage,
  Savant, Sawant, Sayah, Schaetzl, Schale, Scheel, Scheuer, Schindler-Tyka,
  Schmidt, Schnabel, Schofield, Sch{\"{o}}nbeck, Schreiber, Schulte, Schutz,
  Schwarm, Schwartz, Scott, Scott, Seglar-Arroyo, Seidel, Sellers, Sengupta,
  Sennett, Sentenac, Sequino, Sergeev, Setyawati, Shaffer, Shahriar, Sharifi,
  Sharma, Sharma, Shawhan, Shen, Shikauchi, Shink, Shoemaker, Shoemaker,
  Shukla, ShyamSundar, Sieniawska, Sigg, Singer, Singh, Singh, Singha, Singhal,
  Sintes, Sipala, Skliris, Slagmolen, Slaven-Blair, Smetana, Smith, Smith,
  Somala, Son, Soni, Soni, Sorazu, Sordini, Sorrentino, Sorrentino, Soulard,
  Souradeep, Sowell, Spencer, Spera, Srivastava, Srivastava, Staats, Stachie,
  Steer, Steinhoff, Steinke, Steinlechner, Steinlechner, Steinmeyer, Stevenson,
  Stolle-McAllister, Stops, Stover, Strain, Stratta, Strunk, Sturani, Stuver,
  S{\"{u}}dbeck, Sudhagar, Sudhir, Suh, Summerscales, Sun, Sun, Sunil, Sur,
  Suresh, Sutton, Swinkels, Szczepa{\'{n}}czyk, Tacca, Tait, Talbot,
  Tanasijczuk, Tanner, Tao, Tapia, {Tapia San Martin}, Tasson, Taylor, Tenorio,
  Terkowski, Thirugnanasambandam, Thomas, Thomas, Thomas, Thompson, Thondapu,
  Thorne, Thrane, Tiwari, Tiwari, Tiwari, Toland, Tolley, Tonelli, Tornasi,
  Torres-Forn{\'{e}}, Torrie, e~Melo, T{\"{o}}yr{\"{a}}, Tran, Trapananti,
  Travasso, Traylor, Tringali, Tripathee, Trovato, Trudeau, Tsai, Tsang, Tse,
  Tso, Tsukada, Tsuna, Tsutsui, Turconi, Ubhi, Udall, Ueno, Ugolini,
  Unnikrishnan, Urban, Usman, Utina, Vahlbruch, Vajente, Vajpeyi, Valdes,
  Valentini, Valsan, van Bakel, van Beuzekom, van~den Brand, {Van Den Broeck},
  Vander-Hyde, van~der Schaaf, van Heijningen, Vardaro, Vargas, Varma, Vass,
  Vas{\'{u}}th, Vecchio, Vedovato, Veitch, Veitch, Venkateswara, Venneberg,
  Venugopalan, Verkindt, Verma, Veske, Vetrano, Vicer{\'{e}}, Viets,
  Vijaykumar, Villa-Ortega, Vinet, Vitale, Vo, Vocca, Vorvick, Vyatchanin,
  Wade, Wade, Wade, Walet, Walker, Wallace, Wallace, Walsh, Wang, Wang, Wang,
  Wang, Ward, Warner, Was, Washington, Watchi, Weaver, Wei, Weinert, Weinstein,
  Weiss, Wellmann, Wen, We{\ss}els, Westhouse, Wette, Whelan, White, White,
  Whiting, Whittle, Wilken, Williams, Williams, Williamson, Willis, Willke,
  Wilson, Wimmer, Winkler, Wipf, Woan, Woehler, Wofford, Wong, Wrangel, Wright,
  Wu, Wysocki, Xiao, Yamamoto, Yang, Yang, Yang, Yap, Yeeles, Yoon, Yu, Yu,
  Yuen, Zadro{\.{z}}ny, Zanolin, Zelenova, Zendri, Zevin, Zhang, Zhang, Zhang,
  Zhang, Zhao, Zhao, Zheng, Zhou, Zhou, Zhu, Zimmerman, Zlochower, Zucker, \&
  Zweizig}]{Abbot2020_GWTC-2}
Abbott, R., Abbott, T.~D., Abraham, S., {$et~al$.} 2021{\natexlab{a}}, Physical
  Review X, 11, 021053

\bibitem[{Abbott {$et~al$.}(2021{\natexlab{b}})Abbott, Abbott, Acernese,
  Ackley, Adams, Adhikari, Adhikari, Adya, Affeldt, Agarwal, Agathos, Agatsuma,
  Aggarwal, Aguiar, Aiello, Ain, Ajith, Akcay, Akutsu, Albanesi, Allocca,
  Altin, Amato, Anand, Anand, Ananyeva, Anderson, Anderson, Ando, Andrade,
  Andres, Andri{\'{c}}, Angelova, Ansoldi, Antelis, Antier, Appert, Arai, Arai,
  Arai, Araki, Araya, Araya, Areeda, Ar{\`{e}}ne, Aritomi, Arnaud, Arogeti,
  Aronson, Arun, Asada, Asali, Ashton, Aso, Assiduo, Aston, Astone, Aubin,
  Austin, Babak, Badaracco, Bader, Badger, Bae, Bae, Baer, Bagnasco, Bai,
  Baiotti, Baird, Bajpai, Ball, Ballardin, Ballmer, Balsamo, Baltus, Banagiri,
  Bankar, Barayoga, Barbieri, Barish, Barker, Barneo, Barone, Barr, Barsotti,
  Barsuglia, Barta, Bartlett, Barton, Bartos, Bassiri, Basti, Bawaj, Bayley,
  Baylor, Bazzan, B{\'{e}}csy, Bedakihale, Bejger, Belahcene, Benedetto,
  Beniwal, Bennett, Bentley, BenYaala, Bergamin, Berger, Bernuzzi, Berry,
  Bersanetti, Bertolini, Betzwieser, Beveridge, Bhandare, Bhardwaj,
  Bhattacharjee, Bhaumik, Bilenko, Billingsley, Bini, Birney, Birnholtz,
  Biscans, Bischi, Biscoveanu, Bisht, Biswas, Bitossi, Bizouard, Blackburn,
  Blair, Blair, Blair, Bobba, Bode, Boer, Bogaert, Boldrini, Bonavena, Bondu,
  Bonilla, Bonnand, Booker, Boom, Bork, Boschi, Bose, Bose, Bossilkov, Boudart,
  Bouffanais, Bozzi, Bradaschia, Brady, Bramley, Branch, Branchesi, Brandt,
  Brau, Breschi, Briant, Briggs, Brillet, Brinkmann, Brockill, Brooks, Brooks,
  Brown, Brunett, Bruno, Bruntz, Bryant, Bulik, Bulten, Buonanno, Buscicchio,
  Buskulic, Buy, Byer, Davies, Cadonati, Cagnoli, Cahillane, Bustillo,
  Callaghan, Callister, Calloni, Cameron, Camp, Canepa, Canevarolo,
  Cannavacciuolo, Cannon, Cao, Cao, Capocasa, Capote, Carapella, Carbognani,
  Carlin, Carney, Carpinelli, Carrillo, Carullo, Carver, Diaz, Casentini,
  Castaldi, Caudill, Cavagli{\`{a}}, Cavalier, Cavalieri, Ceasar, Cella,
  Cerd{\'{a}}-Dur{\'{a}}n, Cesarini, Chaibi, Chakravarti, Subrahmanya,
  Champion, Chan, Chan, Chan, Chan, Chan, Chandra, Chanial, Chao, Chapman-Bird,
  Charlton, Chase, Chassande-Mottin, Chatterjee, Chatterjee, Chatterjee,
  Chaturvedi, Chaty, Chatziioannou, Chen, Chen, Chen, Chen, Chen, Chen, Chen,
  Chen, Cheng, Cheong, Cheung, Chia, Chiadini, Chiang, Chiarini, Chierici,
  Chincarini, Chiofalo, Chiummo, Cho, Cho, Choudhary, Choudhary, Christensen,
  Chu, Chu, Chu, Chua, Chung, Ciani, Ciecielag, Cie{\'{s}}lar, Cifaldi,
  Ciobanu, Ciolfi, Cipriano, Cirone, Clara, Clark, Clark, Clarke, Clearwater,
  Clesse, Cleva, Coccia, Codazzo, Cohadon, Cohen, Cohen, Colleoni, Collette,
  Colombo, Colpi, Compton, Constancio, Conti, Cooper, Corban, Corbitt,
  Cordero-Carri{\'{o}}n, Corezzi, Corley, Cornish, Corre, Corsi, Cortese,
  Costa, Cotesta, Coughlin, Coulon, Countryman, Cousins, Couvares, Coward,
  Cowart, Coyne, Coyne, Creighton, Creighton, Criswell, Croquette, Crowder,
  Cudell, Cullen, Cumming, Cummings, Cunningham, Cuoco, Cury{\l}o, Dabadie,
  Canton, Dall'Osso, D{\'{a}}lya, Dana, DaneshgaranBajastani, D'Angelo, Danila,
  Danilishin, D'Antonio, Danzmann, Darsow-Fromm, Dasgupta, Datrier, Datta,
  Dattilo, Dave, Davier, Davis, Davis, Daw, de~Alarc{\'{o}}n, Dean, DeBra,
  Deenadayalan, Degallaix, {De Laurentis}, Del{\'{e}}glise, {Del Favero}, {De
  Lillo}, {De Lillo}, {Del Pozzo}, DeMarchi, {De Matteis}, D'Emilio, Demos,
  Dent, Depasse, {De Pietri}, {De Rosa}, {De Rossi}, DeSalvo, {De Simone},
  Dhurandhar, D{\'{i}}az, Diaz-Ortiz, Didio, Dietrich, {Di Fiore}, {Di Fronzo},
  {Di Giorgio}, {Di Giovanni}, {Di Giovanni}, {Di Girolamo}, {Di Lieto}, Ding,
  {Di Pace}, {Di Palma}, {Di Renzo}, Divakarla, Dmitriev, Doctor, D'Onofrio,
  Donovan, Dooley, Doravari, Dorrington, Drago, Driggers, Drori, Ducoin, Dupej,
  Durante, D'Urso, Duverne, Dwyer, Eassa, Easter, Ebersold, Eckhardt, Eddolls,
  Edelman, Edo, Edy, Effler, Eguchi, Eichholz, Eikenberry, Eisenmann,
  Eisenstein, Ejlli, Engelby, Enomoto, Errico, Essick, Estell{\'{e}}s, Estevez,
  Etienne, Etzel, Evans, Evans, Ewing, Fafone, Fair, Fairhurst, Farah, Farinon,
  Farr, Farr, Farrow, Fauchon-Jones, Favaro, Favata, Fays, Fazio, Feicht,
  Fejer, Fenyvesi, Ferguson, Fernandez-Galiana, Ferrante, Ferreira, Fidecaro,
  Figura, Fiori, Fishbach, Fisher, Fittipaldi, Fiumara, Flaminio, Floden, Fong,
  Font, Fornal, Forsyth, Franke, Frasca, Frasconi, Frederick, Freed, Frei,
  Freise, Frey, Fritschel, Frolov, Fronz{\'{e}}, Fujii, Fujikawa, Fukunaga,
  Fukushima, Fulda, Fyffe, Gabbard, Gabella, Gadre, Gair, Gais, Galaudage,
  Gamba, Ganapathy, Ganguly, Gao, Gaonkar, Garaventa, Garc{\'{i}}a,
  Garc{\'{i}}a-N{\'{u}}{\~{n}}ez, Garc{\'{i}}a-Quir{\'{o}}s, Garufi, Gateley,
  Gaudio, Gayathri, Ge, Gemme, Gennai, George, George, Gerberding, Gergely,
  Gewecke, Ghonge, Ghosh, Ghosh, Ghosh, Ghosh, Giacomazzo, Giacoppo, Giaime,
  Giardina, Gibson, Gier, Giesler, Giri, Gissi, Glanzer, Gleckl, Godwin, Goetz,
  Goetz, Gohlke, Golomb, Goncharov, Gonz{\'{a}}lez, Gopakumar, Gosselin,
  Gouaty, Gould, Grace, Grado, Granata, Granata, Grant, Gras, Grassia, Gray,
  Gray, Greco, Green, Green, Gretarsson, Gretarsson, Griffith, Griffiths,
  Griggs, Grignani, Grimaldi, Grimm, Grote, Grunewald, Gruning, Guerra, Guidi,
  Guimaraes, Guix{\'{e}}, Gulati, Guo, Guo, Gupta, Gupta, Gupta, Gustafson,
  Gustafson, Guzman, Ha, Haegel, Hagiwara, Haino, Halim, Hall, Hamilton,
  Hammond, Han, Haney, Hanks, Hanna, Hannam, Hannuksela, Hansen, Hansen,
  Hanson, Harder, Hardwick, Haris, Harms, Harry, Harry, Hartwig, Hasegawa,
  Haskell, Hasskew, Haster, Hattori, Haughian, Hayakawa, Hayama, Hayes, Healy,
  Heidmann, Heidt, Heintze, Heinze, Heinzel, Heitmann, Hellman, Hello,
  Helmling-Cornell, Hemming, Hendry, Heng, Hennes, Hennig, Hennig, Hernandez,
  Vivanco, Heurs, Hild, Hill, Himemoto, Hines, Hiranuma, Hirata, Hirose,
  Hochheim, Hofman, Hohmann, Holcomb, Holland, Holley-Bockelmann, Hollows,
  Holmes, Holt, Holz, Hong, Hopkins, Hough, Hourihane, Howell, Hoy, Hoyland,
  Hreibi, Hsieh, Hsu, Huang, Huang, Huang, Huang, Huang, Huang, H{\"{u}}bner,
  Huddart, Hughey, Hui, Hui, Husa, Huttner, Huxford, Huynh-Dinh, Ide,
  Idzkowski, Iess, Ikenoue, Imam, Inayoshi, Ingram, Inoue, Ioka, Isi, Isleif,
  Ito, Itoh, Iyer, Izumi, JaberianHamedan, Jacqmin, Jadhav, Jadhav, James, Jan,
  Jani, Janquart, Janssens, Janthalur, Jaranowski, Jariwala, Jaume, Jenkins,
  Jenner, Jeon, Jeunon, Jia, Jin, Johns, Johnson-McDaniel, Jones, Jones, Jones,
  Jones, Jones, Jonker, Ju, Jung, Jung, Junker, Juste, Kaihotsu, Kajita,
  Kakizaki, Kalaghatgi, Kalogera, Kamai, Kamiizumi, Kanda, Kandhasamy, Kang,
  Kanner, Kao, Kapadia, Kapasi, Karat, Karathanasis, Karki, Kashyap, Kasprzack,
  Kastaun, Katsanevas, Katsavounidis, Katzman, Kaur, Kawabe, Kawaguchi, Kawai,
  Kawasaki, K{\'{e}}f{\'{e}}lian, Keitel, Key, Khadka, Khalili, Khan, Khazanov,
  Khetan, Khursheed, Kijbunchoo, Kim, Kim, Kim, Kim, Kim, Kim, Kimball, Kimura,
  Kinley-Hanlon, Kirchhoff, Kissel, Kita, Kitazawa, Kleybolte, Klimenko, Knee,
  Knowles, Knyazev, Koch, Koekoek, Kojima, Kokeyama, Koley, Kolitsidou,
  Kolstein, Komori, Kondrashov, Kong, Kontos, Koper, Korobko, Kotake, Kovalam,
  Kozak, Kozakai, Kozu, Kringel, Krishnendu, Kr{\'{o}}lak, Kuehn, Kuei, Kuijer,
  Kulkarni, Kumar, Kumar, Kumar, Kumar, Kume, Kuns, Kuo, Kuo, Kuromiya,
  Kuroyanagi, Kusayanagi, Kuwahara, Kwak, Lagabbe, Laghi, Lalande, Lam,
  Lamberts, Landry, Lane, Lang, Lange, Lantz, {La Rosa}, Lartaux-Vollard,
  Lasky, Laxen, Lazzarini, Lazzaro, Leaci, Leavey, Lecoeuche, Lee, Lee, Lee,
  Lee, Lee, Lee, Lehmann, Lema{\^{i}}tre, Leonardi, Leroy, Letendre, Levesque,
  Levin, Leviton, Leyde, Li, Li, Li, Li, Li, Li, Lin, Lin, Lin, Lin, Lin,
  Linde, Linker, Linley, Littenberg, Liu, Liu, Liu, Liu, Llamas,
  Llorens-Monteagudo, Lo, Lockwood, Loh, London, Longo, Lopez, Portilla,
  Lorenzini, Loriette, Lormand, Losurdo, Lott, Lough, Lousto, Lovelace,
  Lucaccioni, L{\"{u}}ck, Lumaca, Lundgren, Luo, Lynam, Macas, MacInnis,
  Macleod, MacMillan, Macquet, Hernandez, Magazz{\`{u}}, Magee, Maggiore,
  Magnozzi, Mahesh, Majorana, Makarem, Maksimovic, Maliakal, Malik, Man,
  Mandic, Mangano, Mango, Mansell, Manske, Mantovani, Mapelli, Marchesoni,
  Marchio, Marion, Mark, M{\'{a}}rka, M{\'{a}}rka, Markakis, Markosyan,
  Markowitz, Maros, Marquina, Marsat, Martelli, Martin, Martin, Martinez,
  Martinez, Martinez, Martinovic, Martynov, Marx, Masalehdan, Mason, Massera,
  Masserot, Massinger, Masso-Reid, Mastrogiovanni, Matas, Mateu-Lucena,
  Matichard, Matiushechkina, Mavalvala, McCann, McCarthy, McClelland, McClincy,
  McCormick, McCuller, McGhee, McGuire, McIsaac, McIver, McRae, McWilliams,
  Meacher, Mehmet, Mehta, Meijer, Melatos, Melchor, Mendell, Menendez-Vazquez,
  Menoni, Mercer, Mereni, Merfeld, Merilh, Merritt, Merzougui, Meshkov,
  Messenger, Messick, Meyers, Meylahn, Mhaske, Miani, Miao, Michaloliakos,
  Michel, Michimura, Middleton, Milano, Miller, Miller, Miller, Millhouse,
  Mills, Milotti, Minazzoli, Minenkov, Mio, Mir, Miravet-Ten{\'{e}}s, Mishra,
  Mishra, Mistry, Mitra, Mitrofanov, Mitselmakher, Mittleman, Miyakawa,
  Miyamoto, Miyazaki, Miyo, Miyoki, Mo, Modafferi, Moguel, Mogushi, Mohapatra,
  Mohite, Molina, Molina-Ruiz, Mondin, Montani, Moore, Moraru, Morawski, More,
  Moreno, Moreno, Mori, Morisaki, Moriwaki, Morr{\'{a}}s, Mours, Mow-Lowry,
  Mozzon, Muciaccia, Mukherjee, Mukherjee, Mukherjee, Mukherjee, Mukherjee,
  Mukund, Mullavey, Munch, Mu{\~{n}}iz, Murray, Musenich, Muusse, Nadji,
  Nagano, Nagano, Nagar, Nakamura, Nakano, Nakano, Nakashima, Nakayama,
  Napolano, Nardecchia, Narikawa, Naticchioni, Nayak, Nayak, Negishi, Neil,
  Neilson, Nelemans, Nelson, Nery, Neubauer, Neunzert, Ng, Ng, Nguyen, Nguyen,
  Nguyen, Quynh, Ni, Nichols, Nishizawa, Nissanke, Nitoglia, Nocera, Norman,
  North, Nozaki, Siles, Nuttall, Oberling, O'Brien, Obuchi, O'Dell, Oelker,
  Ogaki, Oganesyan, Oh, Oh, Oh, Ohashi, Ohishi, Ohkawa, Ohme, Ohta, Okada,
  Okutani, Okutomi, Olivetto, Oohara, Ooi, Oram, O'Reilly, Ormiston, Ormsby,
  Ortega, O'Shaughnessy, O'Shea, Oshino, Ossokine, Osthelder, Otabe, Ottaway,
  Overmier, Pace, Pagano, Page, Pagliaroli, Pai, Pai, Palamos, Palashov,
  Palomba, Pan, Pan, Panda, Pang, Pang, Pankow, Pannarale, Pant, Panther,
  Paoletti, Paoli, Paolone, Parisi, Park, Park, Parker, Pascucci, Pasqualetti,
  Passaquieti, Passuello, Patel, Pathak, Patricelli, Patron, Paul, Payne,
  Pedraza, Pegoraro, Pele, Arellano, Penn, Perego, Pereira, Pereira, Perez,
  P{\'{e}}rigois, Perkins, Perreca, Perri{\`{e}}s, Petermann, Petterson,
  Pfeiffer, Pham, Phukon, Piccinni, Pichot, Piendibene, Piergiovanni, Pierini,
  Pierro, Pillant, Pillas, Pilo, Pinard, Pinto, Pinto, Piotrzkowski,
  Piotrzkowski, Pirello, Pitkin, Placidi, Planas, Plastino, Pluchar, Poggiani,
  Polini, Pong, Ponrathnam, Popolizio, Porter, Poulton, Powell, Pracchia,
  Pradier, Prajapati, Prasai, Prasanna, Pratten, Principe, Prodi, Prokhorov,
  Prosposito, Prudenzi, Puecher, Punturo, Puosi, Puppo, P{\"{u}}rrer, Qi,
  Quetschke, Quitzow-James, Qutob, Raab, Raaijmakers, Radkins, Radulesco,
  Raffai, Rail, Raja, Rajan, Ramirez, Ramirez, Ramos-Buades, Rana, Rapagnani,
  Rapol, Ray, Raymond, Raza, Razzano, Read, Rees, Regimbau, Rei, Reid, Reid,
  Reitze, Relton, Renzini, Rettegno, Reza, Rezac, Ricci, Richards, Richardson,
  Richardson, Riemenschneider, Riles, Rinaldi, Rink, Rizzo, Robertson, Robie,
  Robinet, Rocchi, Rodriguez, Rolland, Rollins, Romanelli, Romano, Romel,
  Romero-Rodr{\'{i}}guez, Romero-Shaw, Romie, Ronchini, Rosa, Rose,
  Rosi{\'{n}}ska, Ross, Rowan, Rowlinson, Roy, Roy, Roy, Rozza, Ruggi,
  Ruiz-Rocha, Ryan, Sachdev, Sadecki, Sadiq, Sago, Saito, Saito, Sakai, Sakai,
  Sakellariadou, Sakuno, Salafia, Salconi, Saleem, Salemi, Samajdar, Sanchez,
  Sanchez, Sanchez, Sanchis-Gual, Sanders, Sanuy, Saravanan, Sarin, Sassolas,
  Satari, Sathyaprakash, Sato, Sato, Sauter, Savage, Sawada, Sawant, Sawant,
  Sayah, Schaetzl, Scheel, Scheuer, Schiworski, Schmidt, Schmidt, Schnabel,
  Schneewind, Schofield, Sch{\"{o}}nbeck, Schulte, Schutz, Schwartz, Scott,
  Scott, Seglar-Arroyo, Sekiguchi, Sekiguchi, Sellers, Sengupta, Sentenac, Seo,
  Sequino, Sergeev, Setyawati, Shaffer, Shahriar, Shams, Shao, Sharma, Sharma,
  Shawhan, Shcheblanov, Shibagaki, Shikauchi, Shimizu, Shimoda, Shimode,
  Shinkai, Shishido, Shoda, Shoemaker, Shoemaker, ShyamSundar, Sieniawska,
  Sigg, Singer, Singh, Singh, Singha, Sintes, Sipala, Skliris, Slagmolen,
  Slaven-Blair, Smetana, Smith, Smith, Soldateschi, Somala, Somiya, Son, Soni,
  Soni, Sordini, Sorrentino, Sorrentino, Sotani, Soulard, Souradeep, Sowell,
  Spagnuolo, Spencer, Spera, Srinivasan, Srivastava, Srivastava, Staats,
  Stachie, Steer, Steinhoff, Steinlechner, Steinlechner, Stevenson, Stops,
  Stover, Strain, Strang, Stratta, Strunk, Sturani, Stuver, Sudhagar, Sudhir,
  Sugimoto, Suh, Sullivan, Sullivan, Summerscales, Sun, Sun, Sunil, Sur,
  Suresh, Sutton, Suzuki, Suzuki, Swinkels, Szczepa{\'{n}}czyk, Szewczyk,
  Tacca, Tagoshi, Tait, Takahashi, Takahashi, Takamori, Takano, Takeda, Takeda,
  Talbot, Talbot, Tanaka, Tanaka, Tanaka, Tanaka, Tanaka, Tanasijczuk, Tanioka,
  Tanner, Tao, Tao, Mart{\'{i}}n, Taranto, Tasson, Telada, Tenorio, Terhune,
  Terkowski, Thirugnanasambandam, Thomas, Thomas, Thomas, Thompson, Thondapu,
  Thorne, Thrane, Tiwari, Tiwari, Tiwari, Toivonen, Toland, Tolley, Tomaru,
  Tomigami, Tomura, Tonelli, Torres-Forn{\'{e}}, Torrie, Melo,
  T{\"{o}}yr{\"{a}}, Trapananti, Travasso, Traylor, Trevor, Tringali,
  Tripathee, Troiano, Trovato, Trozzo, Trudeau, Tsai, Tsai, Tsang, Tsang, Tsao,
  Tse, Tso, Tsubono, Tsuchida, Tsukada, Tsuna, Tsutsui, Tsuzuki, Turbang,
  Turconi, Tuyenbayev, Ubhi, Uchikata, Uchiyama, Udall, Ueda, Uehara, Ueno,
  Ueshima, Unnikrishnan, Uraguchi, Urban, Ushiba, Utina, Vahlbruch, Vajente,
  Vajpeyi, Valdes, Valentini, Valsan, van Bakel, van Beuzekom, van~den Brand,
  Broeck, Vander-Hyde, van~der Schaaf, van Heijningen, Vanosky, van Putten, van
  Remortel, Vardaro, Vargas, Varma, Vas{\'{u}}th, Vecchio, Vedovato, Veitch,
  Veitch, Venneberg, Venugopalan, Verkindt, Verma, Verma, Veske, Vetrano,
  Vicer{\'{e}}, Vidyant, Viets, Vijaykumar, Villa-Ortega, Vinet, Virtuoso,
  Vitale, Vo, Vocca, von Reis, von Wrangel, Vorvick, Vyatchanin, Wade, Wade,
  Wagner, Walet, Walker, Wallace, Wallace, Walsh, Wang, Wang, Wang, Ward,
  Warner, Was, Washimi, Washington, Watchi, Weaver, Webster, Weinert,
  Weinstein, Weiss, Weller, Weller, Wellmann, Wen, We{\ss}els, Wette, Whelan,
  White, Whiting, Whittle, Wilken, Williams, Williams, Williams, Williamson,
  Willis, Willke, Wilson, Winkler, Wipf, Wlodarczyk, Woan, Woehler, Wofford,
  Wong, Wu, Wu, Wu, Wu, Wysocki, Xiao, Xu, Yamada, Yamamoto, Yamamoto,
  Yamamoto, Yamamoto, Yamashita, Yamazaki, Yang, Yang, Yang, Yang, Yang, Yap,
  Yeeles, Yelikar, Ying, Yokogawa, Yokoyama, Yokozawa, Yoo, Yoshioka, Yu, Yu,
  Yuzurihara, Zadro{\.{z}}ny, Zanolin, Zeidler, Zelenova, Zendri, Zevin, Zhan,
  Zhang, Zhang, Zhang, Zhang, Zhang, Zhao, Zhao, Zhao, Zhao, Zheng, Zhou, Zhou,
  Zhu, Zhu, Zimmerman, Zlochower, Zucker, \& Zweizig}]{Abbot2021_GWTC3}
Abbott, R., Abbott, T.~D., Acernese, F., {$et~al$.} 2021{\natexlab{b}}, arXiv
  e-prints, 2111.03606

\bibitem[{Alam {$et~al$.}(2020)Alam, Arzoumanian, Baker, Blumer, Bohler,
  Brazier, Brook, Burke-Spolaor, Caballero, Camuccio, Chamberlain, Chatterjee,
  Cordes, Cornish, Crawford, Cromartie, DeCesar, Demorest, Dolch, Ellis,
  Ferdman, Ferrara, Fiore, Fonseca, Garcia, Garver-Daniels, Gentile, Good,
  Gusdorff, Halmrast, Hazboun, Islo, Jennings, Jessup, Jones, Kaiser, Kaplan,
  Kelley, Key, Lam, {W. Lazio}, Lorimer, Luo, Lynch, Madison, Maraccini,
  McLaughlin, Mingarelli, Ng, Nguyen, Nice, Pennucci, Pol, Ramette, Ransom,
  Ray, Shapiro-Albert, Siemens, Simon, Spiewak, Stairs, Stinebring, Stovall,
  Swiggum, Taylor, Tripepi, Vallisneri, Vigeland, Witt, \&
  Zhu}]{Alam2020_NANOGrav12.5_narrowband}
Alam, M.~F., Arzoumanian, Z., Baker, P.~T., {$et~al$.} 2020, The Astrophysical
  Journal Supplement Series, 252, 4

\bibitem[{{Alam} {$et~al$.}(2021){Alam}, {Arzoumanian}, {Baker}, {Blumer},
  {Bohler}, {Brazier}, {Brook}, {Burke-Spolaor}, {Caballero}, {Camuccio},
  {Chamberlain}, {Chatterjee}, {Cordes}, {Cornish}, {Crawford}, {Cromartie},
  {Decesar}, {Demorest}, {Dolch}, {Ellis}, {Ferdman}, {Ferrara}, {Fiore},
  {Fonseca}, {Garcia}, {Garver-Daniels}, {Gentile}, {Good}, {Gusdorff},
  {Halmrast}, {Hazboun}, {Islo}, {Jennings}, {Jessup}, {Jones}, {Kaiser},
  {Kaplan}, {Kelley}, {Key}, {Lam}, {Lazio}, {Lorimer}, {Luo}, {Lynch},
  {Madison}, {Maraccini}, {McLaughlin}, {Mingarelli}, {Ng}, {Nguyen}, {Nice},
  {Pennucci}, {Pol}, {Ramette}, {Ransom}, {Ray}, {Shapiro-Albert}, {Siemens},
  {Simon}, {Spiewak}, {Stairs}, {Stinebring}, {Stovall}, {Swiggum}, {Taylor},
  {Tripepi}, {Vallisneri}, {Vigeland}, {Witt}, {Zhu}, \& {Nanograv
  Collaboration}}]{Alam+2021ApJS}
{Alam}, M.~F., {Arzoumanian}, Z., {Baker}, P.~T., {$et~al$.} 2021, \apjs, 252,
  5

\bibitem[{Anholm {$et~al$.}(2009)Anholm, Ballmer, Creighton, Price, \&
  Siemens}]{Anholm2009}
Anholm, M., Ballmer, S., Creighton, J. D.~E., Price, L.~R., \& Siemens, X.
  2009, Physical Review D, 79, 084030

\bibitem[{{Antoniadis} {$et~al$.}(2022){Antoniadis}, {Arzoumanian}, {Babak},
  {Bailes}, {Bak Nielsen}, {Baker}, {Bassa}, {B{\'e}csy}, {Berthereau},
  {Bonetti}, {Brazier}, {Brook}, {Burgay}, {Burke-Spolaor}, {Caballero},
  {Casey-Clyde}, {Chalumeau}, {Champion}, {Charisi}, {Chatterjee}, {Chen},
  {Cognard}, {Cordes}, {Cornish}, {Crawford}, {Cromartie}, {Crowter}, {Dai},
  {DeCesar}, {Demorest}, {Desvignes}, {Dolch}, {Drachler}, {Falxa}, {Ferrara},
  {Fiore}, {Fonseca}, {Gair}, {Garver-Daniels}, {Goncharov}, {Good}, {Graikou},
  {Guillemot}, {Guo}, {Hazboun}, {Hobbs}, {Hu}, {Islo}, {Janssen}, {Jennings},
  {Johnson}, {Jones}, {Kaiser}, {Kaplan}, {Karuppusamy}, {Keith}, {Kelley},
  {Kerr}, {Key}, {Kramer}, {Lam}, {Lamb}, {Lazio}, {Lee}, {Lentati}, {Liu},
  {Luo}, {Lynch}, {Lyne}, {Madison}, {Main}, {Manchester}, {McEwen}, {McKee},
  {McLaughlin}, {Mickaliger}, {Mingarelli}, {Ng}, {Nice}, {Os{\l}owski},
  {Parthasarathy}, {Pennucci}, {Perera}, {Perrodin}, {Petiteau}, {Pol},
  {Porayko}, {Possenti}, {Ransom}, {Ray}, {Reardon}, {Russell}, {Samajdar},
  {Sampson}, {Sanidas}, {Sarkissian}, {Schmitz}, {Schult}, {Sesana},
  {Shaifullah}, {Shannon}, {Shapiro-Albert}, {Siemens}, {Simon}, {Smith},
  {Speri}, {Spiewak}, {Stairs}, {Stappers}, {Stinebring}, {Swiggum}, {Taylor},
  {Theureau}, {Tiburzi}, {Vallisneri}, {van der Wateren}, {Vecchio},
  {Verbiest}, {Vigeland}, {Wahl}, {Wang}, {Wang}, {Wang}, {Witt}, {Zhang}, \&
  {Zhu}}]{IPTA_GWB_2022}
{Antoniadis}, J., {Arzoumanian}, Z., {Babak}, S., {$et~al$.} 2022, \mnras, 510,
  4873

\bibitem[{{Arimoto} {$et~al$.}(2021){Arimoto}, {Asada}, {Cherry}, {Fujii},
  {Fukazawa}, {Harada}, {Hayama}, {Hosokawa}, {Ioka}, {Itoh}, {Kanda},
  {Kawabata}, {Kawaguchi}, {Kawai}, {Kobayashi}, {Kohri}, {Koshio}, {Kotake},
  {Kumamoto}, {Machida}, {Matsufuru}, {Mihara}, {Mori}, {Morokuma},
  {Mukohyama}, {Nakano}, {Narikawa}, {Negoro}, {Nishizawa}, {Ohgami}, {Omukai},
  {Sakamoto}, {Sako}, {Sasada}, {Sekiguchi}, {Serino}, {Soda}, {Sugita},
  {Sumiyoshi}, {Susa}, {Suyama}, {Takahashi}, {Takahashi}, {Takiwaki},
  {Tanaka}, {Tanaka}, {Tanikawa}, {Tominaga}, {Uchikata}, {Utsumi}, {Vagins},
  {Yamada}, \& {Yoshida}}]{Arimoto21}
{Arimoto}, M., {Asada}, H., {Cherry}, M.~L., {$et~al$.} 2021, arXiv e-prints,
  arXiv:2104.02445

\bibitem[{{Arzoumanian} {$et~al$.}(2018){Arzoumanian}, {Baker}, {Brazier},
  {Burke-Spolaor}, {Chamberlin}, {Chatterjee}, {Christy}, {Cordes}, {Cornish},
  {Crawford}, {Thankful Cromartie}, {Crowter}, {DeCesar}, {Demorest}, {Dolch},
  {Ellis}, {Ferdman}, {Ferrara}, {Folkner}, {Fonseca}, {Garver-Daniels},
  {Gentile}, {Haas}, {Hazboun}, {Huerta}, {Islo}, {Jones}, {Jones}, {Kaplan},
  {Kaspi}, {Lam}, {Lazio}, {Levin}, {Lommen}, {Lorimer}, {Luo}, {Lynch},
  {Madison}, {McLaughlin}, {McWilliams}, {Mingarelli}, {Ng}, {Nice}, {Park},
  {Pennucci}, {Pol}, {Ransom}, {Ray}, {Rasskazov}, {Siemens}, {Simon},
  {Spiewak}, {Stairs}, {Stinebring}, {Stovall}, {Swiggum}, {Taylor},
  {Vallisneri}, {van Haasteren}, {Vigeland}, {Zhu}, \& {The NANOGrav
  Collaboration}}]{abb+18}
{Arzoumanian}, Z., {Baker}, P.~T., {Brazier}, A., {$et~al$.} 2018, ApJ, 859, 47

\bibitem[{Arzoumanian {$et~al$.}(2020)Arzoumanian, Baker, Blumer, B{\'{e}}csy,
  Brazier, Brook, Burke-Spolaor, Chatterjee, Chen, Cordes, Cornish, Crawford,
  Cromartie, Decesar, Demorest, Dolch, Ellis, Ferrara, Fiore, Fonseca,
  Garver-Daniels, Gentile, Good, Hazboun, Holgado, Islo, Jennings, Jones,
  Kaiser, Kaplan, Kelley, Key, Laal, Lam, Lazio, Lorimer, Luo, Lynch, Madison,
  McLaughlin, Mingarelli, Ng, Nice, Pennucci, Pol, Ransom, Ray, Shapiro-Albert,
  Siemens, Simon, Spiewak, Stairs, Stinebring, Stovall, Sun, Swiggum, Taylor,
  Turner, Vallisneri, Vigeland, Witt, \& {Nanograv
  Collaboration}}]{Arzoumanian2020_SGWB}
Arzoumanian, Z., Baker, P.~T., Blumer, H., {$et~al$.} 2020, The Astrophysical
  Journal Letters, 905, L34

\bibitem[{{Arzoumanian} {$et~al$.}(2021){Arzoumanian}, {Baker}, {Brazier},
  {Brook}, {Burke-Spolaor}, {Becsy}, {Charisi}, {Chatterjee}, {Cordes},
  {Cornish}, {Crawford}, {Cromartie}, {Decesar}, {Demorest}, {Dolch},
  {Elliott}, {Ellis}, {Ferrara}, {Fonseca}, {Garver-Daniels}, {Gentile},
  {Good}, {Hazboun}, {Islo}, {Jennings}, {Jones}, {Kaiser}, {Kaplan}, {Kelley},
  {Key}, {Lam}, {Lazio}, {Luo}, {Lynch}, {Ma}, {Madison}, {McLaughlin},
  {Mingarelli}, {Ng}, {Nice}, {Pennucci}, {Pol}, {Ransom}, {Ray},
  {Shapiro-Albert}, {Siemens}, {Simon}, {Spiewak}, {Stairs}, {Stinebring},
  {Stovall}, {Swiggum}, {Taylor}, {Vallisneri}, {Vigeland}, {Witt}, \&
  {Nanograv Collaboration}}]{N11_isolated}
{Arzoumanian}, Z., {Baker}, P.~T., {Brazier}, A., {$et~al$.} 2021, \apj, 914,
  121

\bibitem[{Bailes {$et~al$.}(2016)Bailes, Barr, Bhat, Brink, Buchner, Burgay,
  Camilo, Champion, Hessels, Jansseng, Jameson, Johnston, Karastergiou,
  Karuppusamy, Kaspi, Keith, Kramer, McLaughlin, Moodley, Os{\l}owski,
  Possenti, Ransom, Rasio, Sievers, Serylak, Stappers, Stairs, Theureau, van
  Straten, Weltevrede, \& Wex}]{Bailes2018_MeerTime}
Bailes, M., Barr, E., Bhat, N.~D., {$et~al$.} 2016, in Proceedings of Science
  (Trieste, Italy: Sissa Medialab), 011

\bibitem[{{Bailes} {$et~al$.}(2016){Bailes}, {Barr}, {Bhat}, {Brink},
  {Buchner}, {Burgay}, {Camilo}, {Champion}, {Hessels}, {Jameson}, {Johnston},
  {Karastergiou}, {Karuppusamy}, {Kaspi}, {Keith}, {Kramer}, {McLaughlin},
  {Moodley}, {Oslowski}, {Possenti}, {Ransom}, {Rasio}, {Sievers}, {Serylak},
  {Stappers}, {Stairs}, {Theureau}, {van Straten}, {Weltevrede}, \&
  {Wex}}]{SAPTA2016}
{Bailes}, M., {Barr}, E., {Bhat}, N.~D.~R., {$et~al$.} 2016, in MeerKAT
  Science: On the Pathway to the SKA, 11

\bibitem[{{Bailes} {$et~al$.}(2021){Bailes}, {Berger}, {Brady}, {Branchesi},
  {Danzmann}, {Evans}, {Holley-Bockelmann}, {Iyer}, {Kajita}, {Katsanevas},
  {Kramer}, {Lazzarini}, {Lehner}, {Losurdo}, {L{\"u}ck}, {McClelland},
  {McLaughlin}, {Punturo}, {Ransom}, {Raychaudhury}, {Reitze}, {Ricci},
  {Rowan}, {Saito}, {Sanders}, {Sathyaprakash}, {Schutz}, {Sesana}, {Shinkai},
  {Siemens}, {Shoemaker}, {Thorpe}, {van den Brand}, \& {Vitale}}]{Bailes21}
{Bailes}, M., {Berger}, B.~K., {Brady}, P.~R., {$et~al$.} 2021, Nature Reviews
  Physics, 3, 344

\bibitem[{{Banik} \& {Bandyopadhyay}(2017)}]{BB17}
{Banik}, S., \& {Bandyopadhyay}, D. 2017, arXiv e-prints, arXiv:1712.09760

\bibitem[{Bassa {$et~al$.}(2016)Bassa, Janssen, Karuppusamy, Kramer, Lee, Liu,
  McKee, Perrodin, Purver, Sanidas, Smits, \& Stappers}]{Bassa2016}
Bassa, C.~G., Janssen, G.~H., Karuppusamy, R., {$et~al$.} 2016, Monthly Notices
  of the Royal Astronomical Society, 456, 2196

\bibitem[{{Blanchet}(2014)}]{Blanchet2014}
{Blanchet}, L. 2014, Living Reviews in Relativity, 17, 2

\bibitem[{{Boran} {$et~al$.}(2018){Boran}, {Desai}, {Kahya}, \&
  {Woodard}}]{Boran18}
{Boran}, S., {Desai}, S., {Kahya}, E.~O., \& {Woodard}, R.~P. 2018, \prd, 97,
  041501

\bibitem[{{Burke-Spolaor} {$et~al$.}(2019){Burke-Spolaor}, {Taylor}, {Charisi},
  {Dolch}, {Hazboun}, {Holgado}, {Kelley}, {Lazio}, {Madison}, {McMann},
  {Mingarelli}, {Rasskazov}, {Siemens}, {Simon}, \&
  {Smith}}]{Burke-Spolaor2019}
{Burke-Spolaor}, S., {Taylor}, S.~R., {Charisi}, M., {$et~al$.} 2019, \aapr,
  27, 5

\bibitem[{Chen {$et~al$.}(2021)Chen, Caballero, Guo, Chalumeau, Liu,
  Shaifullah, Lee, Babak, Desvignes, Parthasarathy, Hu, van~der Wateren,
  Antoniadis, {Bak Nielsen}, Bassa, Berthereau, Burgay, Champion, Cognard,
  Falxa, Ferdman, Freire, Gair, Graikou, Guillemot, Jang, Janssen, Karuppusamy,
  Keith, Kramer, Liu, Lyne, Main, McKee, Mickaliger, Perera, Perrodin,
  Petiteau, Porayko, Possenti, Samajdar, Sanidas, Sesana, Speri, Stappers,
  Theureau, Tiburzi, Vecchio, Verbiest, Wang, Wang, \& Xu}]{Chen2021_EPTA_6psr}
Chen, S., Caballero, R.~N., Guo, Y.~J., {$et~al$.} 2021, Monthly Notices of the
  Royal Astronomical Society, 508, 4970

\bibitem[{Cho {$et~al$.}(2018)Cho, Gopakumar, Haney, \& Lee}]{Cho2018}
Cho, G., Gopakumar, A., Haney, M., \& Lee, H.~M. 2018, Physical Review D, 98,
  024039

\bibitem[{{Cohen}(2017)}]{Cohen2017}
{Cohen}, M. 2017, Galaxies, 5, 12

\bibitem[{{Damour} \& {Deruelle}(1986)}]{DD1986}
{Damour}, T., \& {Deruelle}, N. 1986, Ann. Inst. Henri Poincar{\'e} Phys.
  Th{\'e}or, 44, 263

\bibitem[{{Damour} {$et~al$.}(2004){Damour}, {Gopakumar}, \& {Iyer}}]{DGI2004}
{Damour}, T., {Gopakumar}, A., \& {Iyer}, B.~R. 2004, \prd, 70, 064028

\bibitem[{{De} \& {Gupta}(2016)}]{dg16}
{De}, K., \& {Gupta}, Y. 2016, Experimental Astronomy, 41, 67

\bibitem[{{Desvignes} {$et~al$.}(2016){Desvignes}, {Caballero}, {Lentati},
  {Verbiest}, {Champion}, {Stappers}, {Janssen}, {Lazarus}, {Os{\l}owski},
  {Babak}, {Bassa}, {Brem}, {Burgay}, {Cognard}, {Gair}, {Graikou},
  {Guillemot}, {Hessels}, {Jessner}, {Jordan}, {Karuppusamy}, {Kramer},
  {Lassus}, {Lazaridis}, {Lee}, {Liu}, {Lyne}, {McKee}, {Mingarelli},
  {Perrodin}, {Petiteau}, {Possenti}, {Purver}, {Rosado}, {Sanidas}, {Sesana},
  {Shaifullah}, {Smits}, {Taylor}, {Theureau}, {Tiburzi}, {van Haasteren}, \&
  {Vecchio}}]{dcl+16}
{Desvignes}, G., {Caballero}, R.~N., {Lentati}, L., {$et~al$.} 2016, MNRAS,
  458, 3341

\bibitem[{Desvignes {$et~al$.}(2016)Desvignes, Caballero, Lentati, Verbiest,
  Champion, Stappers, Janssen, Lazarus, Os{\l}owski, Babak, Bassa, Brem,
  Burgay, Cognard, Gair, Graikou, Guillemot, Hessels, Jessner, Jordan,
  Karuppusamy, Kramer, Lassus, Lazaridis, Lee, Liu, Lyne, McKee, Mingarelli,
  Perrodin, Petiteau, Possenti, Purver, Rosado, Sanidas, Sesana, Shaifullah,
  Smits, Taylor, Theureau, Tiburzi, {Van Haasteren}, \&
  Vecchio}]{Desvignes2016_EPTA}
Desvignes, G., Caballero, R.~N., Lentati, L., {$et~al$.} 2016, Monthly Notices
  of the Royal Astronomical Society, 458, 3341

\bibitem[{{Detweiler}(1979)}]{Detweiler1979}
{Detweiler}, S. 1979, \apj, 234, 1100

\bibitem[{{Dey} {$et~al$.}(2021){Dey}, {Valtonen}, {Gopakumar}, {Lico},
  {G{\'o}mez}, {Susobhanan}, {Komossa}, \& {Pihajoki}}]{Dey2021}
{Dey}, L., {Valtonen}, M.~J., {Gopakumar}, A., {$et~al$.} 2021, \mnras,
  arXiv:2103.05274

\bibitem[{{Dey} {$et~al$.}(2018){Dey}, {Valtonen}, {Gopakumar}, {Zola},
  {Hudec}, {Pihajoki}, {Ciprini}, {Matsumoto}, {Sadakane}, {Kidger}, {Nilsson},
  {Mikkola}, {Sillanp{\"a}{\"a}}, {Takalo}, {Lehto}, {Berdyugin}, {Piirola},
  {Jermak}, {Baliyan}, {Pursimo}, {Caton}, {Alicavus}, {Baransky}, {Blay},
  {Boumis}, {Boyd}, {Campas Torrent}, {Campos}, {Carrillo G{\'o}mez},
  {Chandra}, {Chavushyan}, {Dalessio}, {Debski}, {Drozdz}, {Er}, {Erdem},
  {Escartin P{\'e}rez}, {Fallah Ramazani}, {Filippenko}, {Gafton}, {Ganesh},
  {Garcia}, {Gazeas}, {Godunova}, {G{\'o}mez Pinilla}, {Gopinathan}, {Haislip},
  {Harmanen}, {Hurst}, {Jan{\'\i}k}, {Jelinek}, {Joshi}, {Kagitani},
  {Karjalainen}, {Kaur}, {Keel}, {Kouprianov}, {Kundera}, {Kurowski},
  {Kvammen}, {LaCluyze}, {Lee}, {Liakos}, {Lindfors}, {Lozano de Haro},
  {Mugrauer}, {Naves Nogues}, {Neely}, {Nelson}, {Ogloza}, {Okano},
  {Pajdosz-{\'S}mierciak}, {Pandey}, {Perri}, {Poyner}, {Provencal}, {Raj},
  {Reichart}, {Reinthal}, {Reynolds}, {Saario}, {Sadegi}, {Sakanoi}, {Salto
  Gonz{\'a}lez}, {Sameer}, {Schweyer}, {Simon}, {Siwak}, {Sold{\'a}n Alfaro},
  {Sonbas}, {Steele}, {Stocke}, {Strobl}, {Tomov}, {Tremosa Espasa}, {Valdes},
  {Valero P{\'e}rez}, {Verrecchia}, {Vasylenko}, {Webb}, {Yoneda}, {Zejmo},
  {Zheng}, \& {Zielinski}}]{Dey2018}
---. 2018, \apj, 866, 11

\bibitem[{{Dey} {$et~al$.}(2019){Dey}, {Gopakumar}, {Valtonen}, {Zola},
  {Susobhanan}, {Hudec}, {Pihajoki}, {Pursimo}, {Berdyugin}, {Piirola},
  {Ciprini}, {Nilsson}, {Jermak}, {Kidger}, \& {Komossa}}]{Dey2019}
{Dey}, L., {Gopakumar}, A., {Valtonen}, M., {$et~al$.} 2019, Universe, 5, 108

\bibitem[{{Donner} {$et~al$.}(2020){Donner}, {Verbiest}, {Tiburzi},
  {Os{\l}owski}, {K{\"u}nsem{\"o}ller}, {Bak Nielsen}, {Grie{\ss}meier},
  {Serylak}, {Kramer}, {Anderson}, {Wucknitz}, {Keane}, {Kondratiev}, {Sobey},
  {McKee}, {Bilous}, {Breton}, {Br{\"u}ggen}, {Ciardi}, {Hoeft}, {van Leeuwen},
  \& {Vocks}}]{dvt+20}
{Donner}, J.~Y., {Verbiest}, J.~P.~W., {Tiburzi}, C., {$et~al$.} 2020, \aap,
  644, A153

\bibitem[{Edwards {$et~al$.}(2006)Edwards, Hobbs, \&
  Manchester}]{Edwards2006_tempo2}
Edwards, R.~T., Hobbs, G.~B., \& Manchester, R.~N. 2006, Monthly Notices of the
  Royal Astronomical Society, 372, 1549

\bibitem[{{Einstein}(1918)}]{ein18}
{Einstein}, A. 1918, Sitzungsberichte der K{\"o}niglich Preu{\ss}ischen
  Akademie der Wissenschaften (Berlin), Seite 154-167.

\bibitem[{Estabrook \& Wahlquist(1975)}]{EstabrookWahlquist1975}
Estabrook, F.~B., \& Wahlquist, H.~D. 1975, General Relativity and Gravitation,
  6, 439

\bibitem[{{Fonseca} {$et~al$.}(2021){Fonseca}, {Cromartie}, {Pennucci}, {Ray},
  {Kirichenko}, {Ransom}, {Demorest}, {Stairs}, {Arzoumanian}, {Guillemot},
  {Parthasarathy}, {Kerr}, {Cognard}, {Baker}, {Blumer}, {Brook}, {DeCesar},
  {Dolch}, {Dong}, {Ferrara}, {Fiore}, {Garver-Daniels}, {Good}, {Jennings},
  {Jones}, {Kaspi}, {Lam}, {Lorimer}, {Luo}, {McEwen}, {McKee}, {McLaughlin},
  {McMann}, {Meyers}, {Naidu}, {Ng}, {Nice}, {Pol}, {Radovan},
  {Shapiro-Albert}, {Tan}, {Tendulkar}, {Swiggum}, {Wahl}, \&
  {Zhu}}]{Fonseca+2021ApJ}
{Fonseca}, E., {Cromartie}, H.~T., {Pennucci}, T.~T., {$et~al$.} 2021, \apjl,
  915, L12

\bibitem[{Foster \& Backer(1990)}]{FosterBacker1990}
Foster, R.~S., \& Backer, D.~C. 1990, The Astrophysical Journal, 361, 300

\bibitem[{Goncharov {$et~al$.}(2021)Goncharov, Shannon, Reardon, Hobbs, Zic,
  Bailes, Cury{\l}o, Dai, Kerr, Lower, Manchester, Mandow, Middleton, Miles,
  Parthasarathy, Thrane, Thyagarajan, Xue, Zhu, Cameron, Feng, Luo, Russell,
  Sarkissian, Spiewak, Wang, Wang, Zhang, \& Zhang}]{Goncharov2021}
Goncharov, B., Shannon, R.~M., Reardon, D.~J., {$et~al$.} 2021, The
  Astrophysical Journal Letters, 917, L19

\bibitem[{{Gupta} {$et~al$.}(2017){Gupta}, {Ajithkumar}, {Kale}, {Nayak},
  {Sabhapathy}, {Sureshkumar}, {Swami}, {Chengalur}, {Ghosh},
  {Ishwara-Chandra}, {Joshi}, {Kanekar}, {Lal}, \& {Roy}}]{gak+17}
{Gupta}, Y., {Ajithkumar}, B., {Kale}, H.~S., {$et~al$.} 2017, Current Science,
  113, 707

\bibitem[{{Hellings} \& {Downs}(1983)}]{HellingsDowns1983}
{Hellings}, R.~W., \& {Downs}, G.~S. 1983, \apjl, 265, L39

\bibitem[{Hobbs \& Dai(2017)}]{HobbsDai2017_PTA}
Hobbs, G., \& Dai, S. 2017, National Science Review, 4, 707

\bibitem[{Hobbs {$et~al$.}(2006)Hobbs, Edwards, \&
  Manchester}]{Hobbs2006_tempo2}
Hobbs, G.~B., Edwards, R.~T., \& Manchester, R.~N. 2006, Monthly Notices of the
  Royal Astronomical Society, 369, 655

\bibitem[{{Hodgson} {$et~al$.}(2017){Hodgson}, {Krichbaum}, {Marscher},
  {Jorstad}, {Rani}, {Marti-Vidal}, {Bach}, {Sanchez}, {Bremer}, {Lindqvist},
  {Uunila}, {Kallunki}, {Vicente}, {Fuhrmann}, {Angelakis}, {Karamanavis},
  {Myserlis}, {Nestoras}, {Chidiac}, {Sievers}, {Gurwell}, \&
  {Zensus}}]{Hodgson2017}
{Hodgson}, J.~A., {Krichbaum}, T.~P., {Marscher}, A.~P., {$et~al$.} 2017, \aap,
  597, A80

\bibitem[{Hotan {$et~al$.}(2004)Hotan, {Van Straten}, \&
  Manchester}]{Hotan2004_psrchive}
Hotan, A.~W., {Van Straten}, W., \& Manchester, R.~N. 2004, Publications of the
  Astronomical Society of Australia, 21, 302

\bibitem[{{Janssen} {$et~al$.}(2015){Janssen}, {Hobbs}, {McLaughlin}, {Bassa},
  {Deller}, {Kramer}, {Lee}, {Mingarelli}, {Rosado}, {Sanidas}, {Sesana},
  {Shao}, {Stairs}, {Stappers}, \& {Verbiest}}]{jhm+15}
{Janssen}, G., {Hobbs}, G., {McLaughlin}, M., {$et~al$.} 2015, in Advancing
  Astrophysics with the Square Kilometre Array (AASKA14), 37

\bibitem[{{Johnston} {$et~al$.}(2021){Johnston}, {Sobey}, {Dai}, {Keith},
  {Kerr}, {Manchester}, {Oswald}, {Parthasarathy}, {Shannon}, \&
  {Weltevrede}}]{Johnston+2021}
{Johnston}, S., {Sobey}, C., {Dai}, S., {$et~al$.} 2021, \mnras, 502, 1253

\bibitem[{{Joshi} {$et~al$.}(2018){Joshi}, {Arumugasamy}, {Bagchi},
  {Bandyopadhyay}, {Basu}, {Dhanda Batra}, {Bethapudi}, {Choudhary}, {De},
  {Dey}, {Gopakumar}, {Gupta}, {Krishnakumar}, {Maan}, {Manoharan}, {Naidu},
  {Nandi}, {Pathak}, {Surnis}, \& {Susobhanan}}]{jab+18}
{Joshi}, B.~C., {Arumugasamy}, P., {Bagchi}, M., {$et~al$.} 2018, Journal of
  Astrophysics and Astronomy, 39, 51

\bibitem[{{Keane}(2018)}]{k2018}
{Keane}, E.~F. 2018, in Pulsar Astrophysics the Next Fifty Years, ed.
  P.~{Weltevrede}, B.~B.~P. {Perera}, L.~L. {Preston}, \& S.~{Sanidas}, Vol.
  337, 158--164

\bibitem[{Kerr {$et~al$.}(2020)Kerr, Reardon, Hobbs, Shannon, Manchester, Dai,
  Russell, Zhang, van Straten, Os{\l}owski, Parthasarathy, Spiewak, Bailes,
  Bhat, Cameron, Coles, Dempsey, Deng, Goncharov, Kaczmarek, Keith, Lasky,
  Lower, Preisig, Sarkissian, Toomey, Wang, Wang, Zhang, \&
  Zhu}]{Kerr2020_PPTA_DR2}
Kerr, M., Reardon, D.~J., Hobbs, G., {$et~al$.} 2020, Publications of the
  Astronomical Society of Australia, 37, e020

\bibitem[{K\"onigsd\"orffer \& Gopakumar(2005)}]{KG2005}
K\"onigsd\"orffer, C., \& Gopakumar, A. 2005, Phys. Rev. D, 71, 024039

\bibitem[{{Kramer} \& {Stappers}(2015)}]{ks15}
{Kramer}, M., \& {Stappers}, B. 2015, in Advancing Astrophysics with the Square
  Kilometre Array (AASKA14), 36

\bibitem[{{Krishnakumar} {$et~al$.}(2021){Krishnakumar}, {Manoharan}, {Joshi},
  {Girgaonkar}, {Desai}, {Bagchi}, {Nobleson}, {Dey}, {Susobhanan}, {Susarla},
  {Surnis}, {Maan}, {Gopakumar}, {Basu}, {Batra}, {Choudhary}, {De}, {Gupta},
  {Naidu}, {Pathak}, {Singha}, \& {Prabu}}]{kmj+21}
{Krishnakumar}, M.~A., {Manoharan}, P.~K., {Joshi}, B.~C., {$et~al$.} 2021,
  A\&A, 651, A5

\bibitem[{{Laine} {$et~al$.}(2020){Laine}, {Dey}, {Valtonen}, {Gopakumar},
  {Zola}, {Komossa}, {Kidger}, {Pihajoki}, {G{\'o}mez}, {Caton}, {Ciprini},
  {Drozdz}, {Gazeas}, {Godunova}, {Haque}, {Hildebrand t}, {Hudec}, {Jermak},
  {Kong}, {Lehto}, {Liakos}, {Matsumoto}, {Mugrauer}, {Pursimo}, {Reichart},
  {Simon}, {Siwak}, \& {Sonbas}}]{Laine2020}
{Laine}, S., {Dey}, L., {Valtonen}, M., {$et~al$.} 2020, \apjl, 894, L1

\bibitem[{Lam {$et~al$.}(2018)Lam, Ellis, Grillo, Jones, Hazboun, Brook,
  Turner, Chatterjee, Cordes, Lazio, DeCesar, Arzoumanian, Blumer, Cromartie,
  Demorest, Dolch, Ferdman, Ferrara, Fonseca, Garver-Daniels, Gentile, Gupta,
  Lorimer, Lynch, Madison, McLaughlin, Ng, Nice, Pennucci, Ransom, Spiewak,
  Stairs, Stinebring, Stovall, Swiggum, Vigeland, \& Zhu}]{Lam2018}
Lam, M.~T., Ellis, J.~A., Grillo, G., {$et~al$.} 2018, The Astrophysical
  Journal, 861, 132

\bibitem[{{Lee}(2016)}]{CPTA2016}
{Lee}, K.~J. 2016, in Astronomical Society of the Pacific Conference Series,
  Vol. 502, Frontiers in Radio Astronomy and FAST Early Sciences Symposium
  2015, ed. L.~{Qain} \& D.~{Li}, 19

\bibitem[{Lentati {$et~al$.}(2016)Lentati, Shannon, Coles, Verbiest, van
  Haasteren, Ellis, Caballero, Manchester, Arzoumanian, Babak, Bassa, Bhat,
  Brem, Burgay, Burke-Spolaor, Champion, Chatterjee, Cognard, Cordes, Dai,
  Demorest, Desvignes, Dolch, Ferdman, Fonseca, Gair, Gonzalez, Graikou,
  Guillemot, Hessels, Hobbs, Janssen, Jones, Karuppusamy, Keith, Kerr, Kramer,
  Lam, Lasky, Lassus, Lazarus, Lazio, Lee, Levin, Liu, Lynch, Madison, McKee,
  McLaughlin, McWilliams, Mingarelli, Nice, Os{\l}owski, Pennucci, Perera,
  Perrodin, Petiteau, Possenti, Ransom, Reardon, Rosado, Sanidas, Sesana,
  Shaifullah, Siemens, Smits, Stairs, Stappers, Stinebring, Stovall, Swiggum,
  Taylor, Theureau, Tiburzi, Toomey, Vallisneri, van Straten, Vecchio, Wang,
  Wang, You, Zhu, \& Zhu}]{Lentati2016}
Lentati, L., Shannon, R.~M., Coles, W.~A., {$et~al$.} 2016, Monthly Notices of
  the Royal Astronomical Society, 458, 2161

\bibitem[{{Lorimer}(2011)}]{Lorimer2011_sigproc}
{Lorimer}, D.~R. 2011, {SIGPROC: Pulsar Signal Processing Programs},
  ascl:1107.016

\bibitem[{Maan {$et~al$.}(2021)Maan, van Leeuwen, \& Vohl}]{Maan2020_rfiClean}
Maan, Y., van Leeuwen, J., \& Vohl, D. 2021, Astronomy \& Astrophysics, 650,
  A80

\bibitem[{{Manchester} {$et~al$.}(2005){Manchester}, {Hobbs}, {Teoh}, \&
  {Hobbs}}]{mhth05}
{Manchester}, R.~N., {Hobbs}, G.~B., {Teoh}, A., \& {Hobbs}, M. 2005, \aj, 129,
  1993

\bibitem[{{Manchester} {$et~al$.}(2013){Manchester}, {Hobbs}, {Bailes},
  {Coles}, {van Straten}, {Keith}, {Shannon}, {Bhat}, {Brown}, {Burke-Spolaor},
  {Champion}, {Chaudhary}, {Edwards}, {Hampson}, {Hotan}, {Jameson}, {Jenet},
  {Kesteven}, {Khoo}, {Kocz}, {Maciesiak}, {Os{\l}owski}, {Ravi}, {Reynolds},
  {Sarkissian}, {Verbiest}, {Wen}, {Wilson}, {Yardley}, {Yan}, \&
  {You}}]{mhb+13}
{Manchester}, R.~N., {Hobbs}, G., {Bailes}, M., {$et~al$.} 2013, Proc. Astr.
  Soc. Aust., 30, 17

\bibitem[{{Mapelli}(2020)}]{Mapelli20}
{Mapelli}, M. 2020, Frontiers in Astronomy and Space Sciences, 7, 38

\bibitem[{{Memmesheimer} {$et~al$.}(2004){Memmesheimer}, {Gopakumar}, \&
  {Sch{\"a}fer}}]{MGS2004}
{Memmesheimer}, R.-M., {Gopakumar}, A., \& {Sch{\"a}fer}, G. 2004, \prd, 70,
  104011

\bibitem[{{Naidu} {$et~al$.}(2015){Naidu}, {Joshi}, {Manoharan}, \&
  {Krishnakumar}}]{njmk15}
{Naidu}, A., {Joshi}, B.~C., {Manoharan}, P.~K., \& {Krishnakumar}, M.~A. 2015,
  Experimental Astronomy, 39, 319

\bibitem[{{Nice} {$et~al$.}(2015){Nice}, {Demorest}, {Stairs}, {Manchester},
  {Taylor}, {Peters}, {Weisberg}, {Irwin}, {Wex}, \& {Huang}}]{Nice+2015ascl}
{Nice}, D., {Demorest}, P., {Stairs}, I., {$et~al$.} 2015, {Tempo: Pulsar
  timing data analysis}, ascl:1509.002

\bibitem[{{Nobleson} {$et~al$.}(2022){Nobleson}, {Agarwal}, {Girgaonkar},
  {Pandian}, {Joshi}, {Krishnakumar}, {Susobhanan}, {Desai}, {Prabu},
  {Bathula}, {Pennucci}, {Banik}, {Bagchi}, {Batra}, {Choudhary}, {Dandapat},
  {Dey}, {Gupta}, {Hisano}, {Kato}, {Kharbanda}, {Kikunaga}, {Kolhe}, {Maan},
  {Marmat}, {Arumugam}, {Manoharan}, {Pathak}, {Singha}, {Surnis}, {Susarla},
  \& {Takahashi}}]{Nobleson+2021}
{Nobleson}, K., {Agarwal}, N., {Girgaonkar}, R., {$et~al$.} 2022, \mnras,
  arXiv:2112.06908

\bibitem[{{O'Neill} {$et~al$.}(2022){O'Neill}, {Kiehlmann}, {Readhead},
  {Aller}, {Blandford}, {Liodakis}, {Lister}, {Mr{\'o}z}, {O'Dea}, {Pearson},
  {Ravi}, {Vallisneri}, {Cleary}, {Graham}, {Grainge}, {Hodges}, {Hovatta},
  {L{\"a}hteenm{\"a}ki}, {Lamb}, {Lazio}, {Max-Moerbeck}, {Pavlidou}, {Prince},
  {Reeves}, {Tornikoski}, {Vergara de la Parra}, \& {Zensus}}]{ONeill2022}
{O'Neill}, S., {Kiehlmann}, S., {Readhead}, A.~C.~S., {$et~al$.} 2022, \apjl,
  926, L35

\bibitem[{{Pennucci}(2019)}]{Pennucci2019ApJ}
{Pennucci}, T.~T. 2019, \apj, 871, 34

\bibitem[{{Pennucci} {$et~al$.}(2014){Pennucci}, {Demorest}, \&
  {Ransom}}]{Pennucci+2014}
{Pennucci}, T.~T., {Demorest}, P.~B., \& {Ransom}, S.~M. 2014, \apj, 790, 93

\bibitem[{{Pennucci} {$et~al$.}(2016){Pennucci}, {Demorest}, \&
  {Ransom}}]{2016ascl.soft06013P}
---. 2016, {Pulse Portraiture: Pulsar timing}, ascl:1606.013

\bibitem[{{Perera} {$et~al$.}(2019){Perera}, {DeCesar}, {Demorest}, {Kerr},
  {Lentati}, {Nice}, {Os{\l}owski}, {Ransom}, {Keith}, {Arzoumanian}, {Bailes},
  {Baker}, {Bassa}, {Bhat}, {Brazier}, {Burgay}, {Burke-Spolaor}, {Caballero},
  {Champion}, {Chatterjee}, {Chen}, {Cognard}, {Cordes}, {Crowter}, {Dai},
  {Desvignes}, {Dolch}, {Ferdman}, {Ferrara}, {Fonseca}, {Goldstein},
  {Graikou}, {Guillemot}, {Hazboun}, {Hobbs}, {Hu}, {Islo}, {Janssen},
  {Karuppusamy}, {Kramer}, {Lam}, {Lee}, {Liu}, {Luo}, {Lyne}, {Manchester},
  {McKee}, {McLaughlin}, {Mingarelli}, {Parthasarathy}, {Pennucci}, {Perrodin},
  {Possenti}, {Reardon}, {Russell}, {Sanidas}, {Sesana}, {Shaifullah},
  {Shannon}, {Siemens}, {Simon}, {Spiewak}, {Stairs}, {Stappers}, {Swiggum},
  {Taylor}, {Theureau}, {Tiburzi}, {Vallisneri}, {Vecchio}, {Wang}, {Zhang},
  {Zhang}, {Zhu}, \& {Zhu}}]{IPTA_DR2}
{Perera}, B.~B.~P., {DeCesar}, M.~E., {Demorest}, P.~B., {$et~al$.} 2019,
  \mnras, 490, 4666

\bibitem[{{Phinney}(2001)}]{Phinney2001}
{Phinney}, E.~S. 2001, arXiv e-prints, astro

\bibitem[{{Pol} {$et~al$.}(2021){Pol}, {Taylor}, {Kelley}, {Vigeland}, {Simon},
  {Chen}, {Arzoumanian}, {Baker}, {B{\'e}csy}, {Brazier}, {Brook},
  {Burke-Spolaor}, {Chatterjee}, {Cordes}, {Cornish}, {Crawford}, {Thankful
  Cromartie}, {Decesar}, {Demorest}, {Dolch}, {Ferrara}, {Fiore}, {Fonseca},
  {Garver-Daniels}, {Good}, {Hazboun}, {Jennings}, {Jones}, {Kaiser}, {Kaplan},
  {Shapiro Key}, {Lam}, {Lazio}, {Luo}, {Lynch}, {Madison}, {McEwen},
  {McLaughlin}, {Mingarelli}, {Ng}, {Nice}, {Pennucci}, {Ransom}, {Ray},
  {Shapiro-Albert}, {Siemens}, {Stairs}, {Stinebring}, {Swiggum}, {Vallisneri},
  {Wahl}, {Witt}, \& {Nanograv Collaboration}}]{ptk+21}
{Pol}, N.~S., {Taylor}, S.~R., {Kelley}, L.~Z., {$et~al$.} 2021, \apjl, 911,
  L34

\bibitem[{{Ransom}(2011)}]{Ransom2011_PRESTO}
{Ransom}, S. 2011, {PRESTO: PulsaR Exploration and Search TOolkit},
  ascl:1107.017

\bibitem[{Reddy {$et~al$.}(2017)Reddy, Kudale, Gokhale, Halagalli, Raskar, De,
  Gnanaraj, Ajith~Kumar, \& Gupta}]{rkg+17}
Reddy, S.~H., Kudale, S., Gokhale, U., {$et~al$.} 2017, Journal of Astronomical
  Instrumentation, 06, 1641011

\bibitem[{{Sathyaprakash} \& {Schutz}(2009)}]{SS2009}
{Sathyaprakash}, B.~S., \& {Schutz}, B.~F. 2009, Living Reviews in Relativity,
  12, 2

\bibitem[{{Sazhin}(1978)}]{S78}
{Sazhin}, M.~V. 1978, \sovast, 22, 36

\bibitem[{{Soares-Santos} {$et~al$.}(2019){Soares-Santos}, {Palmese},
  {Hartley}, {Annis}, {Garcia-Bellido}, {Lahav}, {Doctor}, {Fishbach}, {Holz},
  {Lin}, {Pereira}, {Garcia}, {Herner}, {Kessler}, {Peiris}, {Sako}, {Allam},
  {Brout}, {Carnero Rosell}, {Chen}, {Conselice}, {deRose}, {deVicente},
  {Diehl}, {Gill}, {Gschwend}, {Sevilla-Noarbe}, {Tucker}, {Wechsler},
  {Berger}, {Cowperthwaite}, {Metzger}, {Williams}, {Abbott}, {Abdalla},
  {Avila}, {Bechtol}, {Bertin}, {Brooks}, {Buckley-Geer}, {Burke}, {Carrasco
  Kind}, {Carretero}, {Castander}, {Crocce}, {Cunha}, {D'Andrea}, {da Costa},
  {Davis}, {Desai}, {Doel}, {Drlica-Wagner}, {Eifler}, {Evrard}, {Flaugher},
  {Fosalba}, {Frieman}, {Gaztanaga}, {Gerdes}, {Gruen}, {Gruendl}, {Gutierrez},
  {Hollowood}, {Hoyle}, {James}, {Jeltema}, {Kuehn}, {Kuropatkin}, {Li},
  {Lima}, {Maia}, {Marshall}, {Menanteau}, {Miquel}, {Neilsen}, {Ogando},
  {Plazas}, {Romer}, {Roodman}, {Sanchez}, {Scarpine}, {Schindler},
  {Schubnell}, {Serrano}, {Smith}, {Smith}, {Sobreira}, {Suchyta}, {Swanson},
  {Tarle}, {Thomas}, {Walker}, {Wester}, {Zuntz}, {DES Collaboration},
  {Abbott}, {Abbott}, {Abbott}, {Abraham}, {Acernese}, {Ackley}, {Adams},
  {Adhikari}, {Adya}, {Affeldt}, {Agathos}, {Agatsuma}, {Aggarwal}, {Aguiar},
  {Aiello}, {Ain}, {Ajith}, {Allen}, {Allocca}, {Aloy}, {Altin}, {Amato},
  {Ananyeva}, {Anderson}, {Anderson}, {Angelova}, {Appert}, {Arai}, {Araya},
  {Areeda}, {Ar{\`e}ne}, {Ascenzi}, {Ashton}, {Aston}, {Astone}, {Aubin},
  {Aufmuth}, {AultONeal}, {Austin}, {Avendano}, {Avila-Alvarez}, {Babak},
  {Bacon}, {Badaracco}, {Bader}, {Bae}, {Baker}, {Baldaccini}, {Ballardin},
  {Ballmer}, {Banagiri}, {Barayoga}, {Barclay}, {Barish}, {Barker}, {Barkett},
  {Barnum}, {Barone}, {Barr}, {Barsotti}, {Barsuglia}, {Barta}, {Bartlett},
  {Bartos}, {Bassiri}, {Basti}, {Bawaj}, {Bayley}, {Bazzan}, {B{\'e}csy},
  {Bejger}, {Bell}, {Beniwal}, {Bergmann}, {Bernuzzi}, {Bero}, {Berry},
  {Bersanetti}, {Bertolini}, {Betzwieser}, {Bhandare}, {Bidler}, {Bilenko},
  {Bilgili}, {Billingsley}, {Birch}, {Birney}, {Birnholtz}, {Biscans},
  {Biscoveanu}, {Bisht}, {Bitossi}, {Blackburn}, {Blair}, {Blair}, {Blair},
  {Bloemen}, {Bode}, {Boer}, {Boetzel}, {Bogaert}, {Bondu}, {Bonilla},
  {Bonnand}, {Booker}, {Boom}, {Booth}, {Bork}, {Boschi}, {Bose}, {Bossie},
  {Bossilkov}, {Bosveld}, {Bouffanais}, {Bozzi}, {Bradaschia}, {Brady},
  {Bramley}, {Branchesi}, {Brau}, {Briant}, {Briggs}, {Brighenti}, {Brillet},
  {Brinkmann}, {Brockill}, {Brooks}, {Brown}, {Brunett}, {Buikema}, {Bulik},
  {Bulten}, {Buonanno}, {Buskulic}, {Buy}, {Byer}, {Cabero}, {Cadonati},
  {Cagnoli}, {Cahillane}, {Calder{\'o}n Bustillo}, {Callister}, {Calloni},
  {Camp}, {Campbell}, {Cannon}, {Cao}, {Cao}, {Capocasa}, {Carbognani},
  {Caride}, {Carney}, {Carullo}, {Casanueva Diaz}, {Casentini}, {Caudill},
  {Cavagli{\`a}}, {Cavalieri}, {Cella}, {Cerd{\'a}-Dur{\'a}n}, {Cerretani},
  {Cesarini}, {Chaibi}, {Chakravarti}, {Chamberlin}, {Chan}, {Chao},
  {Charlton}, {Chase}, {Chassande-Mottin}, {Chatterjee}, {Chaturvedi},
  {Chatziioannou}, {Cheeseboro}, {Chen}, {Chen}, {Cheng}, {Cheong}, {Chia},
  {Chincarini}, {Chiummo}, {Cho}, {Cho}, {Cho}, {Christensen}, {Chu}, {Chua},
  {Chung}, {Chung}, {Ciani}, {Ciobanu}, {Ciolfi}, {Cipriano}, {Cirone},
  {Clara}, {Clark}, {Clearwater}, {Cleva}, {Cocchieri}, {Coccia}, {Cohadon},
  {Colgan}, {Colleoni}, {Collette}, {Collins}, {Cominsky}, {Constancio},
  {Conti}, {Cooper}, {Corban}, {Corbitt}, {Cordero-Carri{\'o}n}, {Corley},
  {Cornish}, {Corsi}, {Cortese}, {Costa}, {Cotesta}, {Coughlin}, {Coughlin},
  {Coulon}, {Countryman}, {Couvares}, {Covas}, {Cowan}, {Coward}, {Cowart},
  {Coyne}, {Coyne}, {Creighton}, {Creighton}, {Cripe}, {Croquette}, {Crowder},
  {Cullen}, {Cumming}, {Cunningham}, {Cuoco}, {Dal Canton}, {D{\'a}lya},
  {Danilishin}, {D'Antonio}, {Danzmann}, {Dasgupta}, {Da Silva Costa},
  {Datrier}, {Dattilo}, {Dave}, {Davis}, {Daw}, {DeBra}, {Deenadayalan},
  {Degallaix}, {De Laurentis}, {Del{\'e}glise}, {Del Pozzo}, {DeMarchi},
  {Demos}, {Dent}, {De Pietri}, {Derby}, {De Rosa}, {De Rossi}, {DeSalvo}, {de
  Varona}, {Dhurandhar}, {D{\'\i}az}, {Dietrich}, {Di Fiore}, {Di Giovanni},
  {Di Girolamo}, {Di Lieto}, {Ding}, {Di Pace}, {Di Palma}, {Di Renzo},
  {Dmitriev}, {Donovan}, {Dooley}, {Doravari}, {Dorrington}, {Downes}, {Drago},
  {Driggers}, {Du}, {Dupej}, {Dwyer}, {Easter}, {Edo}, {Edwards}, {Effler},
  {Ehrens}, {Eichholz}, {Eikenberry}, {Eisenmann}, {Eisenstein}, {Estelles},
  {Estevez}, {Etienne}, {Etzel}, {Evans}, {Evans}, {Fafone}, {Fair},
  {Fairhurst}, {Fan}, {Farinon}, {Farr}, {Farr}, {Fauchon-Jones}, {Favata},
  {Fays}, {Fazio}, {Fee}, {Feicht}, {Fejer}, {Feng}, {Fernandez-Galiana},
  {Ferrante}, {Ferreira}, {Ferreira}, {Ferrini}, {Fidecaro}, {Fiori},
  {Fiorucci}, {Fisher}, {Fishner}, {Fitz-Axen}, {Flaminio}, {Fletcher},
  {Flynn}, {Fong}, {Font}, {Forsyth}, {Fournier}, {Frasca}, {Frasconi}, {Frei},
  {Freise}, {Frey}, {Fritschel}, {Frolov}, {Fulda}, {Fyffe}, {Gabbard},
  {Gadre}, {Gaebel}, {Gair}, {Gammaitoni}, {Ganija}, {Gaonkar}, {Garcia},
  {Garc{\'\i}a-Quir{\'o}s}, {Garufi}, {Gateley}, {Gaudio}, {Gaur}, {Gayathri},
  {Gemme}, {Genin}, {Gennai}, {George}, {George}, {Gergely}, {Germain},
  {Ghonge}, {Ghosh}, {Ghosh}, {Ghosh}, {Giacomazzo}, {Giaime}, {Giardina},
  {Giazotto}, {Gill}, {Giordano}, {Glover}, {Godwin}, {Goetz}, {Goetz},
  {Goncharov}, {Gonz{\'a}lez}, {Gonzalez Castro}, {Gopakumar}, {Gorodetsky},
  {Gossan}, {Gosselin}, {Gouaty}, {Grado}, {Graef}, {Granata}, {Grant}, {Gras},
  {Grassia}, {Gray}, {Gray}, {Greco}, {Green}, {Green}, {Gretarsson}, {Groot},
  {Grote}, {Grunewald}, {Guidi}, {Gulati}, {Guo}, {Gupta}, {Gupta},
  {Gustafson}, {Gustafson}, {Haegel}, {Halim}, {Hall}, {Hall}, {Hamilton},
  {Hammond}, {Haney}, {Hanke}, {Hanks}, {Hanna}, {Hannuksela}, {Hanson},
  {Hardwick}, {Haris}, {Harms}, {Harry}, {Harry}, {Haster}, {Haughian},
  {Hayes}, {Healy}, {Heidmann}, {Heintze}, {Heitmann}, {Hemming}, {Hendry},
  {Heng}, {Hennig}, {Heptonstall}, {Hernandez Vivanco}, {Heurs}, {Hild},
  {Hinderer}, {Hoak}, {Hochheim}, {Hofman}, {Holgado}, {Holland}, {Holt},
  {Hopkins}, {Horst}, {Hough}, {Howell}, {Hoy}, {Hreibi}, {Huerta}, {Hughey},
  {Hulko}, {Husa}, {Huttner}, {Huynh-Dinh}, {Idzkowski}, {Iess}, {Ingram},
  {Inta}, {Intini}, {Irwin}, {Isa}, {Isac}, {Isi}, {Iyer}, {Izumi}, {Jacqmin},
  {Jadhav}, {Jani}, {Janthalur}, {Jaranowski}, {Jenkins}, {Jiang}, {Johnson},
  {Jones}, {Jones}, {Jones}, {Jonker}, {Ju}, {Junker}, {Kalaghatgi},
  {Kalogera}, {Kamai}, {Kandhasamy}, {Kang}, {Kanner}, {Kapadia}, {Karki},
  {Karvinen}, {Kashyap}, {Kasprzack}, {Katsanevas}, {Katsavounidis}, {Katzman},
  {Kaufer}, {Kawabe}, {Keerthana}, {K{\'e}f{\'e}lian}, {Keitel}, {Kennedy},
  {Key}, {Khalili}, {Khan}, {Khan}, {Khan}, {Khan}, {Khazanov}, {Khursheed},
  {Kijbunchoo}, {Kim}, {Kim}, {Kim}, {Kim}, {Kim}, {Kim}, {Kimball}, {King},
  {King}, {Kinley-Hanlon}, {Kirchhoff}, {Kissel}, {Kleybolte}, {Klika},
  {Klimenko}, {Knowles}, {Koch}, {Koehlenbeck}, {Koekoek}, {Koley},
  {Kondrashov}, {Kontos}, {Koper}, {Korobko}, {Korth}, {Kowalska}, {Kozak},
  {Kringel}, {Krishnendu}, {Kr{\'o}lak}, {Kuehn}, {Kumar}, {Kumar}, {Kumar},
  {Kumar}, {Kuo}, {Kutynia}, {Kwang}, {Lackey}, {Lai}, {Lam}, {Landry}, {Lane},
  {Lang}, {Lange}, {Lantz}, {Lanza}, {Lasky}, {Laxen}, {Lazzarini}, {Lazzaro},
  {Leaci}, {Leavey}, {Lecoeuche}, {Lee}, {Lee}, {Lee}, {Lee}, {Lee}, {Lee},
  {Lehmann}, {Lenon}, {Letendre}, {Levin}, {Li}, {Li}, {Li}, {Li}, {Lin},
  {Linde}, {Linker}, {Littenberg}, {Liu}, {Liu}, {Lo}, {Lockerbie}, {London},
  {Longo}, {Lorenzini}, {Loriette}, {Lormand}, {Losurdo}, {Lough}, {Lousto},
  {Lovelace}, {Lower}, {L{\"u}ck}, {Lumaca}, {Lundgren}, {Lynch}, {Ma},
  {Macas}, {Macfoy}, {MacInnis}, {Macleod}, {Macquet}, {Maga{\~n}a Hernandez},
  {Maga{\~n}a-Sandoval}, {Maga{\~n}a Zertuche}, {Magee}, {Majorana},
  {Maksimovic}, {Malik}, {Man}, {Mandic}, {Mangano}, {Mansell}, {Manske},
  {Mantovani}, {Marchesoni}, {Marion}, {M{\'a}rka}, {M{\'a}rka}, {Markakis},
  {Markosyan}, {Markowitz}, {Maros}, {Marquina}, {Marsat}, {Martelli},
  {Martin}, {Martin}, {Martynov}, {Mason}, {Massera}, {Masserot}, {Massinger},
  {Masso-Reid}, {Mastrogiovanni}, {Matas}, {Matichard}, {Matone}, {Mavalvala},
  {Mazumder}, {McCann}, {McCarthy}, {McClelland}, {McCormick}, {McCuller},
  {McGuire}, {McIver}, {McManus}, {McRae}, {McWilliams}, {Meacher}, {Meadors},
  {Mehmet}, {Mehta}, {Meidam}, {Melatos}, {Mendell}, {Mercer}, {Mereni},
  {Merilh}, {Merzougui}, {Meshkov}, {Messenger}, {Messick}, {Metzdorff},
  {Meyers}, {Miao}, {Michel}, {Middleton}, {Mikhailov}, {Milano}, {Miller},
  {Miller}, {Millhouse}, {Mills}, {Milovich-Goff}, {Minazzoli}, {Minenkov},
  {Mishkin}, {Mishra}, {Mistry}, {Mitra}, {Mitrofanov}, {Mitselmakher},
  {Mittleman}, {Mo}, {Moffa}, {Mogushi}, {Mohapatra}, {Montani}, {Moore},
  {Moraru}, {Moreno}, {Morisaki}, {Mours}, {Mow-Lowry}, {Mukherjee},
  {Mukherjee}, {Mukherjee}, {Mukund}, {Mullavey}, {Munch}, {Mu{\~n}iz},
  {Muratore}, {Murray}, {Nardecchia}, {Naticchioni}, {Nayak}, {Neilson},
  {Nelemans}, {Nelson}, {Nery}, {Neunzert}, {Ng}, {Ng}, {Nguyen}, {Nichols},
  {Nissanke}, {Nocera}, {North}, {Nuttall}, {Obergaulinger}, {Oberling},
  {O'Brien}, {O'Dea}, {Ogin}, {Oh}, {Oh}, {Ohme}, {Ohta}, {Okada}, {Oliver},
  {Oppermann}, {Oram}, {O'Reilly}, {Ormiston}, {Ortega}, {O'Shaughnessy},
  {Ossokine}, {Ottaway}, {Overmier}, {Owen}, {Pace}, {Pagano}, {Page}, {Pai},
  {Pai}, {Palamos}, {Palashov}, {Palomba}, {Pal-Singh}, {Pan}, {Pang}, {Pang},
  {Pankow}, {Pannarale}, {Pant}, {Paoletti}, {Paoli}, {Parida}, {Parker},
  {Pascucci}, {Pasqualetti}, {Passaquieti}, {Passuello}, {Patil}, {Patricelli},
  {Pearlstone}, {Pedersen}, {Pedraza}, {Pedurand}, {Pele}, {Penn}, {Perez},
  {Perreca}, {Pfeiffer}, {Phelps}, {Phukon}, {Piccinni}, {Pichot},
  {Piergiovanni}, {Pillant}, {Pinard}, {Pirello}, {Pitkin}, {Poggiani}, {Pong},
  {Ponrathnam}, {Popolizio}, {Porter}, {Powell}, {Prajapati}, {Prasad},
  {Prasai}, {Prasanna}, {Pratten}, {Prestegard}, {Privitera}, {Prodi},
  {Prokhorov}, {Puncken}, {Punturo}, {Puppo}, {P{\"u}rrer}, {Qi}, {Quetschke},
  {Quinonez}, {Quintero}, {Quitzow-James}, {Radkins}, {Radulescu}, {Raffai},
  {Raja}, {Rajan}, {Rajbhandari}, {Rakhmanov}, {Ramirez}, {Ramos-Buades},
  {Rana}, {Rao}, {Rapagnani}, {Raymond}, {Razzano}, {Read}, {Regimbau}, {Rei},
  {Reid}, {Reitze}, {Ren}, {Ricci}, {Richardson}, {Richardson}, {Ricker},
  {Riles}, {Rizzo}, {Robertson}, {Robie}, {Rocchi}, {Rolland}, {Rollins},
  {Roma}, {Romanelli}, {Romano}, {Romel}, {Romie}, {Rose}, {Rosi{\'n}ska},
  {Rosofsky}, {Ross}, {Rowan}, {R{\"u}diger}, {Ruggi}, {Rutins}, {Ryan},
  {Sachdev}, {Sadecki}, {Sakellariadou}, {Salconi}, {Saleem}, {Samajdar},
  {Sammut}, {Sanchez}, {Sanchez}, {Sanchis-Gual}, {Sandberg}, {Sanders},
  {Santiago}, {Sarin}, {Sassolas}, {Saulson}, {Sauter}, {Savage}, {Schale},
  {Scheel}, {Scheuer}, {Schmidt}, {Schnabel}, {Schofield}, {Sch{\"o}nbeck},
  {Schreiber}, {Schulte}, {Schutz}, {Schwalbe}, {Scott}, {Scott}, {Seidel},
  {Sellers}, {Sengupta}, {Sennett}, {Sentenac}, {Sequino}, {Sergeev},
  {Shaddock}, {Shaffer}, {Shahriar}, {Shaner}, {Shao}, {Sharma}, {Shawhan},
  {Shen}, {Shink}, {Shoemaker}, {Shoemaker}, {ShyamSundar}, {Siellez},
  {Sieniawska}, {Sigg}, {Silva}, {Singer}, {Singh}, {Singhal}, {Sintes},
  {Sitmukhambetov}, {Skliris}, {Slagmolen}, {Slaven-Blair}, {Smith}, {Smith},
  {Somala}, {Son}, {Sorazu}, {Sorrentino}, {Souradeep}, {Sowell}, {Spencer},
  {Srivastava}, {Srivastava}, {Staats}, {Stachie}, {Standke}, {Steer},
  {Steinke}, {Steinlechner}, {Steinlechner}, {Steinmeyer}, {Stevenson},
  {Stocks}, {Stone}, {Stops}, {Strain}, {Stratta}, {Strigin}, {Strunk},
  {Sturani}, {Stuver}, {Sudhir}, {Summerscales}, {Sun}, {Sunil}, {Sur},
  {Suresh}, {Sutton}, {Swinkels}, {Szczepa{\'n}czyk}, {Tacca}, {Tait},
  {Talbot}, {Talukder}, {Tanner}, {T{\'a}pai}, {Taracchini}, {Tasson},
  {Taylor}, {Thies}, {Thomas}, {Thomas}, {Thondapu}, {Thorne}, {Thrane},
  {Tiwari}, {Tiwari}, {Tiwari}, {Toland}, {Tonelli}, {Tornasi},
  {Torres-Forn{\'e}}, {Torrie}, {T{\"o}yr{\"a}}, {Travasso}, {Traylor},
  {Tringali}, {Trovato}, {Trozzo}, {Trudeau}, {Tsang}, {Tse}, {Tso}, {Tsukada},
  {Tsuna}, {Tuyenbayev}, {Ueno}, {Ugolini}, {Unnikrishnan}, {Urban}, {Usman},
  {Vahlbruch}, {Vajente}, {Valdes}, {van Bakel}, {van Beuzekom}, {van den
  Brand}, {Van Den Broeck}, {Vander-Hyde}, {van Heijningen}, {van der Schaaf},
  {van Veggel}, {Vardaro}, {Varma}, {Vass}, {Vas{\'u}th}, {Vecchio},
  {Vedovato}, {Veitch}, {Veitch}, {Venkateswara}, {Venugopalan}, {Verkindt},
  {Vetrano}, {Vicer{\'e}}, {Viets}, {Vine}, {Vinet}, {Vitale}, {Vo}, {Vocca},
  {Vorvick}, {Vyatchanin}, {Wade}, {Wade}, {Wade}, {Walet}, {Walker},
  {Wallace}, {Walsh}, {Wang}, {Wang}, {Wang}, {Wang}, {Wang}, {Ward}, {Warden},
  {Warner}, {Was}, {Watchi}, {Weaver}, {Wei}, {Weinert}, {Weinstein}, {Weiss},
  {Wellmann}, {Wen}, {Wessel}, {We{\ss}els}, {Westhouse}, {Wette}, {Whelan},
  {Whiting}, {Whittle}, {Wilken}, {Williams}, {Williamson}, {Willis}, {Willke},
  {Wimmer}, {Winkler}, {Wipf}, {Wittel}, {Woan}, {Woehler}, {Wofford},
  {Worden}, {Wright}, {Wu}, {Wysocki}, {Xiao}, {Yamamoto}, {Yancey}, {Yang},
  {Yap}, {Yazback}, {Yeeles}, {Yu}, {Yu}, {Yuen}, {Yvert}, {Zadro{\.z}ny},
  {Zanolin}, {Zelenova}, {Zendri}, {Zevin}, {Zhang}, {Zhang}, {Zhang}, {Zhao},
  {Zhou}, {Zhou}, {Zhu}, {Zimmerman}, {Zucker}, {Zweizig}, {LIGO Scientific
  Collaboration}, \& {Virgo Collaboration}}]{DrkSiren19}
{Soares-Santos}, M., {Palmese}, A., {Hartley}, W., {$et~al$.} 2019, \apjl, 876,
  L7

\bibitem[{Susobhanan {$et~al$.}(2020)Susobhanan, Gopakumar, Hobbs, \&
  Taylor}]{Susobhanan2020}
Susobhanan, A., Gopakumar, A., Hobbs, G., \& Taylor, S.~R. 2020, Physical
  Review D, 101, 043022

\bibitem[{{Susobhanan} {$et~al$.}(2021){Susobhanan}, {Maan}, {Joshi}, {Prabu},
  {Desai}, {Nobleson}, {Susarla}, {Girgaonkar}, {Dey}, {Batra}, {Gupta},
  {Gopakumar}, {Bagchi}, {Basu}, {Bethapudi}, {Choudhary}, {De},
  {Krishnakumar}, {Manoharan}, {Naidu}, {Pathak}, {Singha}, \&
  {Surnis}}]{smj+21}
{Susobhanan}, A., {Maan}, Y., {Joshi}, B.~C., {$et~al$.} 2021, Proc. Astr. Soc.
  Aust., 38, e017

\bibitem[{Swarup {$et~al$.}(1991)Swarup, Ananthakrishnan, Kapahi, Rao,
  Subrahmanya, \& Kulkarni}]{Swarup1991}
Swarup, G., Ananthakrishnan, S., Kapahi, V.~K., {$et~al$.} 1991, Current
  Science, 60, 95

\bibitem[{Swarup {$et~al$.}(1971)Swarup, SARMA, JOSHI, KAPAHI, BAGRI, DAMLE,
  ANANTHAKRISHNAN, BALASUBRAMANIAN, BHAVE, \& SINHA}]{Swarup1971}
Swarup, G., SARMA, N. V.~G., JOSHI, M.~N., {$et~al$.} 1971, Nature Physical
  Science, 230, 185

\bibitem[{{Taylor}(1992)}]{Taylor1992}
{Taylor}, J.~H. 1992, Philosophical Transactions of the Royal Society of London
  Series A, 341, 117

\bibitem[{{The LIGO Scientific Collaboration}
  {$et~al$.}(2021{\natexlab{a}}){The LIGO Scientific Collaboration}, {the Virgo
  Collaboration}, {the KAGRA Collaboration}, {Abbott}, {Abbott}, {Acernese},
  {Ackley}, {Adams}, {Adhikari}, {Adhikari}, {Adya}, {Affeldt}, {Agarwal},
  {Agathos}, {Agatsuma}, {Aggarwal}, {Aguiar}, {Aiello}, {Ain}, {Ajith},
  {Akcay}, {Akutsu}, {Albanesi}, {Allocca}, {Altin}, {Amato}, {Anand}, {Anand},
  {Ananyeva}, {Anderson}, {Anderson}, {Ando}, {Andrade}, {Andres},
  {Andri{\'c}}, {Angelova}, {Ansoldi}, {Antelis}, {Antier}, {Appert}, {Arai},
  {Arai}, {Arai}, {Araki}, {Araya}, {Araya}, {Areeda}, {Ar{\`e}ne}, {Aritomi},
  {Arnaud}, {Arogeti}, {Aronson}, {Arun}, {Asada}, {Asali}, {Ashton}, {Aso},
  {Assiduo}, {Aston}, {Astone}, {Aubin}, {Austin}, {Babak}, {Badaracco},
  {Bader}, {Badger}, {Bae}, {Bae}, {Baer}, {Bagnasco}, {Bai}, {Baiotti},
  {Baird}, {Bajpai}, {Ball}, {Ballardin}, {Ballmer}, {Balsamo}, {Baltus},
  {Banagiri}, {Bankar}, {Barayoga}, {Barbieri}, {Barish}, {Barker}, {Barneo},
  {Barone}, {Barr}, {Barsotti}, {Barsuglia}, {Barta}, {Bartlett}, {Barton},
  {Bartos}, {Bassiri}, {Basti}, {Bawaj}, {Bayley}, {Baylor}, {Bazzan},
  {B{\'e}csy}, {Bedakihale}, {Bejger}, {Belahcene}, {Benedetto}, {Beniwal},
  {Bennett}, {Bentley}, {BenYaala}, {Bergamin}, {Berger}, {Bernuzzi}, {Berry},
  {Bersanetti}, {Bertolini}, {Betzwieser}, {Beveridge}, {Bhandare}, {Bhardwaj},
  {Bhattacharjee}, {Bhaumik}, {Bilenko}, {Billingsley}, {Bini}, {Birney},
  {Birnholtz}, {Biscans}, {Bischi}, {Biscoveanu}, {Bisht}, {Biswas}, {Bitossi},
  {Bizouard}, {Blackburn}, {Blair}, {Blair}, {Blair}, {Bobba}, {Bode}, {Boer},
  {Bogaert}, {Boldrini}, {Bonavena}, {Bondu}, {Bonilla}, {Bonnand}, {Booker},
  {Boom}, {Bork}, {Boschi}, {Bose}, {Bose}, {Bossilkov}, {Boudart},
  {Bouffanais}, {Bozzi}, {Bradaschia}, {Brady}, {Bramley}, {Branch},
  {Branchesi}, {Brandt}, {Brau}, {Breschi}, {Briant}, {Briggs}, {Brillet},
  {Brinkmann}, {Brockill}, {Brooks}, {Brooks}, {Brown}, {Brunett}, {Bruno},
  {Bruntz}, {Bryant}, {Bulik}, {Bulten}, {Buonanno}, {Buscicchio}, {Buskulic},
  {Buy}, {Byer}, {Cabourn Davies}, {Cadonati}, {Cagnoli}, {Cahillane},
  {Calder{\'o}n Bustillo}, {Callaghan}, {Callister}, {Calloni}, {Cameron},
  {Camp}, {Canepa}, {Canevarolo}, {Cannavacciuolo}, {Cannon}, {Cao}, {Cao},
  {Capocasa}, {Capote}, {Carapella}, {Carbognani}, {Carlin}, {Carney},
  {Carpinelli}, {Carrillo}, {Carullo}, {Carver}, {Casanueva Diaz}, {Casentini},
  {Castaldi}, {Caudill}, {Cavagli{\`a}}, {Cavalier}, {Cavalieri}, {Ceasar},
  {Cella}, {Cerd{\'a}-Dur{\'a}n}, {Cesarini}, {Chaibi}, {Chakravarti},
  {Chalathadka Subrahmanya}, {Champion}, {Chan}, {Chan}, {Chan}, {Chan},
  {Chan}, {Chandra}, {Chanial}, {Chao}, {Chapman-Bird}, {Charlton}, {Chase},
  {Chassande-Mottin}, {Chatterjee}, {Chatterjee}, {Chatterjee}, {Chaturvedi},
  {Chaty}, {Chatziioannou}, {Chen}, {Chen}, {Chen}, {Chen}, {Chen}, {Chen},
  {Chen}, {Chen}, {Cheng}, {Cheong}, {Cheung}, {Chia}, {Chiadini}, {Chiang},
  {Chiarini}, {Chierici}, {Chincarini}, {Chiofalo}, {Chiummo}, {Cho}, {Cho},
  {Choudhary}, {Choudhary}, {Christensen}, {Chu}, {Chu}, {Chu}, {Chua},
  {Chung}, {Ciani}, {Ciecielag}, {Cie{\'s}lar}, {Cifaldi}, {Ciobanu}, {Ciolfi},
  {Cipriano}, {Cirone}, {Clara}, {Clark}, {Clark}, {Clarke}, {Clearwater},
  {Clesse}, {Cleva}, {Coccia}, {Codazzo}, {Cohadon}, {Cohen}, {Cohen},
  {Colleoni}, {Collette}, {Colombo}, {Colpi}, {Compton}, {Constancio}, {Conti},
  {Cooper}, {Corban}, {Corbitt}, {Cordero-Carri{\'o}n}, {Corezzi}, {Corley},
  {Cornish}, {Corre}, {Corsi}, {Cortese}, {Costa}, {Cotesta}, {Coughlin},
  {Coulon}, {Countryman}, {Cousins}, {Couvares}, {Coward}, {Cowart}, {Coyne},
  {Coyne}, {Creighton}, {Creighton}, {Criswell}, {Croquette}, {Crowder},
  {Cudell}, {Cullen}, {Cumming}, {Cummings}, {Cunningham}, {Cuoco},
  {Cury{\l}o}, {Dabadie}, {Dal Canton}, {Dall'Osso}, {D{\'a}lya}, {Dana},
  {DaneshgaranBajastani}, {D'Angelo}, {Danila}, {Danilishin}, {D'Antonio},
  {Danzmann}, {Darsow-Fromm}, {Dasgupta}, {Datrier}, {Datta}, {Dattilo},
  {Dave}, {Davier}, {Davis}, {Davis}, {Daw}, {de Alarc{\'o}n}, {Dean}, {DeBra},
  {Deenadayalan}, {Degallaix}, {De Laurentis}, {Del{\'e}glise}, {Del Favero},
  {De Lillo}, {De Lillo}, {Del Pozzo}, {DeMarchi}, {De Matteis}, {D'Emilio},
  {Demos}, {Dent}, {Depasse}, {De Pietri}, {De Rosa}, {De Rossi}, {DeSalvo},
  {De Simone}, {Dhurandhar}, {D{\'\i}az}, {Diaz-Ortiz}, {Didio}, {Dietrich},
  {Di Fiore}, {Di Fronzo}, {Di Giorgio}, {Di Giovanni}, {Di Giovanni}, {Di
  Girolamo}, {Di Lieto}, {Ding}, {Di Pace}, {Di Palma}, {Di Renzo},
  {Divakarla}, {Dmitriev}, {Doctor}, {D'Onofrio}, {Donovan}, {Dooley},
  {Doravari}, {Dorrington}, {Drago}, {Driggers}, {Drori}, {Ducoin}, {Dupej},
  {Durante}, {D'Urso}, {Duverne}, {Dwyer}, {Eassa}, {Easter}, {Ebersold},
  {Eckhardt}, {Eddolls}, {Edelman}, {Edo}, {Edy}, {Effler}, {Eguchi},
  {Eichholz}, {Eikenberry}, {Eisenmann}, {Eisenstein}, {Ejlli}, {Engelby},
  {Enomoto}, {Errico}, {Essick}, {Estell{\'e}s}, {Estevez}, {Etienne}, {Etzel},
  {Evans}, {Evans}, {Ewing}, {Fafone}, {Fair}, {Fairhurst}, {Farah}, {Farinon},
  {Farr}, {Farr}, {Farrow}, {Fauchon-Jones}, {Favaro}, {Favata}, {Fays},
  {Fazio}, {Feicht}, {Fejer}, {Fenyvesi}, {Ferguson}, {Fernandez-Galiana},
  {Ferrante}, {Ferreira}, {Fidecaro}, {Figura}, {Fiori}, {Fishbach}, {Fisher},
  {Fittipaldi}, {Fiumara}, {Flaminio}, {Floden}, {Fong}, {Font}, {Fornal},
  {Forsyth}, {Franke}, {Frasca}, {Frasconi}, {Frederick}, {Freed}, {Frei},
  {Freise}, {Frey}, {Fritschel}, {Frolov}, {Fronz{\'e}}, {Fujii}, {Fujikawa},
  {Fukunaga}, {Fukushima}, {Fulda}, {Fyffe}, {Gabbard}, {Gabella}, {Gadre},
  {Gair}, {Gais}, {Galaudage}, {Gamba}, {Ganapathy}, {Ganguly}, {Gao},
  {Gaonkar}, {Garaventa}, {Garc{\'\i}a}, {Garc{\'\i}a-N{\'u}{\~n}ez},
  {Garc{\'\i}a-Quir{\'o}s}, {Garufi}, {Gateley}, {Gaudio}, {Gayathri}, {Ge},
  {Gemme}, {Gennai}, {George}, {George}, {Gerberding}, {Gergely}, {Gewecke},
  {Ghonge}, {Ghosh}, {Ghosh}, {Ghosh}, {Ghosh}, {Giacomazzo}, {Giacoppo},
  {Giaime}, {Giardina}, {Gibson}, {Gier}, {Giesler}, {Giri}, {Gissi},
  {Glanzer}, {Gleckl}, {Godwin}, {Goetz}, {Goetz}, {Gohlke}, {Golomb},
  {Goncharov}, {Gonz{\'a}lez}, {Gopakumar}, {Gosselin}, {Gouaty}, {Gould},
  {Grace}, {Grado}, {Granata}, {Granata}, {Grant}, {Gras}, {Grassia}, {Gray},
  {Gray}, {Greco}, {Green}, {Green}, {Gretarsson}, {Gretarsson}, {Griffith},
  {Griffiths}, {Griggs}, {Grignani}, {Grimaldi}, {Grimm}, {Grote}, {Grunewald},
  {Gruning}, {Guerra}, {Guidi}, {Guimaraes}, {Guix{\'e}}, {Gulati}, {Guo},
  {Guo}, {Gupta}, {Gupta}, {Gupta}, {Gustafson}, {Gustafson}, {Guzman}, {Ha},
  {Haegel}, {Hagiwara}, {Haino}, {Halim}, {Hall}, {Hamilton}, {Hammond}, {Han},
  {Haney}, {Hanks}, {Hanna}, {Hannam}, {Hannuksela}, {Hansen}, {Hansen},
  {Hanson}, {Harder}, {Hardwick}, {Haris}, {Harms}, {Harry}, {Harry},
  {Hartwig}, {Hasegawa}, {Haskell}, {Hasskew}, {Haster}, {Hattori}, {Haughian},
  {Hayakawa}, {Hayama}, {Hayes}, {Healy}, {Heidmann}, {Heidt}, {Heintze},
  {Heinze}, {Heinzel}, {Heitmann}, {Hellman}, {Hello}, {Helmling-Cornell},
  {Hemming}, {Hendry}, {Heng}, {Hennes}, {Hennig}, {Hennig}, {Hernandez},
  {Hernandez Vivanco}, {Heurs}, {Hild}, {Hill}, {Himemoto}, {Hines},
  {Hiranuma}, {Hirata}, {Hirose}, {Hochheim}, {Hofman}, {Hohmann}, {Holcomb},
  {Holland}, {Holley-Bockelmann}, {Hollows}, {Holmes}, {Holt}, {Holz}, {Hong},
  {Hopkins}, {Hough}, {Hourihane}, {Howell}, {Hoy}, {Hoyland}, {Hreibi},
  {Hsieh}, {Hsu}, {Huang}, {Huang}, {Huang}, {Huang}, {Huang}, {Huang},
  {H{\"u}bner}, {Huddart}, {Hughey}, {Hui}, {Hui}, {Husa}, {Huttner},
  {Huxford}, {Huynh-Dinh}, {Ide}, {Idzkowski}, {Iess}, {Ikenoue}, {Imam},
  {Inayoshi}, {Ingram}, {Inoue}, {Ioka}, {Isi}, {Isleif}, {Ito}, {Itoh},
  {Iyer}, {Izumi}, {JaberianHamedan}, {Jacqmin}, {Jadhav}, {Jadhav}, {James},
  {Jan}, {Jani}, {Janquart}, {Janssens}, {Janthalur}, {Jaranowski}, {Jariwala},
  {Jaume}, {Jenkins}, {Jenner}, {Jeon}, {Jeunon}, {Jia}, {Jin}, {Johns},
  {Johnson-McDaniel}, {Jones}, {Jones}, {Jones}, {Jones}, {Jones}, {Jonker},
  {Ju}, {Jung}, {Jung}, {Junker}, {Juste}, {Kaihotsu}, {Kajita}, {Kakizaki},
  {Kalaghatgi}, {Kalogera}, {Kamai}, {Kamiizumi}, {Kanda}, {Kandhasamy},
  {Kang}, {Kanner}, {Kao}, {Kapadia}, {Kapasi}, {Karat}, {Karathanasis},
  {Karki}, {Kashyap}, {Kasprzack}, {Kastaun}, {Katsanevas}, {Katsavounidis},
  {Katzman}, {Kaur}, {Kawabe}, {Kawaguchi}, {Kawai}, {Kawasaki},
  {K{\'e}f{\'e}lian}, {Keitel}, {Key}, {Khadka}, {Khalili}, {Khan}, {Khazanov},
  {Khetan}, {Khursheed}, {Kijbunchoo}, {Kim}, {Kim}, {Kim}, {Kim}, {Kim},
  {Kim}, {Kimball}, {Kimura}, {Kinley-Hanlon}, {Kirchhoff}, {Kissel}, {Kita},
  {Kitazawa}, {Kleybolte}, {Klimenko}, {Knee}, {Knowles}, {Knyazev}, {Koch},
  {Koekoek}, {Kojima}, {Kokeyama}, {Koley}, {Kolitsidou}, {Kolstein}, {Komori},
  {Kondrashov}, {Kong}, {Kontos}, {Koper}, {Korobko}, {Kotake}, {Kovalam},
  {Kozak}, {Kozakai}, {Kozu}, {Kringel}, {Krishnendu}, {Kr{\'o}lak}, {Kuehn},
  {Kuei}, {Kuijer}, {Kulkarni}, {Kumar}, {Kumar}, {Kumar}, {Kumar}, {Kume},
  {Kuns}, {Kuo}, {Kuo}, {Kuromiya}, {Kuroyanagi}, {Kusayanagi}, {Kuwahara},
  {Kwak}, {Lagabbe}, {Laghi}, {Lalande}, {Lam}, {Lamberts}, {Landry}, {Lane},
  {Lang}, {Lange}, {Lantz}, {La Rosa}, {Lartaux-Vollard}, {Lasky}, {Laxen},
  {Lazzarini}, {Lazzaro}, {Leaci}, {Leavey}, {Lecoeuche}, {Lee}, {Lee}, {Lee},
  {Lee}, {Lee}, {Lee}, {Lehmann}, {Lema{\^\i}tre}, {Leonardi}, {Leroy},
  {Letendre}, {Levesque}, {Levin}, {Leviton}, {Leyde}, {Li}, {Li}, {Li}, {Li},
  {Li}, {Li}, {Lin}, {Lin}, {Lin}, {Lin}, {Lin}, {Linde}, {Linker}, {Linley},
  {Littenberg}, {Liu}, {Liu}, {Liu}, {Liu}, {Llamas}, {Llorens-Monteagudo},
  {Lo}, {Lockwood}, {Loh}, {London}, {Longo}, {Lopez}, {Lopez Portilla},
  {Lorenzini}, {Loriette}, {Lormand}, {Losurdo}, {Lott}, {Lough}, {Lousto},
  {Lovelace}, {Lucaccioni}, {L{\"u}ck}, {Lumaca}, {Lundgren}, {Luo}, {Lynam},
  {Macas}, {MacInnis}, {Macleod}, {MacMillan}, {Macquet}, {Maga{\~n}a
  Hernandez}, {Magazz{\`u}}, {Magee}, {Maggiore}, {Magnozzi}, {Mahesh},
  {Majorana}, {Makarem}, {Maksimovic}, {Maliakal}, {Malik}, {Man}, {Mandic},
  {Mangano}, {Mango}, {Mansell}, {Manske}, {Mantovani}, {Mapelli},
  {Marchesoni}, {Marchio}, {Marion}, {Mark}, {M{\'a}rka}, {M{\'a}rka},
  {Markakis}, {Markosyan}, {Markowitz}, {Maros}, {Marquina}, {Marsat},
  {Martelli}, {Martin}, {Martin}, {Martinez}, {Martinez}, {Martinez},
  {Martinovic}, {Martynov}, {Marx}, {Masalehdan}, {Mason}, {Massera},
  {Masserot}, {Massinger}, {Masso-Reid}, {Mastrogiovanni}, {Matas},
  {Mateu-Lucena}, {Matichard}, {Matiushechkina}, {Mavalvala}, {McCann},
  {McCarthy}, {McClelland}, {McClincy}, {McCormick}, {McCuller}, {McGhee},
  {McGuire}, {McIsaac}, {McIver}, {McRae}, {McWilliams}, {Meacher}, {Mehmet},
  {Mehta}, {Meijer}, {Melatos}, {Melchor}, {Mendell}, {Menendez-Vazquez},
  {Menoni}, {Mercer}, {Mereni}, {Merfeld}, {Merilh}, {Merritt}, {Merzougui},
  {Meshkov}, {Messenger}, {Messick}, {Meyers}, {Meylahn}, {Mhaske}, {Miani},
  {Miao}, {Michaloliakos}, {Michel}, {Michimura}, {Middleton}, {Milano},
  {Miller}, {Miller}, {Miller}, {Millhouse}, {Mills}, {Milotti}, {Minazzoli},
  {Minenkov}, {Mio}, {Mir}, {Miravet-Ten{\'e}s}, {Mishra}, {Mishra}, {Mistry},
  {Mitra}, {Mitrofanov}, {Mitselmakher}, {Mittleman}, {Miyakawa}, {Miyamoto},
  {Miyazaki}, {Miyo}, {Miyoki}, {Mo}, {Modafferi}, {Moguel}, {Mogushi},
  {Mohapatra}, {Mohite}, {Molina}, {Molina-Ruiz}, {Mondin}, {Montani}, {Moore},
  {Moraru}, {Morawski}, {More}, {Moreno}, {Moreno}, {Mori}, {Morisaki},
  {Moriwaki}, {Morr{\'a}s}, {Mours}, {Mow-Lowry}, {Mozzon}, {Muciaccia},
  {Mukherjee}, {Mukherjee}, {Mukherjee}, {Mukherjee}, {Mukherjee}, {Mukund},
  {Mullavey}, {Munch}, {Mu{\~n}iz}, {Murray}, {Musenich}, {Muusse}, {Nadji},
  {Nagano}, {Nagano}, {Nagar}, {Nakamura}, {Nakano}, {Nakano}, {Nakashima},
  {Nakayama}, {Napolano}, {Nardecchia}, {Narikawa}, {Naticchioni}, {Nayak},
  {Nayak}, {Negishi}, {Neil}, {Neilson}, {Nelemans}, {Nelson}, {Nery},
  {Neubauer}, {Neunzert}, {Ng}, {Ng}, {Nguyen}, {Nguyen}, {Nguyen}, {Nguyen
  Quynh}, {Ni}, {Nichols}, {Nishizawa}, {Nissanke}, {Nitoglia}, {Nocera},
  {Norman}, {North}, {Nozaki}, {Nu{\~n}o Siles}, {Nuttall}, {Oberling},
  {O'Brien}, {Obuchi}, {O'Dell}, {Oelker}, {Ogaki}, {Oganesyan}, {Oh}, {Oh},
  {Oh}, {Ohashi}, {Ohishi}, {Ohkawa}, {Ohme}, {Ohta}, {Okada}, {Okutani},
  {Okutomi}, {Olivetto}, {Oohara}, {Ooi}, {Oram}, {O'Reilly}, {Ormiston},
  {Ormsby}, {Ortega}, {O'Shaughnessy}, {O'Shea}, {Oshino}, {Ossokine},
  {Osthelder}, {Otabe}, {Ottaway}, {Overmier}, {Pace}, {Pagano}, {Page},
  {Pagliaroli}, {Pai}, {Pai}, {Palamos}, {Palashov}, {Palomba}, {Pan}, {Pan},
  {Panda}, {Pang}, {Pang}, {Pankow}, {Pannarale}, {Pant}, {Panther},
  {Paoletti}, {Paoli}, {Paolone}, {Parisi}, {Park}, {Park}, {Parker},
  {Pascucci}, {Pasqualetti}, {Passaquieti}, {Passuello}, {Patel}, {Pathak},
  {Patricelli}, {Patron}, {Paul}, {Payne}, {Pedraza}, {Pegoraro}, {Pele},
  {Pe{\~n}a Arellano}, {Penn}, {Perego}, {Pereira}, {Pereira}, {Perez},
  {P{\'e}rigois}, {Perkins}, {Perreca}, {Perri{\`e}s}, {Petermann},
  {Petterson}, {Pfeiffer}, {Pham}, {Phukon}, {Piccinni}, {Pichot},
  {Piendibene}, {Piergiovanni}, {Pierini}, {Pierro}, {Pillant}, {Pillas},
  {Pilo}, {Pinard}, {Pinto}, {Pinto}, {Piotrzkowski}, {Piotrzkowski},
  {Pirello}, {Pitkin}, {Placidi}, {Planas}, {Plastino}, {Pluchar}, {Poggiani},
  {Polini}, {Pong}, {Ponrathnam}, {Popolizio}, {Porter}, {Poulton}, {Powell},
  {Pracchia}, {Pradier}, {Prajapati}, {Prasai}, {Prasanna}, {Pratten},
  {Principe}, {Prodi}, {Prokhorov}, {Prosposito}, {Prudenzi}, {Puecher},
  {Punturo}, {Puosi}, {Puppo}, {P{\"u}rrer}, {Qi}, {Quetschke},
  {Quitzow-James}, {Qutob}, {Raab}, {Raaijmakers}, {Radkins}, {Radulesco},
  {Raffai}, {Rail}, {Raja}, {Rajan}, {Ramirez}, {Ramirez}, {Ramos-Buades},
  {Rana}, {Rapagnani}, {Rapol}, {Ray}, {Raymond}, {Raza}, {Razzano}, {Read},
  {Rees}, {Regimbau}, {Rei}, {Reid}, {Reid}, {Reitze}, {Relton}, {Renzini},
  {Rettegno}, {Reza}, {Rezac}, {Ricci}, {Richards}, {Richardson}, {Richardson},
  {Riemenschneider}, {Riles}, {Rinaldi}, {Rink}, {Rizzo}, {Robertson}, {Robie},
  {Robinet}, {Rocchi}, {Rodriguez}, {Rolland}, {Rollins}, {Romanelli},
  {Romano}, {Romel}, {Romero-Rodr{\'\i}guez}, {Romero-Shaw}, {Romie},
  {Ronchini}, {Rosa}, {Rose}, {Rosi{\'n}ska}, {Ross}, {Rowan}, {Rowlinson},
  {Roy}, {Roy}, {Roy}, {Rozza}, {Ruggi}, {Ruiz-Rocha}, {Ryan}, {Sachdev},
  {Sadecki}, {Sadiq}, {Sago}, {Saito}, {Saito}, {Sakai}, {Sakai},
  {Sakellariadou}, {Sakuno}, {Salafia}, {Salconi}, {Saleem}, {Salemi},
  {Samajdar}, {Sanchez}, {Sanchez}, {Sanchez}, {Sanchis-Gual}, {Sanders},
  {Sanuy}, {Saravanan}, {Sarin}, {Sassolas}, {Satari}, {Sathyaprakash}, {Sato},
  {Sato}, {Sauter}, {Savage}, {Sawada}, {Sawant}, {Sawant}, {Sayah},
  {Schaetzl}, {Scheel}, {Scheuer}, {Schiworski}, {Schmidt}, {Schmidt},
  {Schnabel}, {Schneewind}, {Schofield}, {Sch{\"o}nbeck}, {Schulte}, {Schutz},
  {Schwartz}, {Scott}, {Scott}, {Seglar-Arroyo}, {Sekiguchi}, {Sekiguchi},
  {Sellers}, {Sengupta}, {Sentenac}, {Seo}, {Sequino}, {Sergeev}, {Setyawati},
  {Shaffer}, {Shahriar}, {Shams}, {Shao}, {Sharma}, {Sharma}, {Shawhan},
  {Shcheblanov}, {Shibagaki}, {Shikauchi}, {Shimizu}, {Shimoda}, {Shimode},
  {Shinkai}, {Shishido}, {Shoda}, {Shoemaker}, {Shoemaker}, {ShyamSundar},
  {Sieniawska}, {Sigg}, {Singer}, {Singh}, {Singh}, {Singha}, {Sintes},
  {Sipala}, {Skliris}, {Slagmolen}, {Slaven-Blair}, {Smetana}, {Smith},
  {Smith}, {Soldateschi}, {Somala}, {Somiya}, {Son}, {Soni}, {Soni}, {Sordini},
  {Sorrentino}, {Sorrentino}, {Sotani}, {Soulard}, {Souradeep}, {Sowell},
  {Spagnuolo}, {Spencer}, {Spera}, {Srinivasan}, {Srivastava}, {Srivastava},
  {Staats}, {Stachie}, {Steer}, {Steinhoff}, {Steinlechner}, {Steinlechner},
  {Stevenson}, {Stops}, {Stover}, {Strain}, {Strang}, {Stratta}, {Strunk},
  {Sturani}, {Stuver}, {Sudhagar}, {Sudhir}, {Sugimoto}, {Suh}, {Sullivan},
  {Sullivan}, {Summerscales}, {Sun}, {Sun}, {Sunil}, {Sur}, {Suresh}, {Sutton},
  {Suzuki}, {Suzuki}, {Swinkels}, {Szczepa{\'n}czyk}, {Szewczyk}, {Tacca},
  {Tagoshi}, {Tait}, {Takahashi}, {Takahashi}, {Takamori}, {Takano}, {Takeda},
  {Takeda}, {Talbot}, {Talbot}, {Tanaka}, {Tanaka}, {Tanaka}, {Tanaka},
  {Tanaka}, {Tanasijczuk}, {Tanioka}, {Tanner}, {Tao}, {Tao}, {Tapia San
  Mart{\'\i}n}, {Taranto}, {Tasson}, {Telada}, {Tenorio}, {Terhune},
  {Terkowski}, {Thirugnanasambandam}, {Thomas}, {Thomas}, {Thomas}, {Thompson},
  {Thondapu}, {Thorne}, {Thrane}, {Tiwari}, {Tiwari}, {Tiwari}, {Toivonen},
  {Toland}, {Tolley}, {Tomaru}, {Tomigami}, {Tomura}, {Tonelli},
  {Torres-Forn{\'e}}, {Torrie}, {Tosta e Melo}, {T{\"o}yr{\"a}}, {Trapananti},
  {Travasso}, {Traylor}, {Trevor}, {Tringali}, {Tripathee}, {Troiano},
  {Trovato}, {Trozzo}, {Trudeau}, {Tsai}, {Tsai}, {Tsang}, {Tsang}, {Tsao},
  {Tse}, {Tso}, {Tsubono}, {Tsuchida}, {Tsukada}, {Tsuna}, {Tsutsui},
  {Tsuzuki}, {Turbang}, {Turconi}, {Tuyenbayev}, {Ubhi}, {Uchikata},
  {Uchiyama}, {Udall}, {Ueda}, {Uehara}, {Ueno}, {Ueshima}, {Unnikrishnan},
  {Uraguchi}, {Urban}, {Ushiba}, {Utina}, {Vahlbruch}, {Vajente}, {Vajpeyi},
  {Valdes}, {Valentini}, {Valsan}, {van Bakel}, {van Beuzekom}, {van den
  Brand}, {Van Den Broeck}, {Vander-Hyde}, {van der Schaaf}, {van Heijningen},
  {Vanosky}, {van Putten}, {van Remortel}, {Vardaro}, {Vargas}, {Varma},
  {Vas{\'u}th}, {Vecchio}, {Vedovato}, {Veitch}, {Veitch}, {Venneberg},
  {Venugopalan}, {Verkindt}, {Verma}, {Verma}, {Veske}, {Vetrano},
  {Vicer{\'e}}, {Vidyant}, {Viets}, {Vijaykumar}, {Villa-Ortega}, {Vinet},
  {Virtuoso}, {Vitale}, {Vo}, {Vocca}, {von Reis}, {von Wrangel}, {Vorvick},
  {Vyatchanin}, {Wade}, {Wade}, {Wagner}, {Walet}, {Walker}, {Wallace},
  {Wallace}, {Walsh}, {Wang}, {Wang}, {Wang}, {Ward}, {Warner}, {Was},
  {Washimi}, {Washington}, {Watchi}, {Weaver}, {Webster}, {Weinert},
  {Weinstein}, {Weiss}, {Weller}, {Weller}, {Wellmann}, {Wen}, {We{\ss}els},
  {Wette}, {Whelan}, {White}, {Whiting}, {Whittle}, {Wilken}, {Williams},
  {Williams}, {Williams}, {Williamson}, {Willis}, {Willke}, {Wilson},
  {Winkler}, {Wipf}, {Wlodarczyk}, {Woan}, {Woehler}, {Wofford}, {Wong}, {Wu},
  {Wu}, {Wu}, {Wu}, {Wysocki}, {Xiao}, {Xu}, {Yamada}, {Yamamoto}, {Yamamoto},
  {Yamamoto}, {Yamamoto}, {Yamashita}, {Yamazaki}, {Yang}, {Yang}, {Yang},
  {Yang}, {Yang}, {Yap}, {Yeeles}, {Yelikar}, {Ying}, {Yokogawa}, {Yokoyama},
  {Yokozawa}, {Yoo}, {Yoshioka}, {Yu}, {Yu}, {Yuzurihara}, {Zadro{\.z}ny},
  {Zanolin}, {Zeidler}, {Zelenova}, {Zendri}, {Zevin}, {Zhan}, {Zhang},
  {Zhang}, {Zhang}, {Zhang}, {Zhang}, {Zhao}, {Zhao}, {Zhao}, {Zhao}, {Zheng},
  {Zhou}, {Zhou}, {Zhu}, {Zhu}, {Zimmerman}, {Zlochower}, {Zucker}, \&
  {Zweizig}}]{GWTC3_2021}
{The LIGO Scientific Collaboration}, {the Virgo Collaboration}, {the KAGRA
  Collaboration}, {$et~al$.} 2021{\natexlab{a}}, arXiv e-prints,
  arXiv:2111.03606

\bibitem[{{The LIGO Scientific Collaboration}
  {$et~al$.}(2021{\natexlab{b}}){The LIGO Scientific Collaboration}, {the Virgo
  Collaboration}, {the KAGRA Collaboration}, {Abbott}, {Abe}, {Acernese},
  {Ackley}, {Adhikari}, {Adhikari}, {Adkins}, {Adya}, {Affeldt}, {Agarwal},
  {Agathos}, {Agatsuma}, {Aggarwal}, {Aguiar}, {Aiello}, {Ain}, {Ajith},
  {Akutsu}, {de Alarc{\'o}n}, {Albanesi}, {Alfaidi}, {Allocca}, {Altin},
  {Amato}, {Anand}, {Anand}, {Ananyeva}, {Anderson}, {Anderson}, {Ando},
  {Andrade}, {Andres}, {Andr{\'e}s-Carcasona}, {Andri{\'c}}, {Angelova},
  {Ansoldi}, {Antelis}, {Antier}, {Apostolatos}, {Appavuravther}, {Appert},
  {Apple}, {Arai}, {Araya}, {Araya}, {Areeda}, {Ar{\`e}ne}, {Aritomi},
  {Arnaud}, {Arogeti}, {Aronson}, {Arun}, {Asada}, {Asali}, {Ashton}, {Aso},
  {Assiduo}, {Assis de Souza Melo}, {Aston}, {Astone}, {Aubin}, {AultONeal},
  {Austin}, {Babak}, {Badaracco}, {Bader}, {Badger}, {Bae}, {Bae}, {Baer},
  {Bagnasco}, {Bai}, {Baird}, {Bajpai}, {Baka}, {Ball}, {Ballardin}, {Ballmer},
  {Balsamo}, {Baltus}, {Banagiri}, {Banerjee}, {Bankar}, {Barayoga},
  {Barbieri}, {Barish}, {Barker}, {Barneo}, {Barone}, {Barr}, {Barsotti},
  {Barsuglia}, {Barta}, {Bartlett}, {Barton}, {Bartos}, {Basak}, {Bassiri},
  {Basti}, {Bawaj}, {Bayley}, {Bazzan}, {Becher}, {B{\'e}csy}, {Bedakihale},
  {Beirnaert}, {Bejger}, {Belahcene}, {Benedetto}, {Beniwal}, {Benjamin},
  {Bennett}, {Bentley}, {BenYaala}, {Bera}, {Berbel}, {Bergamin}, {Berger},
  {Bernuzzi}, {Berry}, {Bersanetti}, {Bertolini}, {Betzwieser}, {Beveridge},
  {Bhandare}, {Bhandari}, {Bhardwaj}, {Bhatt}, {Bhattacharjee}, {Bhaumik},
  {Bianchi}, {Bilenko}, {Billingsley}, {Bini}, {Birney}, {Birnholtz},
  {Biscans}, {Bischi}, {Biscoveanu}, {Bisht}, {Biswas}, {Bitossi}, {Bizouard},
  {Blackburn}, {Blair}, {Blair}, {Blair}, {Bobba}, {Bode}, {Bo{\"e}r},
  {Bogaert}, {Boldrini}, {Bolingbroke}, {Bonavena}, {Bondu}, {Bonilla},
  {Bonnand}, {Booker}, {Boom}, {Bork}, {Boschi}, {Bose}, {Bose}, {Bossilkov},
  {Boudart}, {Bouffanais}, {Bozzi}, {Bradaschia}, {Brady}, {Bramley}, {Branch},
  {Branchesi}, {Brau}, {Breschi}, {Briant}, {Briggs}, {Brillet}, {Brinkmann},
  {Brockill}, {Brooks}, {Brooks}, {Brown}, {Brunett}, {Bruno}, {Bruntz},
  {Bryant}, {Bucci}, {Bulik}, {Bulten}, {Buonanno}, {Burtnyk}, {Buscicchio},
  {Buskulic}, {Buy}, {Byer}, {Cabourn Davies}, {Cabras}, {Cabrita}, {Cadonati},
  {Caesar}, {Cagnoli}, {Cahillane}, {Calder{\'o}n Bustillo}, {Callaghan},
  {Callister}, {Calloni}, {Cameron}, {Camp}, {Canepa}, {Canevarolo},
  {Cannavacciuolo}, {Cannon}, {Cao}, {Cao}, {Capocasa}, {Capote}, {Carapella},
  {Carbognani}, {Carlassara}, {Carlin}, {Carney}, {Carpinelli}, {Carrillo},
  {Carullo}, {Carver}, {Casanueva Diaz}, {Casentini}, {Castaldi}, {Caudill},
  {Cavagli{\`a}}, {Cavalier}, {Cavalieri}, {Cella}, {Cerd{\'a}-Dur{\'a}n},
  {Cesarini}, {Chaibi}, {Chalathadka Subrahmanya}, {Champion}, {Chan}, {Chan},
  {Chan}, {Chan}, {Chan}, {Chandra}, {Chang}, {Chanial}, {Chao},
  {Chapman-Bird}, {Charlton}, {Chase}, {Chassande-Mottin}, {Chatterjee},
  {Chatterjee}, {Chatterjee}, {Chaturvedi}, {Chaty}, {Chatziioannou}, {Chen},
  {Chen}, {Chen}, {Chen}, {Chen}, {Chen}, {Chen}, {Chen}, {Chen}, {Cheng},
  {Cheong}, {Cheung}, {Chia}, {Chiadini}, {Chiang}, {Chiarini}, {Chierici},
  {Chincarini}, {Chiofalo}, {Chiummo}, {Choudhary}, {Choudhary}, {Christensen},
  {Chu}, {Chu}, {Chua}, {Chung}, {Ciani}, {Ciecielag}, {Cie{\'s}lar},
  {Cifaldi}, {Ciobanu}, {Ciolfi}, {Cipriano}, {Clara}, {Clark}, {Clearwater},
  {Clesse}, {Cleva}, {Coccia}, {Codazzo}, {Cohadon}, {Cohen}, {Colleoni},
  {Collette}, {Colombo}, {Colpi}, {Compton}, {Constancio}, {Conti}, {Cooper},
  {Corban}, {Corbitt}, {Cordero-Carri{\'o}n}, {Corezzi}, {Corley}, {Cornish},
  {Corre}, {Corsi}, {Cortese}, {Costa}, {Cotesta}, {Cottingham}, {Coughlin},
  {Coulon}, {Countryman}, {Cousins}, {Couvares}, {Coward}, {Cowart}, {Coyne},
  {Coyne}, {Creighton}, {Creighton}, {Criswell}, {Croquette}, {Crowder},
  {Cudell}, {Cullen}, {Cumming}, {Cummings}, {Cunningham}, {Cuoco},
  {Cury{\l}o}, {Dabadie}, {Dal Canton}, {Dall'Osso}, {D{\'a}lya}, {Dana},
  {D'Angelo}, {Danilishin}, {D'Antonio}, {Danzmann}, {Darsow-Fromm},
  {Dasgupta}, {Datrier}, {Datta}, {Datta}, {Dattilo}, {Dave}, {Davier},
  {Davis}, {Davis}, {Daw}, {Dean}, {DeBra}, {Deenadayalan}, {Degallaix}, {De
  Laurentis}, {Del{\'e}glise}, {Del Favero}, {De Lillo}, {De Lillo},
  {Dell'Aquila}, {Del Pozzo}, {DeMarchi}, {De Matteis}, {D'Emilio}, {Demos},
  {Dent}, {Depasse}, {De Pietri}, {De Rosa}, {De Rossi}, {DeSalvo}, {De
  Simone}, {Dhurandhar}, {D{\'\i}az}, {Didio}, {Dietrich}, {Di Fiore}, {Di
  Fronzo}, {Di Giorgio}, {Di Giovanni}, {Di Giovanni}, {Di Girolamo}, {Di
  Lieto}, {Di Michele}, {Ding}, {Di Pace}, {Di Palma}, {Di Renzo}, {Divakarla},
  {Divyajyoti}, {Dmitriev}, {Doctor}, {Donahue}, {D'Onofrio}, {Donovan},
  {Dooley}, {Doravari}, {Drago}, {Driggers}, {Drori}, {Ducoin}, {Dupej},
  {Dupletsa}, {Durante}, {D'Urso}, {Duverne}, {Dwyer}, {Eassa}, {Easter},
  {Ebersold}, {Eckhardt}, {Eddolls}, {Edelman}, {Edo}, {Edy}, {Effler},
  {Eguchi}, {Eichholz}, {Eikenberry}, {Eisenmann}, {Eisenstein}, {Ejlli},
  {Engelby}, {Enomoto}, {Errico}, {Essick}, {Estell{\'e}s}, {Estevez},
  {Etienne}, {Etzel}, {Evans}, {Evans}, {Evstafyeva}, {Ewing}, {Fabrizi},
  {Faedi}, {Fafone}, {Fair}, {Fairhurst}, {Fan}, {Farah}, {Farinon}, {Farr},
  {Farr}, {Fauchon-Jones}, {Favaro}, {Favata}, {Fays}, {Fazio}, {Feicht},
  {Fejer}, {Fenyvesi}, {Ferguson}, {Fernandez-Galiana}, {Ferrante}, {Ferreira},
  {Fidecaro}, {Figura}, {Fiori}, {Fiori}, {Fishbach}, {Fisher}, {Fittipaldi},
  {Fiumara}, {Flaminio}, {Floden}, {Fong}, {Font}, {Fornal}, {Forsyth},
  {Franke}, {Frasca}, {Frasconi}, {Freed}, {Frei}, {Freise}, {Freitas}, {Frey},
  {Fritschel}, {Frolov}, {Fronz{\'e}}, {Fujii}, {Fujikawa}, {Fujimoto},
  {Fulda}, {Fyffe}, {Gabbard}, {Gabella}, {Gadre}, {Gair}, {Gais}, {Galaudage},
  {Gamba}, {Ganapathy}, {Ganguly}, {Gao}, {Gaonkar}, {Garaventa}, {Garc{\'\i}a
  N{\'u}{\~n}ez}, {Garc{\'\i}a-Quir{\'o}s}, {Garufi}, {Gateley}, {Gayathri},
  {Ge}, {Gemme}, {Gennai}, {George}, {Gerberding}, {Gergely}, {Gewecke},
  {Ghonge}, {Ghosh}, {Ghosh}, {Ghosh}, {Ghosh}, {Ghosh}, {Giacomazzo},
  {Giacoppo}, {Giaime}, {Giardina}, {Gibson}, {Gier}, {Giesler}, {Giri},
  {Gissi}, {Gkaitatzis}, {Glanzer}, {Gleckl}, {Godwin}, {Goetz}, {Goetz},
  {Gohlke}, {Golomb}, {Goncharov}, {Gonz{\'a}lez}, {Gosselin}, {Gouaty},
  {Gould}, {Goyal}, {Grace}, {Grado}, {Graham}, {Granata}, {Granata}, {Grant},
  {Gras}, {Grassia}, {Gray}, {Gray}, {Greco}, {Green}, {Green}, {Gretarsson},
  {Gretarsson}, {Griffith}, {Griffiths}, {Griggs}, {Grignani}, {Grimaldi},
  {Grimes}, {Grimm}, {Grote}, {Grunewald}, {Gruning}, {Gruson}, {Guerra},
  {Guidi}, {Guimaraes}, {Guix{\'e}}, {Gulati}, {Gunny}, {Guo}, {Guo}, {Gupta},
  {Gupta}, {Gupta}, {Gupta}, {Gupta}, {Gustafson}, {Guzman}, {Ha},
  {Hadiputrawan}, {Haegel}, {Haino}, {Halim}, {Hall}, {Hamilton}, {Hammond},
  {Han}, {Haney}, {Hanks}, {Hanna}, {Hannam}, {Hannuksela}, {Hansen}, {Hansen},
  {Hanson}, {Harder}, {Haris}, {Harms}, {Harry}, {Harry}, {Hartwig},
  {Hasegawa}, {Haskell}, {Haster}, {Hathaway}, {Hattori}, {Haughian},
  {Hayakawa}, {Hayama}, {Hayes}, {Healy}, {Heidmann}, {Heidt}, {Heintze},
  {Heinze}, {Heinzel}, {Heitmann}, {Hellman}, {Hello}, {Helmling-Cornell},
  {Hemming}, {Hendry}, {Heng}, {Hennes}, {Hennig}, {Hennig}, {Henshaw},
  {Hernandez}, {Hernandez Vivanco}, {Heurs}, {Hewitt}, {Higginbotham}, {Hild},
  {Hill}, {Himemoto}, {Hines}, {Hirata}, {Hirose}, {Ho}, {Hochheim}, {Hofman},
  {Hohmann}, {Holcomb}, {Holland}, {Hollows}, {Holmes}, {Holt}, {Holz}, {Hong},
  {Hough}, {Hourihane}, {Howell}, {Hoy}, {Hoyland}, {Hreibi}, {Hsieh}, {Hsieh},
  {Hsiung}, {Hsu}, {Huang}, {Huang}, {Huang}, {Huang}, {Huang}, {Huang},
  {H{\"u}bner}, {Huddart}, {Hughey}, {Hui}, {Hui}, {Husa}, {Huttner},
  {Huxford}, {Huynh-Dinh}, {Ide}, {Idzkowski}, {Iess}, {Inayoshi}, {Inoue},
  {Iosif}, {Isi}, {Isleif}, {Ito}, {Itoh}, {Iyer}, {JaberianHamedan},
  {Jacqmin}, {Jacquet}, {Jadhav}, {Jadhav}, {Jain}, {James}, {Jan}, {Jani},
  {Janquart}, {Janssens}, {Janthalur}, {Jaranowski}, {Jariwala}, {Jaume},
  {Jenkins}, {Jenner}, {Jeon}, {Jia}, {Jiang}, {Jin}, {Johns},
  {Johnson-McDaniel}, {Johnston}, {Jones}, {Jones}, {Jones}, {Jones}, {Joshi},
  {Ju}, {Jue}, {Jung}, {Jung}, {Junker}, {Juste}, {Kaihotsu}, {Kajita},
  {Kakizaki}, {Kalaghatgi}, {Kalogera}, {Kamai}, {Kamiizumi}, {Kanda},
  {Kandhasamy}, {Kang}, {Kanner}, {Kao}, {Kapadia}, {Kapasi}, {Karathanasis},
  {Karki}, {Kashyap}, {Kasprzack}, {Kastaun}, {Kato}, {Katsanevas},
  {Katsavounidis}, {Katzman}, {Kaur}, {Kawabe}, {Kawaguchi},
  {K{\'e}f{\'e}lian}, {Keitel}, {Key}, {Khadka}, {Khalili}, {Khan}, {Khanam},
  {Khazanov}, {Khetan}, {Khursheed}, {Kijbunchoo}, {Kim}, {Kim}, {Kim}, {Kim},
  {Kim}, {Kim}, {Kim}, {Kimball}, {Kimura}, {Kinley-Hanlon}, {Kirchhoff},
  {Kissel}, {Klimenko}, {Klinger}, {Knee}, {Knowles}, {Knust}, {Knyazev},
  {Kobayashi}, {Koch}, {Koekoek}, {Kohri}, {Kokeyama}, {Koley}, {Kolitsidou},
  {Kolstein}, {Komori}, {Kondrashov}, {Kong}, {Kontos}, {Koper}, {Korobko},
  {Kovalam}, {Koyama}, {Kozak}, {Kozakai}, {Kringel}, {Krishnendu},
  {Kr{\'o}lak}, {Kuehn}, {Kuei}, {Kuijer}, {Kulkarni}, {Kumar}, {Kumar},
  {Kumar}, {Kumar}, {Kume}, {Kuns}, {Kuromiya}, {Kuroyanagi}, {Kwak},
  {Lacaille}, {Lagabbe}, {Laghi}, {Lalande}, {Lalleman}, {Lam}, {Lamberts},
  {Landry}, {Lane}, {Lang}, {Lange}, {Lantz}, {La Rosa}, {Lartaux-Vollard},
  {Lasky}, {Laxen}, {Lazzarini}, {Lazzaro}, {Leaci}, {Leavey}, {LeBohec},
  {Lecoeuche}, {Lee}, {Lee}, {Lee}, {Lee}, {Lee}, {Legred}, {Lehmann},
  {Lema{\^\i}tre}, {Lenti}, {Leonardi}, {Leonova}, {Leroy}, {Letendre},
  {Levesque}, {Levin}, {Leviton}, {Leyde}, {Li}, {Li}, {Li}, {Li}, {Li}, {Li},
  {Li}, {Lin}, {Lin}, {Lin}, {Lin}, {Lin}, {Lin}, {Linde}, {Linker}, {Linley},
  {Littenberg}, {Liu}, {Liu}, {Liu}, {Liu}, {Llamas}, {Lo}, {Lo}, {London},
  {Longo}, {Lopez}, {Lopez Portilla}, {Lorenzini}, {Loriette}, {Lormand},
  {Losurdo}, {Lott}, {Lough}, {Lousto}, {Lovelace}, {Lucaccioni}, {L{\"u}ck},
  {Lumaca}, {Lundgren}, {Luo}, {Lynam}, {Ma'arif}, {Macas}, {Machtinger},
  {MacInnis}, {Macleod}, {MacMillan}, {Macquet}, {Maga{\~n}a Hernandez},
  {Magazz{\`u}}, {Magee}, {Maggiore}, {Magnozzi}, {Mahesh}, {Majorana},
  {Maksimovic}, {Maliakal}, {Malik}, {Man}, {Mandic}, {Mangano}, {Mansell},
  {Manske}, {Mantovani}, {Mapelli}, {Marchesoni}, {Mar{\'\i}n Pina}, {Marion},
  {Mark}, {M{\'a}rka}, {M{\'a}rka}, {Markakis}, {Markosyan}, {Markowitz},
  {Maros}, {Marquina}, {Marsat}, {Martelli}, {Martin}, {Martin}, {Martinez},
  {Martinez}, {Martinez}, {Martinovic}, {Martynov}, {Marx}, {Masalehdan},
  {Mason}, {Massera}, {Masserot}, {Masso-Reid}, {Mastrogiovanni}, {Matas},
  {Mateu-Lucena}, {Matichard}, {Matiushechkina}, {Mavalvala}, {McCann},
  {McCarthy}, {McClelland}, {McClincy}, {McCormick}, {McCuller}, {McGhee},
  {McGuire}, {McIsaac}, {McIver}, {McRae}, {McWilliams}, {Meacher}, {Mehmet},
  {Mehta}, {Meijer}, {Melatos}, {Melchor}, {Mendell}, {Menendez-Vazquez},
  {Menoni}, {Mercer}, {Mereni}, {Merfeld}, {Merilh}, {Merritt}, {Merzougui},
  {Meshkov}, {Messenger}, {Messick}, {Meyers}, {Meylahn}, {Mhaske}, {Miani},
  {Miao}, {Michaloliakos}, {Michel}, {Michimura}, {Middleton}, {Mihaylov},
  {Milano}, {Miller}, {Miller}, {Miller}, {Millhouse}, {Mills}, {Milotti},
  {Minenkov}, {Mio}, {Mir}, {Miravet-Ten{\'e}s}, {Mishkin}, {Mishra}, {Mishra},
  {Mistry}, {Mitra}, {Mitrofanov}, {Mitselmakher}, {Mittleman}, {Miyakawa},
  {Miyo}, {Miyoki}, {Mo}, {Modafferi}, {Moguel}, {Mogushi}, {Mohapatra},
  {Mohite}, {Molina}, {Molina-Ruiz}, {Mondin}, {Montani}, {Moore}, {Moragues},
  {Moraru}, {Morawski}, {More}, {Moreno}, {Moreno}, {Mori}, {Morisaki},
  {Morisue}, {Moriwaki}, {Mours}, {Mow-Lowry}, {Mozzon}, {Muciaccia},
  {Mukherjee}, {Mukherjee}, {Mukherjee}, {Mukherjee}, {Mukherjee}, {Mukund},
  {Mullavey}, {Munch}, {Mu{\~n}iz}, {Murray}, {Musenich}, {Muusse}, {Nadji},
  {Nagano}, {Nagar}, {Nakamura}, {Nakano}, {Nakano}, {Nakayama}, {Napolano},
  {Nardecchia}, {Narikawa}, {Narola}, {Naticchioni}, {Nayak}, {Nayak}, {Neil},
  {Neilson}, {Nelson}, {Nelson}, {Nery}, {Neubauer}, {Neunzert}, {Ng}, {Ng},
  {Nguyen}, {Nguyen}, {Nguyen}, {Nguyen Quynh}, {Ni}, {Ni}, {Nichols},
  {Nishimoto}, {Nishizawa}, {Nissanke}, {Nitoglia}, {Nocera}, {Norman},
  {North}, {Nozaki}, {Nurbek}, {Nuttall}, {Obayashi}, {Oberling}, {O'Brien},
  {O'Dell}, {Oelker}, {Ogaki}, {Oganesyan}, {Oh}, {Oh}, {Oh}, {Ohashi},
  {Ohashi}, {Ohkawa}, {Ohme}, {Ohta}, {Okada}, {Okutani}, {Olivetto}, {Oohara},
  {Oram}, {O'Reilly}, {Ormiston}, {Ormsby}, {O'Shaughnessy}, {O'Shea},
  {Oshino}, {Ossokine}, {Osthelder}, {Otabe}, {Ottaway}, {Overmier}, {Pace},
  {Pagano}, {Pagano}, {Page}, {Pagliaroli}, {Pai}, {Pai}, {Pal}, {Palamos},
  {Palashov}, {Palomba}, {Pan}, {Pan}, {Panda}, {Pang}, {Pankow}, {Pannarale},
  {Pant}, {Panther}, {Paoletti}, {Paoli}, {Paolone}, {Pappas}, {Parisi},
  {Park}, {Park}, {Parker}, {Pascucci}, {Pasqualetti}, {Passaquieti},
  {Passuello}, {Patel}, {Pathak}, {Patricelli}, {Patron}, {Paul}, {Payne},
  {Pedraza}, {Pedurand}, {Pegoraro}, {Pele}, {Pe{\~n}a Arellano}, {Penano},
  {Penn}, {Perego}, {Pereira}, {Pereira}, {Perez}, {P{\'e}rigois}, {Perkins},
  {Perreca}, {Perri{\`e}s}, {Pesios}, {Petermann}, {Petterson}, {Pfeiffer},
  {Pham}, {Pham}, {Phukon}, {Phurailatpam}, {Piccinni}, {Pichot}, {Piendibene},
  {Piergiovanni}, {Pierini}, {Pierro}, {Pillant}, {Pillas}, {Pilo}, {Pinard},
  {Pineda-Bosque}, {Pinto}, {Pinto}, {Piotrzkowski}, {Piotrzkowski}, {Pirello},
  {Pitkin}, {Placidi}, {Placidi}, {Planas}, {Plastino}, {Pluchar}, {Poggiani},
  {Polini}, {Pong}, {Ponrathnam}, {Porter}, {Poulton}, {Poverman}, {Powell},
  {Pracchia}, {Pradier}, {Prajapati}, {Prasai}, {Prasanna}, {Pratten},
  {Principe}, {Prodi}, {Prokhorov}, {Prosposito}, {Prudenzi}, {Puecher},
  {Punturo}, {Puosi}, {Puppo}, {P{\"u}rrer}, {Qi}, {Quartey}, {Quetschke},
  {Quinonez}, {Quitzow-James}, {Qutob}, {Raab}, {Raaijmakers}, {Radkins},
  {Radulesco}, {Raffai}, {Rail}, {Raja}, {Rajan}, {Ramirez}, {Ramirez},
  {Ramos-Buades}, {Rana}, {Rapagnani}, {Ray}, {Raymond}, {Raza}, {Razzano},
  {Read}, {Rees}, {Regimbau}, {Rei}, {Reid}, {Reid}, {Reitze}, {Relton},
  {Renzini}, {Rettegno}, {Revenu}, {Reza}, {Rezac}, {Ricci}, {Richards},
  {Richardson}, {Richardson}, {Riemenschneider}, {Riles}, {Rinaldi}, {Rink},
  {Robertson}, {Robie}, {Robinet}, {Rocchi}, {Rodriguez}, {Rolland}, {Rollins},
  {Romanelli}, {Romano}, {Romel}, {Romero}, {Romero-Shaw}, {Romie}, {Ronchini},
  {Rosa}, {Rose}, {Rosi{\'n}ska}, {Ross}, {Rowan}, {Rowlinson}, {Roy}, {Roy},
  {Roy}, {Rozza}, {Ruggi}, {Ruiz-Rocha}, {Ryan}, {Sachdev}, {Sadecki}, {Sadiq},
  {Saha}, {Saito}, {Sakai}, {Sakellariadou}, {Sakon}, {Salafia},
  {Salces-Carcoba}, {Salconi}, {Saleem}, {Salemi}, {Samajdar}, {Sanchez},
  {Sanchez}, {Sanchez}, {Sanchis-Gual}, {Sanders}, {Sanuy}, {Saravanan},
  {Sarin}, {Sassolas}, {Satari}, {Sathyaprakash}, {Sauter}, {Savage}, {Savant},
  {Sawada}, {Sawant}, {Sayah}, {Schaetzl}, {Scheel}, {Scheuer}, {Schiworski},
  {Schmidt}, {Schmidt}, {Schnabel}, {Schneewind}, {Schofield}, {Sch{\"o}nbeck},
  {Schulte}, {Schutz}, {Schwartz}, {Scott}, {Scott}, {Seglar-Arroyo},
  {Sekiguchi}, {Sellers}, {Sengupta}, {Sentenac}, {Seo}, {Sequino}, {Sergeev},
  {Setyawati}, {Shaffer}, {Shahriar}, {Shaikh}, {Shams}, {Shao}, {Sharma},
  {Sharma}, {Shawhan}, {Shcheblanov}, {Sheela}, {Shikano}, {Shikauchi},
  {Shimizu}, {Shimode}, {Shinkai}, {Shishido}, {Shoda}, {Shoemaker},
  {Shoemaker}, {ShyamSundar}, {Sieniawska}, {Sigg}, {Silenzi}, {Singer},
  {Singh}, {Singh}, {Singh}, {Singha}, {Sintes}, {Sipala}, {Skliris},
  {Slagmolen}, {Slaven-Blair}, {Smetana}, {Smith}, {Smith}, {Smith},
  {Soldateschi}, {Somala}, {Somiya}, {Song}, {Soni}, {Soni}, {Sordini},
  {Sorrentino}, {Sorrentino}, {Soulard}, {Souradeep}, {Sowell}, {Spagnuolo},
  {Spencer}, {Spera}, {Spinicelli}, {Srivastava}, {Srivastava}, {Staats},
  {Stachie}, {Stachurski}, {Steer}, {Steinhoff}, {Steinlechner},
  {Steinlechner}, {Stergioulas}, {Stops}, {Stover}, {Strain}, {Strang},
  {Stratta}, {Strong}, {Strunk}, {Sturani}, {Stuver}, {Suchenek}, {Sudhagar},
  {Sudhir}, {Sugimoto}, {Suh}, {Sullivan}, {Sullivan}, {Summerscales}, {Sun},
  {Sunil}, {Sur}, {Suresh}, {Sutton}, {Suzuki}, {Suzuki}, {Suzuki}, {Swinkels},
  {Szczepa{\'n}czyk}, {Szewczyk}, {Tacca}, {Tagoshi}, {Tait}, {Takahashi},
  {Takahashi}, {Takano}, {Takeda}, {Takeda}, {Talbot}, {Talbot}, {Tanaka},
  {Tanaka}, {Tanaka}, {Tanasijczuk}, {Tanioka}, {Tanner}, {Tao}, {Tao},
  {Tapia}, {Tapia San Mart{\'\i}n}, {Taranto}, {Taruya}, {Tasson}, {Tenorio},
  {Terhune}, {Terkowski}, {Thirugnanasambandam}, {Thomas}, {Thomas},
  {Thompson}, {Thompson}, {Thondapu}, {Thorne}, {Thrane}, {Tiwari}, {Tiwari},
  {Tiwari}, {Toivonen}, {Tolley}, {Tomaru}, {Tomura}, {Tonelli}, {Tornasi},
  {Torres-Forn{\'e}}, {Torrie}, {Tosta e Melo}, {T{\"o}yr{\"a}}, {Trapananti},
  {Travasso}, {Traylor}, {Trevor}, {Tringali}, {Tripathee}, {Troiano},
  {Trovato}, {Trozzo}, {Trudeau}, {Tsai}, {Tsang}, {Tsang}, {Tsao}, {Tse},
  {Tso}, {Tsuchida}, {Tsukada}, {Tsuna}, {Tsutsui}, {Turbang}, {Turconi},
  {Tuyenbayev}, {Ubhi}, {Uchikata}, {Uchiyama}, {Udall}, {Ueda}, {Uehara},
  {Ueno}, {Ueshima}, {Unnikrishnan}, {Urban}, {Ushiba}, {Utina}, {Vajente},
  {Vajpeyi}, {Valdes}, {Valentini}, {Valsan}, {van Bakel}, {van Beuzekom}, {van
  Dael}, {van den Brand}, {Van Den Broeck}, {Vander-Hyde}, {van Haevermaet},
  {van Heijningen}, {van Putten}, {van Remortel}, {Vardaro}, {Vargas}, {Varma},
  {Vas{\'u}th}, {Vecchio}, {Vedovato}, {Veitch}, {Veitch}, {Venneberg},
  {Venugopalan}, {Verkindt}, {Verma}, {Verma}, {Vermeulen}, {Veske}, {Vetrano},
  {Vicer{\'e}}, {Vidyant}, {Viets}, {Vijaykumar}, {Villa-Ortega}, {Vinet},
  {Virtuoso}, {Vitale}, {Vocca}, {von Reis}, {von Wrangel}, {Vorvick},
  {Vyatchanin}, {Wade}, {Wade}, {Wagner}, {Wald}, {Walet}, {Walker}, {Wallace},
  {Wallace}, {Wang}, {Wang}, {Wang}, {Ward}, {Warner}, {Was}, {Washimi},
  {Washington}, {Watchi}, {Weaver}, {Weaving}, {Webster}, {Weinert},
  {Weinstein}, {Weiss}, {Weller}, {Weller}, {Wellmann}, {Wen}, {We{\ss}els},
  {Wette}, {Whelan}, {White}, {Whiting}, {Whittle}, {Wilken}, {Williams},
  {Williams}, {Williamson}, {Willis}, {Willke}, {Wilson}, {Wipf}, {Wlodarczyk},
  {Woan}, {Woehler}, {Wofford}, {Wong}, {Wong}, {Wright}, {Wu}, {Wu}, {Wu},
  {Wysocki}, {Xiao}, {Yamada}, {Yamamoto}, {Yamamoto}, {Yamamoto}, {Yamashita},
  {Yamazaki}, {Yang}, {Yang}, {Yang}, {Yang}, {Yang}, {Yang}, {Yap}, {Yeeles},
  {Yeh}, {Yelikar}, {Ying}, {Yokoyama}, {Yokozawa}, {Yoo}, {Yoshioka}, {Yu},
  {Yu}, {Yuzurihara}, {Zadro{\.z}ny}, {Zanolin}, {Zeidler}, {Zelenova},
  {Zendri}, {Zevin}, {Zhan}, {Zhang}, {Zhang}, {Zhang}, {Zhang}, {Zhang},
  {Zhang}, {Zhao}, {Zhao}, {Zhao}, {Zhao}, {Zhou}, {Zhou}, {Zhu}, {Zhu},
  {Zimmerman}, {Zucker}, \& {Zweizig}}]{TGR21}
---. 2021{\natexlab{b}}, arXiv e-prints, arXiv:2112.06861

\bibitem[{{Tiburzi} {$et~al$.}(2021){Tiburzi}, {Shaifullah}, {Bassa}, {Zucca},
  {Verbiest}, {Porayko}, {van der Wateren}, {Fallows}, {Main}, {Janssen},
  {Anderson}, {Bak Nielsen}, {Donner}, {Keane}, {K{\"u}nsem{\"o}ller},
  {Os{\l}owski}, {Grie{\ss}meier}, {Serylak}, {Br{\"u}ggen}, {Ciardi},
  {Dettmar}, {Hoeft}, {Kramer}, {Mann}, \& {Vocks}}]{tsb+21}
{Tiburzi}, C., {Shaifullah}, G.~M., {Bassa}, C.~G., {$et~al$.} 2021, \aap, 647,
  A84

\bibitem[{{Valtonen} {$et~al$.}(2021){Valtonen}, {Dey}, {Gopakumar}, {Zola},
  {Komossa}, {Pursimo}, {Gomez}, {Hudec}, {Jermak}, \& {Berdyugin}}]{VDG2021}
{Valtonen}, M.~J., {Dey}, L., {Gopakumar}, A., {$et~al$.} 2021, Galaxies, 10, 1

\bibitem[{{Van Straten} \& Bailes(2011)}]{vanStraten2011_dspsr}
{Van Straten}, W., \& Bailes, M. 2011, Publications of the Astronomical Society
  of Australia, 28, 1

\bibitem[{{Verbiest} {$et~al$.}(2016){Verbiest}, {Lentati}, {Hobbs}, {van
  Haasteren}, {Demorest}, {Janssen}, {Wang}, {Desvignes}, {Caballero}, {Keith},
  {Champion}, {Arzoumanian}, {Babak}, {Bassa}, {Bhat}, {Brazier}, {Brem},
  {Burgay}, {Burke-Spolaor}, {Chamberlin}, {Chatterjee}, {Christy}, {Cognard},
  {Cordes}, {Dai}, {Dolch}, {Ellis}, {Ferdman}, {Fonseca}, {Gair},
  {Garver-Daniels}, {Gentile}, {Gonzalez}, {Graikou}, {Guillemot}, {Hessels},
  {Jones}, {Karuppusamy}, {Kerr}, {Kramer}, {Lam}, {Lasky}, {Lassus},
  {Lazarus}, {Lazio}, {Lee}, {Levin}, {Liu}, {Lynch}, {Lyne}, {Mckee},
  {McLaughlin}, {McWilliams}, {Madison}, {Manchester}, {Mingarelli}, {Nice},
  {Os{\l}owski}, {Palliyaguru}, {Pennucci}, {Perera}, {Perrodin}, {Possenti},
  {Petiteau}, {Ransom}, {Reardon}, {Rosado}, {Sanidas}, {Sesana}, {Shaifullah},
  {Shannon}, {Siemens}, {Simon}, {Smits}, {Spiewak}, {Stairs}, {Stappers},
  {Stinebring}, {Stovall}, {Swiggum}, {Taylor}, {Theureau}, {Tiburzi},
  {Toomey}, {Vallisneri}, {van Straten}, {Vecchio}, {Wang}, {Wen}, {You},
  {Zhu}, \& {Zhu}}]{IPTA_DR1}
{Verbiest}, J.~P.~W., {Lentati}, L., {Hobbs}, G., {$et~al$.} 2016, \mnras, 458,
  1267

\bibitem[{{Xin} {$et~al$.}(2021){Xin}, {Mingarelli}, \& {Hazboun}}]{Xin2021}
{Xin}, C., {Mingarelli}, C. M.~F., \& {Hazboun}, J.~S. 2021, \apj, 915, 97

\end{thebibliography}

\end{document}